\documentclass[journal,comsoc,letter]{IEEEtran}
\usepackage[T1]{fontenc} 
\ifCLASSINFOpdf
\else
\fi
\usepackage{lipsum,multicol}
\usepackage{url}
\usepackage{amsmath}
\usepackage{pifont}
\usepackage{caption}
\usepackage{color, colortbl}
\usepackage{array}
\usepackage{dingbat}
\usepackage{multirow}
\usepackage[cmintegrals]{newtxmath}
\usepackage{cite}
\usepackage{graphics}
\usepackage{graphicx}
\usepackage{subfigure}
\usepackage{enumitem}
\usepackage{smartdiagram}
\usepackage{algorithm,algpseudocode}
\definecolor{Gray}{gray}{0.8}
\usepackage{stfloats}
\newcolumntype{C}[1]{>{\centering\let\newline\\\arraybackslash\hspace{0pt}}m{#1}}
\newcolumntype{L}[1]{>{\raggedright\let\newline\\\arraybackslash\hspace{0pt}}m{#1}}

\newcommand{\edit}[1]{\textcolor{black}{#1}} 
\newcommand{\rev}[1]{\textcolor{black}{#1}} 
\newcommand{\xmark}{\ding{55}}

\definecolor{mediumaquamarine}{rgb}{0.4, 0.8, 0.67}
\definecolor{bittersweet}{rgb}{1.0, 0.44, 0.37}
\definecolor{sandybrown}{rgb}{0.96, 0.64, 0.38}
\definecolor{saffron}{rgb}{0.96, 0.77, 0.19}
\definecolor{salmonpink}{rgb}{1.0, 0.57, 0.64}
\definecolor{limegreen}{rgb}{0.2, 0.8, 0.2}
\definecolor{persianorange}{rgb}{0.85, 0.56, 0.35}
\definecolor{darkgray}{rgb}{0.66, 0.66, 0.66}
\algnewcommand{\Inputs}[1]{%
  \State \textbf{Inputs:}
  \Statex \hspace*{\algorithmicindent}\parbox[t]{.8\linewidth}{\raggedright #1}
}
\algnewcommand{\Initialize}[1]{%
  \State \textbf{Initialise:}
  \Statex \hspace*{\algorithmicindent}\parbox[t]{.8\linewidth}{\raggedright #1}
}

\begin{document}

\title{Deep Learning in Mobile and Wireless Networking: \mbox{A Survey}}

\author{Chaoyun~Zhang,
        Paul~Patras, and Hamed Haddadi
\thanks{C. Zhang and P. Patras are with the Institute for Computing Systems Architecture (ICSA), School of Informatics, University of Edinburgh, Edinburgh, UK. Emails: \{chaoyun.zhang, paul.patras\}@ed.ac.uk. 

H. Haddadi is with the Dyson School of Design Engineering at Imperial College London. Email: h.haddadi@imperial.ac.uk.}
}

\markboth{IEEE Communications Surveys \& Tutorials}%
{Bare Demo of IEEEtran.cls for IEEE Communications Society Journals}
\maketitle

\begin{abstract}
The rapid uptake of mobile devices and the rising popularity of mobile applications and services pose unprecedented demands on mobile and wireless networking infrastructure. Upcoming 5G systems are evolving to \edit{support exploding mobile traffic volumes, real-time extraction of fine-grained analytics, and agile management of network resources, so as to maximize user experience.} Fulfilling these tasks is challenging, as mobile environments are increasingly complex, heterogeneous, and evolving. One potential solution is to resort to advanced machine learning techniques, \edit{in order to help manage} the rise in data volumes and algorithm-driven applications. The recent success of deep learning underpins new and powerful tools that tackle problems in this space. 

In this paper we bridge the gap between deep learning and mobile and wireless networking research, by presenting a comprehensive survey of the crossovers between the two areas. We first briefly introduce essential background and state-of-the-art in deep learning techniques with potential applications to networking. We then discuss several techniques and platforms that facilitate the efficient deployment of deep learning onto mobile systems. Subsequently, we provide an encyclopedic review of mobile and wireless networking research based on deep learning, which we categorize by different domains. Drawing from our experience, we discuss how to tailor deep learning to mobile environments. We complete this survey by pinpointing current challenges and open future directions for research.  
\end{abstract}


%
\IEEEpeerreviewmaketitle

\section{Introduction}
\IEEEPARstart{I}{nternet} connected mobile devices are penetrating every aspect of individuals' life, work, and entertainment. The increasing number of smartphones and the emergence of evermore diverse applications trigger a surge in mobile data traffic. Indeed, the latest industry forecasts indicate that the annual worldwide IP traffic consumption will reach 3.3 zettabytes (10\textsuperscript{15} MB) by 2021, with smartphone traffic exceeding PC traffic by the same year~\cite{cisco2017}. Given the shift in user preference towards wireless connectivity, current mobile infrastructure faces great capacity demands. In response to this increasing demand, early efforts propose to agilely provision resources~\cite{wang2015backhauling} and tackle mobility management distributively~\cite{giust2015distributed}. In the long run, however, Internet Service Providers (ISPs) must develop \emph{intelligent} heterogeneous architectures and tools that can spawn the 5\textsuperscript{th} generation of mobile systems (5G) and gradually meet more stringent end-user application requirements~\cite{agiwal2016next, gupta2015survey}. 

The growing diversity and complexity of mobile network architectures has made monitoring and managing the multitude of network elements intractable. Therefore, embedding versatile machine intelligence into future mobile networks is drawing unparalleled research interest~\cite{zheng2016big, jiang2017machine}. This trend is reflected in machine learning (ML) based solutions to problems ranging from radio access technology (RAT) selection~\cite{nguyen2017reinforcement} to malware detection~\cite{narudin2016evaluation}, as well as the development of networked systems that support machine learning practices (e.g. \cite{hsieh2017gaia, xiao2017tux2}). ML enables systematic mining of valuable information from traffic data and automatically uncover correlations that would otherwise have been too complex to extract by human experts~\cite{anareport}. 
As the flagship of machine learning, deep learning has achieved remarkable performance in areas  
such as computer vision \cite{zhang2015convolutional} and natural language processing (NLP)~\cite{socher2012deep}. 
Networking researchers are also beginning to recognize the power and importance of deep learning, and are exploring its potential to solve problems specific to the mobile networking domain \cite{specialissue, wang2017machine}. 

Embedding deep learning into the 5G mobile and wireless networks is well justified. In particular, data generated by mobile environments are increasingly heterogeneous, as these are usually collected from various sources, have different formats, and exhibit complex correlations \cite{alsheikh2016mobile}. 
\edit{As a consequence, a range of specific problems become too difficult or impractical for traditional machine learning tools (e.g., shallow neural networks). This is because \emph{(i)} their performance does not improve if provided with more data~\cite{goodfellow2016deep} and \emph{(ii)} they cannot handle highly dimensional state/action spaces in control problems~\cite{mnih2015human}. In contrast, big data fuels the performance of deep learning, as it eliminates domain expertise and instead employs hierarchical feature extraction. In essence this means information can be distilled efficiently and increasingly abstract correlations can be obtained from the data, while reducing the pre-processing effort.} Graphics Processing Unit (GPU)-based parallel computing further enables deep learning to make inferences within milliseconds. This facilitates network analysis and management with high accuracy and in a timely manner, overcoming the run-time limitations of traditional mathematical techniques (e.g. convex optimization, game theory, meta heuristics). 

Despite growing interest in deep learning in the mobile networking domain, existing contributions are scattered across different research areas and a comprehensive survey is lacking. This article fills this gap between deep learning and mobile and wireless networking, by presenting an up-to-date survey of research that lies at the intersection between these two fields. Beyond reviewing the most relevant literature, we discuss the key pros and cons of various deep learning architectures, and outline deep learning model selection strategies, in view of solving mobile networking problems. We further investigate methods that tailor deep learning to individual mobile networking tasks, to achieve the best performance in complex environments. We wrap up this paper by pinpointing future research directions and important problems that remain unsolved and are worth pursing with deep neural networks. Our ultimate goal is to provide a definite guide for networking researchers and practitioners, who intend to employ deep learning to solve problems of interest.
\vspace*{0.25em}

\begin{figure*}[htb]
\begin{center}
\includegraphics[width=1\textwidth]{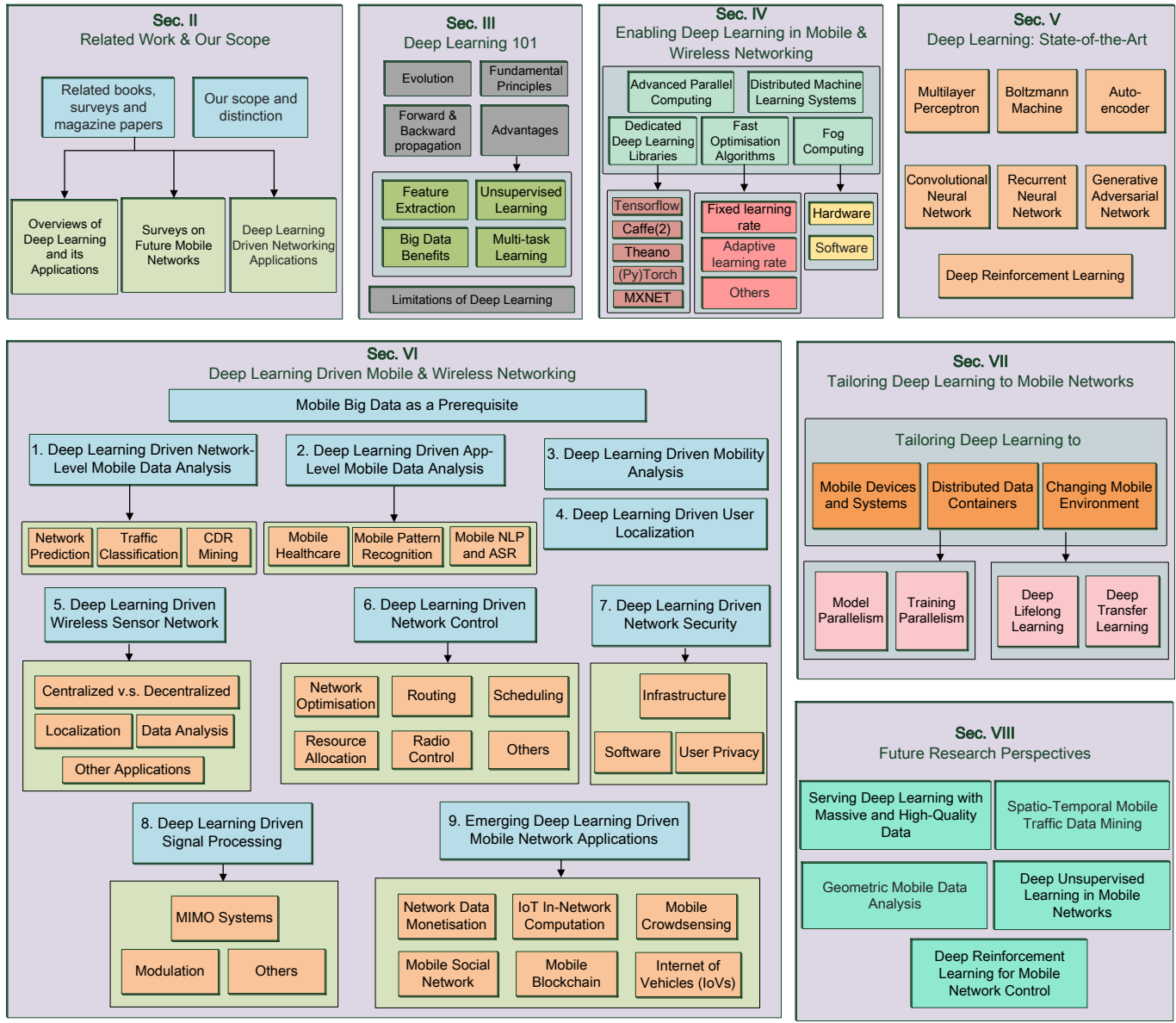}
\end{center}
\caption{\label{fig:org} \edit{Diagramatic view of the organization of this survey.}}
%
%
\end{figure*}

\noindent\textbf{Survey Organization:}  We structure this article in a top-down manner, as shown in Figure~\ref{fig:org}. We begin by discussing work that gives a high-level overview of deep learning, future mobile networks, and networking applications built using deep learning, which help define the scope and contributions of this paper (Section~\ref{sec:related}). Since deep learning techniques are relatively new in the mobile networking community, we provide a basic deep learning background in Section~\ref{sec:back}, highlighting immediate advantages in addressing mobile networking problems. 
There exist many factors that enable implementing deep learning for mobile networking applications (including dedicated deep learning libraries, optimization algorithms, etc.). We discuss these enablers in Section~\ref{sec:stimulator}, aiming to help mobile network researchers and engineers in choosing the right software and hardware platforms for their deep learning deployments.

In Section~\ref{sec:model}, we introduce and compare state-of-the-art \edit{deep learning models and provide guidelines for model selection toward solving networking problems.} 
In Section~\ref{sec:netapp} we review recent deep learning applications to mobile and wireless networking, which we group by different scenarios ranging from mobile traffic analytics to security, and emerging applications.
We then discuss how to tailor deep learning models to mobile networking problems (Section~\ref{sec:tailor}) and \edit{conclude this article with a brief discussion of open challenges, with a view to future research directions (Section~\ref{sec:future})}.\footnote{We list the abbreviations used throughout this paper in Table~\ref{tab:abbreviations}.}

\begin{table}[htpb!]
\centering
\caption{List of abbreviations in alphabetical order.}
\label{tab:abbreviations}
\begin{tabular}{|C{2.1cm}||C{5.7cm}|}
\hline
Acronym & Explanation \\\hline\hline
5G & 5\textsuperscript{th} Generation mobile networks \\\hline
A3C & Asynchronous Advantage Actor-Critic \\\hline
AdaNet & Adaptive learning of neural Network \\\hline
AE & Auto-Encoder \\\hline
AI & Artificial Intelligence \\\hline
AMP &  Approximate Message Passing \\\hline
ANN & Artificial Neural Network \\\hline
ASR & Automatic Speech Recognition \\\hline
BSC & Base Station Controller \\\hline
BP & Back-Propagation \\\hline
CDR & Call Detail Record \\\hline
CNN or ConvNet & Convolutional Neural Network \\\hline
ConvLSTM & Convolutional Long Short-Term Memory \\\hline
CPU & Central Processing Unit \\\hline
CSI & Channel State Information \\\hline
CUDA & Compute Unified Device Architecture \\\hline
cuDNN & CUDA Deep Neural Network library \\\hline
D2D & Device to Device communication \\\hline
DAE & Denoising Auto-Encoder \\\hline
DBN & Deep Belief Network \\\hline
OFDM &  Orthogonal Frequency-Division Multiplexing\\\hline
DPPO & Distributed Proximal Policy Optimization \\\hline
DQN & Deep Q-Network \\\hline
DRL & Deep Reinforcement Learning \\\hline
DT & Decision Tree \\\hline
ELM & Extreme Learning Machine \\\hline
GAN & Generative Adversarial Network \\\hline
GP & Gaussian Process \\\hline
GPS & Global Positioning System \\\hline
GPU & Graphics Processing Unit \\\hline
GRU & Gate Recurrent Unit \\\hline
HMM & Hidden Markov Model \\\hline
HTTP & HyperText Transfer Protocol \\\hline 
IDS & Intrusion Detection System \\\hline 
IoT & Internet of Things \\\hline
IoV & Internet of Vehicle \\\hline
ISP & Internet Service Provider \\\hline 
LAN & Local Area Network \\\hline
LTE & Long-Term Evolution \\\hline 
LSTM & Long Short-Term Memory \\\hline
LSVRC & Large Scale Visual Recognition Challenge\\\hline
MAC & Media Access Control \\\hline
MDP &Markov Decision Process \\\hline
MEC & Mobile Edge Computing  \\\hline
ML & Machine Learning \\\hline
MLP & Multilayer Perceptron \\\hline
MIMO & Multi-Input Multi-Output \\\hline
MTSR & Mobile Traffic Super-Resolution \\\hline
NFL & No Free Lunch theorem \\\hline
NLP & Natural Language Processing \\\hline
NMT & Neural Machine Translation \\\hline
NPU & Neural Processing Unit \\\hline
PCA & Principal Components Analysis \\\hline
PIR & Passive Infra-Red \\\hline
QoE & Quality of Experience \\\hline
RBM & Restricted Boltzmann Machine\\ \hline
ReLU & Rectified Linear Unit\\ \hline
RFID & Radio Frequency Identification \\ \hline
RNC & Radio Network Controller \\ \hline
RNN & Recurrent Neural Network\\ \hline
SARSA & State-Action-Reward-State-Action\\ \hline
SELU & Scaled Exponential Linear Unit \\ \hline
SGD & Stochastic Gradient Descent \\ \hline
SON & Self-Organising Network \\ \hline
SNR & Signal-to-Noise Ratio \\ \hline
SVM & Support Vector Machine \\ \hline
TPU & Tensor Processing Unit \\ \hline
VAE & Variational Auto-Encoder \\ \hline
VR & Virtual Reality \\ \hline
WGAN & Wasserstein Generative Adversarial Network\\ \hline
WSN & Wireless Sensor Network \\\hline
\end{tabular}
\end{table}

\section{Related High-level Articles and \\The Scope of This Survey}\label{sec:related}
Mobile networking and deep learning problems have been researched mostly independently. Only recently crossovers between the two areas have emerged. Several notable works paint a comprehensives picture of the deep learning and/or mobile networking research landscape. We categorize these works into \emph{(i)} pure overviews of deep learning techniques, \emph{(ii)} reviews of analyses and management techniques in modern mobile networks, and \emph{(iii)} reviews of works at the intersection between deep learning and computer networking. We summarize these earlier efforts in Table~\ref{tab:survey} and in this section discuss the most representative publications in each class. 


\begin{table*}[h!]
\centering
\caption{Summary of existing surveys, magazine papers, and books related to deep learning and mobile networking. The symbol \checkmark~indicates a publication is in the scope of a domain; \xmark~marks papers that do not directly cover that area, but from which readers may retrieve some related insights. Publications related to both deep learning and mobile networks are shaded.}
\label{tab:survey}
\setlength\tabcolsep{3pt}
\renewcommand\arraystretch{1.1}
\begin{tabular}{|c|c|C{1.2cm}|C{1.2cm}|C{1.2cm}|C{1.2cm}|}
\hline
\multirow{3}{*}{Publication} & \multirow{3}{*}{One-sentence summary} & \multicolumn{4}{c|}{Scope}                                                   \\ \cline{3-6} 
                             &                          & \multicolumn{2}{c|}{Machine learning} & \multicolumn{2}{c|}{Mobile  networking} \\ \cline{3-6} 
                             &                          & \edit{Deep learning} & Other ML methods & Mobile big data  & 5G technology     \\ \hline
       LeCun \emph{et al.} \cite{lecun2015deep}                      &         A milestone overview of deep learning.                  &        \checkmark          &                     &                     &                \\ 
       \hline
              Schmidhuber \cite{schmidhuber2015deep}                      &         A comprehensive deep learning  survey.          &        \checkmark          &                     &                     &                \\ 
       \hline
               Liu \emph{et al.}  \cite{liu2017survey}                    &         A survey on deep learning and its applications.          &        \checkmark          &                     &                     &                \\ 
       \hline
               Deng \emph{et al.} \cite{deng2014deep}                     &        An overview of deep learning methods and applications.          &        \checkmark          &                     &                     &                \\ 
       \hline
           Deng  \cite{deng2014tutorial}                     &       A tutorial on deep learning.          &        \checkmark          &                     &                     &                \\ 
       \hline 
       Goodfellow \emph{et al.} \cite{goodfellow2016deep}                     &      An essential deep learning textbook.                 &          \checkmark       &        \xmark           &                     &                \\ 
       \hline
 \edit{Pouyanfar \emph{et al.}  \cite{Pouyanfar:2018:SDL:3271482.3234150}}                      &         \edit{A recent survey on deep learning.}                  &        \edit{\checkmark}          &       \edit{\checkmark}              &                     &                \\ 
       \hline
Arulkumaran \emph{et al.} \cite{kai2017brief}                      &         A survey of deep reinforcement learning.          &        \checkmark          &           \xmark          &                     &                \\ 
       \hline
\edit{Hussein \emph{et al.} \cite{hussein2017imitation}}                      &         \edit{A survey of imitation learning.}          &        \edit{\checkmark}          &           \edit{\checkmark}          &                     &                \\ 
       \hline
               Chen \emph{et al.} \cite{chen2014big}                     &       An introduction to deep learning for big data.          &        \checkmark          &        \xmark             &        \xmark                &                \\ 
       \hline      
           Najafabadi  \cite{najafabadi2015deep}                     &        An overview of deep learning applications for big data analytics.          &        \checkmark          &      \xmark               &        \xmark             &                \\ 
       \hline   
Hordri \emph{et al.} \cite{hordri2017systematic}                     &      A brief of survey of deep learning for big data applications.          &        \checkmark          &      \xmark               &        \xmark             &                \\ 
\hline
Gheisari \emph{et al.} \cite{gheisari2017survey}                     &      A high-level literature review on deep learning for big data analytics.          &        \checkmark          &                     &        \xmark             &                \\ 
\hline     
\edit{Zhang \emph{et al.} \cite{zhang2017deep}}                     &      \edit{A survey and outlook of deep learning for recommender systems.}          &        \edit{\checkmark}          &      \edit{\xmark}               &        \edit{\xmark}             &                \\ 
\hline
Yu \emph{et al.}  \cite{yu2017networking}                 &        A survey on networking big data.          &                  &                     &        \checkmark            &                \\ 
\hline  
Alsheikh \emph{et al.}  \cite{alsheikh2014machine}&  A survey on machine learning in wireless sensor networks.          &                  &         \checkmark            &        \checkmark           &                 \\ 
\hline
Tsai \emph{et al.}  \cite{tsai2014data}& A survey on data mining in IoT.          &                  &         \checkmark            &        \checkmark           &                 \\ 
\hline
  Cheng \emph{et al.} \cite{cheng2017exploiting} &      An introductions mobile big data its applications.          &                  &                    &          \checkmark          &        \xmark        \\ 
 \hline  
 Bkassiny \emph{et al.}  \cite{bkassiny2013survey}& A survey on machine learning in  cognitive radios.      &                  &         \checkmark            &        \xmark           &      \xmark           \\ 
\hline
Andrews \emph{et al.}  \cite{andrews2014will}                 &        An introduction and outlook of 5G networks.          &                  &                     &                    &      \checkmark          \\\hline 
Gupta \emph{et al.}  \cite{gupta2015survey}                &        A survey of 5G architecture and technologies.          &                 &                     &                   &       \checkmark          \\ 
\hline
Agiwal \emph{et al.}  \cite{agiwal2016next}                 &        A survey of 5G  mobile networking techniques.          &                  &                     &                    &      \checkmark          \\ 
\hline   
Panwar \emph{et al.}  \cite{panwar2016survey}                 &        A survey of 5G networks features, research progress and open issues.          &                  &                     &                    &      \checkmark          \\\hline 
Elijah \emph{et al.}  \cite{elijah2016comprehensive}                 &        A survey of 5G MIMO systems.          &                  &                     &                    &      \checkmark          \\\hline 
Buzzi \emph{et al.}  \cite{buzzi2016survey}                 &        A survey of 5G energy-efficient techniques.          &                  &                     &                    &      \checkmark          \\\hline 
Peng \emph{et al.}  \cite{peng2015system}                &      An overview of radio access networks in 5G.       &                 &                     &    \xmark              &       \checkmark          \\ 
\hline
Niu \emph{et al.}  \cite{niu2015survey}                 &        A survey of 5G millimeter wave communications.          &                  &                     &                    &      \checkmark          \\\hline 
Wang \emph{et al.}  \cite{wang2015backhauling}                &     5G backhauling techniques and radio resource management.         &                 &                     &                   &       \checkmark          \\ 
\hline
Giust \emph{et al.}  \cite{giust2015distributed}                &        An overview of 5G distributed mobility management.          &                 &                     &                   &       \checkmark          \\ 
\hline
Foukas \emph{et al.}  \cite{foukas2017network}                &        A survey and insights on network slicing in 5G.          &                  &                     &                  &       \checkmark          \\ 
\hline
Taleb \emph{et al.}  \cite{taleb2017multi}                &     A survey on 5G edge architecture and orchestration.       &                 &                     &                &       \checkmark          \\ 
\hline
Mach and Becvar  \cite{mach2017mobile}                &        A survey on MEC.          &                  &                     &                   &       \checkmark          \\ 
\hline
Mao \emph{et al.}  \cite{mao2017survey}                &      A survey on mobile edge computing.        &                 &                     &        \checkmark           &       \checkmark          \\ 
\hline
Wang \emph{et al.}  \cite{wang2017data}                &        An architecture for personalized QoE management in 5G.          &                  &                     &        \checkmark           &       \checkmark          \\ 
\hline
Han \emph{et al.}  \cite{han2015mobile}                &      Insights to mobile cloud sensing, big data, and 5G.           &                  &                    &        \checkmark           &       \checkmark          \\ 
\hline
Singh \emph{et al.}  \cite{singh2017survey}                &      A survey on social networks over 5G.        &                 &     \xmark                &        \checkmark           &       \checkmark          \\ 
\hline
Chen \emph{et al.} \cite{chen20175g} &        An introduction to 5G cognitive systems for healthcare.          &     \xmark             &       \xmark              &        \xmark            &       \checkmark         \\ 
 \hline 
\edit{Chen \emph{et al.}  \cite{chen2015energy}}                &       \edit{Machine learning for traffic offloading in cellular network} &                 &     \edit{\checkmark}                &                  &       \edit{\checkmark}          \\ 
 \hline 
\edit{Wu \emph{et al.}  \cite{wu2016big}}                &       \edit{Big data toward green cellular networks} &                 &     \edit{\checkmark}                &        \edit{\checkmark}           &       \edit{\checkmark}          \\ 
 \hline 
Buda \emph{et al.}  \cite{buda2016can}                &       Machine learning aided use cases and scenarios in 5G.  &                 &     \checkmark                &        \checkmark           &       \checkmark          \\ 
\hline
Imran \emph{et al.} \cite{imran2014challenges} &      An introductions to big data analysis for self-organizing networks (SON) in 5G.          &                  &         \checkmark           &          \checkmark          &        \checkmark        \\ 
 \hline  
Keshavamurthy \emph{et al.}  \cite{keshavamurthy2016conceptual} &        Machine learning perspectives on SON in 5G.          &                  &         \checkmark            &        \checkmark           &       \checkmark          \\ 
\hline
Klaine \emph{et al.}  \cite{valente2017survey}                &      A survey of machine learning applications in SON.       &      \xmark            &         \checkmark            &        \checkmark           &       \checkmark          \\ \hline
Jiang \emph{et al.}  \cite{jiang2017machine}                &        Machine learning paradigms for 5G.          &      \xmark            &         \checkmark            &        \checkmark           &       \checkmark          \\ 
 \hline  
Li \emph{et al.}  \cite{li2017intelligent}                &        Insights into intelligent 5G.          &      \xmark            &         \checkmark            &        \checkmark           &       \checkmark          \\ 
\hline 
 Bui \emph{et al.}  \cite{bui2017survey}                &      A survey of future mobile networks analysis and optimization.       &      \xmark            &         \checkmark            &        \checkmark           &       \checkmark          \\ \hline
\rowcolor{Gray}
  Kasnesis \emph{et al.} \cite{kasnesis2017changing} &        Insights into employing deep learning for mobile data analysis.          &     \checkmark             &                     &        \checkmark            &                \\ 
 \hline  
 \rowcolor{Gray}
 Alsheikh \emph{et al.} \cite{alsheikh2016mobile}
 &        Applying deep learning and Apache Spark for mobile data analytics.     &     \checkmark             &                   &        \checkmark            &                \\
\hline  
  \rowcolor{Gray}
    Cheng \emph{et al.} \cite{cheng2017mobile} &        Survey of mobile big data analysis and outlook.          &     \checkmark             &       \checkmark              &        \checkmark            &      \xmark          \\ 
\hline
 \rowcolor{Gray}
Wang and Jones \cite{wang2017big}                 &        A survey of deep learning-driven network intrusion detection.          &     \checkmark             &       \checkmark              &        \checkmark            &      \xmark          \\ 
\hline
 \rowcolor{Gray}
 Kato \emph{et al.} \cite{kato2017deep} &        Proof-of-concept deep learning for network traffic control.          &     \checkmark             &                    &                    &        \checkmark        \\ 
 \hline 
 \rowcolor{Gray}
  Zorzi \emph{et al.} \cite{zorzi2015cognition} &        An introduction to machine learning driven network optimization.          &     \checkmark             &           \checkmark          &                    &       \checkmark         \\ 
  \hline 
  \rowcolor{Gray}
Fadlullah \emph{et al.}  \cite{fadlullah2017state}
 &        A comprehensive survey of deep learning for network traffic control.     &     \checkmark             &        \checkmark           &        \checkmark            &       \xmark         \\ 
\hline  
\rowcolor{Gray}
 Zheng \emph{et al.}  \cite{zheng2016big}                 &        An introduction to big data-driven 5G optimization.          &     \checkmark             &       \checkmark              &        \checkmark            &      \checkmark          \\ 
 \hline  
\end{tabular}
\end{table*}

\begin{table*}[h!]
\centering
\caption{Continued from Table~\ref{tab:survey}}
\label{tab:survey2}
\setlength\tabcolsep{3pt}
\renewcommand\arraystretch{1.1}
\begin{tabular}{|C{3cm}|C{9.25cm}|C{1.2cm}|C{1.2cm}|C{1.2cm}|C{1.2cm}|}
\hline
\multirow{3}{*}{Publication} & \multirow{3}{*}{One-sentence summary} & \multicolumn{4}{c|}{Scope}                                                   \\ \cline{3-6} 
                             &                          & \multicolumn{2}{c|}{Machine learning} & \multicolumn{2}{c|}{Mobile  networking} \\ \cline{3-6} 
                             &                          & \edit{Deep learning} & Other ML methods & Mobile big data  & 5G technology     \\ \hline
 \rowcolor{Gray}
Mohammadi \emph{et al.}  \cite{mohammadi2018deep}
 &        A survey of deep learning in IoT data analytics.     &     \checkmark             &        \checkmark           &        \checkmark            &                \\ 
\hline     
 \rowcolor{Gray}
Ahad \emph{et al.}  \cite{ahad2016neural}
 &        A survey of neural networks in wireless networks.     &     \checkmark             &        \xmark           &        \xmark            &       \checkmark          \\ 
\hline   
  \rowcolor{Gray}
    \edit{Mao \emph{et al.} \cite{mao2018deep}} &        \edit{A survey of deep learning for wireless networks.}          &     \edit{\checkmark}             &         \edit{\checkmark}            &        \edit{\checkmark}            &               \\ 
  \hline
  \rowcolor{Gray}
    \rev{Luong \emph{et al.} \cite{luong2018applications}} &        \rev{A survey of deep reinforcement learning for networking.}          &     \rev{\checkmark}             &         \rev{\checkmark}            &                    &     \rev{\checkmark}          \\ 
  \hline  
    \rowcolor{Gray}
    \edit{Zhou \emph{et al.} \cite{zhou2017intelligent2}} &        \edit{A survey of ML and cognitive wireless communications.}          &     \edit{\checkmark}             &         \edit{\checkmark}            &        \edit{\checkmark}            &    \edit{\checkmark}           \\ 
  \hline  
    \rowcolor{Gray}
    \edit{Chen \emph{et al.} \cite{chen2017machine2}} &        \edit{A tutorial on neural networks for wireless networks.}          &     \edit{\checkmark}             &         \edit{\checkmark}            &        \edit{\checkmark}            &      \edit{\checkmark}          \\ 
  \hline  
 \rowcolor{Gray}
  Gharaibeh \emph{et al.} \cite{gharaibeh2017smart} &        A survey of smart cities.          &     \checkmark             &        \checkmark             &        \checkmark            &       \checkmark         \\ 
 \hline \rowcolor{Gray}
 Lane \emph{et al.} \cite{lane2015can} &        An overview and introduction of deep learning-driven mobile sensing.          &     \checkmark             &       \checkmark              &        \checkmark            &                \\ 
 \hline   
  \rowcolor{Gray}
  Ota \emph{et al.} \cite{ota2017deep} &        A survey of deep learning for mobile multimedia.          &     \checkmark             &                    &        \checkmark            &       \checkmark         \\ 
  \hline  
  \rowcolor{Gray}
  \edit{Mishra \emph{et al.} \cite{mishra2018detailed}} &        \edit{A survey machine learning driven intrusion detection.}          &     \edit{\checkmark}             &         \edit{\checkmark}            &        \edit{\checkmark}            &       \edit{\checkmark}        \\ 
  \hline  

  \rowcolor{Gray}
\edit{\textbf{Our work}} &        \edit{\textbf{A comprehensive survey of deep learning for mobile and wireless network.}}          &     \edit{\checkmark}             &         \edit{\checkmark}            &        \edit{\checkmark}            &       \edit{\checkmark}        \\ 
  \hline  
\end{tabular}
\end{table*}

\subsection{Overviews of Deep Learning and its Applications}
The era of big data is triggering wide interest in deep learning across different research disciplines~\cite{chen2014big, najafabadi2015deep, gheisari2017survey, hordri2017systematic} and a growing number of surveys and tutorials are emerging (e.g. \cite{deng2014deep, deng2014tutorial}). LeCun \emph{et al.} give a milestone overview of deep learning, introduce several popular models, and look ahead at the potential of deep neural networks \cite{lecun2015deep}. 
Schmidhuber undertakes an encyclopedic survey of deep learning, likely the most comprehensive thus far, covering the evolution, methods, applications, and open research issues~\cite{schmidhuber2015deep}. Liu \emph{et al.} summarize the underlying principles of several deep learning models, and review deep learning developments in selected applications, such as speech processing, pattern
recognition, and computer vision \cite{liu2017survey}. 

Arulkumaran \emph{et al.} present several architectures and core algorithms for deep reinforcement learning, including deep Q-networks, trust region policy optimization, and asynchronous advantage actor-critic~\cite{kai2017brief}. Their survey highlights the remarkable performance of deep neural networks in different control problem (e.g., video gaming, Go board game play, etc.). Similarly, deep reinforcement learning has also been surveyed in~\cite{li2017deeprl}, where the authors shed more light on applications. Zhang \emph{et al.} survey developments in deep learning for recommender systems~\cite{zhang2017deep}, which have potential to play an important role in mobile advertising. As deep learning becomes increasingly popular, Goodfellow \emph{et al.} provide a comprehensive tutorial of deep learning in a book that covers prerequisite knowledge, underlying principles, and popular applications~\cite{goodfellow2016deep}. 

\subsection{Surveys on Future Mobile Networks}
The emerging 5G mobile networks incorporate a host of new techniques to overcome the performance limitations of current deployments and meet new application requirements. Progress to date in this space has been summarized through surveys, tutorials, and magazine papers (e.g. \cite{andrews2014will, agiwal2016next, gupta2015survey, panwar2016survey, mao2017survey}).  
Andrews \emph{et al.} highlight the differences between 5G and prior mobile network architectures, conduct a comprehensive review of 5G techniques, and discuss research challenges facing future developments\edit{\cite{andrews2014will}}. Agiwal \emph{et al.} review new architectures for 5G networks, survey emerging wireless technologies, and point out research problems that remain unsolved~\cite{agiwal2016next}. Gupta \emph{et al.} also review existing work on 5G cellular network architectures, subsequently proposing a framework that incorporates networking ingredients such as Device-to-Device (D2D) communication, small cells, cloud computing, and the IoT \cite{gupta2015survey}.


Intelligent mobile networking is becoming a popular research area and related work has been reviewed in the literature (e.g. \cite{jiang2017machine, buda2016can, keshavamurthy2016conceptual, alsheikh2014machine, li2017intelligent, bkassiny2013survey, bui2017survey, valente2017survey}). Jiang \emph{et al.} discuss the potential of applying machine learning to 5G network applications including massive MIMO and smart grids~\cite{jiang2017machine}. This work further identifies several research gaps between ML and 5G that remain unexplored. Li \emph{et al.} discuss opportunities and challenges of incorporating artificial intelligence (AI) into future network architectures and highlight the significance of AI in the 5G era~\cite{li2017intelligent}. Klaine \emph{et al.} present several successful ML practices in Self-Organizing Networks (SONs), discuss the pros and cons of different algorithms, and identify future research directions in this area~\cite{valente2017survey}. \edit{Potential exists to apply AI and exploit big data for energy efficiency purposes~\cite{wu2016big}. Chen \emph{et al.} survey traffic offloading approaches in wireless networks, and propose a novel reinforcement learning based solution \cite{chen2015energy}. This opens a new research direction toward embedding machine learning towards greening cellular networks.}

\subsection{Deep Learning Driven Networking Applications}
A growing number of papers survey recent works that bring deep learning into the computer networking domain. 
Alsheikh \emph{et al.} identify benefits and challenges of using big data for mobile analytics and propose a Spark based deep learning framework for this purpose~\cite{alsheikh2016mobile}. Wang and Jones discuss evaluation criteria, data streaming and deep learning practices for network intrusion detection, pointing out research challenges inherent to such  applications~\cite{wang2017big}. Zheng \emph{et al.} put forward a big data-driven mobile network optimization framework in 5G networks, to enhance QoE performance~\cite{zheng2016big}. More recently, Fadlullah \emph{et al.} deliver a survey on the progress of deep learning in a board range of areas, highlighting its potential application to network traffic control systems \cite{fadlullah2017state}. Their work also highlights several unsolved research issues worthy of future study. 

Ahad \emph{et al.} introduce techniques, applications, and guidelines on applying neural networks to wireless networking problems~\cite{ahad2016neural}. Despite several limitations of neural networks identified, this article focuses largely on old neural networks models, ignoring recent progress in deep learning and successful applications in current mobile networks. 
Lane \emph{et al.} investigate the suitability and benefits of employing deep learning in mobile sensing, and emphasize on the potential for accurate inference on mobile devices~\cite{lane2015can}.
Ota \emph{et al.} report novel deep learning applications in mobile multimedia. Their survey covers state-of-the-art deep learning practices in mobile health and wellbeing, mobile security, mobile ambient intelligence, language translation, and speech recognition. Mohammadi \emph{et al.} survey recent deep learning techniques for Internet of Things (IoT) data analytics \cite{mohammadi2018deep}. They overview comprehensively existing efforts that incorporate deep learning into the IoT domain and shed light on current research challenges and future directions. \edit{Mao \emph{et al.} focus on deep learning in wireless networking~\cite{mao2018deep}. Their work surveys state-of-the-art deep learning applications in wireless networks, and discusses research challenges to be solved in the future.}

\subsection{Our Scope}
The objective of this survey is to provide a comprehensive view on state-of-the-art deep learning practices in the mobile networking area. By this we aim to answer the following key questions: 
\begin{enumerate}
\item Why is deep learning promising for solving mobile networking problems? 
\item What are the cutting-edge deep learning models relevant to mobile and wireless networking?
\item What are the most recent successful deep learning applications in the mobile networking domain?
\item How can researchers tailor deep learning to specific mobile networking problems?
\item Which are the most important and promising directions worthy of further study?
\end{enumerate}

The research papers and books we mentioned previously only partially answer these questions.  This article goes beyond these previous works and specifically focuses on the crossovers between deep learning and mobile networking. \edit{We cover a range of neural network (NN) structures that are increasingly important and have not been explicitly discussed in earlier tutorials, e.g.,~\cite{chen2018deep0}. This includes auto-encoders and Generative Adversarial Networks. Unlike such existing tutorials, we also review open-source libraries for deploying and training neural networks, a range of optimization algorithms, and the parallelization of neural networks models and training across large numbers of mobile devices. We also review applications not looked at in other related surveys, including traffic/user analytics, security and privacy, mobile health, etc.} 

While our main scope remains the mobile networking domain, for completeness we also discuss deep learning applications to wireless networks, and identify emerging application domains intimately connected to these areas. \edit{We differentiate between mobile networking, which refers to scenarios where devices are portable, battery powered, potentially wearable, and routinely connected to cellular infrastructure, and wireless networking, where devices are mostly fixed, and part of a distributed infrastructure (including WLANs and WSNs), and serve a single application.} Overall, \textbf{our paper distinguishes itself from earlier surveys from the following perspectives}: 

\begin{enumerate}[label=\emph{(\roman*)}]
\item We particularly focus on deep learning applications for mobile network analysis and management, instead of broadly discussing deep learning methods (as, e.g., in ~\cite{lecun2015deep, schmidhuber2015deep}) or centering on a single application domain, e.g. mobile big data analysis with a specific platform~\cite{alsheikh2016mobile}.

\item We discuss cutting-edge deep learning techniques from the perspective of mobile networks (e.g., \cite{mnih2016asynchronous, arjovsky2017wasserstein}), focusing on their applicability to this area, whilst giving less attention to conventional deep learning models that may be out-of-date.


\item We analyze similarities between existing non-networking problems and those specific to mobile networks; based on this analysis we provide insights into both best deep learning architecture selection strategies and adaptation approaches, so as to exploit the characteristics of mobile networks for analysis and management tasks.
\end{enumerate}

To the best of our knowledge, \textbf{this is the first time that mobile network analysis and management are jointly reviewed from a deep learning angle. We also provide for the first time insights into how to tailor deep learning to mobile networking problems.}

\section{Deep Learning 101}\label{sec:back}
\begin{figure}[htb]
\begin{center}
\includegraphics[width=0.5\textwidth]{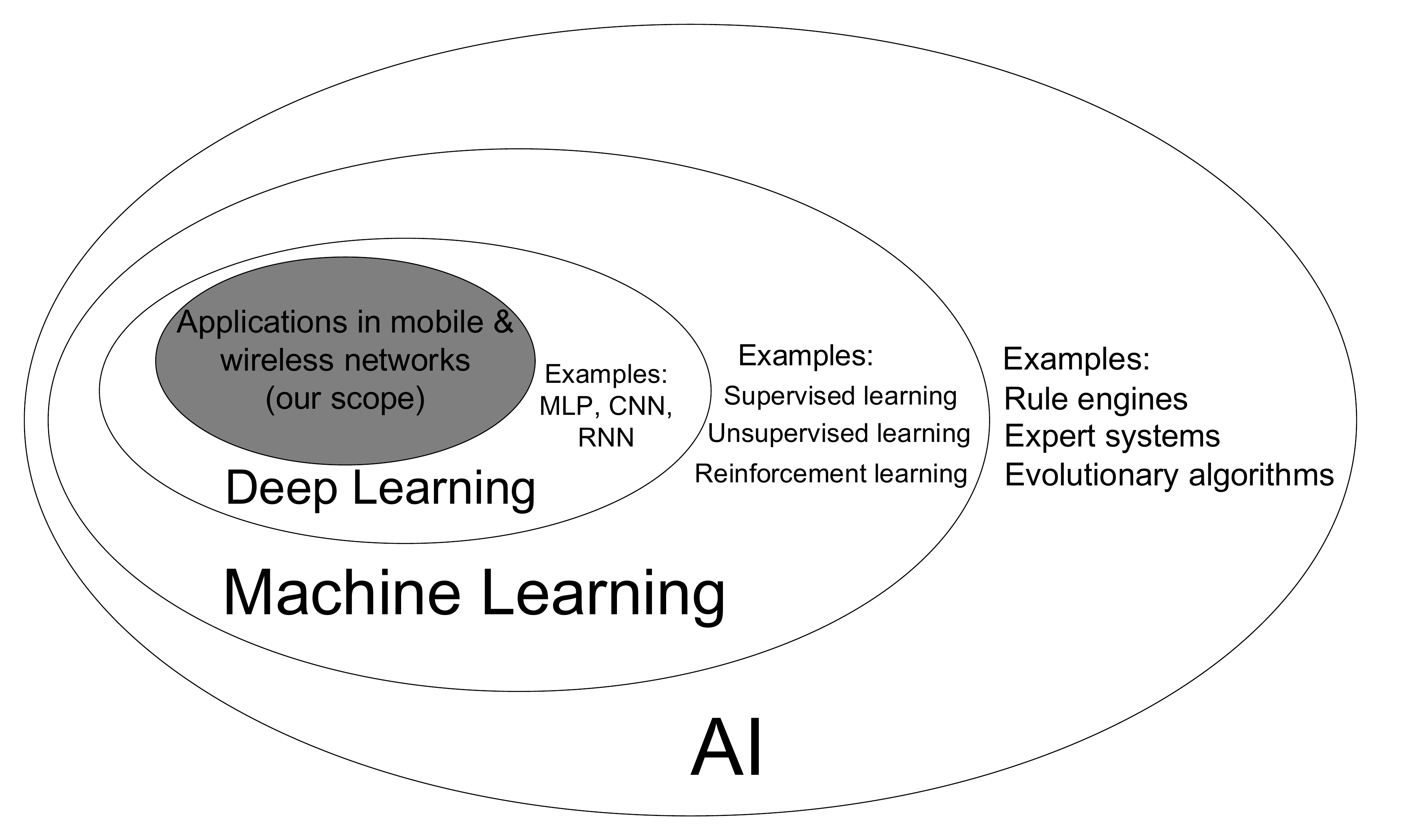}
\end{center}
\caption{\label{fig:venn} Venn diagram of the relation between deep learning, machine learning, and AI. This survey particularly focuses on deep learning applications in mobile and wireless networks.}
%
%
\end{figure}

We begin with a brief introduction to deep learning, highlighting the basic principles behind computation techniques in this field, as well as key advantages that lead to their success. Deep learning is essentially a sub-branch of ML, \edit{which essentially enables an algorithm to make predictions, classifications, or decisions based on data, without being explicitly programmed. Classic examples include linear regression, the k-nearest neighbors classifier, and Q-learning. In contrast to traditional ML tools that rely heavily on features defined by domain experts, deep learning} algorithms hierarchically extract knowledge from raw data through multiple layers of nonlinear processing units, in order to make predictions or take actions according to some target objective. 
\edit{The most well-known deep learning models are neural networks (NNs), but only NNs that have a sufficient number of hidden layers (usually more than one) can be regarded as `deep' models. Besides deep NNs, other architectures have multiple layers, such as deep Gaussian processes \cite{damianou2013deep2}, neural processes \cite{garnelo2018neural}, and deep random forests~\cite{zhou2017deep2}, and can also be regarded as deep learning structures.
The major benefit of deep learning over traditional ML is thus the automatic feature extraction, by which expensive hand-crafted feature engineering can be circumvented.} 
We illustrate the relation between deep learning, machine learning, and artificial intelligence (AI) at a high level in Fig.~\ref{fig:venn}. 

\rev{In general, AI is a computation paradigm that endows machines with intelligence, aiming to teach them how to work, react, and learn like humans. Many techniques fall under this broad umbrella, including machine learning, expert systems, and evolutionary algorithms. Among these, machine learning enables the artificial processes to absorb knowledge from data and make decisions without being explicitly programmed. Machine learning algorithms are typically categorized into supervised, unsupervised, and reinforcement learning. Deep learning is a family of machine learning techniques that mimic biological nervous systems and perform representation learning through multi-layer transformations, extending across all three learning paradigms mentioned before. As deep learning has growing number of applications in mobile an wireless networking, the crossovers between these domains make the scope of this manuscript.}
 \subsection{The Evolution of Deep Learning}
The discipline traces its origins 75 years back, when threshold logic was employed to produce a computational model for neural networks~\cite{mcculloch:1943}. However, it was only in the late 1980s that neural networks (NNs) gained interest, as Rumelhart \emph{et al.} showed that multi-layer NNs could be trained effectively by back-propagating errors \cite{williams1986learning}. LeCun and Bengio subsequently proposed the now popular Convolutional Neural Network (CNN) architecture \cite{lecun1995convolutional}, but progress stalled due to computing power limitations of systems available at that time. Following the recent success of GPUs, CNNs have been employed to dramatically reduce the error rate in the Large Scale Visual Recognition Challenge (LSVRC)~\cite{krizhevsky2012imagenet}.~This has drawn unprecedented interest in deep learning and breakthroughs continue to appear in a wide range of computer science areas.
 
\begin{figure}[b]
\begin{center}
\includegraphics[width=0.5\textwidth]{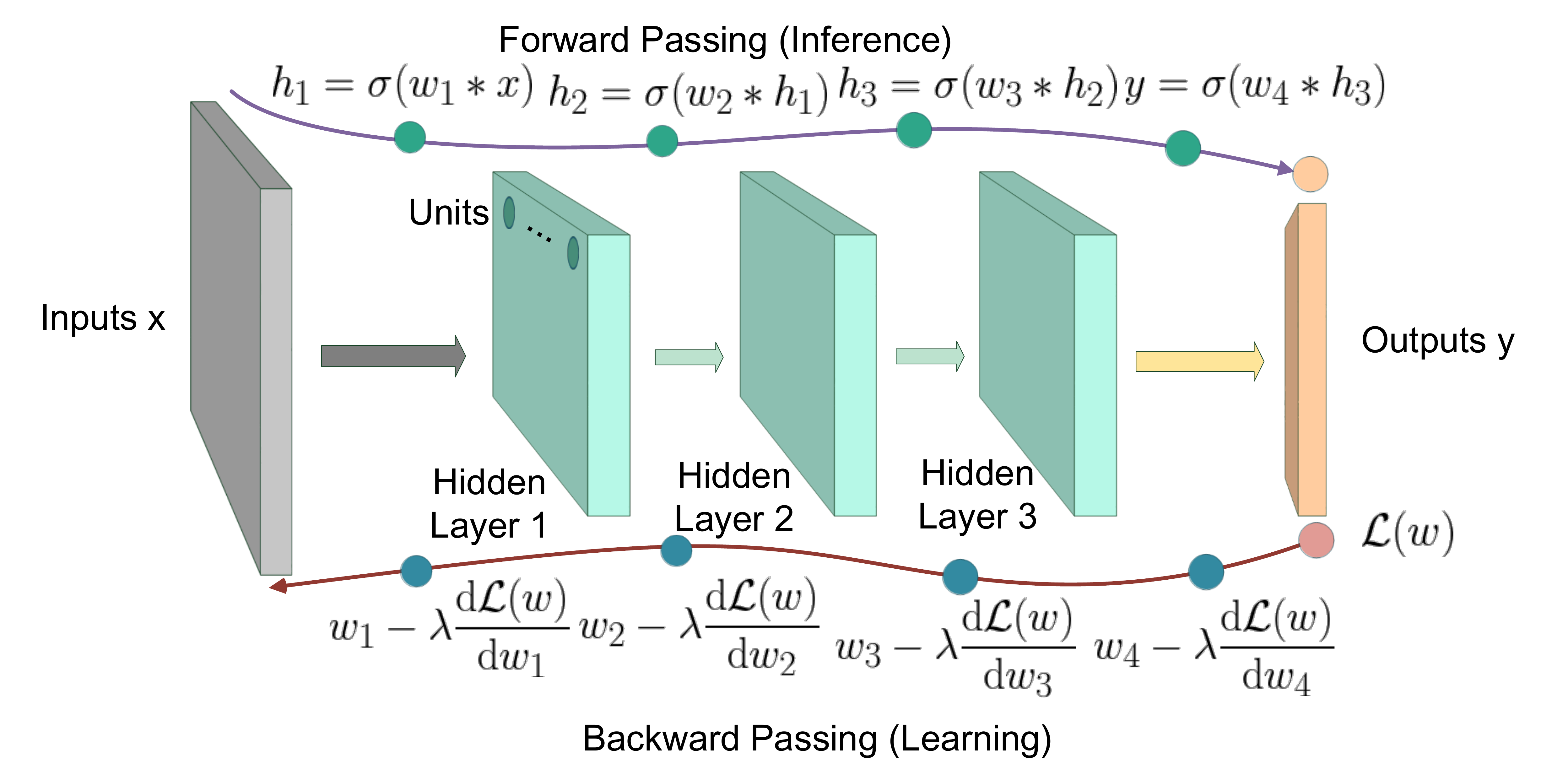}
\end{center}
\caption{\label{fig:bp} Illustration of the learning and inference processes of a 4-layer CNN. $w_{(\cdot)}$ denote weights of each hidden layer, $\sigma(\cdot)$ is an activation function, $\lambda$ refers to the learning rate, $*(\cdot)$ denotes the convolution operation and $\mathcal{L}(w)$ is the loss function to be optimized.}
%
%
\end{figure}

 \subsection{Fundamental Principles of Deep Learning}
The key aim of deep neural networks is to approximate complex functions through a composition of simple and predefined operations of units (or neurons). Such an objective function can be almost of any type, such as a mapping between images and their class labels (classification), computing future stock prices based on historical values (regression), or even deciding the next optimal chess move given the current status on the board (control). The operations performed are usually defined by a weighted combination of a specific group of hidden units with a non-linear activation function, depending on the structure of the model. Such operations along with the output units are named ``layers''. The neural network architecture resembles the perception process in a brain, where a specific set of units are activated given the current environment, influencing the output of the neural network model.

\subsection{\edit{Forward and Backward Propagation}}
In mathematical terms, the architecture of deep neural networks is usually differentiable, therefore the weights (or parameters) of the model can be learned by minimizing a loss function using gradient descent methods through back-propagation, following the fundamental chain rule \cite{williams1986learning}. We illustrate the principles of the learning and inference processes of a deep neural network in Fig.~\ref{fig:bp}, where we use a two-dimensional (2D) Convolutional Neural Network (CNN) as an example. \\

\noindent\edit{\textbf{Forward propagation:} The figure shows a CNN with 5 layers, i.e., an input layer (grey), 3 hidden layers (blue) and an output layer (orange). In forward propagation,  A 2D input $\mathbf{x}$ (e.g images) is first processed by a convolutional layer, which perform the following convolutional operation:
\begin{equation}
h_1 = \sigma(w_1\ast \mathbf{x}).
\end{equation}
Here $h_1$ is the output of the first hidden layer, $w_1$ is the convolutional filter and $\sigma(\cdot)$ is the activation function, aiming at improving the non-linearity and representability of the model. The output $h_1$ is subsequently provided as input to and processed by the following two convolutional layers, which eventually produces a final output $\mathbf{y}$. This could be for instance vector of probabilities for different possible patterns (shapes) discovered in the (image) input. To train the CNN appropriately, one uses a loss function $\mathcal{L}(w)$ to measure the distance between the output $\mathbf{y}$ and the ground truth $\mathbf{y^*}$. The purpose of training is to find the best weights $\mathbf{w}$, so as to minimize the loss function $\mathcal{L}(w)$. This can be achieved by the back propagation through gradient descent.}\\

\noindent\edit{\textbf{Backward propagation:} During backward propagation, one computes the gradient of the loss function $\mathcal{L}(w)$ over the weight of the last hidden layer, and updates the weight by computing:
\begin{equation}
w_4 = w_4-\lambda\frac{\mathrm{d} \mathcal{L}(w)}{\mathrm{d} w_4}.
\end{equation}}
\noindent\edit{Here $\lambda$ denotes the learning rate, which controls the step size of moving in the direction indicated by the gradient. The same operation is performed for each weight, following the chain rule. The process is repeated and eventually the gradient descent will lead to a set $w$ that minimizes the $\mathcal{L}(w)$.}

\edit{For other NN structures, the training and inference processes are similar. To help less expert readers we detail the principles and computational details of various deep learning techniques in Sec.\ref{sec:model}.}

\begin{table}[ht]
\centering
\caption{\edit{Summary of the benefits of applying deep learning to solve problems in mobile and wireless networks.}}
\label{tab:advantage}
\color{black}
\begin{tabular}{|C{1.5cm}|C{3cm}|C{3cm}|}
\hline
\textbf{Key aspect}    & \textbf{Description}                                                                                                     & \textbf{Benefits}                                                                                         \\ \hline
Feature extraction    & Deep neural networks can automatically extract high-level features through layers of different depths.                    & Reduce expensive hand-crafted feature engineering in processing heterogeneous and noisy mobile big data. \\ \hline
Big data exploitation      & Unlike traditional ML tools, the performance of deep learning usually grow significantly with the size of training data. & Efficiently utilize huge amounts of mobile data generated at high rates.                          \\ \hline
Unsupervised learning & Deep learning is effective in processing un-/semi- labeled data, enabling unsupervised learning.    & Handling large amounts of unlabeled data, which are common in mobile system.                      \\ \hline
Multi-task learning   & Features learned by neural networks through hidden layers can be applied to different tasks by transfer learning.              & Reduce computational and memory requirements when performing multi-task learning in mobile systems.      \\ \hline
\rev{Geometric mobile data learning}   & \rev{Dedicated deep learning architectures exist to model geometric mobile data}              & \rev{Revolutionize geometric mobile data analysis}      \\ \hline
\end{tabular}
\end{table}

\subsection{Advantages of Deep Learning \rev{in Mobile and Wireless Networking} \label{sec:adv}}
We recognize several benefits of employing deep learning to address network engineering problems, \edit{as summarized in Table~\ref{tab:advantage}. Specifically:}
\begin{enumerate}
\item It is widely acknowledged that, while vital to the performance of traditional ML algorithms, feature engineering is costly \cite{Domingos:2012}. A key advantage of deep learning is that it can automatically extract high-level features from data that has complex structure and inner correlations. The learning process does not need to be designed by a human, which tremendously simplifies prior feature handcrafting~\cite{lecun2015deep}. The importance of this is amplified in the context of mobile networks, as mobile data is usually generated by heterogeneous sources, is often noisy, and exhibits non-trivial spatial/temporal patterns~\cite{alsheikh2016mobile}, whose labeling would otherwise require outstanding \mbox{human effort.} 
\item Secondly, deep learning is capable of handling large amounts of data. Mobile networks generate high volumes of different types of data at fast pace. Training traditional ML algorithms (e.g., Support Vector Machine (SVM) \cite{tsang2005core} and Gaussian Process (GP)~\cite{rasmussen2006gaussian}) sometimes requires to store all the data in memory, which is computationally infeasible under big data scenarios. Furthermore, the performance of ML does not grow significantly with large volumes of data and plateaus relatively fast~\cite{goodfellow2016deep}. In contrast, Stochastic Gradient Descent (SGD) employed to train NNs only requires sub-sets of data at each training step, which guarantees deep learning's scalability with big data. Deep neural networks further benefit as training with big data prevents model over-fitting.

\item Traditional supervised learning is only effective when sufficient labeled data is available. However, most current mobile systems generate unlabeled or semi-labeled data~\cite{alsheikh2016mobile}. Deep learning provides a variety of methods that allow exploiting unlabeled data to learn useful patterns in an unsupervised manner, e.g., Restricted Boltzmann Machine (RBM)~\cite{le2008representational}, Generative Adversarial Network (GAN)~\cite{goodfellow2014generative}. Applications include clustering~\cite{schroff2015facenet}, data distributions approximation~\cite{goodfellow2014generative}, un/semi-supervised learning~\cite{kingma2014semi, stewart2017label}, and one/zero shot learning~\cite{rezende2016one, socher2013zero}, among others. 

\item Compressive representations learned by deep neural networks can be shared across different tasks, while this is limited or difficult to achieve in other ML paradigms (e.g., linear regression, random forest, etc.). Therefore, a single model can be trained to fulfill multiple objectives, without requiring complete model retraining for different tasks. We argue that this is essential for mobile network engineering, as it reduces computational and memory requirements of mobile systems when performing multi-task learning applications~\cite{georgiev2017low}. 

\item \rev{Deep learning is effective in handing geometric mobile data \cite{monti2017geometric}, while this is a conundrum for other ML approaches. Geometric data refers to multivariate data represented by coordinates, topology, metrics and order \cite{le2004geometric}. Mobile data, such as mobile user location and network connectivity can be naturally represented by point clouds and graphs, which have important geometric properties. These data can be effectively modelled by dedicated deep learning architectures, such as PointNet++ \cite{qi2017pointnet} and Graph CNN \cite{kipf2016semi}. Employing these architectures has great potential to revolutionize the geometric mobile data analysis \cite{wang2018spatio}.}
\end{enumerate}

\subsection{\rev{Limitations of Deep Learning in Mobile and Wireless Networking} \label{sec:limit}}
\rev{Although deep learning has unique advantages when addressing mobile network problems, it also has several shortcomings, which partially restricts its applicability in this domain. Specifically, }
\begin{enumerate}
    \item \rev{In general, deep learning (including deep reinforcement learning) is vulnerable to adversarial examples \cite{nguyen2015deep, behzadan2017vulnerability}. These refer to artifact inputs that are intentionally designed by an attacker to fool machine learning models into making mistakes \cite{nguyen2015deep}. While it is difficult to distinguish such samples from genuine ones, they can trigger mis-adjustments of a model with high likelihood. We illustrate an example of such an adversarial attack in Fig.~\ref{fig:adver_attack}. Deep learning, especially CNNs are vulnerable to these types of attacks. This may also affect the applicability of deep learning in mobile systems. For instance, hackers may exploit this vulnerability and construct cyber attacks that subvert deep learning based detectors \cite{madani2018robustness}. Constructing deep models that are robust to adversarial examples is imperative, but remains challenging. }
\begin{figure}[htb]
\begin{center}
\includegraphics[width=0.5\textwidth]{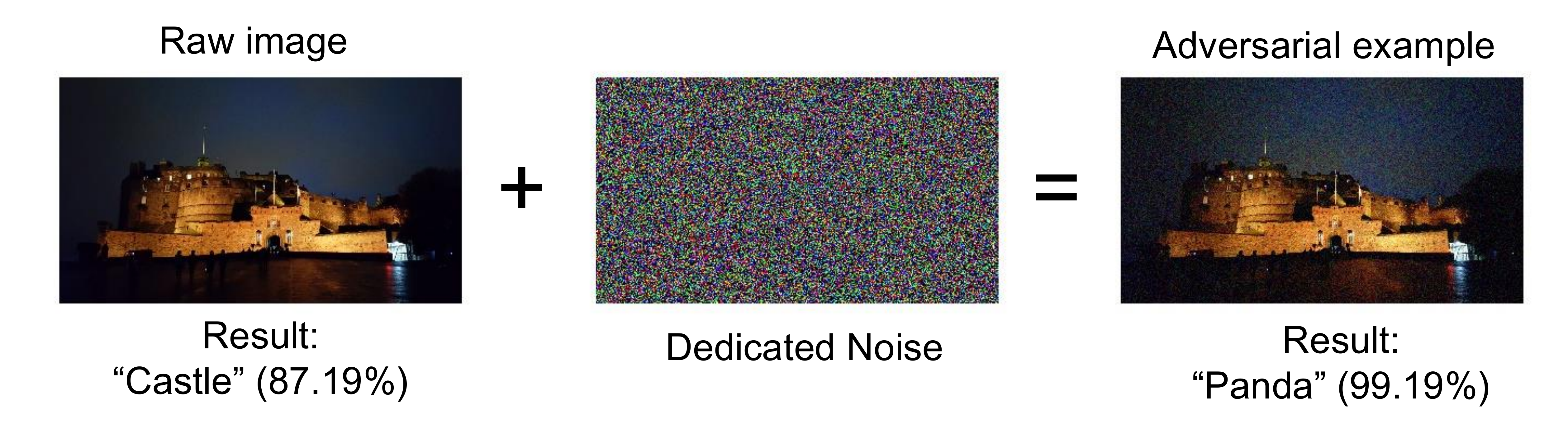}
\end{center}
\caption{\label{fig:adver_attack} \rev{An example of an adversarial attack on deep learning.}}
\end{figure}
    \item \rev{Deep learning algorithms are largely black boxes and have low interpretability. Their major breakthroughs are in terms of accuracy, as they significantly improve performance of many tasks in different areas. However, although deep learning enables creating ``machines'' that have high accuracy in specific tasks, we still have limited knowledge as of why NNs make certain decisions. This limits the applicability of deep learning, e.g. in network economics. Therefore, businesses would rather continue to employ statistical methods that have high interpretability, whilst sacrificing on accuracy. Researchers have recognized this problem and investing continuous efforts to address this limitation of deep learning (e.g. \cite{bau2017network, wu2017beyond, chakraborty2017interpretability}).}
    \item \rev{Deep learning is heavily reliant on data, which sometimes can be more important  than the model itself. Deep models can further benefit from training data augmentation \cite{perez2017effectiveness}. This is indeed an opportunity for mobile networking, as networks generates tremendous amounts of data. However, data collection may be costly, and face privacy concern, therefore it may be difficult to obtain sufficient information for model training. In such scenarios, the benefits of employing deep learning may be outweigth by the costs.}
    \item \rev{Deep learning can be computationally demanding. Advanced parallel computing (e.g. GPUs, high-performance chips) fostered the development and popularity of deep learning, yet deep learning also heavily relies on these. Deep NNs usually require complex structures to obtain satisfactory accuracy performance. However, when deploying NNs on embedded and mobile devices, energy and capability constraints have to be considered. Very deep NNs may not be suitable for such scenario and this would inevitably compromise accuracy. Solutions are being developed to mitigate this problem and we will dive deeper into these in Sec. \ref{sec:stimulator} and \ref{sec:tailor}.}
    \item \rev{Deep neural networks usually have many hyper-parameters and finding their optimal configuration can be difficult. For a single convolutional layer, we need to configure at least hyper-parameters for the number, shape, stride, and dilation of filters, as well as for the residual connections. The number of such hyper-parameters grows exponentially with the depth of the model and can highly influence its performance. Finding a good set of hyper-parameters can be similar to looking for a needle in a haystack. The AutoML platform\footnote{\rev{AutoML -- training high-quality custom machine learning models with minimum effort and machine learning expertise. \url{https://cloud.google.com/automl/}}} provides a first solution to this problem, by employing progressive neural architecture search \cite{liu2017progressive}. This task, however, remains costly.}
\end{enumerate}

\rev{To circumvent some of the aforementioned problems and allow for effective deployment in mobile networks, deep learning requires certain system and software support. We review and discuss such enablers in the next section.}

\section{Enabling Deep Learning in Mobile Networking}\label{sec:stimulator}
\begin{figure}[htb]
\begin{center}
\includegraphics[width=0.85\columnwidth]{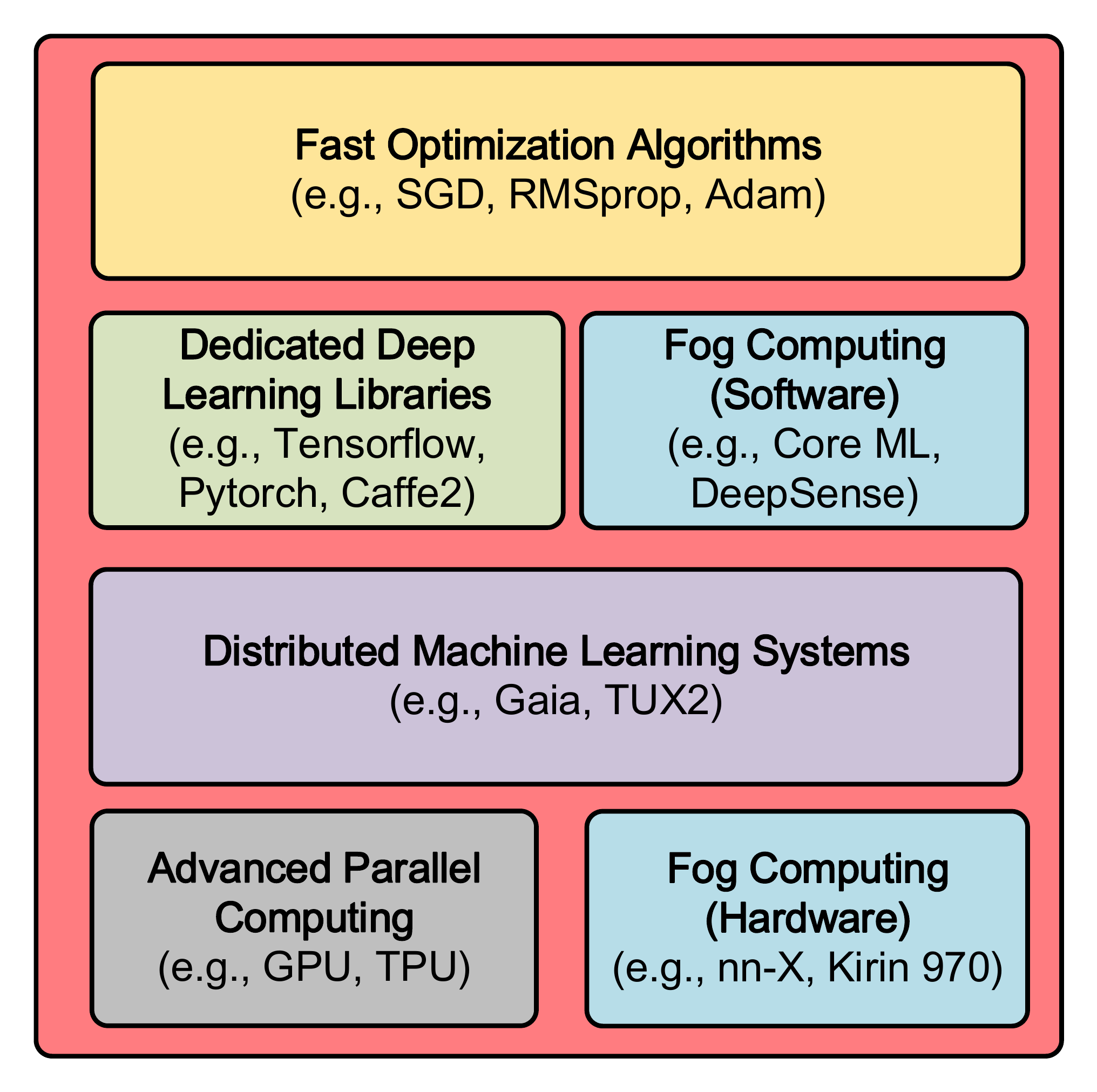}
\end{center}
\caption{\label{fig:enable} \edit{Hierarchical view of deep learning enablers. Parallel computing and hardware in fog computing lay foundations for deep learning. Distributed machine learning systems can build upon them, to support large-scale deployment of deep learning. Deep learning libraries run at the software level, to enable fast deep learning implementation. Higher-level optimizers are used to train the NN, to fulfill specific objectives.}}
\end{figure}

5G systems seek to provide high throughput and ultra-low latency communication services, to improve users' QoE~\cite{agiwal2016next}. 
Implementing deep learning to build intelligence into 5G systems, so as to meet these objectives is expensive. This is because powerful hardware and software is required to support training and inference in complex settings. Fortunately, several tools are emerging, which make deep learning in mobile networks tangible; namely, \emph{(i)} advanced parallel computing, \emph{(ii)} distributed machine learning systems,  \emph{(iii)} dedicated deep learning libraries, \emph{(iv)} fast optimization algorithms, and \emph{(v)} fog computing. \edit{These tools can be seen as forming a hierarchical structure, as illustrated in Fig.~\ref{fig:enable}; synergies between them exist that makes networking problem amenable to deep learning based solutions. By employing these tools, once the training is completed, inferences can be made within millisecond timescales, as already reported by a number of papers for a range of tasks (e.g., \cite{zhang2016deep3, ordonez2016deep, de2016distributed} ).} We summarize these advances in Table~\ref{tab:enabler} and review them in what follows.

\begin{table*}[htb]
\centering
\caption{Summary of tools and techniques that enable deploying deep learning in mobile systems.}
\label{tab:enabler}
\begin{tabular}{|C{2cm}|C{2.5cm}|C{2cm}|C{3.5cm}|C{2cm}|C{1.5cm}|C{1.5cm}|}
\hline
\textbf{Technique} & \textbf{Examples} & \textbf{Scope} & \textbf{Functionality} & \textbf{Performance improvement}& \textbf{Energy consumption}& \textbf{Economic cost}\\ \hline
Advanced parallel computing      &   GPU, TPU \cite{jouppi2017datacenter}, CUDA \cite{nickolls2008scalable}, cuDNN~\cite{chetlur2014cudnn}       &     Mobile servers, workstations               &   Enable fast, parallel training/inference of deep learning models in mobile applications    &   High &High &Medium (hardware) \\ \hline
Dedicated deep learning library     &    TensorFlow \cite{tensorflow2015-whitepaper}, Theano \cite{2016arXiv160502688short}, Caffe~\cite{jia2014caffe}, Torch~\cite{torch}      &       Mobile servers and devices                &   High-level toolboxes that enable network engineers to build purpose-specific deep learning architectures    &   Medium &Associated with hardware & Low (software) \\ \hline
Fog computing &nn-X \cite{gokhale2014240}, ncnn \cite{ncnn}, Kirin 970 \cite{huawei2017kirin}, Core~ML~\cite{coreml}& Mobile devices & Support edge-based deep learning computing & Medium & Low & Medium (hardware) \\ \hline
Fast optimization algorithms     &    Nesterov \cite{sutskever2013importance}, Adagrad \cite{dean2012large}, RMSprop,~Adam~\cite{kingma2015adam}      &       Training deep architectures                &   Accelerate and stabilize the model optimization process &   Medium & Associated with hardware & Low (software)   \\ \hline
Distributed machine learning systems    &   MLbase \cite{kraska2013mlbase}, Gaia~\cite{hsieh2017gaia}, Tux\textsuperscript{2} \cite{xiao2017tux2}, Adam \cite{chilimbi2014project}, GeePS~\cite{cui2016geeps}       &      Distributed data centers, cross-server  &     Support deep learning frameworks in mobile systems across data centers    &   High &High&High (hardware) \\ \hline
\end{tabular}
\end{table*}

\subsection{Advanced Parallel Computing}
Compared to traditional machine learning models, deep neural networks have significantly larger parameters spaces, intermediate outputs, and number of gradient values. Each of these need to be updated during every training step, requiring powerful computation resources. The training and inference processes involve huge amounts of matrix multiplications and other operations, though they could be massively parallelized. Traditional Central Processing Units (CPUs) have a limited number of cores, thus they only support restricted computing parallelism. Employing CPUs for deep learning implementations is highly inefficient and will not satisfy the low-latency requirements of mobile systems. 

Engineers address these issues by exploiting the power of GPUs. GPUs were originally designed for high performance video games and graphical rendering, but new techniques such as Compute Unified Device Architecture (CUDA)~\cite{nickolls2008scalable} and the CUDA Deep Neural Network library (cuDNN)~\cite{chetlur2014cudnn} developed by NVIDIA add flexibility to this type of hardware, allowing users to customize their usage for specific purposes. GPUs usually incorporate thousand of cores and perform exceptionally in fast matrix multiplications required for training neural networks. This provides higher memory bandwidth over CPUs and dramatically speeds up the learning process. Recent advanced Tensor Processing Units (TPUs) developed by Google even demonstrate 15-30$\times$ higher processing speeds and 30-80$\times$ higher performance-per-watt, as compared to CPUs and GPUs~\cite{jouppi2017datacenter}. 

\edit{Diffractive neural networks (D\textsuperscript{2}NNs) that completely rely on light communication were recently introduced in \cite{Lineaat8084}, to enable zero-consumption and zero-delay deep learning. The D\textsuperscript{2}NN is composed of several transmissive layers, where points on these layers act as neurons in a NN. The structure is trained to optimize the transmission/reflection coefficients, which are equivalent to weights in a NN. Once trained, transmissive layers will be materialized via 3D printing and they can subsequently be used for inference.}

There are also a number of toolboxes that can assist the computational optimization of deep learning on the server side. Spring and Shrivastava introduce a hashing based technique that substantially reduces computation requirements of deep network implementations~\cite{spring2017scalable}. Mirhoseini \emph{et al.} employ a reinforcement learning scheme to enable machines to learn the optimal operation placement over mixture hardware for deep neural networks.  Their solution achieves up to 20\% faster computation speed than human experts' designs of such placements~\cite{mirhoseini2017device}.

Importantly, these systems are easy to deploy, therefore mobile network engineers do not need to rebuild mobile servers from scratch to support deep learning computing. This makes implementing deep learning in mobile systems feasible and accelerates the processing of mobile data streams.

\vspace*{-0.5em}
\subsection{Distributed Machine Learning Systems}
Mobile data is collected from heterogeneous sources (e.g., mobile devices, network probes, etc.), and stored in multiple distributed data centers. With the increase of data volumes, it is impractical to move all mobile data to a central data center to run deep learning applications~\cite{hsieh2017gaia}. Running network-wide deep learning algorithms would therefore require distributed machine learning systems that support different interfaces (e.g., operating systems, programming language, libraries), so as to enable training and evaluation of deep models across geographically distributed servers simultaneously, with high efficiency and low overhead.

Deploying deep learning in a distributed fashion will inevitably introduce several system-level problems, which \mbox{require} satisfying the following properties:

\noindent\edit{
\textbf{Consistency} -- Guaranteeing that model parameters and computational processes are consistent across all  machines.\\
\textbf{Fault tolerance} -- Effectively dealing with equipment breakdowns in large-scale distributed machine learning systems.\\
\textbf{Communication} -- Optimizing communication between nodes in a cluster and  to avoid congestion.\\
\textbf{Storage} -- Designing efficient storage mechanisms tailored to different environments (e.g., distributed clusters, single machines, GPUs), given I/O and data processing diversity.\\
\textbf{Resource management} --  Assigning workloads and ensuring that nodes work well-coordinated.\\
\textbf{Programming model} -- Designing programming interfaces to support multiple programming languages.
\vspace*{0.5em}
}

There exist several distributed machine learning systems that facilitate deep learning in mobile networking applications. Kraska \emph{et al.} introduce a distributed system named MLbase, which enables to intelligently specify, select, optimize, and parallelize ML algorithms~\cite{kraska2013mlbase}. Their system helps non-experts deploy a wide range of ML methods, allowing optimization and running ML applications across different servers. Hsieh \emph{et al.} develop a geography-distributed ML system called Gaia, which breaks the throughput bottleneck by employing an advanced communication mechanism over Wide Area Networks, while preserving the accuracy of ML algorithms~\cite{hsieh2017gaia}. Their proposal supports versatile ML interfaces (e.g. TensorFlow, Caffe), without requiring significant changes to the ML algorithm itself. This system enables deployments of complex deep learning applications over large-scale mobile networks. 

Xing \emph{et al.} develop a large-scale machine learning platform to support big data applications \cite{xing2015petuum}. Their architecture achieves efficient model and data parallelization, enabling parameter state synchronization with low communication cost. Xiao \emph{et al.} propose a distributed graph engine for ML named TUX\textsuperscript{2}, to support data layout optimization across machines and reduce cross-machine communication~\cite{xiao2017tux2}. 
They demonstrate remarkable performance in terms of runtime and convergence on a large dataset with up to 64 billion edges. Chilimbi~\emph{et al.} build a distributed, efficient, and scalable system named ``Adam"\footnote{Note that this is distinct from the Adam optimizer discussed in Sec.~\ref{set:opt}} tailored to the training of deep models~\cite{chilimbi2014project}. Their architecture demonstrates impressive performance in terms of throughput, delay, and fault tolerance. Another dedicated distributed deep learning system called GeePS is developed by Cui~\emph{et al.}~\cite{cui2016geeps}. Their framework allows data parallelization on distributed GPUs, and demonstrates higher training throughput and faster convergence rate.
\rev{More recently, Moritz \emph{et al.} designed a dedicated distributed framework named Ray to underpin reinforcement learning applications \cite{moritz2018ray}. Their framework is supported by an dynamic task execution engine, which incorporates the actor and task-parallel abstractions. They further introduce a bottom-up distributed scheduling strategy and a dedicated state storage scheme, to improve scalability and fault tolerance.}

\subsection{Dedicated Deep Learning Libraries}
\begin{table*}[t!]
\centering
\caption{\edit{Summary and comparison of mainstream deep learning libraries.}}
\label{tab:library}
\color{black}
\begin{tabular}{|C{1.3cm}|C{1.5cm}|C{1.3cm}|l|l|C{1cm}|C{1cm}|C{2.1cm}|}
\hline
\textbf{Library} & \textbf{Low-Layer Language} & \textbf{Available Interface}                                       & \textbf{Pros}                                                                                                                                                                                                                                                                                                & \textbf{Cons}                                                                                                                                                  & \textbf{Mobile Supported} & \textbf{Popularity} & \textbf{Upper-Level Library}                                                               \\ \hline
TensorFlow       & C++                         & \begin{tabular}[c]{@{}l@{}}Python, Java,\\ C, C++, Go\end{tabular} & \begin{tabular}[c]{@{}l@{}}$\bullet$ Large user community\\ $\bullet$ Well-written document\\ $\bullet$ Complete functionality\\ $\bullet$ Provides visualization \\ tool (TensorBoard)\\ $\bullet$ Multiple interfaces support\\ \rev{$\bullet$ Allows distributed training} \\ \rev{and model serving}\end{tabular} & \begin{tabular}[c]{@{}l@{}}$\bullet$ Difficult to debug\\ $\bullet$ The package is heavy\\ \rev{$\bullet$ Higher entry barrier} \\ \rev{for beginners}\end{tabular} & Yes                       & High                & \begin{tabular}[c]{@{}l@{}}Keras, TensorLayer, \\ Luminoth\end{tabular} \\ \hline
Theano           & Python                      & Python                                                             & \begin{tabular}[c]{@{}l@{}}$\bullet$ Flexible\\ $\bullet$ Good running speed\end{tabular}                                                                                                                                                                                                                    & \begin{tabular}[c]{@{}l@{}}$\bullet$ Difficult to learn\\ $\bullet$ Long compilation time\\ $\bullet$ No longer maintained\end{tabular}                        & No                        & Low                 & Keras, Blocks, Lasagne                                                                     \\ \hline
Caffe(2)         & C++                         & Python, Matlab                                                     & \begin{tabular}[c]{@{}l@{}}$\bullet$ Fast runtime\\ $\bullet$ Multiple platform\\ s support\end{tabular}                                                                                                                                                                                                     & \begin{tabular}[c]{@{}l@{}}$\bullet$ Small user base\\ $\bullet$ Modest documentation\end{tabular}                                                             & Yes                       & Medium              & None                                                                                       \\ \hline
(Py)Torch        & Lua, C++                    & Lua, Python, C, C++                                                & \begin{tabular}[c]{@{}l@{}}$\bullet$ Easy to build models\\ $\bullet$ Flexible\\ $\bullet$ Well documented\\ $\bullet$ Easy to debug\\ \rev{$\bullet$ Rich pretrained models}\\ \rev{available}\\ \rev{$\bullet$ Declarative data parallelism}\end{tabular}                                                            & \begin{tabular}[c]{@{}l@{}}$\bullet$ Has limited resource\\ \rev{$\bullet$ Lacks model serving}\\ \rev{$\bullet$ Lacks visualization tools}\end{tabular}         & Yes                       & High                & None                                                                                       \\ \hline
MXNET            & C++                         & C++, Python, Matlab, R                                             & \begin{tabular}[c]{@{}l@{}}$\bullet$ Lightweight\\ $\bullet$ Memory-efficient\\ $\bullet$ Fast training\\ \rev{$\bullet$ Simple model serving}\\ \rev{$\bullet$ Highly scalable}\end{tabular}                                                                                                          & \begin{tabular}[c]{@{}l@{}}$\bullet$ Small user base\\ $\bullet$ Difficult to learn\end{tabular}                                                               & Yes                       & Low                 & Gluon                                                                                      \\ \hline
\end{tabular}
\end{table*}

Building a deep learning model from scratch can prove complicated to engineers, as this requires definitions of forwarding behaviors and gradient propagation operations at each layer, in addition to CUDA coding for GPU parallelization. With the growing popularity of deep learning, several dedicated libraries simplify this process. Most of these toolboxes work with multiple programming languages, and are built with GPU acceleration and automatic differentiation support. This eliminates the need of hand-crafted definition of gradient propagation. We summarize these libraries below, \edit{and give a comparison among them in Table~\ref{tab:library}}.\\

\noindent \textbf{TensorFlow}\footnote{TensorFlow, \url{https://www.tensorflow.org/}} is a machine learning library developed by Google~\cite{tensorflow2015-whitepaper}. It enables deploying computation graphs on CPUs, GPUs, and even mobile devices \cite{alzantot2017rstensorflow}, allowing ML implementation on both single and distributed architectures. \rev{This permits fast implementation of deep NNs on both cloud and fog services.} Although originally designed for ML and deep neural networks applications, TensorFlow is also suitable for other data-driven research purposes. \rev{It provides TensorBoard,\footnote{\rev{TensorBoard -- A visualization tool for TensorFlow, \url{https://www.tensorflow.org/guide/summaries_and_tensorboard}.}} a sophisticated visualization tool, to help users understand model structures and data flows, and perform debugging. } Detailed documentation and tutorials for Python exist, while other programming languages such as C, Java, and Go are also supported. \edit{currently it is the most popular deep learning library.} Building upon TensorFlow, several dedicated deep learning toolboxes were released to provide higher-level programming interfaces, including Keras\footnote{Keras deep learning library, \url{https://github.com/fchollet/keras}}, Luminoth \footnote{Luminoth deep learning library for computer vision, \url{https://github.com/tryolabs/luminoth}} and TensorLayer \cite{tensorlayer}. \\

\noindent \textbf{Theano} is a Python library that allows to efficiently define, optimize, and evaluate numerical computations involving multi-dimensional data  \cite{2016arXiv160502688short}. It provides both GPU and CPU modes, which enables users to tailor their programs to individual machines. \edit{Learning Theano is however difficult and building a NNs with it involves substantial compiling time.} Though Theano has a large user base and a support community, and at some stage was one of the most popular deep learning tools, its popularity is decreasing rapidly, as core ideas and attributes are absorbed by TensorFlow.\\

\noindent \textbf{Caffe(2)} is a dedicated deep learning framework developed by Berkeley AI Research \cite{jia2014caffe} and the latest version, Caffe2,\footnote{Caffe2, \url{https://caffe2.ai/}} was recently released by Facebook. \edit{Inheriting all the advantages of the old version, Caffe2 has become a very flexible framework that enables users to build their models efficiently}.  It also allows to train neural networks on multiple GPUs within distributed systems, and supports deep learning implementations on mobile operating systems, such as iOS and Android. Therefore, it has the potential to play an important role in the future mobile edge computing.\\

\noindent \textbf{(Py)Torch} is a scientific computing framework with wide support for machine learning models and algorithms \cite{torch}. It was originally  developed in the Lua language, but developers later released an improved Python version \cite{paszke2017automatic}. In essence PyTorch is a lightweight toolbox that can run on embedded systems such as smart phones, but lacks comprehensive documentations. \edit{Since building NNs in PyTorch is straightforward,  the popularity of this library is growing rapidly.} \rev{It also offers rich pretrained models and modules that are easy to reuse and combine.} PyTorch is now officially maintained by Facebook and mainly employed for research purposes.\\ 

\noindent \textbf{MXNET} is a flexible and scalable deep learning library that provides interfaces for multiple languages (e.g., C++, Python, Matlab, R, etc.) \cite{chen2015mxnet}. It supports different levels of machine learning models, from logistic regression to GANs. MXNET provides fast numerical computation for both single machine and distributed ecosystems. It wraps workflows commonly used in deep learning into high-level functions, such that standard neural networks can be easily constructed without substantial coding effort. \edit{However, learning how to work with this toolbox in short time frame is difficult, hence the number of users who prefer this library is relatively small.}  MXNET is the official deep learning framework in Amazon. 

Although less popular, there are other excellent deep learning libraries, such as CNTK,\footnote{MS Cognitive Toolkit, \url{https://www.microsoft.com/en-us/cognitive-toolkit/}} Deeplearning4j,\footnote{Deeplearning4j, \url{http://deeplearning4j.org}} Blocks,\footnote{Blocks, A Theano framework for building and training neural networks \url{https://github.com/mila-udem/blocks}} Gluon,\footnote{Gluon, A deep learning library \url{https://gluon.mxnet.io/}} and Lasagne,\footnote{Lasagne, \url{https://github.com/Lasagne}} which can also be employed in mobile systems. Selecting among these varies according to specific applications. \rev{For AI beginners who intend to employ deep learning for the networking domain, PyTorch is a good candidate, as it is easy to build neural networks in this environment and the library is well optimized for GPUs. On the other hand, if for people who pursue advanced operations and large-scale implementation, Tensorflow might be a better choice, as it is well-established, under good maintainance and has standed the test of many Google industrial projects.}

\subsection{Fast Optimization Algorithms}\label{set:opt}
The objective functions to be optimized in deep learning are usually complex, as they involve sums of extremely large numbers of data-wise likelihood functions. As the depth of the model increases, such functions usually exhibit high non-convexity with multiple local minima, critical points, and saddle points. In this case, conventional Stochastic Gradient Descent (SGD) algorithms \cite{ruder2016overview} are slow in terms of convergence, which will restrict their applicability to latency constrained mobile systems. \edit{To overcome this problem and stabilize the optimization process, many algorithms evolve the traditional SGD, allowing NN models to be trained faster for mobile applications. We summarize the key principles behind these optimizers and make a comparison between them in Table~\ref{tab:opt}. We delve into the details of their operation next.} 
\begin{table*}[t!]
\centering
\caption{\edit{Summary and comparison of different optimization algorithms.}}
\label{tab:opt}
\color{black}
\begin{tabular}{|c|C{3.2cm}|L{4.6cm}|L{4.6cm}|}
\hline
\textbf{Optimization algorithm}                    & \textbf{Core idea}                                                                         & \textbf{Pros}                                                                                                                                                                                                                            & \textbf{Cons}                                                                                                                                                                                                         \\ \hline
SGD \cite{ruder2016overview}                       & Computes the gradient of mini-batches iteratively and updates the parameters                 & $\bullet$ Easy to implement                                                                                                                                                                                                              & \begin{tabular}[c]{@{}l@{}}$\bullet$ Setting a global learning rate required\\ $\bullet$ Algorithm may get stuck on saddle \\points or local minima\\ $\bullet$ Slow in terms of convergence\\ $\bullet$ Unstable\end{tabular} \\ \hline
Nesterov's momentum \cite{sutskever2013importance} & Introduces momentum to maintain the last gradient direction for the next update        & \begin{tabular}[c]{@{}l@{}}$\bullet$ Stable\\ $\bullet$ Faster learning\\ $\bullet$ Can escape local minima\end{tabular}                                                                                                                 & $\bullet$ Setting a learning rate needed                                                                                                                                                                                 \\ \hline
Adagrad \cite{dean2012large}                       & Applies different learning rates to different parameters                                       & \begin{tabular}[c]{@{}l@{}}$\bullet$ Learning rate tailored to each\\ parameter\\ $\bullet$ Handle sparse gradients well\end{tabular}                                                                                             & \begin{tabular}[c]{@{}l@{}}$\bullet$ Still requires setting a global \\learning rate\\ $\bullet$ Gradients sensitive to the regularizer\\ $\bullet$ Learning rate becomes very slow in \\the late stages\end{tabular}   \\ \hline
Adadelta \cite{zeiler2012adadelta}                 & Improves Adagrad, by applying a self-adaptive learning rate                               & \begin{tabular}[c]{@{}l@{}}
$\bullet$ Does not rely on a global learning rate\\ $\bullet$ Faster speed of convergence\\ $\bullet$ Fewer hyper-parameters to adjust\end{tabular}  & $\bullet$ May get stuck in a local minima at late training                                                                                                                                                        \\ \hline
RMSprop \cite{ruder2016overview}                   & Employs root mean square as a constraint of the learning rate                             & \begin{tabular}[c]{@{}l@{}}$\bullet$ Learning rate tailored to each \\parameter\\ $\bullet$ Learning rate do not decrease \\dramatically at late training\\ $\bullet$ Works well in RNN training\end{tabular}                      & \begin{tabular}[c]{@{}l@{}}$\bullet$ Still requires a global learning rate\\ $\bullet$ Not good at handling sparse gradients\end{tabular}                                                                              \\ \hline
Adam \cite{kingma2015adam}                         & Employs a momentum mechanism to store an exponentially decaying average of past gradients & \begin{tabular}[c]{@{}l@{}}$\bullet$ Learning rate stailored to each\\ parameter\\ $\bullet$ Good at handling sparse gradients and\\ non-stationary problems\\ $\bullet$ Memory-efficient\\ $\bullet$ Fast convergence\end{tabular} & $\bullet$ It may turn unstable during training                                                                                                                                                                              \\ \hline
Nadam \cite{dozat2016incorporating}                & Incorporates Nesterov accelerated gradients into Adam                                           & $\bullet$ Works well in RNN training                                                                                                                                                                                                      & ---                                                                                                                                                                                                                     \\ \hline
Learn to optimize \cite{andrychowicz2016learning}  & Casts the optimization problem as a learning problem using a RNN                          & $\bullet$ Does not require to design the learning by hand                                                                                                                                                                                     & $\bullet$ Require an additional RNN for learning in the optimizer                                                                                                                                                         \\ \hline
Quantized training \cite{szegedy2015going}         & Quantizes the gradients into \{-1, 0, 1\} for training                                     & \begin{tabular}[c]{@{}l@{}}$\bullet$ Good for distributed training\\ $\bullet$ Memory-efficient\end{tabular}                                                                                                                             & $\bullet$ Loses training accuracy                                                                                                                                                                                      \\ \hline
Stable gradient descent~\cite{stable2018zhou} & Employs a differential private mechanism to compare training and validation gradients, to reuse samples and keep them fresh. &\begin{tabular}[c]{@{}l@{}}$\bullet$ More stable\\ $\bullet$ Less overfitting \\$\bullet$ Converges faster than SGD \end{tabular} & $\bullet$ Only validated on convex functions \\ \hline
\end{tabular}
\end{table*}

\edit{\textbf{Fixed Learning Rate SGD Algorithms}}: Suskever \emph{et al.} introduce a variant of the SGD optimizer with Nesterov's momentum, which evaluates gradients after the current velocity is applied \cite{sutskever2013importance}. Their method demonstrates faster convergence rate when optimizing convex functions. Another approach is Adagrad, which  performs adaptive learning to model parameters according to their update frequency. This is suitable for handling sparse data and significantly outperforms SGD in terms of robustness \cite{dean2012large}. \edit{Adadelta improves the traditional Adagrad algorithm, enabling it to converge faster, and does not rely on a global learning rate~\cite{zeiler2012adadelta}.} RMSprop is a popular SGD based method introduced by G. Hinton. RMSprop divides the learning rate by an exponential smoothing the average of gradients and does not require one to set the learning rate for each training step \cite{ruder2016overview}. 

\edit{\textbf{Adaptive Learning Rate SGD Algorithms}}: Kingma and Ba propose an adaptive learning rate optimizer named Adam, which incorporates momentum by the first-order moment of the gradient \cite{kingma2015adam}. This algorithm is fast in terms of convergence, highly robust to model structures, and is considered as the first choice if one cannot decide what algorithm to use. \edit{By incorporating the  momentum into Adam, Nadam applies stronger constraints to the gradients, which enables faster convergence \cite{dozat2016incorporating}.} 

\edit{\textbf{Other Optimizers}}: Andrychowicz \emph{et al.} suggest that the optimization process can be even learned dynamically~\cite{andrychowicz2016learning}. They pose the gradient descent as a trainable learning problem, which demonstrates good generalization ability in neural network training. Wen \emph{et al.} propose a training algorithm tailored to distributed systems \cite{wen2017terngrad}. They quantize float gradient values to \{-1, 0 and +1\} in the training processing, which theoretically require 20 times less gradient communications between nodes. The authors prove that such gradient approximation mechanism allows the objective function to converge to optima with probability 1, where in their experiments only a 2\% accuracy loss is observed on average on GoogleLeNet~\cite{szegedy2015going} training. \edit{Zhou \emph{et al.} employ a differential private mechanism to compare training and validation gradients, to reuse samples and keep them fresh \cite{stable2018zhou}. This can dramatically reduce overfitting during training.}

\subsection{Fog Computing}
The fog computing paradigm presents a new opportunity to implement deep learning in mobile systems. Fog computing refers to a set of techniques that permit deploying applications or data storage at the edge of networks~\cite{bonomi2014fog}, e.g., on individual mobile devices. This reduces the communications overhead, offloads data traffic, reduces user-side latency, and lightens the sever-side computational burdens~\cite{mao2017modnn, mukherjee2018survey}. \edit{A formal definition of fog computing is given in \cite{vaquero2014finding}, where this is interpreted as \emph{'a  huge  number  of  heterogeneous  (wireless  and  sometimes  autonomous) ubiquitous and decentralized devices [that] communicate and potentially cooperate among them and with the network to perform storage and processing tasks without the intervention of third parties.'} To be more concrete, it can refer to smart phones, wearables devices and vehicles which store, analyze and exchange data, to offload the burden from cloud and perform more delay-sensitive tasks~\rev{\cite{aazam2018offloading, Buyya2018Manifesto}}. Since fog computing involves deployment at the edge, participating devices usually have limited computing resource and battery power. Therefore, special hardware and software are required for deep learning implementation, as we explain next.}

\edit{\textbf{Hardware:}} There exist several efforts that attempt to shift deep learning computing from the cloud side to mobile devices \cite{sze2017efficient}.  For example, Gokhale \emph{et al.} develop a mobile coprocessor named neural network neXt (nn-X), which accelerates the deep neural networks execution in mobile devices, while retaining low energy consumption~\cite{gokhale2014240}. Bang \emph{et al.} introduce a low-power and programmable deep learning processor to deploy mobile intelligence on edge devices \cite{bang201714}. Their hardware only consumes 288 $\mu$W  but  achieves  374 GOPS/W efficiency. A Neurosynaptic Chip called TrueNorth is proposed by IBM \cite{akopyan2016design}. Their solution seeks to support computationally intensive applications on embedded battery-powered mobile devices. Qualcomm introduces a Snapdragon neural processing engine to enable deep learning computational optimization tailored to mobile devices.\footnote{Qualcomm Helps Make Your Mobile Devices Smarter With New Snapdragon Machine Learning Software Development Kit: \url{https://www.qualcomm.com/news/releases/2016/05/02/qualcomm-helps-make-your-mobile-devices-smarter-new-snapdragon-machine}} Their hardware allows developers to execute neural network models on Snapdragon 820 boards to serve a variety of applications. In close collaboration with Google, Movidius\footnote{\rev{Movidius, an Intel company, provides cutting edge solutions for deploying deep learning and computer vision algorithms on ultra-low power devices. \url{https://www.movidius.com/}}} develops an embedded neural network computing framework that allows user-customized deep learning deployments at the edge of mobile networks. Their products can achieve satisfying runtime efficiency, while operating with ultra-low power requirements. \rev{It further supports difference frameworks, such as TensorFlow and Caffe, providing users with flexibility in choosing among toolkits.} More recently, Huawei officially announced the Kirin 970 as a mobile AI computing system on chip.\footnote{Huawei announces the {Kirin} 970 -- new flagship {SoC} with {AI} capabilities \url{http://www.androidauthority.com/huawei-announces-kirin-970-797788/}} Their innovative framework incorporates dedicated Neural Processing Units (NPUs), which dramatically accelerates neural network computing, enabling classification of 2,000 images per second on mobile devices. 

\edit{\textbf{Software:}} Beyond these hardware advances, there are also software platforms that seek to optimize deep learning on mobile devices (e.g., \cite{latifi2016cnndroid}). We compare and summarize all these platforms in Table~\ref{tab:platform}.\footnote{Adapted from \url{https://mp.weixin.qq.com/s/3gTp1kqkiGwdq5olrpOvKw}} In addition to the mobile version of TensorFlow and Caffe, Tencent released a lightweight, high-performance neural network inference framework tailored to mobile platforms, which relies on CPU computing.\footnote{ncnn is a high-performance neural network inference framework optimized for the mobile platform, \url{https://github.com/Tencent/ncnn}} This toolbox performs better than all known CPU-based open source frameworks in terms of inference speed. Apple has developed ``Core ML", a private ML framework to facilitate mobile deep learning implementation on iOS 11.\footnote{Core ML: Integrate machine learning models into your app, \url{https://developer.apple.com/documentation/coreml}} This lowers the entry barrier for developers wishing to deploy ML models on Apple equipment. Yao \emph{et al.} develop a deep learning framework called DeepSense dedicated to mobile sensing related data processing, which provides a general machine learning toolbox that accommodates a wide range of edge applications. It has moderate energy consumption and low latency, thus being amenable to deployment on smartphones.

\begin{table*}[htb]
\centering
\caption{Comparison of mobile deep learning platform.}
\label{tab:platform}
\begin{tabular}{|c|c|c|c|c|c|c|}
\hline
\textbf{Platform}   & \textbf{Developer} & \textbf{Mobile hardware supported} & \textbf{Speed}  & \textbf{Code size} & \textbf{Mobile compatibility} & \textbf{Open-sourced} \\ \hline
TensorFlow & Google    & CPU                       & Slow   & Medium    & Medium               & Yes        \\ \hline
Caffe      & Facebook  & CPU                       & Slow   & Large     & Medium               & Yes        \\ \hline
ncnn       & Tencent   & CPU                       & Medium & Small     & Good                 & Yes        \\ \hline
CoreML     & Apple     & CPU/GPU                   & Fast   & Small     & Only iOS 11+ supported  & No  \\ \hline
DeepSense     & Yao \emph{et al.}     & CPU                   & Medium   & Unknown     & Medium  & No  \\ \hline
\end{tabular}
\end{table*}

The techniques and toolboxes mentioned above make the deployment of deep learning practices in mobile network applications feasible. In what follows, we briefly introduce several representative deep learning architectures and discuss their applicability to mobile networking problems.

\section{Deep Learning: State-of-the-Art}\label{sec:model}
\begin{figure*}[h!]
\centering
\setcounter{subfigure}{0}
\subfigure[Structure of an MLP with 2 hidden layers (blue circles).]{
\label{fig:mlp} 
\includegraphics[width=3in]{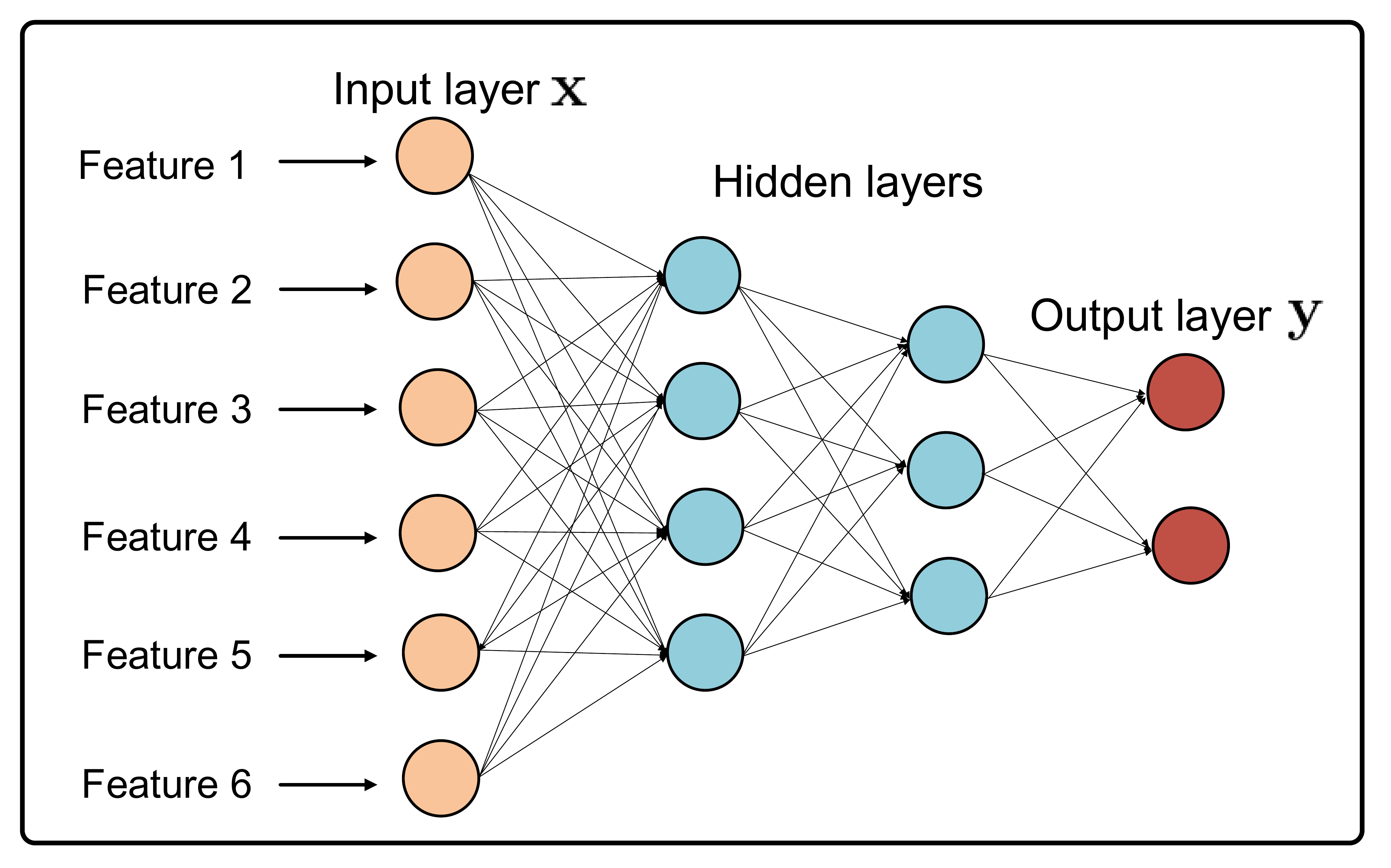}}
\subfigure[Graphical model and training process of an RBM. $\mathbf{v}$ and $\mathbf{h}$ denote visible and hidden variables, respectively.]{
\label{fig:rbm} 
\includegraphics[width=3in]{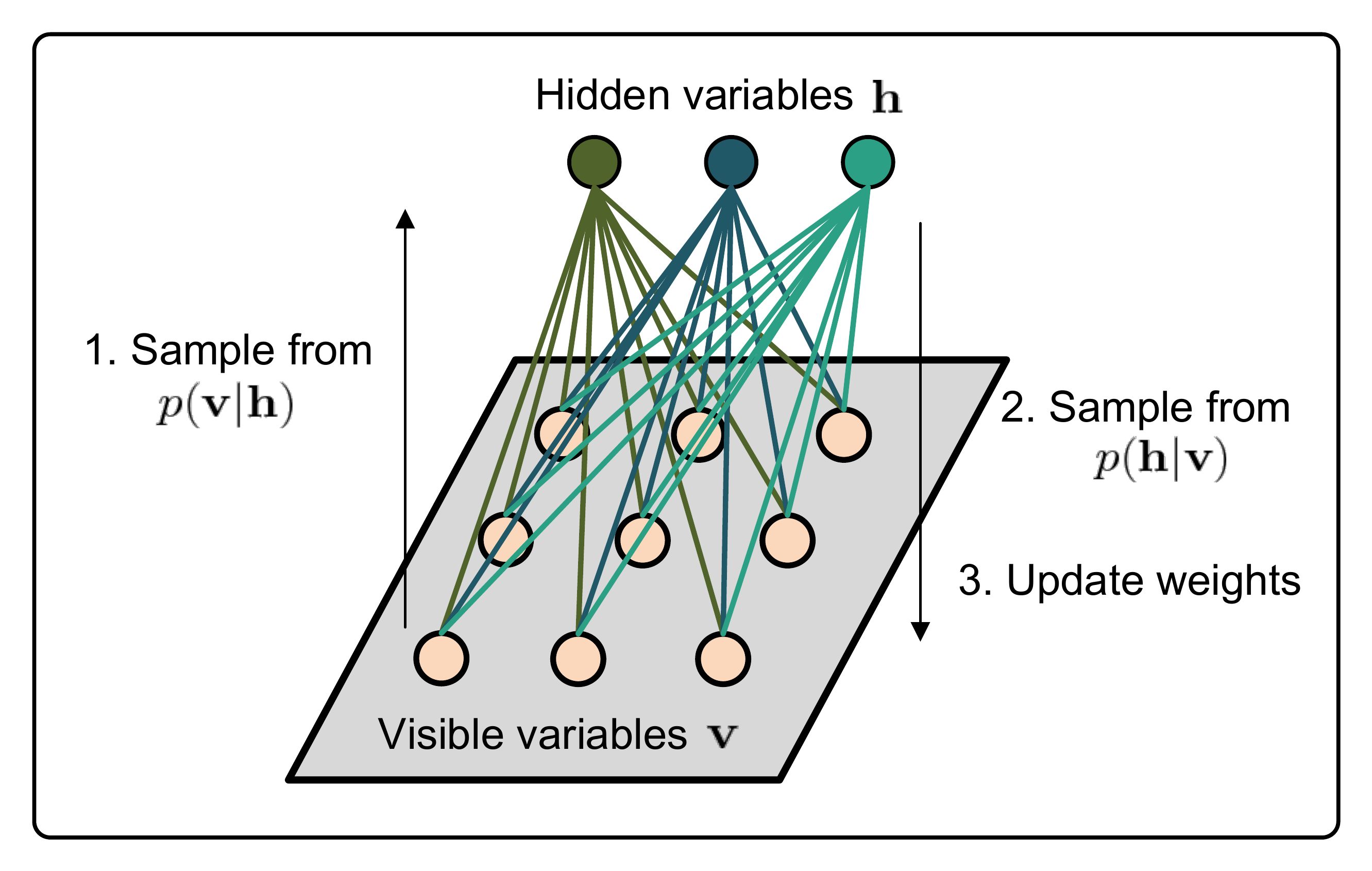}}
\subfigure[Operating principle of an auto-encoder, which seeks to reconstruct the input from the hidden layer.]{
\label{fig:auto} 
\includegraphics[width=3in]{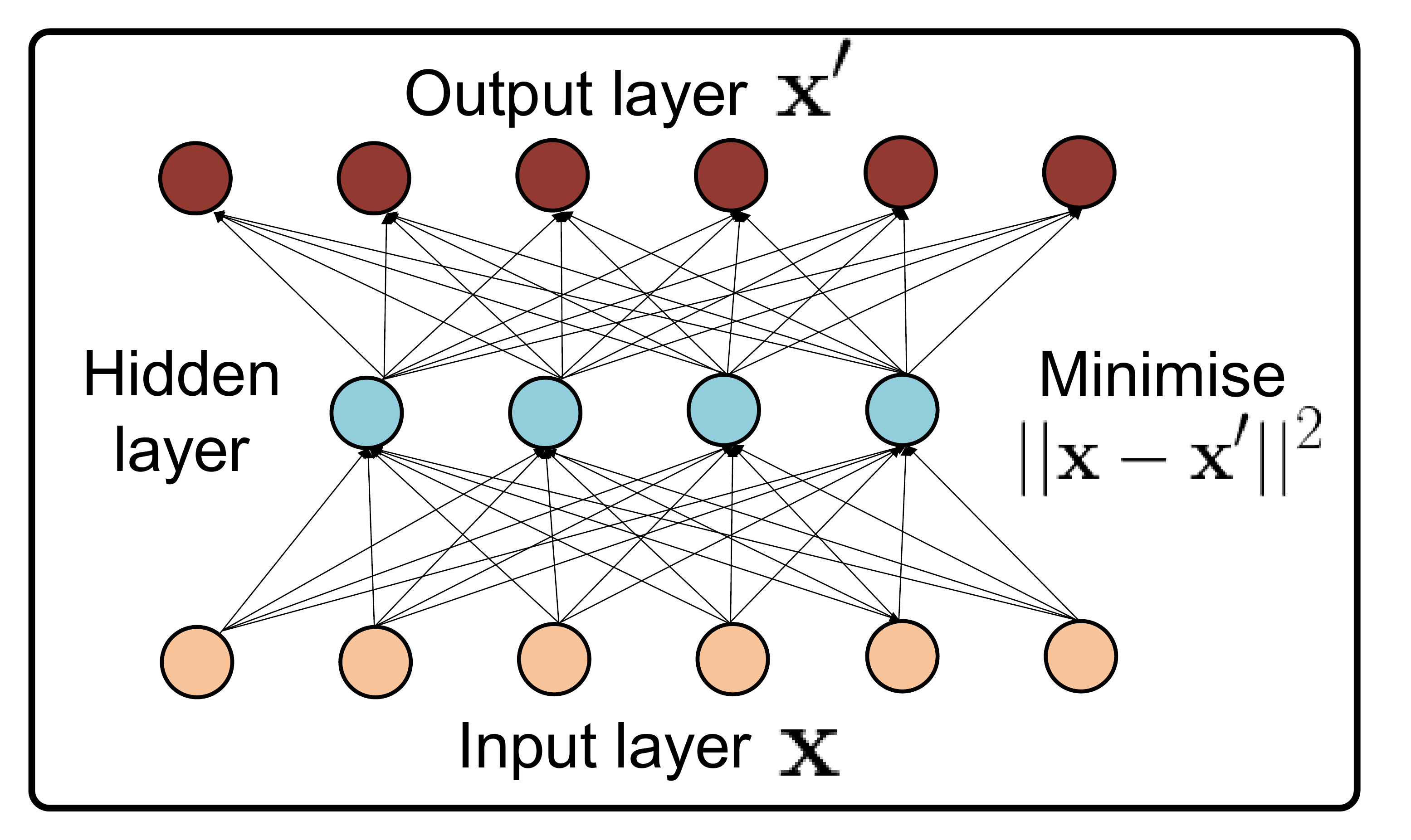}}
\subfigure[Operating principle of a convolutional layer. ]{
\label{fig:cnn} 
\includegraphics[width=3in]{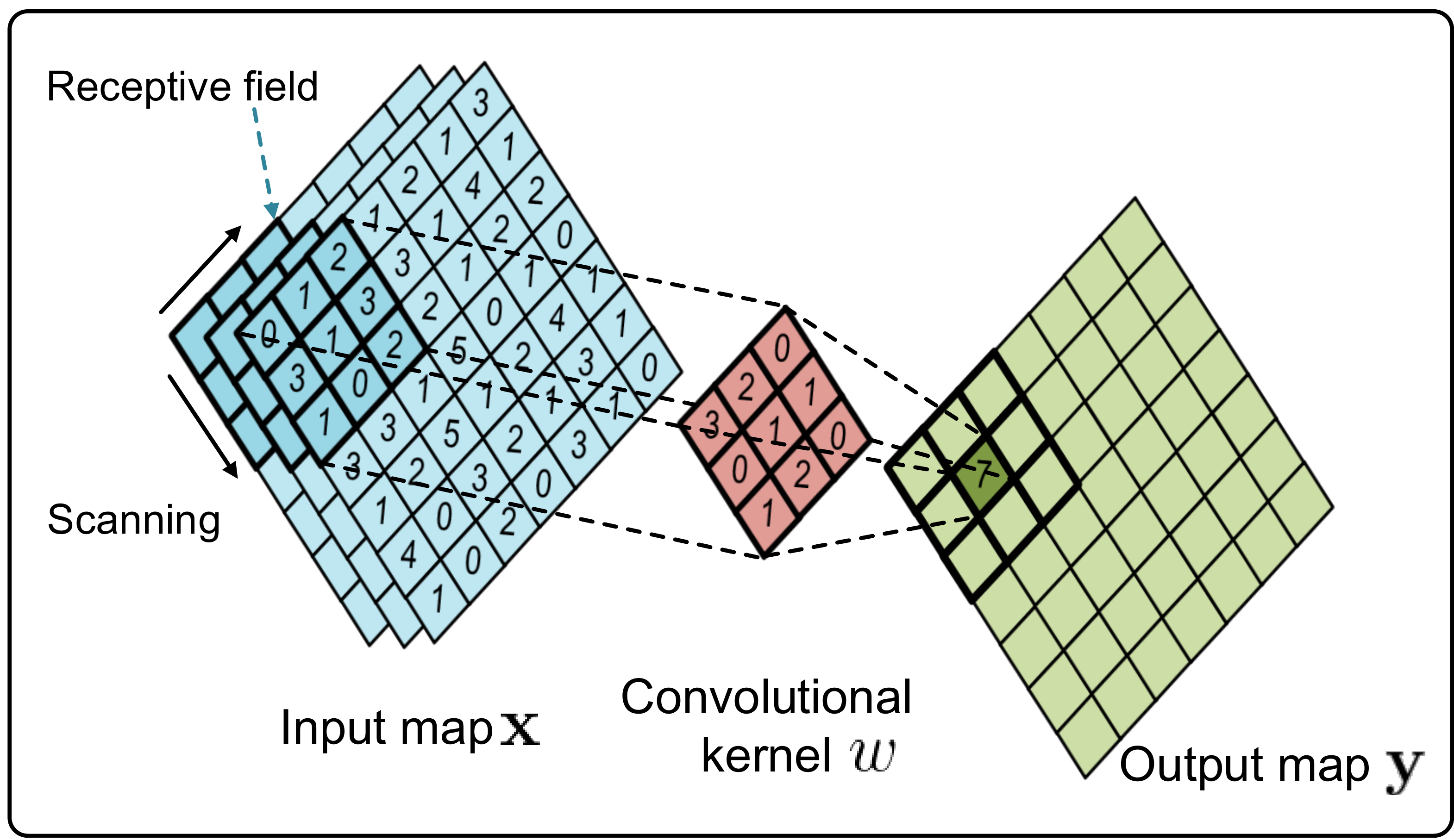}}
\subfigure[Recurrent layer -- $x_{1:t}$ is the input sequence, indexed by~time~$t$, $s_t$ denotes the state vector and $h_t$ the hidden outputs.]{
\label{fig:rnn} 
\includegraphics[width=3in]{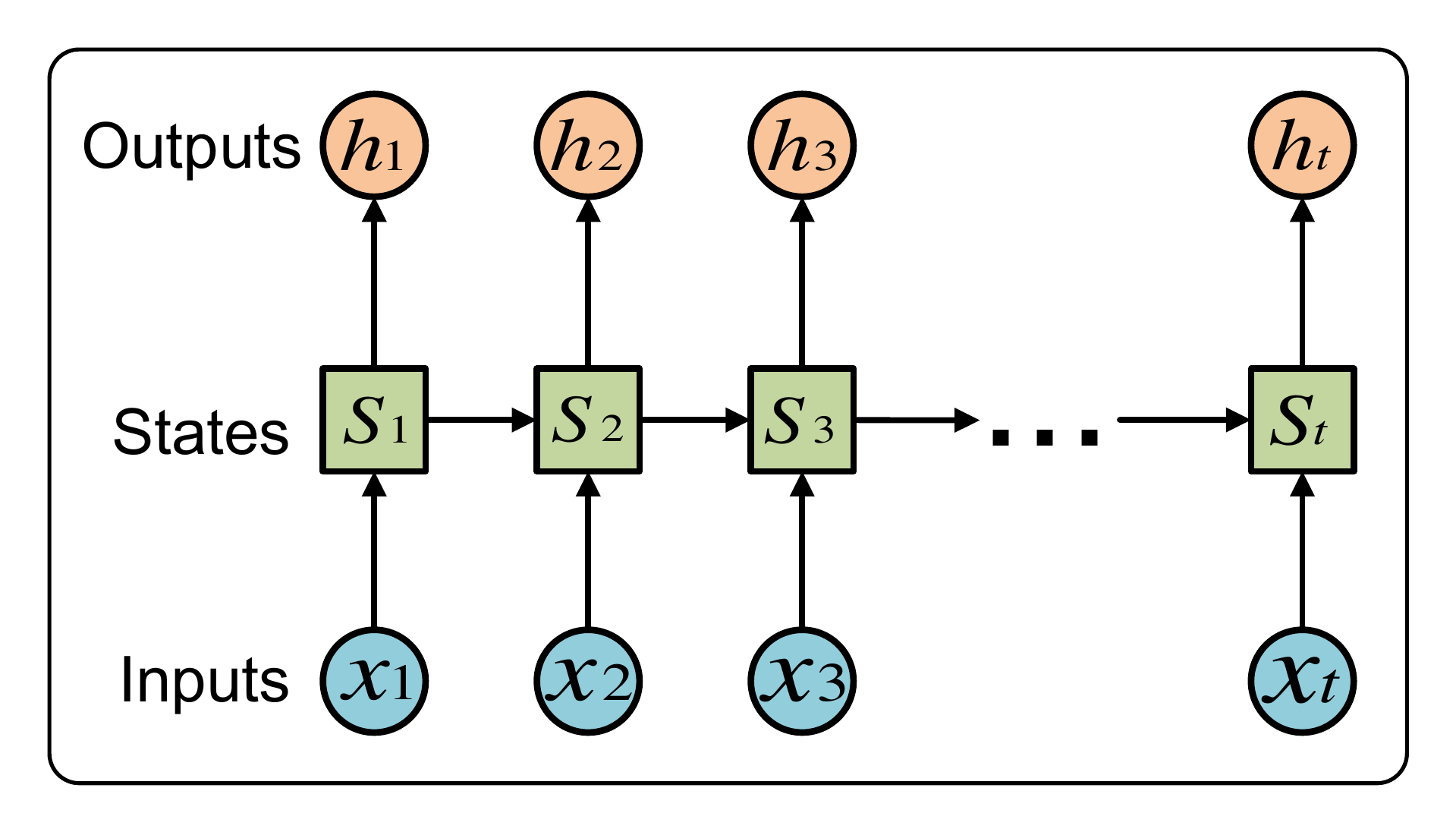}}
\subfigure[\rev{The inner structure of an LSTM layer.}]{
\label{fig:lstm} 
\includegraphics[width=3in]{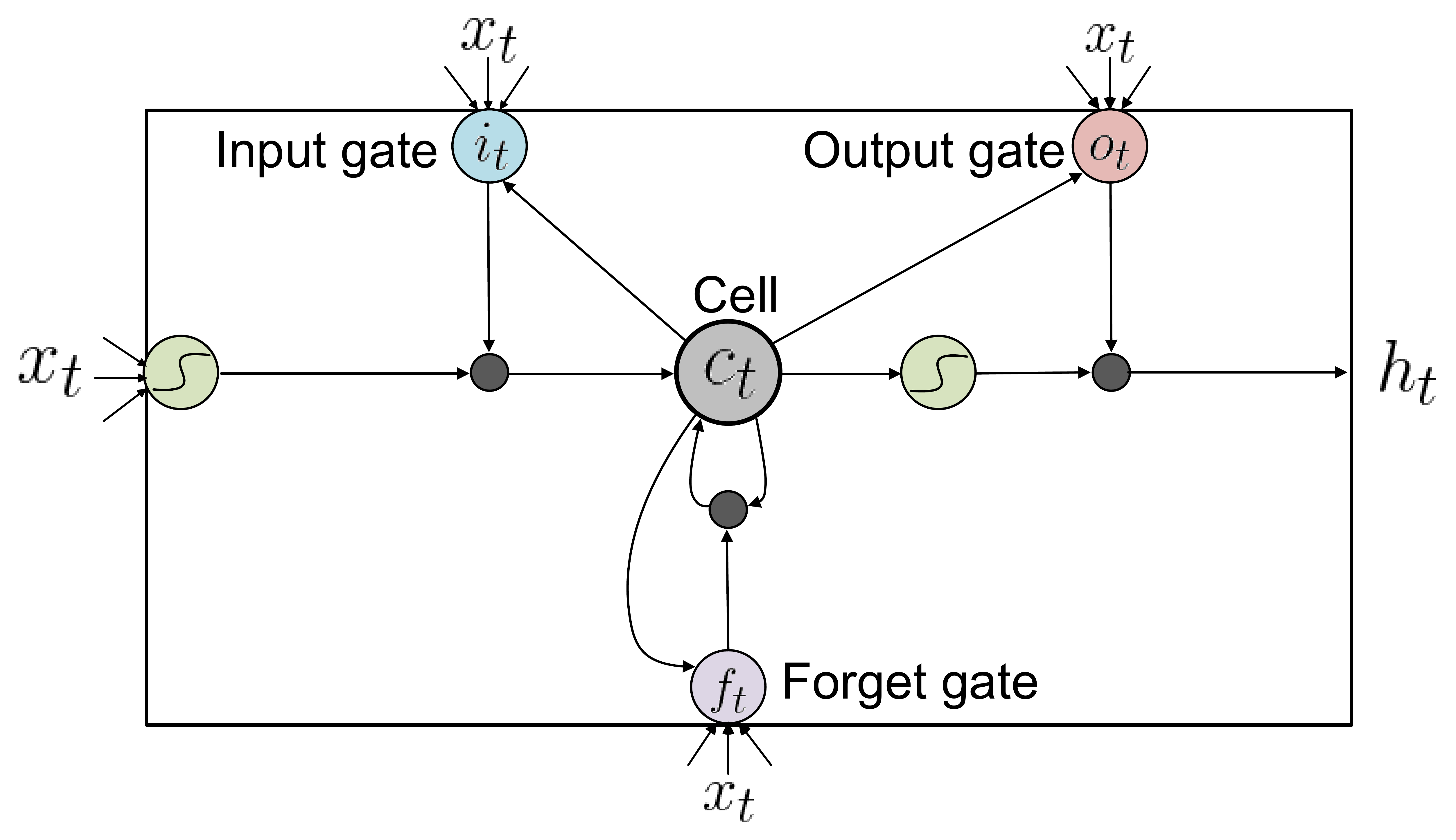}}
\subfigure[Underlying principle of a generative adversarial network (GAN).]{
\label{fig:gan}
\includegraphics[width=3in]{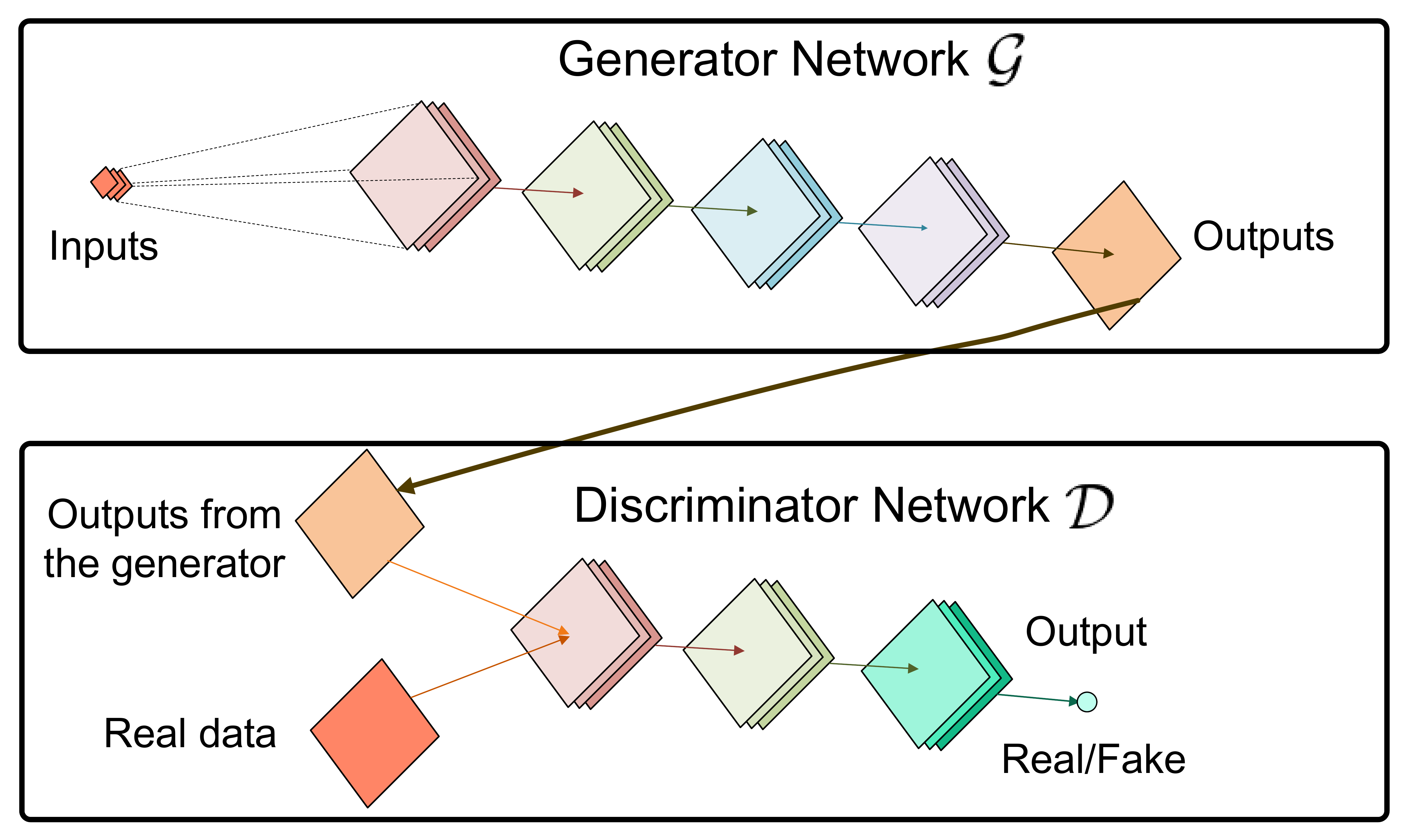}}
\subfigure[Typical deep reinforcement learning architecture. The agent is a neural network model that approximates the required function.]{
\label{fig:drl}
\includegraphics[width=3in]{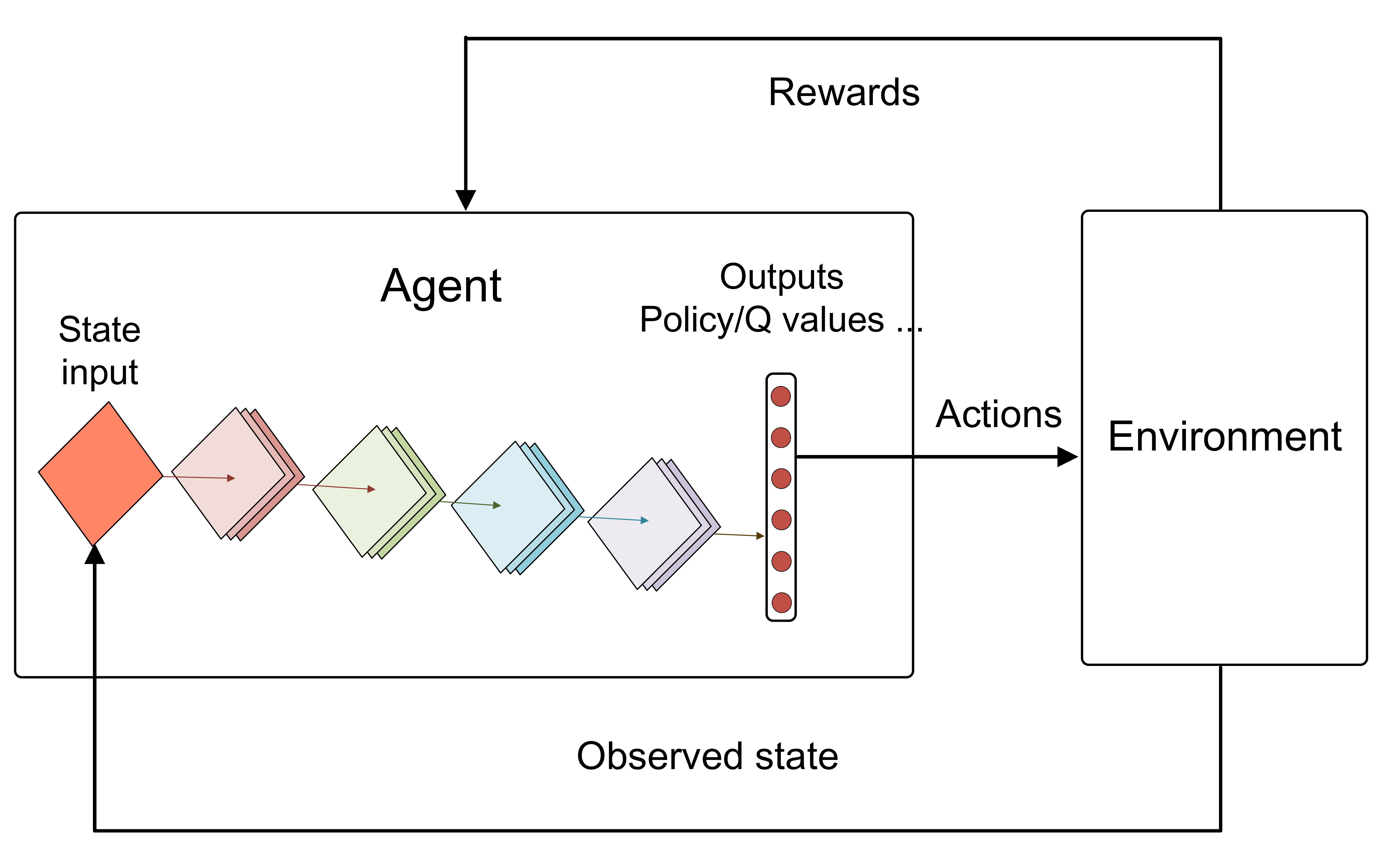}}
\caption{\label{Fig:deeplearning} Typical structure and operation principles of MLP, RBM, AE, CNN, RNN, LSTM, GAN, and DRL.}
\end{figure*}

Revisiting Fig. \ref{fig:venn}, machine learning methods can be naturally categorized into three classes, namely supervised learning, unsupervised learning, and reinforcement learning. Deep learning architectures have achieved remarkable performance in all these areas. In this section, we introduce the key principles underpinning several deep learning models and discuss their largely unexplored potential to solve mobile networking problems. \edit{Technical details of classical models are provided to readers who seek to obtain a deeper understanding of neural networks. The more experienced can continue reading with Sec.~\ref{sec:netapp}.} We illustrate and summarize the most salient architectures that we present in Fig. \ref{Fig:deeplearning} and Table~\ref{tab:model}, respectively.

\begin{table*}[h]
\centering
\caption{Summary of different deep learning architectures. GAN and DRL are shaded, since they are built upon other~models.}
\label{tab:model}
\begin{tabular}{|C{1cm}|C{1.8cm}|C{2.3cm}|C{2cm}|C{2.5cm}|C{2.5cm}|C{2.5cm}|}
\hline
\textbf{Model}     & \textbf{Learning scenarios} & \textbf{Example architectures} & \textbf{Suitable problems} & \textbf{Pros} & \textbf{Cons} & \textbf{Potential applications in mobile networks} \\ \hline
MLP       &    Supervised, unsupervised, reinforcement    &        ANN, AdaNet~\cite{cortes2017adanet}      &   Modeling data with simple correlations          &  Naive structure and straightforward to build     &   High complexity, modest performance and slow convergence   &          Modeling multi-attribute mobile data; auxiliary or component of other deep architectures        \\ \hline

RBM     &       Unsupervised            &          DBN \cite{hinton2006fast}, Convolutional DBN~\cite{lee2009convolutional}       &     Extracting robust representations     &    Can generate virtual samples     &   Difficult to train well   &             Learning representations from unlabeled mobile data; model weight initialization; network flow prediction                 \\ \hline

AE     &       Unsupervised            &  DAE \cite{vincent2010stacked}, VAE~\cite{kingma2014auto}        &     Learning  sparse and compact representations         &    Powerful and effective unsupervised learning    &   Expensive to pretrain with big data   &          model weight initialization; mobile data dimension reduction; mobile anomaly detection                               \\ \hline

CNN       &        Supervised, unsupervised, reinforcement            &      AlexNet \cite{krizhevsky2012imagenet}, ResNet \cite{he2016deep}, 3D-ConvNet \cite{ji20133d}, GoogLeNet \cite{szegedy2015going}, DenseNet \cite{huang2017densely}               &  Spatial data modeling                 & Weight sharing; affine invariance &   High computational cost; challenging to find optimal hyper-parameters; requires deep structures for complex tasks   &      Spatial mobile data analysis                              \\ \hline
RNN       &        Supervised, unsupervised, reinforcement            &      LSTM \cite{gers1999learning}, Attention based RNN \cite{sutskever2014sequence}, ConvLSTM \cite{xingjian2015convolutional}               &        Sequential data modeling          &   Expertise in capturing temporal dependencies  & High model complexity; gradient vanishing and exploding problems     &          Individual traffic flow analysis; network- wide (spatio-) temporal data modeling                                  \\ \hline \rowcolor{Gray}
GAN       &      Unsupervised             &          WGAN \cite{arjovsky2017wasserstein}, LS-GAN \cite{qi2017loss}, BigGAN \cite{brock2018large}         &   Data generation    &   Can produce lifelike artifacts from a target distribution  &  Training process is unstable (convergence difficult)   &   Virtual mobile data generation; assisting supervised learning tasks in network data analysis                                     \\ \hline \rowcolor{Gray}
DRL       &      Reinforcement        &         DQN \cite{mnih2015human}, Deep~Policy Gradient~\cite{silver2016mastering}, A3C~\cite{mnih2016asynchronous}, Rainbow~\cite{hessel2017rainbow}, DPPO \cite{schulman2017proximal}             &    Control problems with high- dimensional inputs              &  Ideal for high-dimensional environment  modeling   &  Slow in terms of convergence    &     Mobile network control and management.                                       \\ \hline
\end{tabular}
\end{table*}

\subsection{Multilayer Perceptron}
The Multilayer Perceptrons (MLPs) is the initial Artificial Neural Network (ANN) design, which consists of at least three layers of operations \cite{collobert2004links}. Units in each layer are densely connected, hence require to configure a substantial number of weights. We show an MLP with two hidden layers in Fig. \ref{fig:mlp}. \rev{Note that usually only MLPs containing more than one hidden layer are regarded as deep learning structures.} 

Given an input vector $\mathbf{x}$, a standard MLP layer performs the following operation:
\begin{equation}
\mathbf{y} = \sigma(W\cdot \mathbf{x} + b).
\end{equation}
Here $\mathbf{y}$ denotes the output of the layer, $W$ are the weights and $b$ the biases. $\sigma(\cdot)$ is an activation function, which aims at improving the non-linearity of the model. Commonly used activation function are the sigmoid,
\[
\textbf{sigmoid}(\mathbf{x}) = \frac{1}{1+e^{-\mathbf{x}}},
\]
the Rectified Linear Unit (ReLU) \cite{glorot2011deep},
\[
\textbf{ReLU}(\mathbf{x}) = \max(\mathbf{x}, 0),
\]
tanh,
\[
\textbf{tanh}(\mathbf{x}) = \frac{e^{\mathbf{x}}-e^{\mathbf{-x}}}{e^{\mathbf{x}}+e^{\mathbf{-x}}},
\]
and the Scaled Exponential Linear Units (SELUs) \cite{klambauer2017self}, 
$$\textbf{SELU}(\mathbf{x})=\lambda
\begin{cases}
\mathbf{x},& \text{if}\:  \mathbf{x}>0;\\
\alpha e^{\mathbf{x}}-\alpha, & \text{if}\:  \mathbf{x}\leq 0,
\end{cases}$$ 
where the parameters $\lambda=1.0507$ and $\alpha= 1.6733$  are frequently used. In addition, the softmax function is typically employed in the last layer when performing classification:
\[
\textbf{softmax}(\mathbf{x}_i) = \frac{e^{\mathbf{x}_i}}{\sum_{j=0}^k e^{\mathbf{x}_k}},
\]
where $k$ is the number of labels involved in classification. \rev{Until recently, \textbf{sigmoid} and \textbf{tanh} have been the activation functions most widely used. However, they suffer from a known gradient vanishing problem, which hinders gradient propagation through layers. Therefore these functions are increasingly more often replaced by ReLU or SELU. SELU enables to normalize the output of each layer, which dramatically accelerates the training convergence, and can be viewed as a replacement of Batch Normalization~\cite{ioffe2015batch}.}

The MLP can be employed for supervised, unsupervised, and even reinforcement learning purposes. Although this structure was the most popular neural network in the past, its popularity is decreasing because it entails high complexity (fully-connected structure), modest performance, and low convergence efficiency. MLPs are mostly used as a baseline or integrated into more complex architectures (e.g., the final layer in CNNs used for classification). Building an MLP is straightforward, and it can be employed, e.g., to assist with feature extraction in models built for specific objectives in mobile network applications. The advanced Adaptive learning of neural Network (AdaNet) enables MLPs to dynamically train their structures to adapt to the input~\cite{cortes2017adanet}. This new architecture can be potentially explored for analyzing continuously changing mobile environments.

\subsection{Boltzmann Machine}
Restricted Boltzmann Machines (RBMs) \cite{le2008representational} were originally designed for unsupervised learning purposes. They are essentially a type of energy-based undirected graphical models, and include a visible layer and a hidden layer, and where each unit can only \edit{assume binary values (i.e., 0 and 1).
The probabilities of these values are given by:
\begin{flalign*}
& P(h_j=1|v) = \frac{1}{1+e^{-\mathbf{W\cdot v + b_j}}}\\
& P(v_j=1|h) = \frac{1}{1+e^{-\mathbf{W^T\cdot h + a_j}}},
\end{flalign*}
where $h, v$ are the hidden and visible units respectively, and $\mathbf{W}$ are weights and $\mathbf{a}, \mathbf{b}$ are biases. The visible units are conditional independent to the hidden units, and \emph{vice versa}. 
} \rev{A typical structure of an RBM is shown in Fig.~\ref{fig:rbm}. In general, input data are assigned to visible units $v$. Hidden units $h$ are invisible and they fully connect to all $v$ through weights $W$, which is similar to a standard feed forward neural network. However, unlike in MLPs where only the input vector can affect the hidden units, with RBMs the state of $v$ can affect the state of $h$, and \emph{vice versa}.}

RBMs can be effectively trained using the contrastive divergence algorithm~\cite{hinton2002training} through multiple steps of Gibbs sampling~\cite{casella1992explaining}. We illustrate the structure and the training process of an RBM in Fig.~\ref{fig:rbm}. RBM-based models are usually employed to initialize the weights of a neural network in more recent applications. The pre-trained model can be subsequently fine-tuned for supervised learning purposes using a standard back-propagation algorithm. A stack of RBMs is called a Deep Belief Network (DBN)~\cite{hinton2006fast}, which performs layer-wise training and achieves superior performance as compared to MLPs in many applications, including time series forecasting~\cite{kuremoto2014forecast}, ratio matching~\cite{dauphin2013stochastic}, and speech recognition \cite{sainath2011making}. Such structures can be even extended to a convolutional architecture, to learn hierarchical spatial representations~\cite{lee2009convolutional}.

\subsection{Auto-Encoders}
Auto-Encoders (AEs) are also designed for unsupervised learning and attempt to copy inputs to outputs. The underlying principle of an AE is shown in Fig.~\ref{fig:auto}. AEs are frequently used to learn compact representation of data for dimension reduction \cite{bengio2009learning}. Extended versions can be further employed to initialize the weights of a deep architecture, e.g., the Denoising Auto-Encoder (DAE)~\cite{vincent2010stacked}), and generate virtual examples from a target data distribution, e.g. Variational Auto-Encoders (VAEs)~\cite{kingma2014auto}. 

\edit{
A VAE typically comprises two neural networks -- an encoder and a decoder. The input of the encoder is a data point $\mathbf{x}$ (e.g., images) and its functionality is to encode this input into a latent representation space $\mathbf{z}$. Let $f_\Theta(\mathbf{z}|\mathbf{x})$ be an encoder parameterized by $\Theta$ and $\mathbf{z}$ is sampled from a Gaussian distribution, the objective of the encoder is to output the mean and variance of the Gaussian distribution.}
\edit{
Similarly, denoting $g_\Omega(\mathbf{x}|\mathbf{z})$ the decoder parameterized by $\Omega$, this accepts the latent representation $\mathbf{z}$ as input, and outputs the parameter of the distribution of $\mathbf{x}$. The objective of the VAE is to minimize the reconstruction error of the data and the  Kullback-Leibler (KL) divergence between $p(\mathbf{z})$ and $f_\Theta(\mathbf{z}|\mathbf{x})$. Once trained, the VAE can generate new data point samples by \emph{(i)} drawing latent variables $z_i \sim p(\mathbf{z})$ and \emph{(ii)} drawing a new data point $x_i \sim p(\mathbf{x}\vert \mathbf{z})$.}

AEs can be employed to address network security problems, as several research papers confirm their effectiveness in detecting anomalies under different circumstances~\cite{sakurada2014anomaly, nicolau2016hybrid, thing2017ieee}, which we will further discuss in subsection~\ref{sec:security}.
The structures of RBMs and AEs are based upon MLPs, CNNs or RNNs. Their goals are similar, while their learning processes are different. Both  can be exploited to extract patterns from unlabeled mobile data, which may be subsequently employed for various supervised learning tasks, e.g., routing~\cite{mao2017routing}, mobile activity recognition~\cite{radu2016towards, radu2018multimodal}, periocular verification~\cite{raghavendra2016learning} and base station user number prediction~\cite{li2016wavelet}.

\subsection{Convolutional Neural Network}
Instead of employing full connections between layers, Convolutional Neural Networks (CNNs or ConvNets) employ a set of locally connected kernels (filters) to capture correlations between different data regions. Mathematically, for each location $\boldsymbol{p}_{y}$ of the output $\mathbf{y}$, the standard convolution performs the following operation:
\begin{equation}\label{eq:conv}
    \mathbf{y}(\boldsymbol{p}_y) = \sum_{\boldsymbol{p}_G \in \mathbb{G}} \mathbf{w}(\boldsymbol{p}_G)\cdot \mathbf{x}(\boldsymbol{p}_y + \boldsymbol{p}_G),
\end{equation}
where $\boldsymbol{p}_G$ denotes all positions in the receptive field $\mathbb{G}$ of the convolutional filter $W$, \rev{effectively representing the receptive range of each neuron to inputs in a convolutional layer}. Here the weights $W$ are shared across different locations of the input map. We illustrate the operation of one 2D convolutional layer in Fig.~\ref{fig:cnn}. \rev{Specifically, the inputs of a 2D CNN layer are multiple 2D matrices with different channels (e.g. the RGB representation of images). A convolutional layer employs multiple filters shared across different locations, to ``scan'' the inputs and produce output maps. In general, if the inputs and outputs have $M$ and $N$ filters respectively, the convolutional layer will require $M\times N$ filters to perform the convolution operation.}

CNNs improve traditional MLPs by leveraging three important ideas, \edit{namely, \emph{(i)} sparse interactions, \emph{(ii)} parameter sharing, and \emph{(iii)} equivariant representations~\cite{goodfellow2016deep}. This reduces the number} of model parameters significantly and maintains the affine invariance (i.e., \rev{recognition results} are robust to the affine transformation of objects). 
\edit{
Specifically, The sparse interactions imply that the weight kernel has smaller size than the input. It performs moving filtering to produce outputs (with roughly the same size as the inputs) for the current layer. Parameter sharing refers to employing the same kernel to scan the whole input map. This significantly reduces the number of parameters needed, which mitigates the risk of over-fitting. Equivariant representations indicate that convolution operations are invariant in terms of translation, scale, and shape. This is particularly useful for image processing, since essential features may show up at different locations in the image, with various affine patterns. }

Owing to the properties mentioned above, CNNs achieve remarkable performance in imaging applications. Krizhevsky \emph{et al.} \cite{krizhevsky2012imagenet} exploit a CNN to classify images on the ImageNet dataset \cite{ILSVRC15}. Their method reduces the top-5 error by 39.7\% and revolutionizes the imaging classification field. GoogLeNet \cite{szegedy2015going} and ResNet \cite{he2016deep} significantly increase the depth of CNN structures, and propose inception and residual learning techniques to address problems such as over-fitting and gradient vanishing introduced by ``depth''. Their structure is further improved by the Dense Convolutional Network (DenseNet)~\cite{huang2017densely}, which reuses feature maps from each layer, thereby achieving significant accuracy improvements over other CNN based models, while requiring fewer layers. CNNs have also been extended to video applications. Ji \emph{et al.} propose 3D convolutional neural networks for video activity recognition~\cite{ji20133d}, demonstrating superior accuracy as compared to 2D CNN. More recent research focuses on learning the shape of convolutional kernels \cite{active2017yunho, dai2017deformable, Zhu2018morel}. These dynamic architectures allow to automatically focus on important regions in input maps. Such properties are particularly important in analyzing large-scale mobile environments exhibiting clustering behaviors (e.g., surge of mobile traffic associated with a popular event). 

Given the high similarity between image and spatial mobile data (e.g., mobile traffic snapshots, users' mobility, etc.), CNN-based models have huge potential for network-wide mobile data analysis. This is a promising future direction that we further discuss in Sec.~\ref{sec:future}.

\subsection{Recurrent Neural Network}
Recurrent Neural Networks (RNNs) are designed for modeling sequential data, \rev{where sequential correlations exist between samples}. At each time step, they produce output via recurrent connections between hidden units \cite{goodfellow2016deep}, as shown in Fig.~\ref{fig:rnn}. Given a sequence of inputs $\mathbf{x} = \{x_1, x_2,\cdots, x_T\}$, a \rev{standard RNN} performs the following operations:
\begin{flalign*}
& s_t = \sigma_s(W_x x_t + W_s s_{t-1} + b_s)\\
& h_t = \sigma_h(W_h s_t + b_h),
\end{flalign*}
where \rev{$s_t$ represents the state of the network at time $t$ and it constructs a memory unit for the network. Its values are computed by a function of the input $x_t$ and previous state $s_{t-1}$. $h_t$ is the output of the network at time $t$. In natural language processing applications, this usually represents a language vector and becomes the input at $t+1$ after being processed by an embedding layer. The weights $W_x, W_h$ and biases $b_s, b_h$ are shared across different temporal locations. This reduces the model complexity and the degree of over-fitting.}

The RNN is trained via a Backpropagation Through Time (BPTT) algorithm. However, gradient vanishing and exploding problems are frequently reported in traditional RNNs, which make them particularly hard to train~\cite{bengio1994learning}. The Long Short-Term Memory (LSTM) mitigates these issues by introducing a set of ``gates''~\cite{gers1999learning}, which has been proven successful in many applications (e.g., speech recognition~\cite{graves2013hybrid}, text categorization~\cite{johnson2016supervised}, and wearable activity recognition~\cite{ordonez2016deep}). \edit{A standard LSTM performs the following operations:
\begin{flalign*}
&i_t = \sigma(W_{xi}X_t + W_{hi}H_{t-1}+W_{ci}\odot C_{t-1} + b_i),\\
&f_t = \sigma(W_{xf}X_t + W_{hf}H_{t-1}+W_{cf}\odot C_{t-1}+b_f),\\
&C_t = f_t\odot C_{t-1} + i_t \odot \tanh(W_{xc}X_t + W_{hc}H_{t-1}+b_c),\\
&o_t = \sigma(W_{xo}X_t + W_{ho}H_{t-1}+W_{co} \odot C_t +b_o),\\
&H_t = o_t\odot \tanh(C_t).
\end{flalign*}
Here, `$\odot$' denotes the Hadamard product, $C_t$ denotes the cell outputs,  $H_t$ are the hidden states,  $i_t$,  $f_t$, and  $o_t$ are input gates, forget gates, and output gates, respectively. These gates mitigate the gradient issues and significantly improve the RNN. We illustrated the structure of an LSTM in Fig. \ref{fig:lstm}.}

Sutskever \emph{et al.} introduce attention mechanisms to RNNs, which achieves outstanding accuracy in tokenized predictions~\cite{sutskever2014sequence}. Shi \emph{et al.} substitute the dense matrix multiplication in LSTMs with convolution operations, designing a Convolutional Long Short-Term Memory (ConvLSTM)~\cite{xingjian2015convolutional}. Their proposal reduces the complexity of traditional LSTM and demonstrates significantly lower prediction errors in precipitation \mbox{nowcasting} \edit{(i.e., forecasting the volume of precipitation).}

Mobile networks produce massive sequential data from various sources, such as data traffic flows, and the evolution of mobile network subscribers' trajectories and application latencies. Exploring the RNN family is promising to enhance the analysis of time series data in mobile networks.  

\subsection{Generative Adversarial Network}
\begin{algorithm}[t]
  \caption{\rev{Typical GAN training algorithm. \label{alg:gan}}}
  \label{tra_gan}
  \begin{algorithmic}[1]
  \color{black}
    \Inputs{Batch size $m$. \\
    The number of steps for the discriminator $K$. \\
    Learning rate $\lambda$ and an optimizer Opt($\cdot$)\\
    Noise vector $z\sim p_g(z)$. \\
    Target data set $x\sim p_{data}(x)$.}
    \Initialize{Generative and discriminative models, $\mathcal{G}$~and $\mathcal{D}$,  parameterized by $\Theta_\mathcal{G}$ and $\Theta_\mathcal{D}$.}
    \While{$\Theta_\mathcal{G}$ and $\Theta_\mathcal{D}$ have not converged}\label{ln:bigloop}
    \For{$k = 1$ to $K$} \label{ln:loopd}
        	\State{Sample $m$-element noise vector $\{z^{(1)},\cdots, z^{(m)}\}$ } \Statex{\hspace*{3em}from the noise prior $p_g(z)$ }\label{ln:sampling1}
        	\State{Sample $m$ data points $\{x^{(1)},\cdots, x^{(m)}\}$ from the} \Statex{\hspace*{3em}target data distribution $p_{data}(x)$ }\label{ln:sampling1}
               \State{$g_\mathcal{D} \leftarrow \Delta_{\Theta_\mathcal{D}}[\frac{1}{m}\sum_{i=1}^m \log \mathcal{D}(x^{(i)}) +$}\label{ln:grad1}
               \Statex{\hspace*{6em}$+ \frac{1}{m}\sum_{i=1}^m \log (1-\mathcal{D}(\mathcal{G}(z^{(i)})))]$.}               
               \State{$\Theta_\mathcal{D}\leftarrow \Theta_\mathcal{D} + \lambda \cdot \text{Opt}(\Theta_\mathcal{D}, g_\mathcal{D})$.} \label{ln:adam1}
	\EndFor
    	\State{Sample $m$-element noise vector $\{z^{(1)},\cdots, z^{(m)}\}$ from} 
    	\Statex{\hspace*{1em} the noise prior $p_g(z)$ }
           \State{$g_\mathcal{G} \leftarrow \frac{1}{m}\sum_{i=1}^m \log (1 - \mathcal{D}(\mathcal{G}(z^{(i)})))$}
           \State{$\Theta_\mathcal{G}\leftarrow \Theta_\mathcal{G} - \lambda \cdot \text{Opt}(\Theta_\mathcal{G}, g_\mathcal{G})$.}\label{ln:adam2}
    \EndWhile
    \end{algorithmic}
\end{algorithm}
The Generative Adversarial Network (GAN) is a framework that trains generative models using the following adversarial process. It simultaneously trains two models: a generative one $\mathcal{G}$ that seeks to approximate the target data distribution from training data, and a discriminative model $\mathcal{D}$ that estimates the probability that a sample comes from the real training data rather than the output of $\mathcal{G}$~\cite{goodfellow2014generative}.  Both of $\mathcal{G}$ and  $\mathcal{D}$ are normally neural networks. The training procedure for $\mathcal{G}$ aims to maximize the probability of $\mathcal{D}$ making a mistake. 
\edit{The overall objective is solving the following minimax problem~\cite{goodfellow2014generative}:
\begin{equation*}
\label{eq:minmax}
\begin{aligned}
\min\limits_{\mathcal{G}}\max\limits_{\mathcal{D}} \mathbb{E}_{x\sim P_r(x)}[\log \mathcal{D}(x)]+\mathbb{E}_{z\sim P_n(z)}[\log(1- \mathcal{D}(\mathcal{G}(z)))].
\end{aligned}
\end{equation*}}\rev{Algorithm~\ref{alg:gan} shows the typical routine used to train a simple GAN.} Both the generators and the discriminator are trained iteratively while fixing the other one. Finally $\mathcal{G}$ can produce data close to a target distribution (the same with training examples), if the model converges. We show the overall structure of a GAN in Fig. \ref{fig:gan}. \rev{In practice, the generator $\mathcal{G}$ takes a noise vector $z$ as input, and generates an output $\mathcal{G}(z)$ that follows the target distribution. $\mathcal{D}$ will try to discriminate whether $\mathcal{G}(z)$ is a real sample or an artifact \cite{goodfellow2016nips}. This effectively constructs a dynamic game, for which a Nash Equilibrium is reached if both $\mathcal{G}$ and $\mathcal{D}$ become optimal, and $\mathcal{G}$ can produce lifelike data that $\mathcal{D}$ can no longer discriminate, i.e. $\mathcal{D}(\mathcal{G}(z))=0.5, \forall z$.}

The training process of traditional GANs is highly sensitive to model structures, learning rates, and other hyper-parameters. Researchers are usually required to employ numerous ad hoc `tricks' to achieve convergence and improve the fidelity of data generated. There exist several solutions for mitigating this problem, e.g., Wasserstein Generative Adversarial Network (WGAN) \cite{arjovsky2017wasserstein}, Loss-Sensitive Generative Adversarial Network (LS-GAN) \cite{qi2017loss} and BigGAN \cite{brock2018large}, but research on the theory of GANs remains shallow.  Recent work confirms that GANs can promote the performance of some supervised tasks (e.g., super-resolution \cite{ledig2017photo}, object detection \cite{li2017perceptual}, and face completion \cite{li2017generative}) by minimizing the divergence between inferred and real data distributions. Exploiting the unsupervised learning abilities of GANs is promising in terms of generating \edit{synthetic mobile data} for simulations, or assisting specific supervised tasks in mobile network applications. This becomes more important in tasks where appropriate datasets are lacking, given that operators are generally reluctant to share their network data.

\subsection{Deep Reinforcement Learning}
Deep Reinforcement Learning (DRL) refers to a set of methods that approximate value functions (deep Q learning) or policy functions (policy gradient method) through deep neural networks. An agent (neural network) continuously interacts with an environment and receives reward signals as feedback. The agent selects an action at each step, which will change the state of the environment. The training goal of the neural network is to optimize its parameters, such that it can select actions that potentially lead to the best future return. We illustrate this principle in Fig. \ref{fig:drl}. DRL is well-suited to problems that have a huge number of possible states (i.e., environments are high-dimensional). Representative DRL methods include Deep Q-Networks (DQNs)~\cite{mnih2015human}, deep policy gradient methods~\cite{silver2016mastering}, Asynchronous Advantage Actor-Critic~\cite{mnih2016asynchronous}, {Rainbow \cite{hessel2017rainbow} and Distributed Proximal Policy Optimization (DPPO) \cite{schulman2017proximal}}. These perform remarkably in AI gaming (e.g., Gym\footnote{Gym is a toolkit for developing and comparing reinforcement learning algorithms. It supports teaching agents everything from walking to playing games like Pong or Pinball. \rev{In combination with the NS3 simulator Gym becomes applicable to networking research. \cite{gawlowicz2018ns3}} \url{https://gym.openai.com/}}), robotics, and autonomous driving~\cite{gu2016continuous, moravvcik2017deepstack, levine2016learning, sallab2017deep}, and have made inspiring deep learning breakthroughs recently. 

\edit{In particular, the DQN~\cite{mnih2015human} is first proposed by DeepMind to play Atari video games. However, traditional DQN requires several important adjustments to work well. The A3C \cite{mnih2016asynchronous} employs an actor-critic mechanism, where the actor selects the action given the state of the environment, and the critic estimates the value given the state and the action, then delivers feedback to the actor. The A3C deploys different actors and critics on different threads of a CPU to break the dependency of data. This significantly improves training convergence, enabling fast training of DRL agents on CPUs. Rainbow~\cite{hessel2017rainbow} combines different variants of DQNs, and discovers that these are complementary to some extent. This insight improved performance in many Atari games. To solve the step size problem in  policy gradients methods, Schulman \emph{et al.} propose a Distributed Proximal Policy Optimization (DPPO) method to constrain the update step of new policies, and implement this on multi-threaded CPUs in a distributed manner~\cite{schulman2017proximal}. Based on this method, an agent developed by OpenAI defeated a human expert in Dota2 team in a 5v5 match.\footnote{\edit{Dota2 is a popular multiplayer online battle arena video game.}}}

Many mobile networking problems can be formulated as Markov Decision Processes (MDPs), where reinforcement learning can play an important role (e.g., base station on-off switching strategies~\cite{li2014tact}, routing~\cite{al2015application}, and adaptive tracking control~\cite{liu2015reinforcement}). Some of these problems nevertheless involve high-dimensional inputs, which limits the applicability of traditional reinforcement learning algorithms. DRL techniques broaden the ability of traditional reinforcement learning algorithms to handle high dimensionality, in scenarios previously considered intractable. Employing DRL is thus promising to address network management and control problems under complex, changeable, and heterogeneous mobile environments. We further discuss this potential in Sec.~\ref{sec:future}.

\section{Deep Learning Driven Mobile and Wireless Networks}\label{sec:netapp}

Deep learning has a wide range of applications in mobile and wireless networks. 
In what follows, we present the most important research contributions across different mobile networking areas and compare their design and principles. \edit{In particular, we first discuss a key prerequisite, that of mobile big data, then organize the review of relevant works into nine subsections, focusing on specific domains where deep learning has made advances. Specifically, 
\begin{enumerate}
\item \textbf{Deep Learning Driven Network-Level Mobile Data Analysis} focuses on deep learning applications built on mobile big data collected within the network, including network prediction, traffic classification, and Call Detail Record (CDR) mining.
\item \textbf{Deep Learning Driven App-Level Mobile Data Analysis} shifts the attention towards mobile data analytics on edge devices.
\item \rev{\textbf{Deep Learning Driven User Mobility Analysis} sheds light on the benefits of employing deep neural networks to understand the movement patterns of mobile users, either at group or individual levels.}
\item \rev{\textbf{Deep Learning Driven User Localization} reviews literature that employ deep neural networks to localize users in indoor or outdoor environments, based on different signals received from mobile devices or wireless channels.}
\item \textbf{Deep Learning Driven Wireless Sensor Networks} discusses important work on deep learning applications in WSNs \rev{from four different perspectives, namely centralized vs. decentralized sensing, WSN data analysis, WSN localization and other applications. }
\item \textbf{Deep Learning Driven Network Control} investigate the usage of deep reinforcement learning and deep imitation learning on network optimization, routing, scheduling, resource allocation, and radio control.
\item \textbf{Deep Learning Driven Network Security} presents work that leverages deep learning to improve  network security, which we cluster by focus as infrastructure, software, and privacy related.
\item \textbf{Deep Learning Driven Signal Processing} scrutinizes physical layer aspects that benefit from deep learning and reviews relevant work on signal processing.
\item \textbf{Emerging Deep Learning Driven Mobile Network Application} warps up this section, presenting other interesting deep learning applications in mobile networking.
\end{enumerate} 
For each domain, we summarize work broadly in tabular form, providing readers with a general picture of individual topics. Most important works in each domain are discussed in more details in text. Lessons learned are also discussed at the end of each subsection. We give a diagramatic view of the topics dealt with by the literature reviewed in this section in Fig.~\ref{fig:bubble}.
}
\begin{figure}[t]
\begin{center}
\smartdiagramset{
    distance text center bubble=0.15cm,
    bubble center node size=3cm,
    bubble node size=2cm,
    distance center/other bubbles=1.2cm,
    bubble center node font=\normalsize,
    bubble node font=\scriptsize,
    bubble center node color=bittersweet,
    module y sep=13,
    set color list = {saffron, mediumaquamarine, salmonpink, limegreen, persianorange, violet, cyan, darkgray}
}%
\smartdiagram[bubble diagram]{%
    \textbf{Deep Learning Driven} \\ \textbf{Mobile and Wireless} \\ \textbf{Networks},
    \textbf{Network-Level Mobile} \\\textbf{Data Analysis} \\\cite{pierucci2016neural, gwon2014inferring, nie2017network, moyo2015generalization, wangspatiotemporal, zhang2017long, chaoyun2017zipnet, huang2017study, zhang2018citywide, chen2018deep0, navabi2018predicting, wang2015applications, wang2017end, lotfollahi2017deep, wang2017malware, liang2016mercury, felbo2016using, chen2017comprehensive, lin2017deep, xu2017large, meng2018qoe, wang2018spatio, fang2018mobile, luo2018channel, 8553650, feng2018deeptp, zhu2018deep},
    \textbf{App-Level Mobile Data} \\\textbf{Analysis}\\\cite{liu2016poster, sicong2017ubiear, jindal2016integrating, kim2016deep, sathyanarayana2016sleep, li2017personal, hosseini2017deep, stamate2017deep, quisel2017collecting, khan2017deep, li2016deepcham, tobias2016convolutional, pouladzadeh2017mobile, tanno2016deepfoodcam, kuhad2015using, teng2016facial, rao2017mobile,  zeng2014convolutional, almaslukh2017effective, li2016deep22, bhattacharya2016smart, antoniou2016general, ordonez2016deep, wang2016interacting, gao2016ihear, zhu2015using, sundsoy2016deep, chen2015deep, ha2016convolutional, edel2016binarized, xue2018appdna, liu2018finding, okita2017recognition, alsheikh2016mobile}\\ \cite{mittal2016spotgarbage, seidenari2017deep, zeng2017mobiledeeppill, zoudeepsense, zeng2017mobile, radu2016towards, wang2015phasefi, wang2016csi, feng2018evaluation, cao2017deepmood, ran2018deepdecision, siri, mcgraw2016personalized, prabhavalkar2016compression, yoshioka2015ntt, ruan2016speech, georgiev2017low, ignatov2017dslr, lu2017demo, lee2016reducing, vu2016transportation, fang2017learning, zhao2018rf, katevas2017practical, yao2017deepsense, lane2015can, ohara2017detecting, liu2017deep},
    \textbf{Mobility Analysis} \\\cite{ouyang2016deepspace, yang2017neural, song2016deeptransport, zhang2017deep123, lin2017deep, subramanian2014implementation, ezema2017artificial, shao2018depedo, yayeh2018mobility, chen2016learning2, song2017deepmob, yao2017trajectory, liu2018urban, wickramasuriya2017base, tkavcik2016neural, kim2018method, jiang2018deepurbanmomentum, wang2018deep1231, jiang2018deep, feng2018deepmove},
    \textbf{User Localization} \\\cite{wang2015deepfi, wang2015phasefi, wang2016csi, wang2017cifi, wang2017biloc, nowicki2017low, zhang2016device}\\\cite{wang2017device, mohammadi2017semi, anzum2018zone, wang2018deepml, kumar2016indoor, zhengj2016mobile, vieira2017deep, zhang2016deep3, wang2017csi, chen2017confi, shokry2018deeploc, zhou2018device, zhang2017deeppositioning, adege2018applying, ibrahim2018cnn, niitsoo2018convolutional, wang2018deep6, xiao20173, Hsu:2017:ZIS:3139486.3130924, guan2017high},
    \textbf{Wireless Sensor Networks}\\\cite{chuang2014effective, bernas2015fully, payal2015analysis, dong2017range, yan2016real, wang2017temperature, lee2017deep222, li2016adaptive, khorasani2017energy, li2015distributed, luo2018distributed, kumar2019machine, heydari2017reduce, phoemphon2018hybrid, banihashemian2018new, sun2017wnn, kang2018novel, mehmood2017eldc, alsheikh2016rate, kumar2019machine, el2016robust, wang2017deep3, jia2018continuous},
    \textbf{Network Control}\\ \cite{liu2017deep222, subramanian2016poster, he2017optimization, he2017deep3, mismar2017deep, wang2018handover, chen2018heterogeneous, chen2018auto, lee2017classification, yang2017neural, mao2017routing, tang2017removing, zhang2017energy, atallah2017deep} \\\cite{chinchali2018, atallah2017deep, wei2018joint, sun2017learning, xu2017deep3, ferreira2017multi, ye2018deep12, challita2018proactive, naparstek2017deep, o2016deep, wijaya2015intercell, rutagemwa2018dynamic, wijaya2016neural, mao2017neural, oda2017design, oda2017performance, kim2017load, challita2018deep, luo2018online, yu2018deep2, xu2018experience, liu2018deepnap, zhao2018deep, li2018deep, mennes2018neural, 8553651, zhu2018deep, mao2017tensor, geyer2018learning, luong2018joint, li2017intelligent2, lee2018deep, liu2018energy, he2017software, liu2018anti, pham2018deep, ferreira2016multi},
    \textbf{Network Security}\\ \cite{yousefi2017autoencoder, thing2017ieee, aminanto2016detecting, feng2016anomaly, khan2016distributed, diro2017distributed, saied2016detection, lopez2017conditional, hamedani2018reservoir, luo2018distributed, das2018deep, jiang2018virtual, yuan2014droid, yuan2016droiddetector, su2016deep, hou2016deep4maldroid, martinelli2017evaluating, nguyen2018cyberattack}\\ \cite{mclaughlin2017deep, chen2017deep123, wang2017malware, oulehla2016detection, torres2016analysis, eslahi2016mobile, alauthaman2016p2p, shokri2015privacy, aono2017privacy, ossia2017hybrid, abadi2016deep, osia2017privacy, servia2017personal, hitaj2017deep, abadi2016deep, greydanus2017learning, maghrebi2016breaking, liu2018genpass, ning2018deepmag},
    \textbf{Signal Processing}\\ \cite{wijaya2015intercell, wijaya2016neural, o2017deep, borgerding2017amp, fujihashi2018nonlinear, rajendran2017distributed, west2017deep, o2016radio, gante2018beamformed, alkhateeb2018deep}\\
    \cite{vieira2017deep, neumann2017deep, samuel2017deep, yan2017signal, timothy2017introduction, jagannath2018artificial, o2016end, o2016learning, ye2018power, liang2018exploiting, lyu2018performance, dorner2017deep, liao2018rayleigh, huang2018deep, huang2018fully},
\textbf{Emerging Applications}\\ \cite{gonzalez2017network, kaminski2017neural, xiao2017secure, luong2018optimal, gulati2018deep}
}%
\end{center}
\caption{\edit{Classification of the literature reviewed in Sec.~\ref{sec:netapp}. }\label{fig:bubble}}
\end{figure}

\subsection{Mobile Big Data as a Prerequisite}
\begin{figure*}[htb]
\begin{center}
\includegraphics[width=1\textwidth]{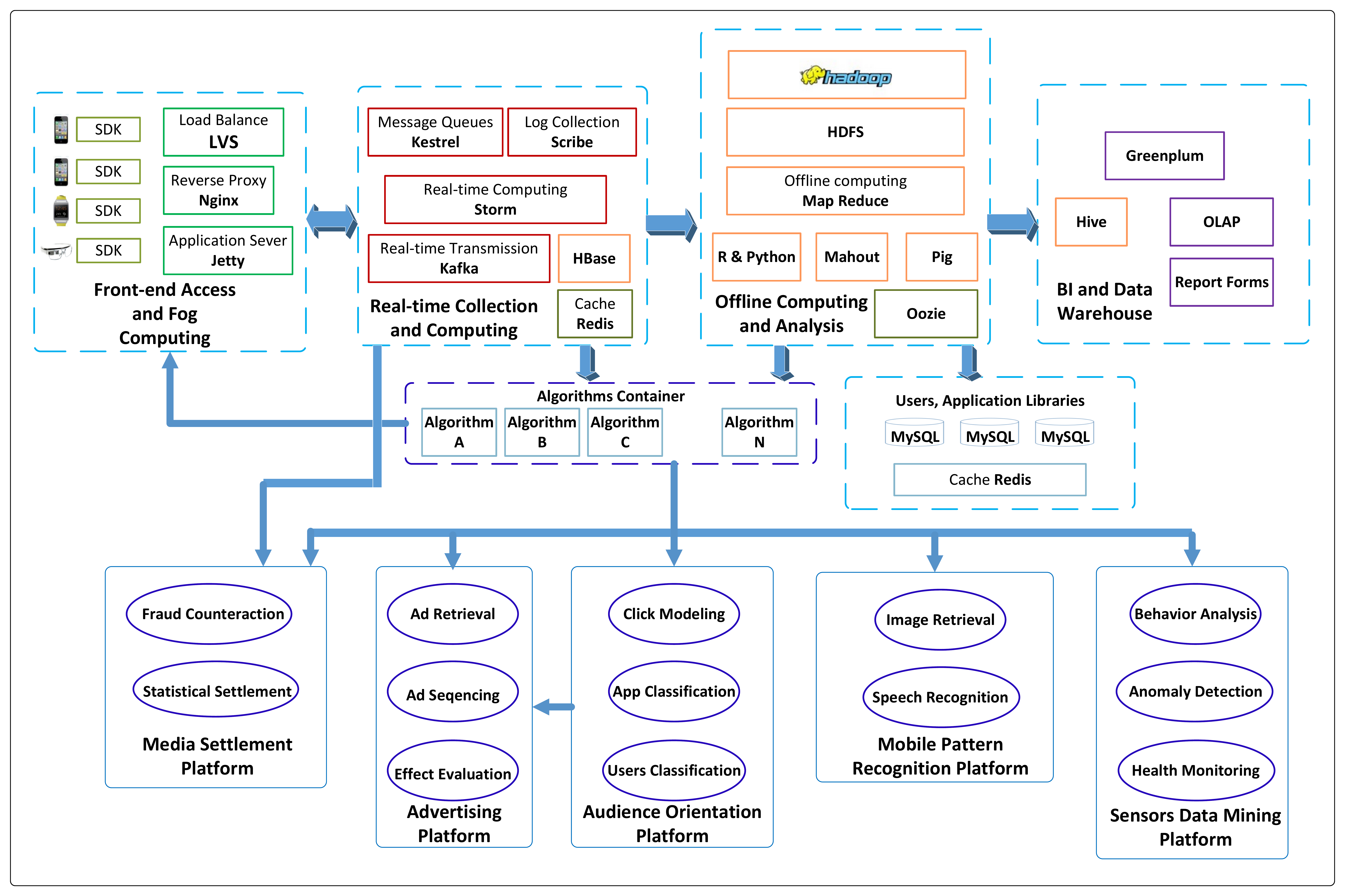}
\caption{\label{fig:app} \rev{Typical pipeline of an app-level mobile data processing system.} }
\end{center}
\end{figure*}
The development of mobile technology (e.g. smartphones, augmented reality, etc.) are forcing mobile operators to evolve mobile network infrastructures. As a consequence, both the cloud and edge side of mobile networks are becoming increasingly sophisticated to cater for users who produce and consume huge amounts of mobile data daily. These data can be either generated by the sensors of mobile devices that record individual user behaviors, or from the mobile network infrastructure, which reflects dynamics in urban environments. Appropriately mining these data can benefit multidisciplinary research fields and the industry in areas such mobile network management, social analysis, public transportation, personal services provision, and so on~\cite{cheng2017exploiting}.
Network operators, however, could become overwhelmed when managing and analyzing massive amounts of heterogeneous mobile data \cite{ahmed2018recent}. Deep learning is probably the most powerful methodology that can overcoming this burden. We begin therefore by introducing characteristics of mobile big data, then present a holistic review of deep learning driven mobile data analysis research.

\begin{figure}[t]
\begin{center}
\includegraphics[width=0.5\textwidth]{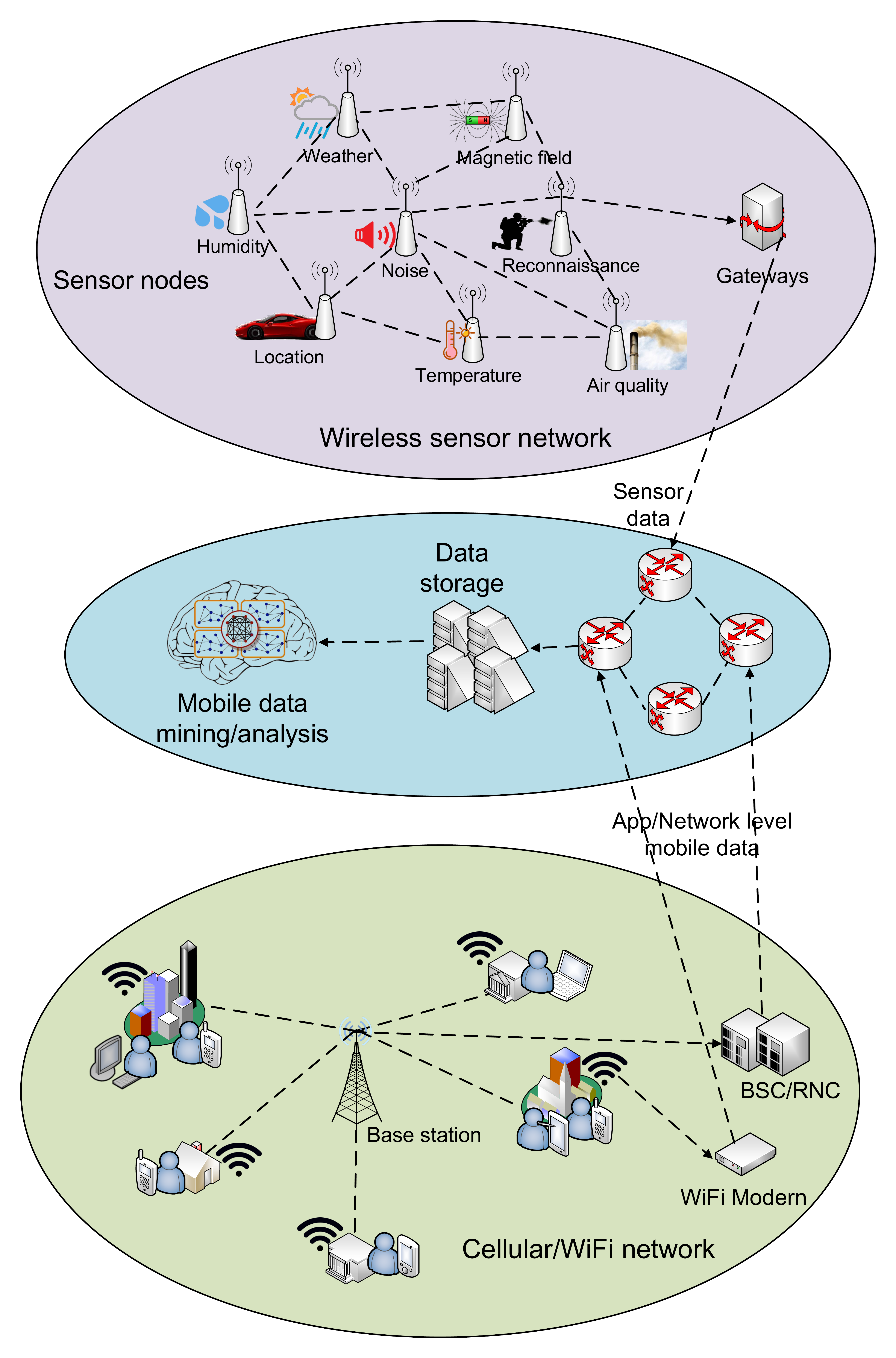}
\caption{\label{fig:collection} Illustration of the mobile data collection process in cellular, WiFi and wireless sensor networks. BSC: Base Station Controller; RNC: Radio Network Controller.}
\end{center}
\end{figure}

Yazti and Krishnaswamy propose to categorize mobile data into two groups, namely \emph{network-level} data and \emph{app-level} data~\cite{yazti2014mobile}. The key difference between them is that in the former data is usually collected by the edge mobile devices, while in the latter obtained throughout network infrastructure. We summarize these two types of data and their information comprised in Table \ref{tab:mbd2}. Before delving into mobile data analytics, we illustrate the typical data collection process in Figure~\ref{fig:collection}.

\begin{table}[htb]
\centering
\caption{The taxonomy of mobile big data.}
\label{tab:mbd2}
  \begin{tabular}{|c|C{1.8cm}|C{3cm}|}
\hline
\textbf{Mobile data}                         & \textbf{Source}                & \textbf{Information}           

\\ \hline
\multirow{4}{*}{Network-level data} & Infrastructure        & Infrastructure locations, capability, equipment  holders, etc                           \\ \cline{2-3} 
                                    & Performance indicators & Data traffic, end-to-end delay, QoE, jitter, etc.                            \\ \cline{2-3} 
                                    & Call detail records (CDR)    & Session start and end times, type, sender and receiver, etc. 
\\ \cline{2-3} 
                                    & Radio information    & Signal power, frequency, spectrum, modulation etc.
\\ \hline
\multirow{5}{*}{App-level data}     & Device                & Device type, usage, Media Access Control (MAC) address, etc.                                        \\ \cline{2-3} 
                                    & Profile               & User settings, personal information, etc                                      \\ \cline{2-3} 
                                    & Sensors                & Mobility, temperature, magnetic field, movement, etc                         \\ \cline{2-3} 
                                    & Application           & Picture, video, voice, health condition, preference, etc.                    \\ \cline{2-3} 
                                    & System log            & Software and hardware failure logs, etc.  
 \\ \hline
\end{tabular}
\end{table}

\emph{Network-level} mobile data generated by the networking infrastructure not only deliver a global view of mobile network performance (e.g. throughput, end-to-end delay, jitter, etc.), but also log individual session times, communication types, sender and receiver information, through Call Detail Records (CDRs). Network-level data usually exhibit significant spatio-temporal variations resulting from users' behaviors~\cite{naboulsi2016large}, which can be utilized for network diagnosis and management, user mobility analysis and public transportation planning~\cite{chaoyun2017zipnet}. Some network-level data (e.g. mobile traffic snapshots) can be viewed as pictures taken by `panoramic cameras', which provide a city-scale sensing system for urban sensing. 

On the other hand, \emph{App-level} data is directly recorded by sensors or mobile applications installed in various mobile devices. These data are frequently collected through crowd-sourcing schemes from heterogeneous sources, such as Global Positioning Systems (GPS), mobile cameras and video recorders, and portable medical monitors. Mobile devices act as sensor hubs, which are responsible for data gathering and preprocessing, and subsequently distributing such data to specific locations, as required \cite{cheng2017exploiting}.  \rev{We show a typical app-level data processing system in Fig.~\ref{fig:app}. App-level mobile data is generated and collected by a Software Development Kit (SDK) installed on mobile devices. Such data is subsequently processed by real-time collection and computing services (e.g., Storm,\footnote{\rev{Storm is a free and open-source distributed real-time computation system, \url{http://storm.apache.org/}}} Kafka,\footnote{\rev{Kafka\textsuperscript{\textregistered} is used for building real-time data pipelines and streaming apps, \url{https://kafka.apache.org/}}} HBase,\footnote{\rev{Apache HBase\texttrademark ~is the Hadoop database, a distributed, scalable, big data store, \url{https://hbase.apache.org/}}} Redis,\footnote{\rev{Redis is an open source, in-memory data structure store, used as a database, cache and message broker, \url{https://redis.io/}}} etc.) as required. Further offline storage and computing with mobile data can be performed with various tools, such as Hadoop Distribute File System (HDFS),\footnote{\rev{The Hadoop Distributed File System (HDFS) is a distributed file system designed to run on commodity hardware, \url{https://hadoop.apache.org/docs/r1.2.1/hdfs_design.html}}} Python, Mahout,\footnote{\rev{Apache Mahout\texttrademark ~is a distributed linear algebra framework, \url{https://mahout.apache.org/}}} Pig,\footnote{\rev{Apache Pig is a high-level platform for creating programs that run on Apache Hadoop, \url{https://pig.apache.org/}}} or Oozie.\footnote{\rev{Oozie is a workflow scheduler system to manage Apache Hadoop jobs, \url{http://oozie.apache.org/}}} The raw data and analysis results will be further transferred to databases (e.g., MySQL\footnote{\rev{MySQL is the open source database, \url{https://www.oracle.com/technetwork/database/mysql/index.html}}}) Business Intelligence -- BI (e.g. Online Analytical Processing -- OLAP), and data warehousing (e.g., Hive\footnote{\rev{The Apache Hive\texttrademark ~is a data warehouse software, \url{https://hive.apache.org/}}}). Among these, the algorithms container is the core of the entire system as it connects to front-end access and fog computing, real-time collection and computing, and offline computing and analysis modules, while it links directly to mobile applications, such as mobile healthcare, pattern recognition, and advertising platforms. Deep learning logic can be placed within the algorithms container.}

App-level data may directly or indirectly reflect users' behaviors, such as mobility, preferences, and social links \cite{cheng2017mobile}. Analyzing app-level data from individuals can help reconstructing one's personality and preferences, which can be used in recommender systems and users targeted advertising. Some of these data comprise explicit information about individuals' identities. Inappropriate sharing and use can raise significant privacy issues. Therefore, extracting useful patterns from multi-modal sensing devices without compromising user's privacy remains a challenging endeavor.

Compared to traditional data analysis techniques, deep learning embraces several unique features to address the aforementioned challenges \cite{alsheikh2016mobile}. Namely:
\begin{enumerate}
\item Deep learning achieves remarkable performance in various data analysis tasks, on both structured and unstructured data. Some types of mobile data can be represented as image-like (e.g. \cite{chaoyun2017zipnet}) or sequential data \cite{liang2016mercury}.
\item Deep learning performs remarkably well in feature extraction from raw data. This saves tremendous effort of hand-crafted feature engineering, which allows spending more time on model design and less on sorting through the data itself.
\item Deep learning offers excellent tools (e.g. RBM, AE, GAN) for handing unlabeled data, which is common in mobile network logs. 
\item Multi-modal deep learning allows to learn features over multiple modalities \cite{ngiam2011multimodal}, which makes it powerful in modeling with data collected from heterogeneous sensors and data sources.
\end{enumerate}
These advantages make deep learning as a powerful tool for mobile data analysis.

\begin{table*}[h!]
\centering
\caption{A summary of work on network-level mobile data analysis.}
\label{tab:netdata}
\begin{tabular}{|c|C{1.5cm}|C{3cm}|C{2cm}|>{\color{black}}C{1.5cm}|>{\color{black}}C{5.5cm}|}
\hline
\textbf{Domain}                         & \textbf{Reference}                                    & \textbf{Applications}                                                  & \textbf{Model}              & \textbf{Optimizer}                & \textbf{Key contribution}                                                                                 \\ \hline
\multirow{14}{*}{Network prediction}    & Pierucci and Micheli \cite{pierucci2016neural}        & QoE prediction                                                         & MLP                         & Unknown                           & Uses NNs to correlate Quality of Service parameters and QoE estimations.                                       \\ \cline{2-6} 
                                        & Gwon and Kung \cite{gwon2014inferring}                & Inferring Wi-Fi flow patterns                                          & Sparse coding + Max pooling & SGD                               & Semi-supervised learning.                                                                                      \\ \cline{2-6} 
                                        & Nie \emph{et al.} \cite{nie2017network}               & Wireless mesh network traffic prediction                               & DBN + Gaussian models       & SGD                               & Considers both long-term dependency and short-term fluctuations.                                               \\ \cline{2-6} 
                                        & Moyo and Sibanda \cite{moyo2015generalization}        & TCP/IP traffic prediction                                              & MLP                         & Unknown                           & Investigates the impact of learning in traffic forecasting.                                                     \\ \cline{2-6} 
                                        & Wang \emph{et al.} \cite{wangspatiotemporal}          & Mobile traffic forecasting                                             & AE + LSTM                   & SGD                               & Uses an AE to model spatial correlations and an LSTM to model temporal correlation                       \\ \cline{2-6} 
                                        & Zhang and Patras \cite{zhang2017long}                 & Long-term mobile traffic forecasting                                   & ConvLSTM + 3D-CNN           & Adam                              & Combines 3D-CNNs and ConvLSTMs to perform long-term forecasting                   \\ \cline{2-6} 
                                        & Zhang \emph{et al.} \cite{chaoyun2017zipnet}          & Mobile traffic super-resolution                                        & CNN + GAN                   & Adam                              & Introduces the MTSR concept and applies image processing techniques for mobile traffic analysis                  \\ \cline{2-6} 
                                        & Huang \emph{et al.} \cite{huang2017study}             & Mobile traffic forecasting                                            & LSTM + 3D-CNN               & Unknown                           & Combines CNNs and RNNs to extract geographical and temporal features from mobile traffic.                        \\ \cline{2-6} 
                                        & \edit{Zhang \emph{et al.} \cite{zhang2018citywide}}     & \edit{Cellular traffic prediction}                                         & \edit{Densely connected CNN}      & Adam                              & Uses separate CNNs to model closeness and periods in temporal dependency.                                        \\ \cline{2-6} 
                                        & Chen \emph{et al.} \cite{chen2018deep0}               & Cloud RAN optimization & Multivariate LSTM           & Unknown                           & Uses mobile traffic forecasting to aid cloud radio access network optimization                                 \\ \cline{2-6} 
                                        & \edit{Navabi \emph{et al.} \cite{navabi2018predicting}} & \edit{Wireless WiFi channel feature prediction}                          & \edit{MLP}                    & SGD                               & Infers non-observable channel information from observable features.                                            \\\cline{2-6} 
                                        & \rev{Feng \emph{et al.} \cite{feng2018deeptp}} & \rev{Mobile cellular traffic prediction}              & \rev{LSTM}           & \rev{Adam}                           & \rev{Extracts spatial and temporal dependencies using separate modules.}                                 \\ \cline{2-6} 
                                        & \rev{Alawe \emph{et al.} \cite{alawe2018improving}} & \rev{Mobile traffic load forecasting}              & \rev{MLP, LSTM}           & \rev{Unknown}                           & \rev{Employs traffic forecasting to improve 5G network scalability.}                                 \\ \cline{2-6} 
                                        & \rev{Wang \emph{et al.} \cite{wang2018spatio}} & \rev{Cellular traffic prediction}                          & \rev{Graph neural networks}                    & \rev{Pineda algorithm}                               & \rev{Represents spatio-temporal dependency via graphs and first work employing Graph neural networks for traffic forecasting.}                                           \\ \cline{2-6} 
                                        & \rev{Fang \emph{et al.} \cite{fang2018mobile}} & \rev{Mobile demand forecasting}                          & \rev{Graph CNN, LSTM, Spatio-temporal graph ConvLSTM}                    & \rev{Unknown}                               & \rev{Modelling the spatial correlations between cells using a dependency graph.}                                           \\ \cline{2-6} 
                                        & \rev{Luo \emph{et al.} \cite{luo2018channel}} & \rev{Channel state information prediction}                          & \rev{CNN and LSTM}                    & \rev{RMSprop}                               & \rev{Employing a two-stage offline-online training scheme to improve the stability of framework.}                                           \\ \hline
\multirow{7}{*}{Traffic classification} & Wang \cite{wang2015applications}                      & Traffic classification                                                 & MLP, stacked AE             & Unknown                           & Performs feature learning, protocol identification and anomalous protocol detection simultaneously.                       \\ \cline{2-6} 
                                        & Wang \emph{et al.} \cite{wang2017end}                 & Encrypted traffic classification                                       & CNN                         & SGD                               & Employs an end-to-end deep learning approach to perform encrypted traffic classification.                       \\ \cline{2-6} 
                                        & Lotfollahi \emph{et al.} \cite{lotfollahi2017deep}    & Encrypted traffic classification                                       & CNN                         & Adam                              & Can perform both traffic characterization and application identification.                                      \\ \cline{2-6} 
                                        & Wang \emph{et al.} \cite{wang2017malware}             & Malware traffic classification                                         & CNN                         & SGD                               & First work to use representation learning for malware classification from raw traffic.                  \\ \cline{2-6} 
                                        & \edit{Aceto \emph{et al.} \cite{aceto2018mobile}}       & \edit{Mobile encrypted traffic classification}                           & \edit{MLP, CNN, LSTM}         & SGD, Adam                         & Comprehensive evaluations of different NN architectures and excellent performance.                            \\
                                        \cline{2-6} 
                                        & \rev{Li \emph{et al.} \cite{8553650}}       & \rev{Network traffic classification}                           & \rev{Bayesian auto-encoder}         & \rev{SGD}                         & \rev{Applying Bayesian probability theory to obtain the a posteriori distribution of model parameters.}                            \\
                                        \hline
\multirow{4}{*}{CDR mining}             & Liang \emph{et al.} \cite{liang2016mercury}           & Metro density prediction                                               & RNN                         & SGD                               & Employs geo-spatial data processing, a weight-sharing RNN and parallel stream analytic programming.    \\ \cline{2-6} 
                                        & Felbo \emph{et al.} \cite{felbo2016using}             & Demographics prediction                                                & CNN                         & Adam                              & Exploits the temporal correlation inherent to mobile phone metadata.                                            \\ \cline{2-6} 
                                        & Chen \emph{et al.} \cite{chen2017comprehensive}       & Tourists' next visit location prediction                               & MLP, RNN                    & Scaled conjugate gradient descent & LSTM that performs significantly better than other ML approaches.                                      \\ \cline{2-6} 
                                        & Lin \emph{et al.} \cite{lin2017deep}                  & Human activity chains generation                                       & Input-Output HMM + LSTM     & Adam                              & First work that uses an RNN to generate human activity chains.                                                      \\ \hline
\multirow{2}{*}{Others}                 & Xu \emph{et al.} \cite{xu2017large}                   & Wi-Fi hotpot classification                                            & CNN                         & Unknown                           & Combining deep learning with frequency analysis.                                                                 \\ \cline{2-6} 
                                        & \edit{Meng \emph{et al.} \cite{meng2018qoe}}            & \edit{QoE-driven big data analysis}                                      & \edit{CNN}                    & SGD                               & Investigates trade-off between accuracy of high-dimensional big data analysis and model training speed. \\ \hline
\end{tabular}
\end{table*}

\subsection{Deep Learning Driven Network-level Mobile Data Analysis}\label{sec:netdata}
Network-level mobile data refers broadly to logs recorded by Internet service providers, including infrastructure metadata, network performance indicators and call detail records (CDRs) (see Table. \ref{tab:netdata}). The recent remarkable success of deep learning ignites global interests in exploiting this methodology for mobile network-level data analysis, so as to optimize mobile networks configurations, thereby improving end-uses' QoE. These work can be categorized into four types: network state prediction, network traffic classification, CDR mining and radio analysis. In what follows, we review work in these directions, which we first summarize and compare in Table~\ref{tab:netdata}.\\

\noindent \textbf{Network State Prediction} refers to inferring mobile network traffic or performance indicators, given historical measurements or related data. Pierucci and Micheli investigate the relationship between key objective metrics and QoE \cite{pierucci2016neural}. They employ MLPs to predict users' QoE in mobile communications, based on average user throughput, number of active users in a cells, average data volume per user, and channel quality indicators, demonstrating high prediction accuracy. Network traffic forecasting is another field where deep learning is gaining importance. By leveraging sparse coding and max-pooling, Gwon and Kung develop a semi-supervised deep learning model to classify received frame/packet patterns and infer the original properties of flows in a WiFi network \cite{gwon2014inferring}. Their proposal demonstrates superior performance over traditional ML techniques. Nie~\emph{et al.} investigate the traffic demand patterns in wireless mesh network \cite{nie2017network}. They design a DBN along with Gaussian models to precisely estimate traffic distributions. 

\begin{figure}[htb]
\begin{center}
\includegraphics[width=0.5\textwidth]{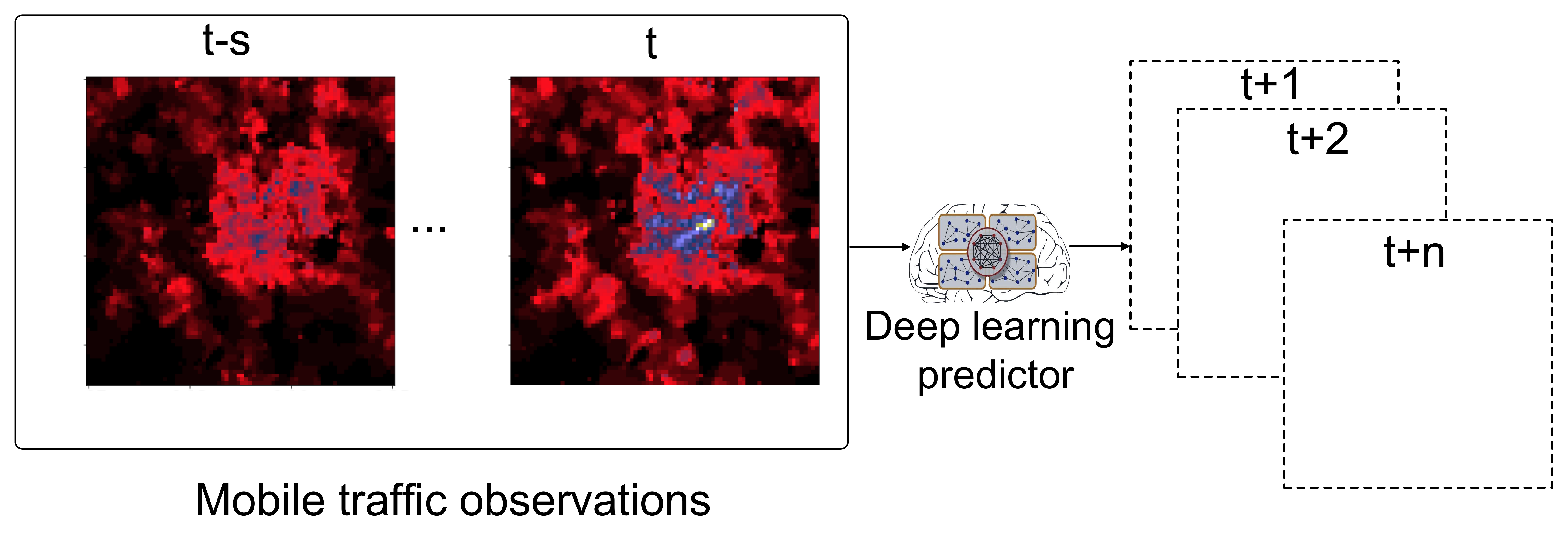}
\end{center}
\caption{\label{fig:forecasting} \rev{The underlying principle of city-scale mobile traffic forecasting. The deep learning predictor takes as input a sequence of mobile traffic measurements in a region (snapshots $t-s$ to $t$), and forecasts how much mobile traffic will be consumed in the same areas in the future $t+1$ to $t+n$ instances.} }
\end{figure}

\rev{In addition to the above, several researchers employ deep learning to forecast mobile traffic at city scale, by considering spatio-temporal correlations of geographic mobile traffic measurements. We illustrate the underlying principle in Fig.~\ref{fig:forecasting}.} In \cite{wangspatiotemporal}, Wang \emph{et al.} propose to use an AE-based architecture and LSTMs to model spatial and temporal correlations of mobile traffic distribution, respectively. In particular, the authors use a global and multiple local stacked AEs for spatial feature extraction, dimension reduction and training parallelism. Compressed representations extracted are subsequently processed by LSTMs, to perform final forecasting. Experiments with a real-world dataset demonstrate superior performance over SVM and the Autoregressive Integrated Moving Average (ARIMA) model. The work in \cite{zhang2017long} extends mobile traffic forecasting to long time frames. The authors combine ConvLSTMs and 3D CNNs to construct spatio-temporal neural networks that capture the complex spatio-temporal features at city scale. They further introduce a fine-tuning scheme and lightweight approach to blend predictions with historical means, which significantly extends the length of reliable prediction steps. \edit{Deep learning was also employed in \cite{huang2017study, alawe2018improving, feng2018deeptp} and \cite{chen2018deep0}, where the authors employ CNNs and LSTMs to perform mobile traffic forecasting. By effectively extracting spatio-temporal features, their proposals gain significantly higher accuracy than traditional approaches, such as ARIMA.} \rev{Wang \emph{et al.} represent spatio-temporal dependencies in mobile traffic using graphs, and learn such dependencies using Graph Neural Networks \cite{wang2018spatio}. Beyond the accurate inference achieved in their study, this work also demonstrates potential for precise social events inference.}

\begin{figure}[htb]
\begin{center}
\includegraphics[width=0.5\textwidth]{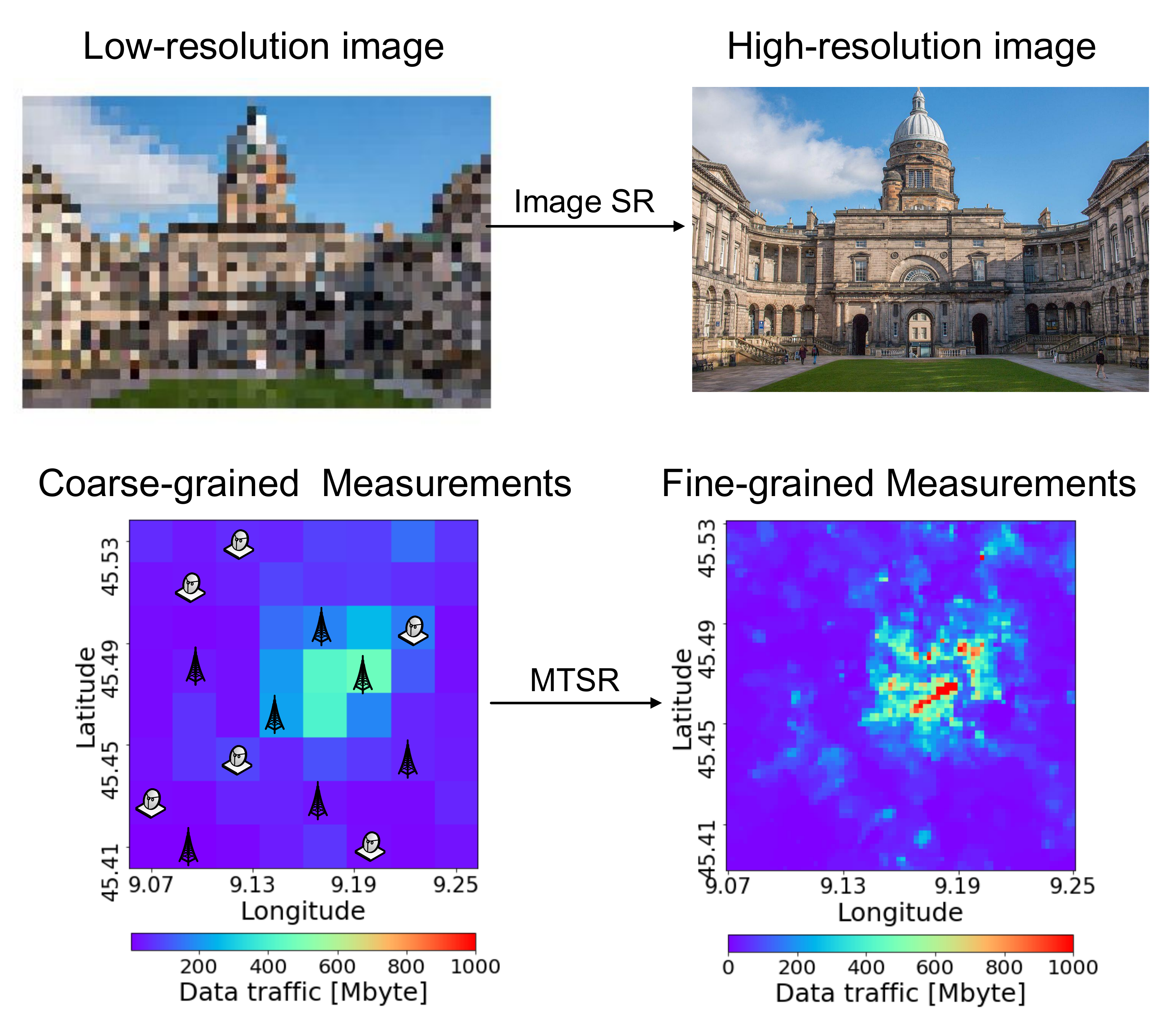}
\end{center}
\caption{\label{fig:mtsr} \rev{Illustration of the image super-resolution (SR) principle (above) and the mobile traffic super-resolution (MTSR) technique (below). Figure adapted from \cite{chaoyun2017zipnet}.} }
\end{figure}

More recently, Zhang \emph{et al.} propose an original Mobile Traffic Super-Resolution (MTSR) technique to infer network-wide fine-grained mobile traffic consumption given coarse-grained counterparts obtained by probing, thereby reducing traffic measurement overheads \cite{chaoyun2017zipnet}. \rev{We illustrate the principle of MTSR in Fig.~\ref{fig:mtsr}.} Inspired by image super-resolution techniques, they design \edit{a dedicated CNN with multiple skip connections between layers, named deep zipper network, along with a Generative Adversarial Network (GAN) to perform precise MTSR and improve the fidelity of inferred traffic snapshots. Experiments with a real-world dataset show that this architecture can improve the granularity of mobile traffic measurements over a city by up to 100$\times$, while significantly outperforming other interpolation techniques.}\\

\noindent \textbf{Traffic Classification}
is aimed at identifying specific applications or protocols among the traffic in networks. Wang recognizes the powerful feature learning ability of deep neural networks and uses \rev{a deep AE} to identify protocols in a TCP flow dataset, achieving excellent precision and recall rates~\cite{wang2015applications}. Work in~\cite{wang2017end} proposes to use a 1D CNN for encrypted traffic classification. The authors suggest that this structure works well for modeling sequential data and has lower complexity, thus being promising in addressing the traffic classification problem. Similarly, Lotfollahi \emph{et al.} present Deep Packet, which is based on a CNN, for encrypted traffic classification \cite{lotfollahi2017deep}. Their framework reduces the amount of hand-crafted feature engineering and achieves great accuracy. \rev{An improved stacked AE is employed in~\cite{8553650}, where Li \emph{et al.} incorporate Bayesian methods into AEs to enhance the inference accuracy in network traffic classification.} \edit{More recently, Aceto \emph{et al.} employ MLPs, CNNs, and LSTMs to perform encrypted mobile traffic classification \cite{aceto2018mobile}, arguing that deep NNs can automatically extract complex features present in mobile traffic. As reflected by their results, deep learning based solutions obtain superior accuracy over RFs in classifying Android, IOS and Facebook traffic.}
CNNs have also been used to identify malware traffic, where work in \cite{wang2017malware} regards traffic data as images and unusual patterns that malware traffic exhibit are classified by representation learning. Similar work on mobile malware detection will be further discussed in subsection \ref{sec:security}. 
\\

\noindent\textbf{CDR Mining} involves extracting knowledge from specific instances of telecommunication transactions such as phone number, cell ID, session start/end time, traffic consumption, etc. Using deep learning to mine useful information from CDR data can serve a variety of functions. For example, Liang \emph{et al.} propose Mercury to estimate metro density from streaming CDR data, using RNNs \cite{liang2016mercury}. They take the trajectory of a mobile phone user as a sequence of locations; RNN-based models work well in handling such sequential data. Likewise, Felbo \emph{et al.} use CDR data to study demographics \cite{felbo2016using}. They employ a CNN to predict the age and gender of mobile users, demonstrating the superior accuracy of these structures over other ML tools. More recently, Chen \emph{et al.} compare different ML models to predict tourists' next locations of visit by analyzing CDR data \cite{chen2017comprehensive}. Their experiments suggest that RNN-based predictors significantly outperform traditional ML methods, including Naive Bayes, SVM, RF, and MLP.\\

\textbf{Lessons learned}: Network-level mobile data, such as mobile traffic, usually involves essential spatio-temporal correlations. These correlations can be effectively learned by CNNs and RNNs, as they are specialized in modeling spatial and temporal data (e.g., images, traffic series). An important observation is that large-scale mobile network traffic can be processed as sequential snapshots, as suggested in \cite{chaoyun2017zipnet, zhang2017long}, which resemble images and videos. Therefore, potential exists to exploit image processing techniques for network-level analysis. Techniques previously used for imaging usually, however, cannot be directly employed with mobile data. Efforts must be made to adapt them to the particularities of the mobile networking domain. We expand on this future research direction in Sec.~\ref{sec:st-traffic}. 

\rev{On the other hand, although deep learning brings precision in network-level mobile data analysis, making causal inference remains challenging, due to limited model interpretability. For example, a NN may predict there will be a traffic surge in a certain region in the near future, but it is hard to explain why this will happen and what triggers such a surge. Additional efforts are required to enable explanation and confident decision making. At this stage, the community should rather use deep learning algorithms as intelligent assistants that can make accurate inferences and reduce human effort, instead of relying exclusively on these. }

\subsection{Deep Learning Driven App-level Mobile Data Analysis}\label{sec:appdata}
Triggered by the increasing popularity of Internet of Things (IoT), current mobile devices bundle increasing numbers of applications and sensors that can collect massive amounts of app-level mobile data \cite{al2015internet}.  Employing artificial intelligence to extract useful information from these data can extend the capability of devices \cite{ota2017deep, seneviratne2017survey, li2018learning}, thus greatly benefiting users themselves, mobile operators, and indirectly device manufacturers. Analysis of mobile data therefore becomes an important and popular research direction in the mobile networking domain. Nonetheless, mobile devices usually operate in noisy, uncertain and unstable environments, where their users move fast and change their location and activity contexts frequently. \edit{As a result, app-level mobile data analysis becomes difficult for traditional machine learning tools, which performs relatively poorly.} Advanced deep learning practices provide a powerful solution for app-level data mining, as they demonstrate better precision and higher robustness in IoT applications \cite{lane2015early}.

\begin{figure*}[htb]
\begin{center}
\includegraphics[width=1\textwidth]{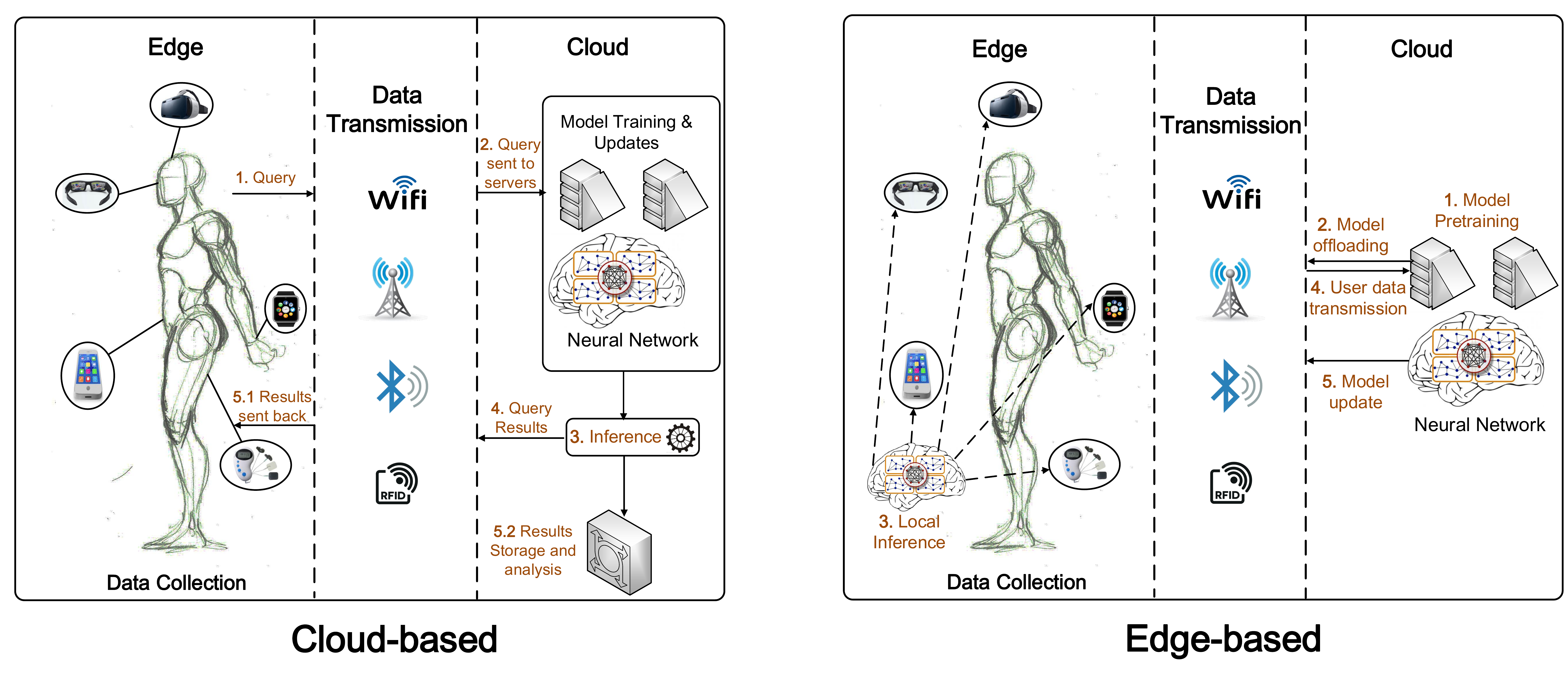}
\caption{\label{fig:scenario} Illustration of two deployment approaches for app-level mobile data analysis, namely cloud-based (left) and edge-based (right). The cloud-based approach makes inference on clouds and send results to edge devices. On the contrary, the edge-based approach deploys models on edge devices which can make local inference. }
\end{center}
\end{figure*}

There exist two approaches to app-level mobile data analysis, namely \emph{(i)} cloud-based computing and \emph{(ii)} edge-based computing. We illustrate the difference between these scenarios in Fig.~\ref{fig:scenario}.\footnotetext{Human profile source: \url{https://lekeart.deviantart.com/art/male-body-profile-251793336}} \edit{As shown in the left part of the figure, the cloud-based computing treats mobile devices as data collectors and messengers that constantly send data to cloud servers, via local points of access with limited data preprocessing capabilities. This scenario typically includes the following steps: \emph{(i)} users query on/interact with local mobile devices; \emph{(ii)} queries are transmitted to severs in the cloud; \emph{(iii)} servers gather the data received for model training and inference; \emph{(iv)} query results are subsequently sent back to each device, or stored and analyzed without further dissemination, depending on specific application requirements. The drawback of this scenario is that constantly sending and receiving messages to/from servers over the Internet introduces overhead and may result in severe latency. In contrast, in the edge-based computing scenario pre-trained models are offloaded from the cloud to individual mobile devices, such that they can make inferences locally. As illustrated in the right part of  Fig.~\ref{fig:scenario}, this scenario typically consists of the following: \emph{(i)} servers use offline datasets to per-train a model; \emph{(ii)} the pre-trained model is offloaded to edge devices; \emph{(iii)} mobile devices perform inferences locally using the model; \emph{(iv)} cloud servers accept data from local devices; \emph{(v)} the model is updated using these data whenever necessary. While this scenario requires less interactions with the cloud, its applicability is limited by the computing and battery capabilities of edge hardware. Therefore, it can only support tasks that require light computations. }

Many researchers employ deep learning for app-level mobile data analysis. We group the works reviewed according to their application domains, namely mobile healthcare, mobile pattern recognition, and mobile Natural Language Processing (NLP) and Automatic Speech Recognition (ASR). Table \ref{tab:mbd} gives a high-level summary of existing research efforts and we discuss representative work next.\\

\begin{table*}[t!]
\centering
\caption{A summary of works on app-level mobile data analysis.}
\label{tab:mbd}
\begin{tabular}{|c|C{3.5cm}|C{5cm}|C{1.8cm}|C{2.3cm}|}
\hline
\textbf{Subject}                             & \textbf{Reference}                                            & \textbf{Application}                                                          & \textbf{Deployment}            & \textbf{Model}           \\ \hline
\multirow{10}{*}{Mobile Healthcare}          & Liu and Du \cite{liu2016poster}                               & Mobile ear                                                                    & Edge-based                     & CNN                      \\ \cline{2-5} 
                                             & Liu \emph{at al.} \cite{sicong2017ubiear}                     & Mobile ear                                                                    & Edge-based                     & CNN                      \\ \cline{2-5} 
                                             & Jindal \cite{jindal2016integrating}                           & Heart rate prediction                                                         & Cloud-based                    & DBN                      \\ \cline{2-5} 
                                             & Kim \emph{et al.} \cite{kim2016deep}                          & Cytopathology classification                                                  & Cloud-based                    & CNN                      \\ \cline{2-5} 
                                             & Sathyanarayana \emph{et al.} \cite{sathyanarayana2016sleep}   & Sleep quality prediction                                                      & Cloud-based                    & MLP, CNN, LSTM           \\ \cline{2-5} 
                                             & Li and Trocan \cite{li2017personal}                           & Health conditions analysis                                                    & Cloud-based                    & Stacked AE               \\ \cline{2-5} 
                                             & Hosseini \emph{et al.} \cite{hosseini2017deep}                & Epileptogenicity localisation                                                 & Cloud-based                    & CNN                      \\ \cline{2-5} 
                                             & Stamate \emph{et al.} \cite{stamate2017deep}                  & Parkinson's symptoms management                                               & Cloud-based                    & MLP                      \\ \cline{2-5} 
                                             & Quisel \emph{et al.} \cite{quisel2017collecting}              & Mobile health data analysis                                                   & Cloud-based                    & CNN, RNN                 \\ \cline{2-5} 
                                             & Khan \emph{et al.}\cite{khan2017deep}                         & Respiration surveillance                                                      & Cloud-based                    & CNN                      \\ \hline
\multirow{38}{*}{Mobile Pattern Recognition} & Li \emph{et al.} \cite{li2016deepcham}                        & Mobile object recognition                                                     & Edge-based                     & CNN                      \\ \cline{2-5} 
                                             & Tob{\'\i}as \emph{et al.} \cite{tobias2016convolutional}      & Mobile object recognition                                                     & Edge-based \& Cloud based      & CNN                      \\ \cline{2-5} 
                                             & Pouladzadeh and Shirmohammadi \cite{pouladzadeh2017mobile}    & Food recognition system                                                       & Cloud-based                    & CNN                      \\ \cline{2-5} 
                                            & Tanno \emph{et al.} \cite{tanno2016deepfoodcam}               & Food recognition system                                                       & Edge-based                     & CNN                      \\ \cline{2-5} 
                                             & Kuhad \emph{et al.} \cite{kuhad2015using}                     & Food recognition system                                                       & Cloud-based                    & MLP                      \\ \cline{2-5} 
                                             & Teng and Yang \cite{teng2016facial}                           & Facial recognition                                                            & Cloud-based                    & CNN                      \\ \cline{2-5} 
                                            & Wu \emph{et al.} \cite{liu2017deep}                           & Mobile visual search                                                          & Edge-based                     & CNN                      \\ \cline{2-5} 
                                             & Rao \emph{et al.} \cite{rao2017mobile}                        & Mobile augmented reality                                                      & Edge-based                     & CNN                      \\ \cline{2-5} 
                                            & Ohara \emph{et al.} \cite{ohara2017detecting}                 & WiFi-driven indoor change detection                                           & Cloud-based                    & CNN,LSTM                 \\ \cline{2-5} 
                                             & Zeng \emph{et al.} \cite{zeng2014convolutional}               & Activity recognition                                                          & Cloud-based                    & CNN, RBM                 \\ \cline{2-5} 
                                             & Almaslukh \emph{et al.} \cite{almaslukh2017effective}         & Activity recognition                                                          & Cloud-based                    & AE                       \\ \cline{2-5} 
                                             & Li \emph{et al.} \cite{li2016deep22}                          & RFID-based activity recognition                                               & Cloud-based                    & CNN                      \\ \cline{2-5} 
                                             & Bhattacharya and Lane \cite{bhattacharya2016smart}            & Smart watch-based activity recognition                                        & Edge-based                     & RBM                      \\ \cline{2-5} 
                                             & Antreas and Angelov \cite{antoniou2016general}                & Mobile surveillance system                                                    & Edge-based \& Cloud based      & CNN                      \\ \cline{2-5} 
                                             & Ord{\'o}{\~n}ez and Roggen \cite{ordonez2016deep}             & Activity recognition                                                          & Cloud-based                    & ConvLSTM                 \\ \cline{2-5} 
                                             & Wang \emph{et al.} \cite{wang2016interacting}                 & Gesture recognition                                                           & Edge-based                     & CNN, RNN                 \\ \cline{2-5} 
                                             & Gao \emph{et al.} \cite{gao2016ihear}                         & Eating detection                                                              & Cloud-based                    & DBM, MLP                 \\ \cline{2-5} 
                                             & Zhu \emph{et al.} \cite{zhu2015using}                         & User energy expenditure estimation                                            & Cloud-based                    & CNN, MLP                 \\ \cline{2-5} 
                                             & Sunds{\o}y \emph{et al.} \cite{sundsoy2016deep}               & Individual income classification                                              & Cloud-based                    & MLP                      \\ \cline{2-5} 
                                             & Chen and Xue \cite{chen2015deep}                              & Activity recognition                                                          & Cloud-based                    & CNN                      \\ \cline{2-5} 
                                             & Ha and Choi \cite{ha2016convolutional}                        & Activity recognition                                                          & Cloud-based                    & CNN                      \\ \cline{2-5} 
                                             & Edel and K{\"o}ppe \cite{edel2016binarized}                   & Activity recognition                                                          & Edge-based                     & Binarized-LSTM           \\ \cline{2-5} 
                                             & Okita and Inoue \cite{okita2017recognition}                   & Multiple overlapping activities recognition                                   & Cloud-based                    & CNN+LSTM                 \\ \cline{2-5} 
                                             & Alsheikh \emph{et al.} \cite{alsheikh2016mobile}              & Activity recognition using Apache Spark                                       & Cloud-based                    & MLP                      \\ \cline{2-5} 
                                             & Mittal \emph{et al.} \cite{mittal2016spotgarbage}             & Garbage detection                                                             & Edge-based \& Cloud based      & CNN                      \\ \cline{2-5} 
                                             & Seidenari \emph{et al.} \cite{seidenari2017deep}              & Artwork detection and retrieval                                               & Edge-based                     & CNN                      \\ \cline{2-5} 
                                             & Zeng \emph{et al.} \cite{zeng2017mobiledeeppill}              & Mobile pill classification                                                    & Edge-based                     & CNN                      \\ \cline{2-5} 
                                            & Lane and Georgiev \cite{lane2015can}                          & Mobile activity recognition, emotion recognition and speaker identification   & Edge-based                     & MLP                      \\ \cline{2-5} 
                                            & Yao \emph{et al.} \cite{yao2017deepsense}                     & Car tracking,heterogeneous human activity recognition and user identification & Edge-based                     & CNN, RNN                 \\ \cline{2-5} 
                                            & \edit{Zou \emph{et al.} \cite{zoudeepsense}}                    & \edit{IoT human activity recognition}                                           & \edit{Cloud-based}               & \edit{AE, CNN, LSTM}       \\ \cline{2-5} 
                                             & Zeng \cite{zeng2017mobile}                                    & Mobile object recognition                                                     & Edge-based                     & Unknown                  \\ \cline{2-5} 
                                             & Katevas \emph{et al.} \cite{katevas2017practical}             & Notification attendance prediction                                            & Edge-based                     & RNN                      \\ \cline{2-5} 
                                             & Radu \emph{et al.} \cite{radu2016towards}                     & Activity recognition                                                          & Edge-based                     & RBM, CNN                 \\ \cline{2-5} 
                                             & Wang \emph{et al.} \cite{wang2015phasefi, wang2016csi}        & Activity and gesture recognition                                              & Cloud-based                    & Stacked AE               \\ \cline{2-5} 
                                            & \edit{Feng \emph{et al.} \cite{feng2018evaluation}}             & \edit{Activity detection}                                                       & \edit{Cloud-based}               & \edit{LSTM}                \\ \cline{2-5} 
                                             & Cao \emph{et al.} \cite{cao2017deepmood}                      & Mood detection                                                                & Cloud-based                    & GRU                      \\ \cline{2-5} 
                                             & \edit{Ran \emph{et al.} \cite{ran2018deepdecision}}             & \edit{Object detection for AR applications.}                                    & \edit{Edge-based \& cloud-based} & \edit{CNN}                 \\ \cline{2-5} 
                                             & \edit{Zhao \emph{et al.} \cite{zhao2018rf}}                     & \edit{Estimating 3D human skeleton from radio frequently signal}                & \edit{Cloud-based}               & \edit{CNN}                 \\ \hline
\multirow{7}{*}{Mobile NLP and ASR}          
                                             & Siri \cite{siri}                                              & Speech synthesis                                                              & Edge-based                     & Mixture density networks \\ \cline{2-5} 
                                             & McGraw \emph{et al.} \cite{mcgraw2016personalized}            & Personalised speech recognition                                               & Edge-based                     & LSTM                     \\ \cline{2-5} 
                                             & Prabhavalkar \emph{et al.} \cite{prabhavalkar2016compression} & Embedded speech recognition                                                   & Edge-based                     & LSTM                     \\ \cline{2-5} 
                                             & Yoshioka \emph{et al.} \cite{yoshioka2015ntt}                 & Mobile speech recognition                                                     & Cloud-based                    & CNN                      \\ \cline{2-5} 
                                             & Ruan \emph{et al.} \cite{ruan2016speech}                      & Shifting from typing to speech                                                & Cloud-based                    & Unknown                  \\ \cline{2-5} 
                                             & Georgiev \emph{et al.} \cite{georgiev2017low}                 & Multi-task mobile audio sensing                                               & Edge-based                     & MLP                      \\ \hline
\multirow{7}{*}{Others}                      & Ignatov \emph{et al.} \cite{ignatov2017dslr}                  & Mobile images quality enhancement                                             & Cloud-based                    & CNN                      \\ \cline{2-5} 
                                             & Lu \emph{et al.} \cite{lu2017demo}                            & Information retrieval from videos in wireless network                         & Cloud-based                    & CNN                      \\ \cline{2-5} 
                                             & Lee \emph{et al.} \cite{lee2016reducing}                      & Reducing distraction for smartwatch users                                     & Cloud-based                    & MLP                      \\ \cline{2-5} 
                                             & Vu \emph{et al.} \cite{vu2016transportation}                  & Transportation mode detection                                                 & Cloud-based                    & RNN                      \\ \cline{2-5} 
                                             & Fang \emph{et al.} \cite{fang2017learning}                    & Transportation mode detection                                                 & Cloud-based                    & MLP                      \\
                                             \cline{2-5} 
                                             &\rev{Xue \emph{et al.} \cite{xue2018appdna}}                    & \rev{Mobile App classification}                                                 & \rev{Cloud-based}                    & \rev{AE, MLP, CNN, and LSTM}                      \\
                                             \cline{2-5} 
                                             & \rev{Liu \emph{et al.} \cite{liu2018finding}}                    & \rev{Mobile motion sensor fingerprinting}                                                 & \rev{Cloud-based}                    & \rev{LSTM}                      \\\hline
\end{tabular}
\end{table*}

\noindent\textbf{Mobile Health.}
There is an increasing variety of wearable health monitoring devices being introduced to the market. By incorporating medical sensors, these devices can capture the physical conditions of their carriers and provide real-time feedback (e.g. heart rate, blood pressure, breath status etc.), or trigger alarms to remind users of taking medical actions \cite{ravi2017deep}. 

Liu and Du design a deep learning-driven MobiEar to aid deaf people's awareness of emergencies \cite{liu2016poster}. Their proposal accepts acoustic signals as input, allowing users to register different acoustic events of interest. MobiEar operates efficiently on smart phones and only requires infrequent communications with servers for updates. Likewise, Liu \emph{et al.} develop a UbiEar, which is operated on the Android platform to assist hard-to-hear sufferers in recognizing acoustic events, without requiring location information \cite{sicong2017ubiear}. Their design adopts a lightweight CNN architecture for inference acceleration and demonstrates comparable accuracy over traditional CNN models.

Hosseini \emph{et al.} design an edge computing system for health monitoring and treatment \cite{hosseini2017deep}. They use CNNs to extract features from mobile sensor data, which plays an important role in their epileptogenicity localization application. Stamate \emph{et al.} develop a mobile Android app called cloudUPDRS to manage Parkinson's symptoms \cite{stamate2017deep}. In their work, MLPs are employed to determine the acceptance of data collected by smart phones, to maintain high-quality data samples. The proposed method outperforms other ML methods such as GPs and RFs. Quisel \emph{et al.} suggest that deep learning can be effectively used for mobile health data analysis \cite{quisel2017collecting}. They exploit CNNs and RNNs to classify lifestyle and environmental traits of volunteers. Their models demonstrate superior prediction accuracy over RFs and logistic regression, over six datasets. 

As deep learning performs remarkably in medical data analysis \cite{miotto2017deep}, we expect more and more deep learning powered health care devices will emerge to improve physical monitoring and illness diagnosis.\\

\noindent\textbf{Mobile Pattern Recognition.} Recent advanced mobile devices offer people a \edit{portable intelligent assistant, which fosters a diverse set of applications that can classify surrounding objects (e.g. \cite{li2016deepcham, tobias2016convolutional, teng2016facial, pouladzadeh2017mobile}) or users' behaviors (e.g. \cite{zeng2014convolutional, ronao2016human, chen2015deep, bhattacharya2016smart, ordonez2016deep,  ha2016convolutional, wang2017deep}) based on patterns observed in the output of the mobile camera or other sensors.} We review and compare recent works on mobile pattern recognition in this part. 

\emph{Object classification} in pictures taken by mobile devices is drawing increasing research interest. Li \emph{et al.} develop DeepCham as a mobile object recognition framework \cite{li2016deepcham}. Their architecture involves a crowd-sourcing labeling process, which aims to reduce the hand-labeling effort, and a collaborative training instance generation pipeline that is built for deployment on mobile devices. Evaluations of the prototype system suggest that this framework is efficient and effective in terms of training and inference. Tob{\'\i}as \emph{et al.} investigate the applicability of employing CNN schemes on mobile devices for objection recognition tasks \cite{tobias2016convolutional}. They conduct experiments on three different model deployment scenarios, i.e., on GPU, CPU, and respectively on mobile devices, with two benchmark datasets. The results obtained suggest that deep learning models can be efficiently embedded in mobile devices to perform real-time inference.

Mobile classifiers can also assist Virtual Reality (VR) applications. A CNN framework is proposed in \cite{teng2016facial} for facial expressions recognition when users are wearing head-mounted displays in the VR environment. Rao \emph{et al.} incorporate a deep learning object detector into a mobile augmented reality (AR) system \cite{rao2017mobile}. Their system achieves outstanding performance in detecting and enhancing geographic objects in outdoor environments. Further work focusing on mobile AR applications is introduced in \cite{ran2017delivering}, where the authors characterize the tradeoffs between accuracy, latency, and energy efficiency of object detection. 


\emph{Activity recognition} is another interesting area that relies on data collected by mobile motion sensors \cite{wang2017deep, vyas2017survey}. \edit{This refers to the ability to classify based on data collected via, e.g., video capture, accelerometer readings, motion -- Passive Infra-Red (PIR) sensing, specific actions and activities that a human subject performs. Data collected will be delivered to servers for model training and the model will be subsequently deployed for domain-specific tasks.} 

Essential features of sensor data can be automatically extracted by neural networks.
The first work in this space that is based on deep learning employs a CNN to capture local dependencies and preserve scale invariance in motion sensor data \cite{zeng2014convolutional}. The authors evaluate their proposal on 3 offline datasets, demonstrating their proposal yields higher accuracy over statistical methods and \edit{Principal Components Analysis (PCA)}. Almaslukh \emph{et al.} employ a deep AE to perform human activity recognition by analyzing an offline smart phone dataset gathered from accelerometers and gyroscope sensors \cite{almaslukh2017effective}. Li \emph{et al.} consider different scenarios for activity recognition~\cite{li2016deep22}. In their implementation, Radio Frequency Identification (RFID) data is directly sent to a CNN model for recognizing human activities. While their mechanism achieves high accuracy in different applications, experiments suggest that the RFID-based method does not work well with metal objects or liquid containers.

\cite{bhattacharya2016smart} exploits an RBM to predict human activities, given 7 types of sensor data collected by a smart watch. Experiments on prototype devices show that this approach can efficiently fulfill the recognition objective under tolerable power requirements. Ord{\'o}{\~n}ez and Roggen architect an advanced ConvLSTM to fuse data gathered from multiple sensors and perform activity recognition \cite{ordonez2016deep}. By leveraging CNN and LSTM structures, ConvLSTMs can automatically compress spatio-temporal sensor data into low-dimensional representations, without heavy data post-processing effort. Wang \emph{et al.} exploit Google Soli to architect a mobile user-machine interaction platform \cite{wang2016interacting}. By analyzing radio frequency signals captured by millimeter-wave radars, their architecture is able to recognize 11 types of gestures with high accuracy. Their models are trained on the server side, and inferences are performed locally on mobile devices. \edit{More recently, Zhao \emph{et al.} design a 4D CNN framework (3D for the spatial dimension + 1D for the temporal dimension) to reconstruct human skeletons using radio frequency signals \cite{zhao2018rf}. This novel approach resembles virtual ``X-ray'', enabling to accurately estimate human poses, without requiring an actual camera.}\\



\noindent\textbf{Mobile NLP and ASR.}
Recent remarkable achievements obtained by deep learning in Natural Language Processing (NLP) and Automatic Speech Recognition (ASR) are also embraced by applications for mobile devices. 


Powered by deep learning, the intelligent personal assistant Siri, developed by Apple, employs a deep mixture density networks \cite{zen2014deep} to fix typical robotic voice issues and synthesize more human-like voice \cite{siri}. An Android app released by Google supports mobile personalized speech recognition \cite{mcgraw2016personalized}; this quantizes the parameters in LSTM model compression, allowing the app to run on low-power mobile phones. Likewise, Prabhavalkar \emph{et al.} propose a mathematical RNN compression technique that reduces two thirds of an LSTM acoustic model size, while only compromising negligible accuracy \cite{prabhavalkar2016compression}. This allows building both memory- and energy-efficient ASR applications on mobile devices.

Yoshioka \emph{et al.} present a framework that incorporates a network-in-network architecture into a CNN model, which allows to perform ASR with mobile multi-microphone devices used in noisy environments \cite{yoshioka2015ntt}. Mobile ASR can also accelerate text input on mobile devices, Ruan \emph{et al.}'s study showing that with the help of ASR, the input rates of English and Mandarin are 3.0 and 2.8 times faster over standard typing on keyboards \cite{ruan2016speech}. 
\edit{More recently, the applicability of deep learning to multi-task audio sensing is investigated in \cite{georgiev2017low}, where Georgiev \emph{et al.} propose and evaluate a novel deep learning modelling and optimization framework tailored to embedded audio sensing tasks. To this end, they selectively share compressed representations between different tasks, which reduces training and data storage overhead, without significantly compromising accuracy of an individual task. The authors evaluate their framework on a memory-constrained smartphone performing four audio tasks (i.e., speaker identification, emotion recognition, stress detection, and ambient scene analysis). Experiments suggest this proposal can achieve high efficiency in terms of energy, runtime and memory, while maintaining excellent accuracy.}\\


\noindent\edit{\textbf{Other applications.}} Deep learning also plays an important role in other applications that involve app-level data analysis. For instance, Ignatov \emph{et al.} show that deep learning can enhance the quality of pictures taken by mobile phones. By employing a CNN, they successfully improve the quality of images obtained by different mobile devices, to a digital single-lens reflex camera level \cite{ignatov2017dslr}. Lu \emph{et al.} focus on video post-processing under wireless networks \cite{lu2017demo}, where their framework exploits a customized AlexNet to answer queries about detected objects. This framework further involves an optimizer, which instructs mobile devices to offload videos, in order to reduce query response time. 

Another interesting application is presented in \cite{lee2016reducing}, where Lee \emph{et al.} show that deep learning can help smartwatch users reduce distraction by eliminating unnecessary notifications. Specifically, the authors use an 11-layer MLP to predict the importance of a notification.  Fang \emph{et al.} exploit an MLP to extract features from high-dimensional and heterogeneous sensor data, \edit{including accelerometer,  magnetometer,  and  gyroscope  measurements \cite{fang2017learning}. Their architecture achieves 95\%  accuracy in recognizing human transportation modes, i.e., still, walking, running, biking, and on vehicle.}\\

\noindent \edit{\textbf{Lessons Learned:} App-level data is heterogeneous and generated from distributed mobile devices, and there is a trend to offload the inference process to these devices. However, due to computational and battery power limitations, models employed in the edge-based scenario are constrained to light-weight architectures, which are less suitable for complex tasks. Therefore, the trade-off between model complexity and accuracy should be carefully considered \cite{mohammadi2018deep}. \rev{Numerous efforts were made towards tailoring deep learning to mobile devices, in order to make algorithms faster and less energy-consuming on embedded equipment. For example, model compression, pruning, and quantization are commonly used for this purpose. Mobile device manufacturers are also developing new software and hardware to support deep learning based applications. We will discuss this work in more detail in Sec.~\ref{sec:tailor}. }}

\edit{
At the same time, app-level data usually contains important users information and processing this poses significant privacy concerns. Although there have been efforts that commit to preserve user privacy, as we discuss in Sec.\ref{sec:security}, research efforts in this direction are new, especially in terms of protecting user information in distributed training. We expect more efforts in this direction in the future.}

\subsection{Deep Learning Driven Mobility Analysis}
\begin{figure}[htb]
\begin{center}
\includegraphics[width=0.5\textwidth]{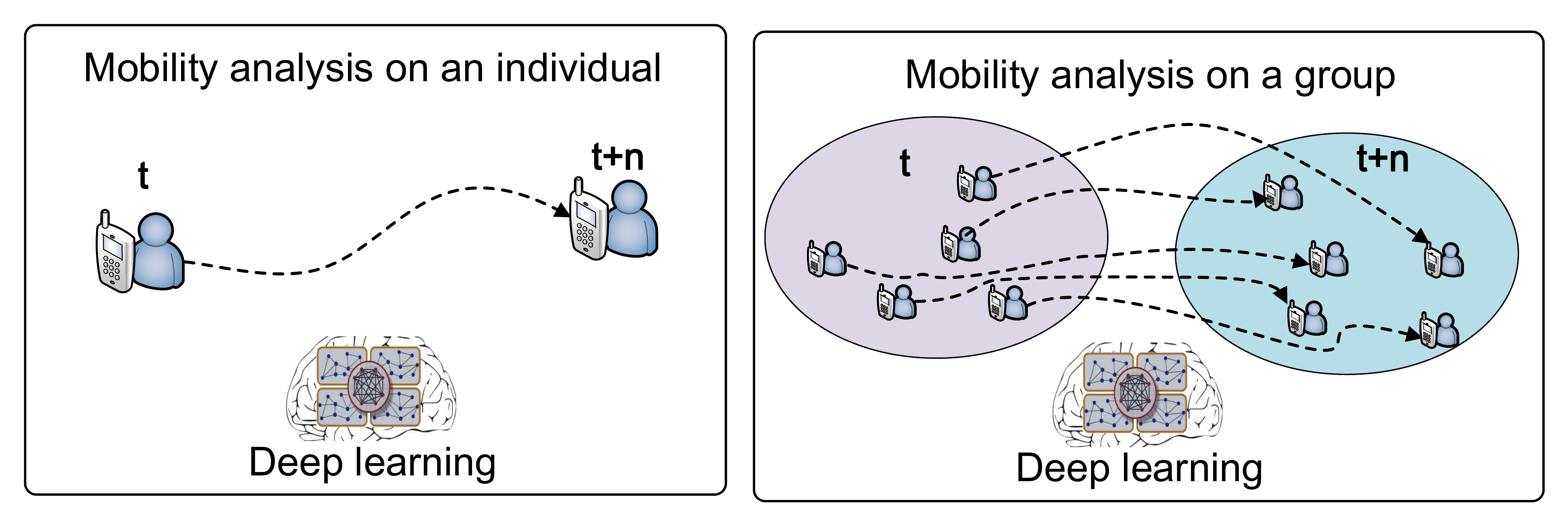}
\caption{\label{fig:mobility} \rev{Illustration of mobility analysis paradigms at individual (left) and group  (right) levels.} }
\end{center}
\end{figure}

Understanding movement patterns of groups of human beings and individuals is becoming crucial for epidemiology, urban planning, public service provisioning, and mobile network resource management \cite{zhao2016urban}. Deep learning is gaining increasing attention in this area, both from a group and individual level perspective (see  Fig.~\ref{fig:mobility}). In this subsection, we thus discuss research using deep learning in this space, which we summarize in Table \ref{tab:mobility}.

\begin{table*}[htb]
\centering
\caption{A summary of work on deep learning driven mobility analysis.}
\label{tab:mobility}
\begin{tabular}{|C{2cm}|C{3.5cm}|>{\color{black}}C{2cm}|C{2.5cm}|C{5.5cm}|}
\hline
\textbf{Reference}                                                            & \textbf{Application}                                            & \textbf{Mobility level} & \textbf{Model}                                     & \textbf{Key contribution}                                                                                                                          \\ \hline
Ouyang \emph{et al.}  \cite{ouyang2016deepspace}                              & Mobile user trajectory prediction                               & Individual              & CNN                                                & Online framework for data stream processing.                                                                                                        \\ \hline
Yang \emph{et al.} \cite{yang2017neural}                                      & Social networks and mobile trajectories modeling                & Mobile Ad-hoc Network   & RNN, GRU                                           & Multi-task learning.                                                                                                                                \\ \hline
\rev{Tka{\v{c}}{\'\i}k and Kord{\'\i}k \cite{tkavcik2016neural}} & \rev{Mobility modelling and prediction}                          & Individual              & \rev{Neural Turing Machine}                         & \rev{The Neural Turing Machine can store historical data and perform ``read'' and ``write'' operations automatically.}                               \\ \hline
Song \emph{et al.} \cite{song2016deeptransport}                               & City-wide mobility prediction and transportation modeling       & City-wide               & Multi-task LSTM                                    & Multi-task learning.                                                                                                                                \\ \hline
Zhang \emph{et al.} \cite{zhang2017deep123}                                   & City-wide crowd flows prediction                                & City-wide               & Deep spatio-temporal residual networks (CNN-based) & Exploitation of spatio-temporal characteristics of mobility events.                                                                                 \\ \hline
Lin \emph{et al.} \cite{lin2017deep}                                          & Human activity chains generation                                & User group              & Input-Output HMM + LSTM                            & Generative model.                                                                                                                                   \\ \hline
Subramanian and Sadiq \cite{subramanian2014implementation}                    & Mobile movement prediction                                      & Individual              & MLP                                                & Fewer location updates and lower paging signaling costs.                                                                                            \\ \hline
Ezema and Ani \cite{ezema2017artificial}                                      & Mobile location estimation                                      & Individual              & MLP                                                & Operates with received signal strength in GSM.                                                                                                     \\ \hline
\edit{Shao \emph{et al.} \cite{shao2018depedo}}                               & \edit{CNN driven Pedometer}                                     & Individual              & \edit{CNN}                                         & \edit{Reduced false negatives  caused by periodic movements and lower initial response time.}                                                       \\ \hline
\rev{Yayeh \emph{et al.} \cite{yayeh2018mobility}}                             & \rev{Mobility prediction in mobile Ad-hoc network}               & Individual              & \rev{MLP}                                           & \rev{Achieving high prediction accuracy under random waypoint mobility model.}                                                                         \\ \hline
\rev{Chen \emph{et al.} \cite{chen2016learning2}}                              & \rev{Mobility driven traffic accident risk prediction}           & City-wide               & \rev{Stacked denoising AE}                            & \rev{Automatically learning correlation between human mobility and traffic accident risk.}                                                              \\ \hline
\rev{Song \emph{et al.} \cite{song2017deepmob}}                                & \rev{Human emergency behavior and mobility modelling}            & City-wide               & \rev{DBN}                                           & \rev{Achieves accurate prediction over various disaster events, including earthquake, tsunami and nuclear accidents.}                                 \\ \hline
\rev{Yao \emph{et al.} \cite{yao2017trajectory}}                               & \rev{Trajectory clustering}                                    & User group              & \rev{sequence-to-sequence AE with RNNs}             & \rev{The learned representations can robustly encode the movement characteristics of the objects and generate spatio-temporally invariant clusters.} \\ \hline
\rev{Liu \emph{et al.} \cite{liu2018urban}}                                    & \rev{Urban traffic prediction}                                   & City-wide               & \rev{CNN, RNN, LSTM, AE, and RBM}                    & \rev{Reveals the potential of employing deep learning to urban traffic prediction with mobility data.}                                                 \\ \hline
\rev{Wickramasuriya \emph{et al.} \cite{wickramasuriya2017base}}               & \rev{Base station prediction with proactive mobility management} & Individual              & \rev{RNN}                                           & \rev{Employs proactive and   anticipatory mobility management for dynamic base station selection.}                                                    \\ \hline
\rev{Kim and Song \cite{kim2018method}}               & \rev{User mobility and personality modelling} & Individual              & \rev{MLP, RBM}                                           & \rev{Foundation for the customization of location-based services.}                                                    \\ \hline
\rev{Jiang \emph{et al.} \cite{jiang2018deepurbanmomentum}}               & \rev{Short-term urban mobility prediction} & City-wide              & \rev{RNN}                                           & \rev{Superior prediction accuracy and verified as highly deployable prototype system.}                                                    \\ \hline
\rev{Wang \emph{et al.} \cite{wang2018deep1231}}               & \rev{Mobility management in dense networks} & Individual              & \rev{LSTM}                                           & \rev{Improves the quality of service of mobile users in the handover process, while maintaining network energy efficiency.}                                                    \\ \hline
\rev{Jiang \emph{et al.} \cite{jiang2018deep}}               & \rev{Urban human mobility prediction} & City-wide              & \rev{RNN}                                           & \rev{First work that utilizes urban region of interest to model human mobility city-wide.}                                                    \\ \hline
\rev{Feng \emph{et al.} \cite{feng2018deepmove}}               & \rev{Human mobility forecasting} & Individual              & \rev{Attention RNN}                                           & \rev{Employs the attention mechanism and combines heterogeneous transition regularity and multi-level periodicity.}                                                    \\ \hline
\end{tabular}
\end{table*}

Since deep learning is able to capture spatial dependencies in sequential data, it is becoming a powerful tool for mobility analysis. The applicability of deep learning for trajectory prediction is studied in \cite{yang2017neural2}. By sharing representations learned by RNN and Gate Recurrent Unit (GRU), the framework can perform multi-task learning on both social networks and mobile trajectories modeling. Specifically, the authors first use deep learning to reconstruct social network representations of users, subsequently employing RNN and GRU models to learn patterns of mobile trajectories with different time granularity. Importantly, these two components jointly share representations learned, which tightens the overall architecture and enables efficient implementation. Ouyang \emph{et al.} argue that mobility data are normally high-dimensional, which may be problematic for traditional ML models. Therefore, they build upon deep learning advances and propose an online learning scheme to train a hierarchical CNN architecture, allowing model parallelization for data stream processing \cite{ouyang2016deepspace}. By analyzing usage records, their framework ``DeepSpace'' predicts individuals' trajectories with much higher accuracy as compared to naive CNNs, as shown with experiments on a real-world dataset. \rev{In \cite{tkavcik2016neural}, Tka{\v{c}}{\'\i}k and Kord{\'\i}k design a Neural Turing Machine \cite{graves2014neural} to predict trajectories of individuals using mobile phone data. The Neural Turing Machine embraces two major components: a memory module to store the historical trajectories, and a controller to manage the ``read'' and ``write'' operations over the memory. Experiments show that their architecture achieves superior generalization over stacked RNN and LSTM, while also delivering more precise trajectory prediction than the $n$-grams and $k$ nearest neighbor methods.}

Instead of focusing on individual trajectories, Song \emph{et al.} shed light on the mobility analysis at a larger scale \cite{song2016deeptransport}. In their work, LSTM networks are exploited to jointly model the city-wide movement patterns of a large group of people and vehicles. Their multi-task architecture demonstrates superior prediction accuracy over a standard LSTM. City-wide mobile patterns is also researched in \cite{zhang2017deep123}, where the authors architect deep spatio-temporal residual networks to forecast the movements of crowds. In order to capture the unique characteristics of spatio-temporal correlations associated with human mobility, their framework abandons RNN-based models and constructs three ResNets to extract nearby and distant spatial dependencies within a city. This scheme learns temporal features and fuses representations extracted by all models for the final prediction. By incorporating external events information, their proposal achieves the highest accuracy among all deep learning and non-deep learning methods studied. \rev{An RNN is also employed in \cite{jiang2018deepurbanmomentum}, where Jiang \emph{et al.} perform short-term urban mobility forecasting on a huge dataset collected from a real-world deployment. Their model delivers superior accuracy over the $n$-gram and Markovian approaches.}

Lin \emph{et al.} consider generating human movement chains from cellular data, to support transportation planning \cite{lin2017deep}. In particular, they first employ an input-output Hidden Markov Model (HMM) to label activity profiles for CDR data pre-processing. Subsequently, an LSTM is designed for activity chain generation, given the labeled activity sequences. They further synthesize urban mobility plans using the generative model and the simulation results reveal reasonable fit accuracy. \edit{Jiang \emph{et al.} design 24-h mobility prediction system base on RNN mdoels \cite{jiang2018deep}. They employ dynamic Region of Interests (ROIs) for each hour to discovered through divide-and-merge mining from raw trajectory database, which leads to high prediction accuracy. Feng \emph{et al.} incorporate attention mechanisms on RNN \cite{feng2018deepmove}, to capture the complicated sequential transitions of human mobility. By combining the heterogeneous transition regularity and multi-level periodicity, their model delivers up to 10\% of accuracy improvement compared to state-of-the-art forecasting models.}

\rev{Yayeh \emph{et al.} employ an MLP to predict the mobility of mobile devices in mobile ad-hoc networks, given previously observed pause time, speed, and movement direction \cite{yayeh2018mobility}. Simulations conducted using the random waypoint mobility model show that their proposal achieves high prediction accuracy. An MLP is also adopted in \cite{kim2018method}, where Kim and Song model the relationship between human mobility and personality, and achieve high prediction accuracy. Yao \emph{et al.} discover groups of similar trajectories to facilitate higher-level mobility driven applications using RNNs \cite{yao2017trajectory}. Particularly, a sequence-to-sequence AE is adopted to learn fixed-length representations of mobile users' trajectories. Experiments show that their method can effectively capture spatio-temporal patterns in both real and synthetic datasets. Shao \emph{et al.} design a sophisticated pedometer using a CNN \cite{shao2018depedo}. By reducing false negative steps caused by periodic movements, their proposal significantly improves the robustness of the pedometer.}

\rev{In \cite{chen2016learning2}, Chen \emph{et al.} combine GPS records and traffic accident data to understand the correlation between human mobility and traffic accidents. To this end, they design a stacked denoising AE to learn a compact representation of the human mobility, and subsequently use that to predict the traffic accident risk. Their proposal can deliver accurate, real-time prediction across large regions. GPS records are also used in other mobility-driven applications. Song \emph{et al.} employ DBNs to predict and simulate human emergency behavior and mobility in natural disaster, learning from GPS records of 1.6 million users \cite{song2017deepmob}. Their proposal yields accurate predictions in different disaster scenarios such as earthquakes, tsunamis, and nuclear accidents. GPS data is also utilized in \cite{liu2018urban}, where Liu \emph{et al.} study the potential of employing deep learning for urban traffic prediction using mobility data.
}\\

\noindent\textbf{Lessons learned:} Mobility analysis is concerned with the movement trajectory of a single user or large groups of users. The data of interest are essential time series, but have an additional spatial dimension. \rev{Mobility data is usually subject to stochasticity, loss, and noise; therefore precise modelling is not straightforward. As deep learning is able to perform automatic feature extraction, it becomes a strong candidate for human mobility modelling.} Among them, CNNs and RNNs are the most successful architectures in such applications (e.g., \cite{ouyang2016deepspace, yang2017neural, song2016deeptransport, zhang2017deep123, lin2017deep}), as they can effectively exploit spatial and temporal correlations.

\subsection{\rev{Deep Learning Driven User Localization}}
\begin{figure}[t]
\begin{center}
\includegraphics[width=0.45\textwidth]{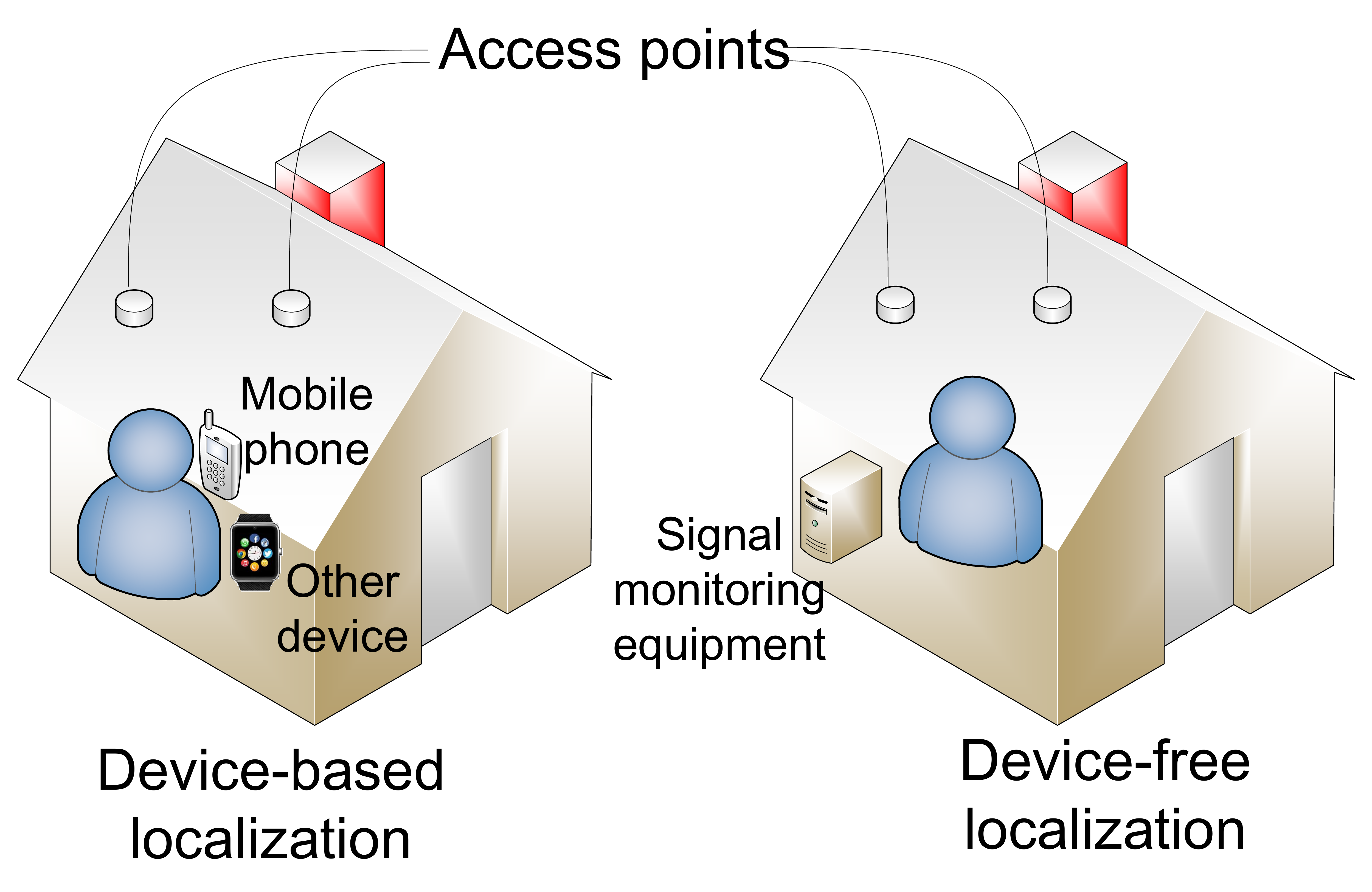}
\caption{\label{fig:localization} \rev{An illustration of device-based (left) and device-free (right) indoor localization systems.}}
\end{center}
\end{figure}

\begin{table*}[h!]
\centering
\caption{Leveraging Deep learning in user localization}
\label{tab:localization}
\begin{tabular}{|C{2cm}|C{3.5cm}|>{\color{black}}C{2cm}|C{3cm}|C{5cm}|}
\hline
\textbf{Reference}                                        & \textbf{Application}                       & \textbf{Input data}                   & \textbf{Model}                            & \textbf{Key contribution}                                                     \\ \hline
Wang \emph{et al.} \cite{wang2015deepfi}                  & Indoor fingerprinting                      & CSI                                   & RBM                                       & First deep learning driven indoor localization based on CSI                   \\ \hline
Wang \emph{et al.} \cite{wang2015phasefi, wang2016csi}    & Indoor localization                        & CSI                                   & RBM                                       & Works with calibrated phase information of CSI                                \\ \hline
Wang \emph{et al.} \cite{wang2017cifi}                    & Indoor localization                        & CSI                                   & CNN                                       & Uses more robust angle of arrival for estimation                              \\ \hline
Wang \emph{et al.} \cite{wang2017biloc}                   & Indoor localization                        & CSI                                   & RBM                                       & Bi-modal framework using both angle of arrival and average amplitudes of CSI  \\ \hline
Nowicki and Wietrzykowski \cite{nowicki2017low}           & Indoor localization                        & WiFi scans                            & Stacked AE                                & Requires less system tuning or filtering effort                               \\ \hline
Wang \emph{et al.} \cite{zhang2016device, wang2017device} & Indoor localization                        & Received signal strength              & Stacked AE                                & Device-free framework, multi-task learning                                    \\ \hline
Mohammadi\emph{et al.} \cite{mohammadi2017semi}           & Indoor localization                        & Received signal strength              & VAE+DQN                                   & Handles unlabeled data; reinforcement learning aided semi-supervised learning \\ \hline
\edit{Anzum \emph{et al.} \cite{anzum2018zone}}           & \edit{Indoor localization}                 & Received signal strength              & \edit{Counter propagation neural network} & \edit{Solves the ambiguity among zones}                                       \\ \hline
\edit{Wang \emph{et al.} \cite{wang2018deepml}}           & \edit{Indoor localization}                 & Smartphone magnetic and light sensors & \edit{LSTM}                               & \edit{Employ bimodal magnetic field and light intensity data}                 \\ \hline
Kumar \emph{et al.} \cite{kumar2016indoor}                & Indoor vehicles localization               & Camera images                         & CNN                                       & Focus on vehicles applications                                                \\ \hline
Zheng and Weng \cite{zhengj2016mobile}                    & Outdoor navigation                         & Camera images ad GPS                  & Developmental network                     & Online learning scheme; edge-based                                            \\ \hline
Zhang \emph{et al.} \cite{zhang2016deep3}                 & Indoor and outdoor localization            & Received signal strength              & Stacked AE                                & Operates under both indoor and outdoor environments                           \\ \hline
Vieira \emph{et al.} \cite{vieira2017deep}                & Massive MIMO fingerprint-based positioning & Associated channel fingerprint         & CNN                                       & Operates with massive MIMO channels                                           \\ \hline
\rev{Hsu \emph{et al.} \cite{Hsu:2017:ZIS:3139486.3130924}}                & \rev{Activity recognition, localization and sleep monitoring} & \rev{RF Signal}         & \rev{CNN}      & \rev{Multi-user, device-free localization and sleep monitoring }                                          \\ \hline
\rev{Wang \emph{et al.} \cite{wang2017csi}}                & \rev{Indoor localization} & \rev{CSI}         & \rev{RBM}      & \rev{Explores features of wireless channel data
                                    and obtains optimal weights as fingerprints} \\ \hline
\rev{Wang \emph{et al.} \cite{wang2018deep6}}                & \rev{Indoor localization} & \rev{CSI}         & \rev{CNN}      & \rev{Exploits the angle of arrival for stable indoor localization} \\ \hline
\rev{Xiao \emph{et al.} \cite{xiao20173}}                & \rev{3D Indoor localization} & \rev{Bluetooth relative received signal strength}         & \rev{Denoising AE}      & \rev{Low-cost and robust localization} \\ \hline
\rev{Niitsoo \emph{et al.} \cite{niitsoo2018convolutional}}                & \rev{Indoor localization} & \rev{Raw channel impulse response data}         & \rev{CNN}      & \rev{Robust to multipath propagation environments} \\ \hline
\rev{Ibrahim \emph{et al.} \cite{ibrahim2018cnn}}                & \rev{Indoor localization} & \rev{Received signal strength}         & \rev{CNN}      & \rev{100\% accuracy in building and floor identification} \\ \hline
\rev{Adege \emph{et al.} \cite{adege2018applying}}                & \rev{Indoor localization} & \rev{Received signal strength}         & \rev{MLP + linear discriminant analysis}      & \rev{Accurate in multi-building environments} \\ \hline
\rev{Zhang \emph{et al.} \cite{zhang2017deeppositioning}}                & \rev{Indoor localization} & \rev{Pervasive magnetic field and CSI}         & \rev{MLP}      & \rev{Using the magnetic field to improve the positioning accuracy} \\ \hline
\rev{Zhou \emph{et al.} \cite{zhou2018device}}                & \rev{Indoor localization} & \rev{CSI}         & \rev{MLP}      & \rev{Device-free localization } \\ \hline
\rev{Shokry \emph{et al.} \cite{shokry2018deeploc}}                & \rev{Outdoor localization} & \rev{Crowd-sensed received signal strength information}         & \rev{MLP}      & \rev{More accurate than cellular localization while requiring less energy} \\ \hline 
\rev{Chen \emph{et al.} \cite{chen2017confi}}                & \rev{Indoor localization} & \rev{CSI}         & \rev{CNN}      & \rev{Represents CSI as feature images} \\ \hline
\rev{Guan \emph{et al.} \cite{guan2017high}}                & \rev{Indoor localization} & \rev{Line-of-sight and non line-of-sight radio signals}         & \rev{MLP}      & \rev{Combining deep learning with genetic
algorithms} \\ \hline
\end{tabular}
\end{table*}
Location-based services and applications (e.g. mobile AR, GPS) demand precise individual positioning technology \cite{xia2017indoor}. As a result, research on user localization is evolving rapidly and numerous techniques are emerging \cite{davidson2016survey}. \rev{In general, user localization methods can be categorized as device-based and device-free \cite{xiao2016survey}. We illustrate the two different paradigms in Fig.~\ref{fig:localization}. Specifically, in the first category specific devices carried by users become prerequisites for fulfilling the applications' localization function. This type of approaches rely on signals from the device to identify the location. Conversely, approaches that require no device pertain to the device-free category. Instead these employ special equipment to monitor signal changes, in order to localize the entities of interest. Deep learning can enable high localization accuracy with both paradigms. We summarize the most notable contributions in Table~\ref{tab:localization} and delve into the details of these works next.} 

To overcome the variability and coarse-granularity limitations of signal strength based methods, Wang \emph{et al.}  propose a deep learning driven fingerprinting  system name ``DeepFi'' to perform indoor localization based on Channel State Information (CSI) \cite{wang2015deepfi}. Their toolbox yields much higher accuracy as compared to traditional methods, including including FIFS \cite{xiao2012fifs}, Horus \cite{youssef2005horus}, and Maximum  Likelihood \cite{brunato2005statistical}. The same group of authors extend their work in \cite{wang2015phasefi, wang2016csi} and \cite{wang2017cifi, wang2017biloc}, where they update the localization system, such that it can work with calibrated phase information of CSI \cite{wang2015phasefi, wang2016csi, wang2017csi}. They further use more sophisticated CNN \cite{wang2017cifi, wang2018deep6} and bi-modal structures \cite{wang2017biloc} to improve the accuracy. 

Nowicki and Wietrzykowski propose a localization framework that reduces significantly the effort of system tuning or filtering and obtains satisfactory prediction performance \cite{nowicki2017low}. Wang \emph{et al.} suggest that the objective of indoor localization can be achieved without the help of mobile devices. In \cite{wang2017device}, the authors employ an AE to learn useful patterns from WiFi signals. By automatic feature extraction, they produce a predictor that can fulfill multi-tasks simultaneously, including indoor localization, activity, and gesture recognition. \rev{A similar work in presented in \cite{zhou2018device}, where Zhou \emph{et al.} employ an MLP structure to perform device-free indoor localization using CSI.} Kumar \emph{et al.} use deep learning to address the problem of indoor vehicles localization \cite{kumar2016indoor}. They employ CNNs to analyze visual signal and localize vehicles in a car park. This can help driver assistance systems operate in underground environments where the system has limited vision ability. 

\rev{In \cite{xiao20173}, Xiao \emph{et al.} achieve low cost indoor localization with Bluetooth technology. The authors design a denosing AE to extract fingerprint features from the received signal strength of Bluetooth Low Energy beacon and subsequently project that to the exact position in 3D space. Experiments conducted in a conference room demonstrate that the proposed framework can perform precise positioning in both vertical and horizontal dimensions in real-time. Niitsoo \emph{et al.} employ a CNN to perform localization given raw channel impulse response data \cite{niitsoo2018convolutional}. Their framework is robust to multipath propagation environments and more precise than signal processing based approaches. A CNN is also adopted in \cite{ibrahim2018cnn}, where the authors work with received signal strength series and achieve 100\% prediction accuracy in terms of building and floor identification. The work in \cite{adege2018applying} combines deep learning with linear discriminant analysis for feature reduction, achieving low positioning errors in multi-building environments. Zhang \emph{et al.} combine pervasive magnetic field and WiFi fingerprinting for indoor localization using an MLP \cite{zhang2017deeppositioning}. Experiments show that adding magnetic field information to the input of the model can improve the prediction accuracy, compared to solutions based soley on WiFi fingerprinting.}

\begin{figure}[t]
\begin{center}
\includegraphics[width=0.5\textwidth]{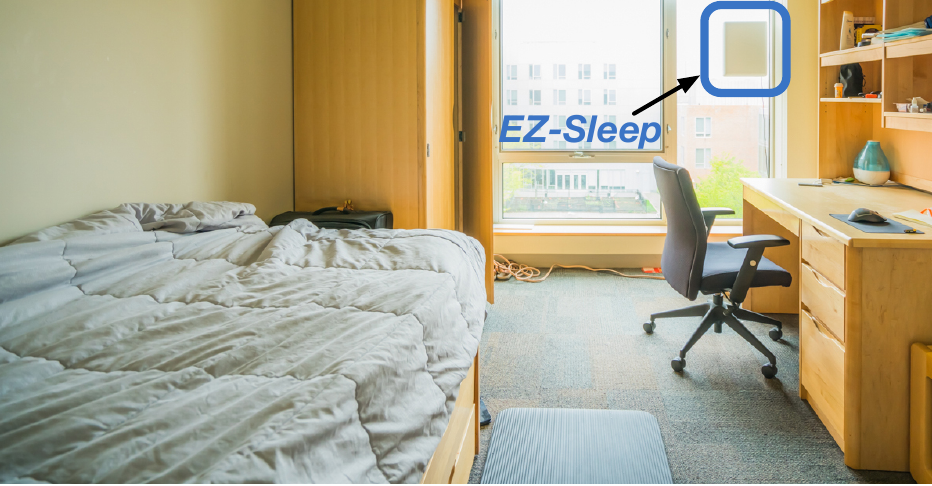}
\caption{\label{fig:ezsleep} \rev{EZ-Sleep setup in a subject's bedroom. Figure adopted from~\cite{Hsu:2017:ZIS:3139486.3130924}.}}
\end{center}
\end{figure}

\rev{Hsu \emph{et al.} use deep learning to provide Radio Frequency-based user localization, sleep monitoring, and insomnia analysis in multi-user home scenarios where individual sleep monitoring devices might not be available~\cite{Hsu:2017:ZIS:3139486.3130924}. They use a CNN classifier with a 14-layer residual network model for sleep monitoring, in addition to Hidden Markov Models, to accurately track when the user enters or leaves the bed. By deploying sleep sensors called EZ-Sleep in 8 homes (see Fig.~\ref{fig:ezsleep}), collecting data for 100 nights of sleep over a month, and cross-validating this using an electroencephalography-based sleep monitor, the authors demonstrate the performance of their solution is comparable to that of individual, electroencephalography-based devices. }

Most mobile devices can only produce unlabeled position data, therefore unsupervised and semi-supervised learning become essential. Mohammadi \emph{et al.} address this problem by leveraging DRL and VAE. In particular, their framework envisions a virtual agent in indoor environments \cite{mohammadi2017semi}, which can constantly receive state information during training, including signal strength indicators, current agent location, and the real (labeled data) and inferred (via a VAE) distance to the target. The agent can virtually move in eight directions at each time step. Each time it takes an action, the agent receives an reward signal, identifying whether it moves to a correct direction. By employing deep Q learning, the agent can finally localize accurately a user, given both labeled and unlabeled data. 

Beyond indoor localization, there also exist several research works that apply deep learning in outdoor scenarios. For example, Zheng and Weng introduce a lightweight developmental network for outdoor navigation applications on mobile devices \cite{zhengj2016mobile}. Compared to CNNs, their architecture requires 100 times fewer weights to be updated, while maintaining decent accuracy. This enables efficient outdoor navigation on mobile devices. Work in \cite{zhang2016deep3} studies localization under both indoor and outdoor environments. They use an AE to pre-train a four-layer MLP, in order to avoid hand-crafted feature engineering. The MLP is subsequently used to estimate the coarse position of targets. The authors further introduce an HMM to fine-tune the predictions based on temporal properties of data. This improves the accuracy  estimation in both in-/out-door positioning with Wi-Fi signals. \rev{More recently, Shokry \emph{et al.} propose DeepLoc, a deep learning-based outdoor localization system using crowdsensed geo-tagged received signal strength information \cite{shokry2018deeploc}. By using an MLP to learn the correlation between cellular signal and users' locations, their framework can deliver median localization accuracy within 18.8m in urban areas and within 15.7m in rural areas on Android devices, while requiring modest energy budgets.}\\

\noindent\edit{\textbf{Lessons learned:} Localization relies on sensorial output, signal strength, or CSI.  These data usually have complex features, therefore large amounts of data are required for learning \cite{nowicki2017low}. As deep learning can extract features in an unsupervised manner, it has become a strong candidate for localization tasks.} \rev{On the other hand, it can be observed that positioning accuracy and system robustness can be improved by fusing multiple types of signals when providing these as the input (see, e.g., \cite{zhang2017deeppositioning}). Using deep learning to automatically extract features and correlate information from different sources for localization purposes is becoming a trend.}

\subsection{Deep Learning Driven Wireless Sensor Networks}
\begin{table*}[h!]
\centering
\caption{A summary of work on deep learning driven WSNs.}
\label{tab:wsn}
\begin{tabular}{|c|C{2CM}|C{3CM}|C{1.5CM}|C{2CM}|C{3.8CM}|}
\hline
\rev{\textbf{Perspective}}                            & \textbf{Reference}                                           & \textbf{Application}                                                                                   & \textbf{Model}                               & \textbf{Optimizer}                                            & \textbf{Key contribution}                                                                                                                           \\ \hline
\multirow{2}{*}{\rev{Centralized vs. Decentralized}} & Khorasani and Naji \cite{khorasani2017energy}                & Data aggregation                                                                                       & MLP                                          & Unknown                                                       & Improves the energy efficiency in the aggregation process                                                                                           \\ \cline{2-6} 
                                                     & Li \emph{et al.} \cite{li2015distributed}                    & Distributed data mining                                                                                & MLP                                          & Unknown                                                       & Performing data analysis at distributed nodes, reducing by 58.31\% the energy consumption                                                           \\ \hline
\multirow{7}{*}{\rev{WSN Localization}}               & Bernas and P{\l}aczek \cite{bernas2015fully}                 & Indoor localization                                                                                    & MLP                                          & Resilient backpropagation                                     & Dramatically reduces the memory consumption of received signal strength map storage                                                                 \\ \cline{2-6} 
                                                     & Payal \emph{et al.} \cite{payal2015analysis}                 & Node localization                                                                                      & MLP                                          & First-order and second-order gradient descent algorithm       & Compares different training algorithms of MLP for WSN localization                                                                                  \\ \cline{2-6} 
                                                     & Dong \emph{et al.} \cite{dong2017range}                      & Underwater localization                                                                                & MLP                                          & RMSprop                                                       & Performs WSN localization in underwater environments                                                                                                \\ \cline{2-6} 
                                                     & \rev{Phoemphon \emph{et al.} \cite{phoemphon2018hybrid}}      & \rev{WSN node localization}                                                                             & \rev{Logic fuzzy system + ELM}                & \rev{Algorithm 4 described in \cite{phoemphon2018hybrid}}      & \rev{Combining logic fuzzy system and ELM to achieve robust range-free WSN node localization}                                                        \\ \cline{2-6} 
                                                     & \rev{Banihashemian \emph{et al.} \cite{banihashemian2018new}} & \rev{Range-free WSN node localization}                                                                  & \rev{MLP, ELM}                                & \rev{Conjugate gradient based method }                         & \rev{Employs a particle swarm optimization algorithm to simultaneously optimize the neural network based on storage cost and localization accuracy} \\ \cline{2-6} 
                                                     & \rev{Kang \emph{et al.} \cite{kang2018novel}}                 & \rev{Water leakage detection and localization}                                                          & \rev{CNN + SVM}                               & \rev{SGD}                                                      & \rev{Propose an enhanced graph-based local search algorithm using a virtual node scheme to select the nearest leakage location}            \\ \cline{2-6} 
                                                     & \rev{El \emph{et al.} \cite{el2016robust}}                    & \rev{WSN localization in the presence of anisotropic signal attenuation}                                & \rev{MLP}                                     & \rev{Unknown}                                                  & \rev{Robust against anisotropic signal attenuation}                                                                                              \\ \hline
\multirow{7}{*}{\rev{WSN Data Analysis}}              & Yan \emph{et al.} \cite{yan2016real}                         & Smoldering and flaming combustion identification                                                       & MLP                                          & SGD                                                           & Achieves high accuracy in detecting fire in forests using smoke, $CO_2$ and temperature sensors                                                     \\ \cline{2-6} 
                                                     & Wang \emph{et al.} \cite{wang2017temperature}                & Temperature correction                                                                                 & MLP                                          & SGD                                                           & Employs deep learning to learn correlation between polar radiation and air temperature error                                                        \\ \cline{2-6} 
                                                     & Lee \emph{et al.} \cite{lee2017deep222}                      & Online query processing                                                                                & CNN                                          & Unknown                                                       & Employs adaptive query refinement to enable real-time analysis                                                                                      \\ \cline{2-6} 
                                                     & Li and Serpen \cite{li2016adaptive}                          & Self-adaptive WSN                                                                                      & Hopfield network                             & Unknown                                                       & Embedding Hopfield NNs as a static optimizer for the weakly-connected dominating problem                                                            \\ \cline{2-6} 
                                                     & Khorasani and Naji \cite{khorasani2017energy}                & Data aggregation                                                                                       & MLP                                          & Unknown                                                       & Improves the energy efficiency in the aggregation process                                                                                           \\ \cline{2-6} 
                                                     & Li \emph{et al.} \cite{li2015distributed}                    & Distributed data mining                                                                                & MLP                                          & Unknown                                                       & Distributed data mining                                                                                                                             \\ \cline{2-6} 
                                                     & Luo and Nagarajany \cite{luo2018distributed}                 & Distributed WSN anomaly detection                                                                      & AE                                           & SGD                                                           & Employs distributed anomaly detection techniques to offload computations from the cloud                                                             \\ \hline
\multirow{4}{*}{\rev{Other}}                              & \rev{Heydari \emph{et al.} \cite{heydari2017reduce}}          & \rev{Energy consumption optimization and secure communication in wireless multimedia sensor networks} & \rev{Stacked AE}                              & \rev{Unknown}                                                  & \rev{Use deep learning to enable fast data transfer and reduce energy consumption}                                                                   \\ \cline{2-6} 
                                                     & \rev{Mehmood \emph{et al.} \cite{mehmood2017eldc}}            & \rev{Robust routing for pollution monitoring in WSNs}                                            & \rev{MLP}                                     & \rev{SGD}                                                      & \rev{Highly energy-efficient}                                                                                                                        \\ \cline{2-6} 
                                                     & \rev{Alsheikh \emph{et al.} \cite{alsheikh2016rate}}          & \rev{Rate-distortion balanced data compression for WSNs}                                                & \rev{AE}                                      & \rev{Limited memory Broyden Fletcher Goldfarb Shann algorithm} & \rev{Energy-efficient and bounded reconstruction errors}                                                                              \\ \cline{2-6} 
                                                     & \rev{Wang \emph{et al.} \cite{wang2017deep3}}                 & \rev{Blind drift calibration for WSNs}                                                                  & \rev{Projection-recovery network (CNN based)} & \rev{Adam}                                                     & \rev{Exploits spatial and temporal correlations of data from all sensors; first work that adopts deep learning in WSN data calibration}       \\ \cline{2-6} 
                                                     & \rev{Jia \emph{et al.} \cite{jia2018continuous}}                 & \rev{Ammonia monitoring}                                                                  & \rev{LSTM} & \rev{Adam}                                                     & \rev{Low-power and accurate ammonia monitoring}       \\\hline
\end{tabular}
\end{table*}

\begin{figure}[t]
\begin{center}
\includegraphics[width=0.5\textwidth]{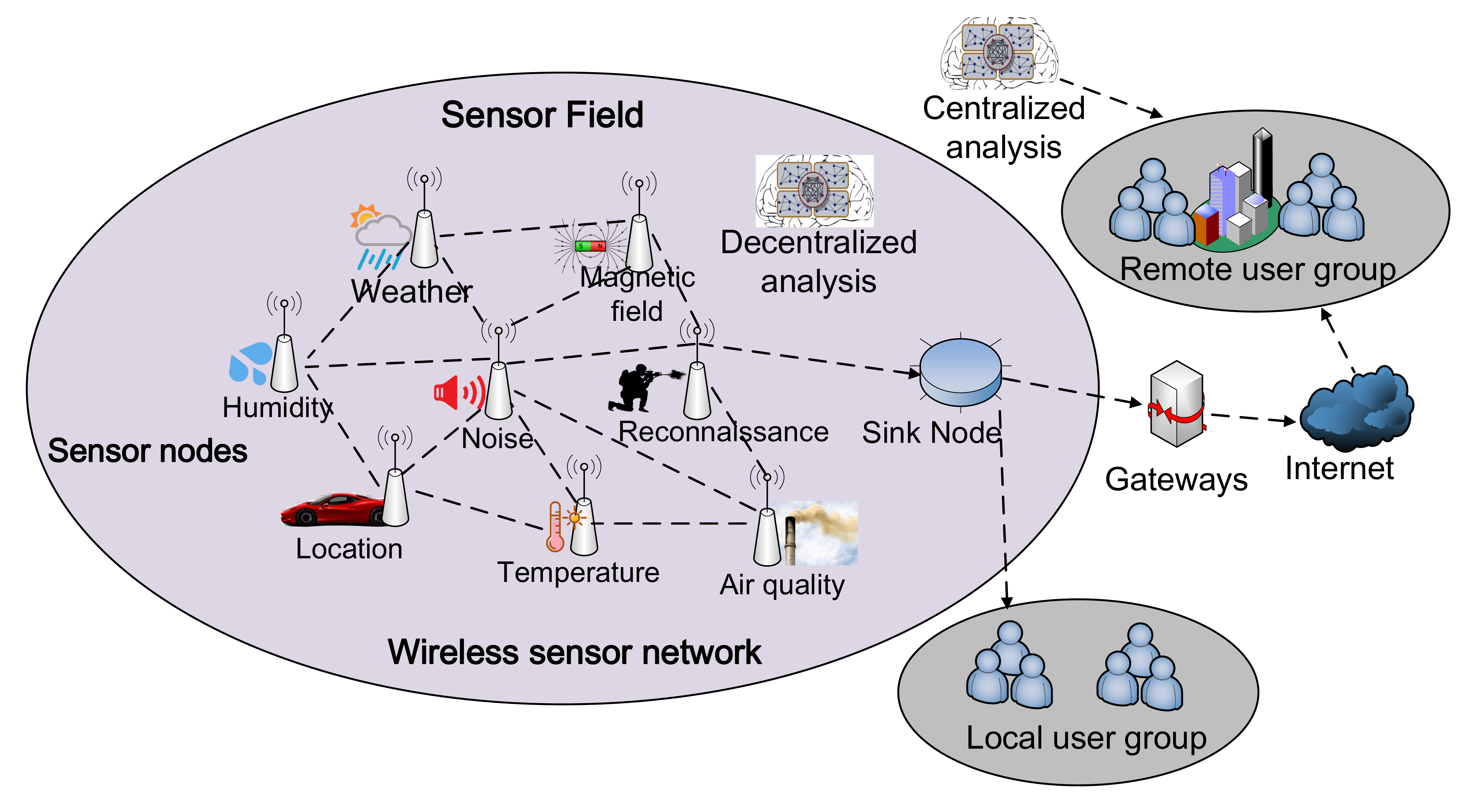}
\caption{\label{fig:wsn} \rev{An example framework for WSN data collection and (centralized and decentralized) analysis.}}
\end{center}
\end{figure}

Wireless Sensor Networks (WSNs) consist of a set of unique or heterogeneous sensors that are distributed over geographical regions. Theses sensors collaboratively monitor physical or environment status (e.g. temperature, pressure, motion, pollution, etc.) and transmit the data collected to centralized servers through wireless channels (see top circle in Fig. \ref{fig:collection} for an illustration). A WSN typically involves three key core tasks, namely sensing, communication and analysis. 
Deep learning is becoming increasingly popular also for WSN applications \cite{kumar2019machine}. In what follows, \rev{we review works adopting deep learning in this domain, covering different angles, namely: centralized vs. decentralized analysis paradigms, WSN data analysis per se, WSN localization, and other applications}. Note that the contributions of these works are distinct from mobile data analysis discussed in subsections \ref{sec:netdata} and \ref{sec:appdata}, as in this subsection we only focus on WSN applications.  We begin by summarizing the most important works in Table \ref{tab:wsn}.

\rev{\noindent \textbf{Centralized vs Decentralized Analysis Approaches:}}
There exist two data processing scenarios in WSNs, namely centralized and decentralized. The former simply takes sensors as data collectors, which are only responsible for gathering data and sending these to a central location for processing. The latter assumes sensors have some computational ability and the main server offloads part of the jobs to the edge, each sensor performing data processing individually. 
\rev{We show an example framework for WSN data collection and analysis in Fig.~\ref{fig:wsn}, where sensor data is collected via various nodes in a field of interest. Such data is delivered to a sink node, which aggregates and optionally further processes this.} 
Work in \cite{khorasani2017energy} focuses on the centralized approach and the authors apply a 3-layer MLP to reduce data redundancy while maintaining essential points for data aggregation. These data are sent to a central server for analysis. In contrast, Li \emph{et al.} propose to distribute data mining to individual sensors \cite{li2015distributed}. They partition a deep neural network into different layers and offload layer operations to sensor nodes. Simulations conducted suggest that, by pre-processing with NNs, their framework obtains high fault detection accuracy, while reducing power consumption at the central server. \\

\rev{\noindent \textbf{WSN Localization:}}
Localization is also an important and challeging task in WSNs. Chuang and Jiang exploit neural networks to localize sensor nodes in WSNs \cite{chuang2014effective}. To adapt deep learning models to specific network topology, they employ an online training scheme and correlated topology-trained data, enabling efficient model implementations and accurate location estimation. Based on this, Bernas and P{\l}aczek architect an ensemble system that involves multiple MLPs for location estimation in different regions of interest \cite{bernas2015fully}. In this scenario, node locations inferred by multiple MLPs are fused by a fusion algorithm, which improves the localization accuracy, particularly benefiting sensor nodes that are around the boundaries of regions. A comprehensive comparison of different training algorithms that apply MLP-based node localization is presented in \cite{payal2015analysis}. Experiments suggest that the Bayesian regularization algorithm in general yields the best performance. Dong \emph{et al.} consider an underwater node localization scenario \cite{dong2017range}. Since acoustic signals are subject to loss caused by absorption, scattering, noise, and interference, underwater localization is not straightforward. By adopting a deep neural network, their framework successfully addresses the aforementioned challenges and achieves higher inference accuracy as compared to SVM and generalized least square methods.

Phoemphon \emph{et al.} \cite{phoemphon2018hybrid} combine a fuzzy logic system and an ELM via a particle swarm optimization technique to achieve robust range-free location estimation for sensor nodes. In particular, the fuzzy logic system is employed for adjusting the weight of traditional centroids, while the ELM is used for optimization for the localization precision. Their method achieves superior accuracy over other soft computing-based approaches. Similarly, Banihashemian \emph{et al.} employ the particle swarm optimization technique combining with MLPs to perform range-free WSN localization, which achieves low localization error \cite{banihashemian2018new}. Kang \emph{et al.} shed light water leakage and localization in water distribution systems \cite{kang2018novel}. They represent the water pipeline network as a graph and assume leakage events occur at vertices. They combine CNN with SVM to perform detection and localization on wireless sensor network testbed, achieving 99.3\% leakage detection accuracy and localization error for less than 3 meters.\\

\rev{\noindent \textbf{WSN Data Analysis:}} Deep learning has also been exploited for identification of smoldering and flaming combustion phases in forests. In \cite{yan2016real}, Yan \emph{et al.} embed a set of sensors into a forest to monitor CO$_2$, smoke, and temperature.  They suggest that various burning scenarios will emit different gases, which can be taken into account when classifying smoldering and flaming combustion. Wang \emph{et al.} consider deep learning to correct inaccurate measurements of air temperature \cite{wang2017temperature}. They discover a close relationship between solar radiation and actual air temperature, which can be effectively learned by neural networks. \rev{In \cite{sun2017wnn}, Sun \emph{et al.} employ a Wavelet neural network based solution to evaluate radio link quality in WSNs on smart grids. Their proposal is more precise than traditional approaches and can provide end-to-end reliability guarantees to smart grid applications. }

Missing data or de-synchronization are common in WSN data collection. These may lead to serious problems in analysis due to inconsistency. Lee \emph{et al.} address this problem by plugging a query refinement component in deep learning based WSN analysis systems \cite{lee2017deep222}. They employ exponential smoothing to infer missing data, thereby maintaining the integrity of data for deep learning analysis without significantly compromising  accuracy. To enhance the intelligence of WSNs, Li and Serpen embed an artificial neural network into a WSN, allowing it to agilely react to potential changes and following deployment in the field~\cite{li2016adaptive}. To this end, they employ a minimum weakly-connected dominating set to represent the WSN topology, and subsequently use a Hopfield recurrent neural network as a static optimizer, to adapt network infrastructure to potential changes as necessary. This work represents an important step towards embedding machine \mbox{intelligence in WSNs}. \\

\rev{\noindent \textbf{Other Applications:}} \rev{The benefits of deep learning have also been demonstrated in other WSN applications. The work in \cite{heydari2017reduce} focuses on reducing energy consumption while maintaining security in wireless multimedia sensor networks. A stacked AE is employed to categorize images in the form of continuous pieces, and subsequently send the data over the network. This enables faster data transfer rates and lower energy consumption. Mehmood \emph{et al.} employ MLPs to achieve robust routing in WSNs, so as to facilitate  pollution monitoring \cite{mehmood2017eldc}. Their proposal use the NN to provide an efficiency threshold value and switch nodes that consume less energy than this threshold, thereby improving energy efficiency. Alsheikh \emph{et al.} introduce an algorithm for WSNs that uses AEs to minimize the energy expenditure \cite{alsheikh2016rate}. Their architecture exploits spatio-temporal correlations to reduce the dimensions of raw data and provides reconstruction error bound guarantees.}

\rev{Wang \emph{et al.} design a dedicated projection-recovery neural network to blindly calibrate sensor measurements in an online manner \cite{wang2017deep3}. Their proposal can automatically extract features from sensor data and exploit spatial and temporal correlations among information from all sensors, to achieve high accuracy. This is the first effort that adopts deep learning in WSN data calibration. Jia \emph{et al.} shed light on ammonia monitoring using deep learning \cite{jia2018continuous}. In their design, an LSTM is employed to predict the sensors' electrical resistance during a very short heating pulse, without waiting for settling in an equilibrium state. This dramatically reduces the energy consumption of sensors in the waiting process. Experiments with 38 prototype sensors and a home-built gas flow system show that the proposed LSTM can deliver precise prediction of equilibrium state resistance under different ammonia concentrations, cutting down the overall energy consumption by approximately 99.6\%.}


\textbf{Lessons learned:} \rev{The centralized and decentralized WSN data analysis paradigms resemble the cloud and fog computing philosophies in other areas. Decentralized methods exploit the computing ability of sensor nodes and perform light processing and analysis locally. This offloads the burden on the cloud and significantly reduces the data transmission overheads and storage requirements. However, at the moment, the centralized approach dominates the WSN data analysis landscape. As deep learning implementation on embedded devices becomes more accessible, in the future we expect to witness a grow in the popularity of the decentralized schemes.}

On the other hand, looking at Table \ref{tab:wsn}, it is interesting to see that the majority of deep learning practices in WSNs employ MLP models. Since MLP is straightforward to architect and performs reasonably well, it remains a good candidate for WSN applications. However, since most sensor data collected is sequential, we expect RNN-based models will play a more important role in this area. 

\subsection{Deep Learning Driven Network Control \label{sec:control}}
In this part, we turn our attention to mobile network control problems. Due to powerful function approximation mechanism, deep learning has made remarkable breakthroughs in improving traditional reinforcement learning \cite{kai2017brief} and imitation learning \cite{ho2016generative}. These advances have potential to solve mobile network control problems which are complex and previously considered intractable \cite{zorzi2016cobanets, roopaei2017deep}. Recall that in reinforcement learning, an agent continuously interacts with the environment to learn the best action. With constant exploration and exploitation, the agent learns to maximize its expected return. Imitation learning follows a different learning paradigm called ``learning by demonstration''. This learning paradigm relies on a `teacher' who tells the agent what action should be executed under certain observations during the training. After sufficient demonstrations, the agent learns a policy that imitates the behavior of the teacher and can operate standalone without supervision. \edit{For instance, an agent is trained to mimic human behaviour (e.g., in applications such as game play, self-driving vehicles, or robotics), instead of learning by interacting with the environment, as in the case of pure reinforcement learning. This is because in such applications, making mistakes can have fatal consequences~\cite{hussein2017imitation}.}

\begin{figure}[t]
\begin{center}
\includegraphics[width=0.5\textwidth]{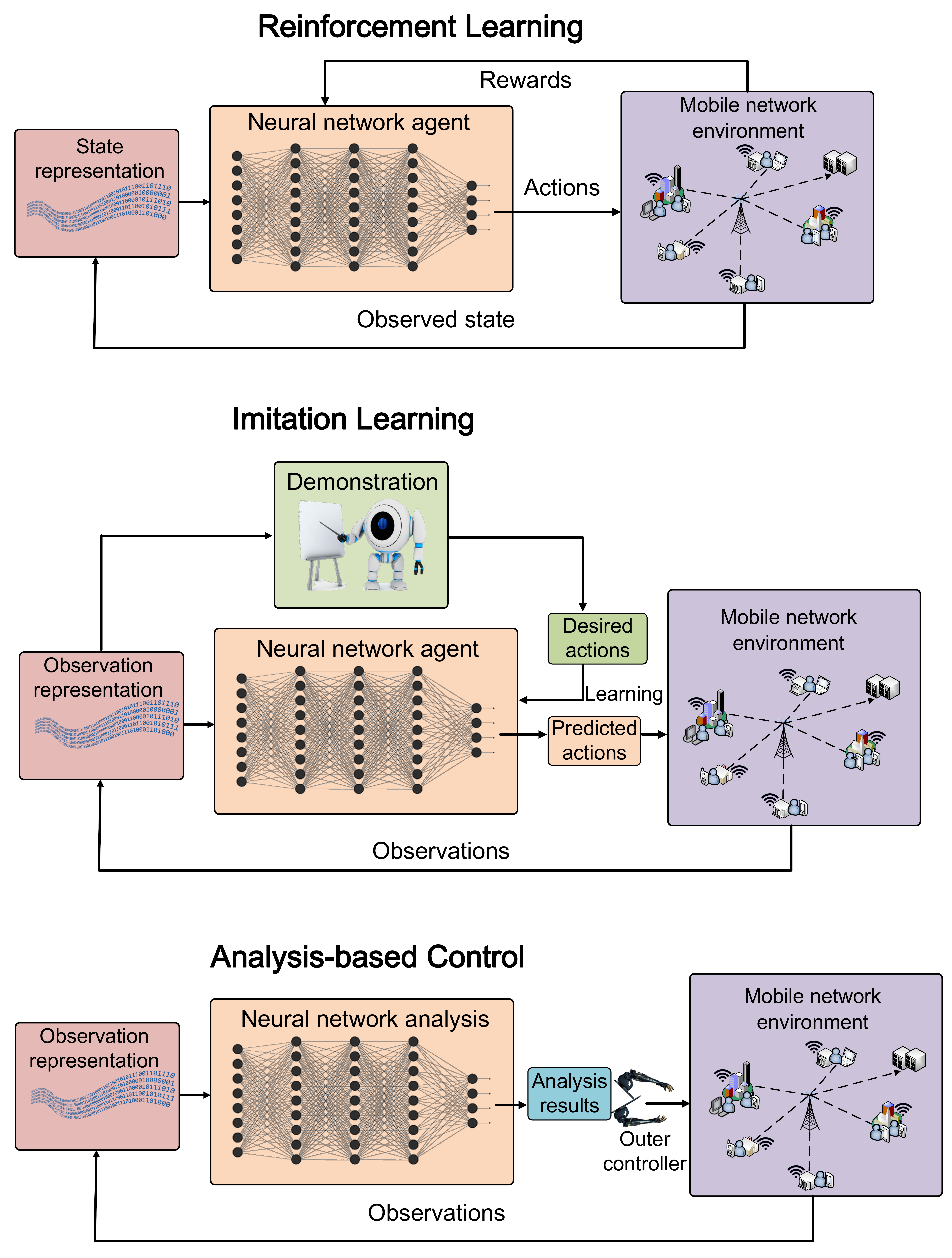}
\end{center}
\caption{\label{fig:control} Principles of three control approaches applied in mobile and wireless networks control, namely reinforcement learning (above), imitation learning (middle), and analysis-based control (below).}
\end{figure}

\begin{table*}[h!]
\centering
\caption{A summary of work on deep learning driven network control.}
\label{tab:control}
\begin{tabular}{|c|C{3CM}|C{5CM}|c|C{3CM}|}
\hline
\textbf{Domain}                       & \textbf{Reference}                                                & \textbf{Application}                                                & \textbf{Control approach}   & \textbf{Model}                       \\ \hline
\multirow{7}{*}{Network optimization} & Liu \emph{et al.} \cite{liu2017deep222}                           & Demand constrained energy minimization                              & Analysis-based              & DBN                                  \\ \cline{2-5} 
                                      & Subramanian and Banerjee \cite{subramanian2016poster}             & Machine to machine system optimization                              & Analysis-based              & Deep multi-modal network             \\ \cline{2-5} 
                                      & He \emph{et al.} \cite{he2017optimization, he2017deep3}           & Caching and interference alignment                                  & Reinforcement learning      & Deep Q learning                      \\ \cline{2-5} 
                                      & Masmar and Evans \cite{mismar2017deep}                            & mmWave Communication performance optimization                       & Reinforcement learning      & Deep Q learning                      \\ \cline{2-5} 
                                      & \edit{Wang \emph{et al.} \cite{wang2018handover}}                   & \edit{Handover optimization in wireless systems}                      & \edit{Reinforcement learning} & \edit{Deep Q learning}                 \\ \cline{2-5} 
                                      & \edit{Chen and Smith \cite{chen2018heterogeneous}}                  & \edit{Cellular network random access optimization}                   & \edit{Reinforcement learning} & \edit{Deep Q learning}                 \\ \cline{2-5} 
                                      & \edit{Chen \emph{et al.} \cite{chen2018auto}}                       & \multicolumn{1}{l|}{\edit{Automatic traffic optimization}}            & \edit{Reinforcement learning} & \edit{Deep policy gradient}            \\ \hline
\multirow{4}{*}{Routing}              & Lee \emph{et al.} \cite{lee2017classification}                    & Virtual route assignment                                            & Analysis-based              & MLP                                  \\ \cline{2-5} 
                                      & Yang \emph{et al.} \cite{yang2017neural}                          & Routing optimization                                                & Analysis-based              & Hopfield neural networks             \\ \cline{2-5} 
                                      & Mao \emph{et al.} \cite{mao2017routing}                           & Software defined routing                                            & Imitation learning          & DBN                                  \\ \cline{2-5} 
                                      & Tang \emph{et al.} \cite{tang2017removing}                        & Wireless network routing                                            & Imitation learning          & CNN                                  \\ \cline{2-5} 
                                      & \rev{Mao \emph{et al.} \cite{mao2017tensor}}                        & \rev{Intelligent packet routing}                                            & \rev{Imitation learning}          & \rev{Tensor-based DBN}                                  \\ \cline{2-5} 
                                      & \rev{Geyer \emph{et al.} \cite{geyer2018learning}}                        & \rev{Distributed routing}                                            & \rev{Imitation learning}          & \rev{Graph-query neural network}                                  \\ \cline{2-5} 
                                      & \rev{Pham \emph{et al.} \cite{pham2018deep}}                        & \rev{Routing for knowledge-defined networking}                                            & \rev{Reinforcement learning}          & \rev{Deep Deterministic Policy Gradient} \\
                                      \hline
\multirow{5}{*}{Scheduling}           & Zhang \emph{et al.} \cite{zhang2017energy}                        & Hybrid dynamic voltage and frequency scaling scheduling             & Reinforcement learning      & Deep Q learning                      \\ \cline{2-5} 
                                      & Atallah \emph{et al.} \cite{atallah2017deep}                      & Roadside communication network scheduling                           & Reinforcement learning      & Deep Q learning                      \\ \cline{2-5} 
                                      & Chinchali \emph{et al.} \cite{chinchali2018}                      & Cellular network traffic scheduling                                 & Reinforcement learning      & Policy gradient                      \\ \cline{2-5} 
                                      & Atallah \emph{et al.} \cite{atallah2017deep}                      & Roadside communications networks scheduling                          & Reinforcement learning      & Deep Q learning                      \\ \cline{2-5} 
                                      & \edit{Wei \emph{et al.} \cite{wei2018joint}}                        & \edit{User scheduling and content caching for mobile edge networks}   & \edit{Reinforcement learning} & \edit{Deep policy gradient}            \\ \cline{2-5} 
                                      & \rev{Mennes \emph{et al.} \cite{mennes2018neural}}                        & \rev{Predicting free slots in a Multiple Frequencies Time Division Multiple Access
                                      (MF-TDMA) network. }   & \rev{Imitation learning} & \rev{MLP}            \\ \hline
\multirow{6}{*}{Resource allocation}  & Sun \emph{et al.} \cite{sun2017learning}                          & Resource management over wireless networks                          & Imitation learning          & MLP                                  \\ \cline{2-5} 
                                      & Xu \emph{et al.} \cite{xu2017deep3}                               & Resource allocation in cloud radio access networks                  & Reinforcement learning      & Deep Q learning                      \\ \cline{2-5} 
                                      & Ferreira \emph{et al.} \cite{ferreira2017multi}                   & Resource management in cognitive communications               & Reinforcement learning      & Deep SARSA                           \\ \cline{2-5} 
                                      & \edit{Challita \emph{et al.} \cite{challita2018proactive}}          & \edit{Proactive resource management for LTE}                         & \edit{Reinforcement learning} & \edit{Deep policy gradient}            \\ \cline{2-5} 
                                      & \edit{Ye and Li \cite{ye2018deep12}}                                & \edit{Resource allocation in vehicle-to-vehicle communication}          & \edit{Reinforcement learning} & \edit{Deep Q learning}                 \\ \cline{2-5} 
                                      & \rev{Li \emph{et al.} \cite{li2018deep}}                                & \rev{Computation
                                      offloading and resource allocation for mobile edge computing}          & \rev{Reinforcement learning} & \rev{Deep Q learning}                 \\ \cline{2-5} 
                                      & \rev{Zhou \emph{et al.} \cite{8553651}}                                & \rev{Radio resource assignment }          & \rev{Analysis-based} & \rev{LSTM}                 \\ \hline
\multirow{4}{*}{Radio control}        & Naparstek and Cohen \cite{naparstek2017deep}                      & Dynamic spectrum access                                             & Reinforcement learning      & Deep Q learning                      \\ \cline{2-5} 
                                      & O'Shea and Clancy \cite{o2016deep}                                & Radio control and signal detection                                  & Reinforcement learning      & Deep Q learning                      \\ \cline{2-5} 
                                      & Wijaya \emph{et al.} \cite{wijaya2015intercell, wijaya2016neural} & Intercell-interference cancellation and transmit power optimization & Imitation learning          & RBM                                  \\ \cline{2-5} 
                                      & \edit{Rutagemwa \emph {et al.} \cite{rutagemwa2018dynamic}}         & \edit{Dynamic spectrum alignment}                                   & \edit{Analysis-based}         & \edit{RNN}                             \\ \cline{2-5} 
                                      & \edit{Yu \emph{et al.} \cite{yu2018deep2}}                          & \edit{Multiple access for wireless network}                           & \edit{Reinforcement learning} & \edit{Deep Q learning}                 \\ \cline{2-5} 
                                      & \rev{Li \emph {et al.} \cite{li2017intelligent2}}         & \rev{Power control for spectrum sharing in cognitive radios}                                   & \rev{Reinforcement learning}         & \rev{DQN}                             \\ \cline{2-5} 
                                      & \rev{Liu \emph {et al.} \cite{liu2018anti}}         & \rev{Anti-jamming communications in dynamic and unknown environment}                                   & \rev{Reinforcement learning}         & \rev{DQN}                             \\ \cline{2-5} 
                                      & \rev{Luong \emph {et al.} \cite{luong2018joint}}         & \rev{ Transaction transmission and channel selection in cognitive radio blockchain}                                   & \rev{Reinforcement learning}         & \rev{Double DQN}                             \\ \cline{2-5} 
                                      & \rev{Ferreira \emph {et al.} \cite{ferreira2018multi}}         & \rev{
Radio  transmitter  settings selection in satellite  communications
}                                   & \rev{Reinforcement learning}         & \rev{Deep multi-objective reinforcement learning}                             \\ \hline
\multirow{9}{*}{Other}               & Mao \emph{et al.} \cite{mao2017neural}                            & Adaptive video bitrate                                              & Reinforcement learning      & A3C                                  \\ \cline{2-5} 
                                      & Oda \emph{et al.} \cite{oda2017design, oda2017performance}        & Mobile actor node control                                           & Reinforcement learning      & Deep Q learning                      \\ \cline{2-5} 
                                      & Kim \cite{kim2017load}                                            & IoT load balancing                                                  & Analysis-based              & DBN                                  \\ \cline{2-5} 
                                      & \edit{Challita \emph{et al.} \cite{challita2018deep}}               & \edit{Path planning for aerial vehicle networking}                    & \edit{Reinforcement learning} & \edit{Multi-agent echo state networks} \\ \cline{2-5} 
                                      & \edit{Luo \emph{et al.} \cite{luo2018online}}                       & \edit{Wireless online power control}                                  & \edit{Reinforcement learning} & \edit{Deep Q learning}                 \\  \cline{2-5} 
                                      & \edit{Xu \emph{et al.} \cite{xu2018experience}}                     & \edit{Traffic engineering}                                            & \edit{Reinforcement learning} & \edit{Deep policy gradient}            \\ \cline{2-5} 
                                      & \edit{Liu \emph{et al.} \cite{liu2018deepnap}}                      & \edit{Base station sleep control}                                & \edit{Reinforcement learning} & \edit{Deep Q learning}                 \\ \cline{2-5} 
                                      & \rev{Zhao \emph{et al.} \cite{zhao2018deep}}                      & \rev{Network slicing}                                & \rev{Reinforcement learning} & \rev{Deep Q learning} \\ \cline{2-5} 
                                      & \rev{Zhu \emph{et al.} \cite{zhu2018deep}}                      & \rev{Mobile edge caching}                                & \rev{Reinforcement learning} & \rev{A3C} \\ \cline{2-5} 
                                      & \rev{Liu \emph{et al.} \cite{liu2018energy}}                      & \rev{Unmanned aerial vehicles control}                                & \rev{Reinforcement learning} & \rev{DQN} \\ \cline{2-5} 
                                      & \rev{Lee \emph{et al.} \cite{lee2018deep}}                      & \rev{Transmit power Control in device-to-device communications}                                & \rev{Imitation learning} & \rev{MLP} \\ \cline{2-5} 
                                      & \rev{He \emph{et al.} \cite{he2017software}}                      & \rev{Dynamic orchestration of networking, caching, and computing }                                & \rev{Reinforcement learning} & \rev{DQN} \\\hline
\end{tabular}
\end{table*}

Beyond these two approaches, analysis-based control is gaining traction in mobile networking. Specifically, this scheme uses ML models for network data analysis, and subsequently exploits the results to aid network control. Unlike reinforcement/imitation learning, analysis-based control does not directly output actions. Instead, it extract useful information and delivers this to an agent, to execute the actions. We illustrate the principles between the three control paradigms in Fig. \ref{fig:control}. We review works proposed so far in this space next, and summarize these efforts in Table~\ref{tab:control}.\\

\noindent\textbf{Network Optimization} refers to the management of network resources and functions in a given environment, with the goal of improving the network performance. Deep learning has recently achieved several successful results in this area. For example,  Liu \emph{et al.} exploit a DBN to discover the correlations between multi-commodity flow demand information and link usage in wireless networks \cite{liu2017deep222}. Based on the predictions made, they remove the links that are unlikely to be scheduled, so as to reduce the size of data for the demand constrained energy minimization. Their method reduces runtime by up to 50\%, without compromising optimality. Subramanian and Banerjee propose to use deep learning to predict the health condition of heterogeneous devices in machine to machine communications \cite{subramanian2016poster}. The results obtained are subsequently exploited for optimizing health aware policy change decisions. 

He \emph{et al.} employ deep reinforcement learning to address caching and interference alignment problems in wireless networks \cite{he2017optimization, he2017deep3}. In particular, they treat time-varying channels as finite-state Markov channels and apply deep Q networks to learn the best user selection policy. This novel framework demonstrates significantly higher sum rate and energy efficiency over existing approaches. \edit{Chen \emph{et al.} shed light on automatic traffic optimization using a deep reinforcement learning approach \cite{chen2018auto}. Specifically, they architect a two-level DRL framework, which imitates the Peripheral and Central Nervous Systems in animals, to address scalability problems at datacenter scale. In their design, multiple peripheral systems are deployed on all end-hosts, so as to make decisions locally for short traffic flows. A central system is further employed to decide on the optimization with long traffic flows, which are more tolerant to longer delay. Experiments in a testbed with 32 severs suggest that the proposed design  reduces the traffic optimization turn-around time and flow completion time significantly, compared to existing approaches.}\\

\noindent\textbf{Routing:} Deep learning can also improve the efficiency of routing rules. Lee \emph{et al.} exploit a 3-layer deep neural network to classify node degree, given detailed information of the routing nodes \cite{lee2017classification}. The classification results along with temporary routes are exploited for subsequent virtual route generation using the Viterbi algorithm.  Mao \emph{et al.} employ a DBN to decide the next routing node and construct a software defined router \cite{mao2017routing}. By considering Open Shortest Path First as the optimal routing strategy, their method achieves up to 95\% accuracy, while reducing significantly the overhead and delay, and achieving higher throughput with a signaling interval of 240 milliseconds. \rev{In follow up work, the authors use tensors to represent hidden layers, weights and biases in DBNs, which further improves the routing performance \cite{mao2017tensor}.} 

A similar outcome is obtained in \cite{yang2017neural}, where the authors employ Hopfield neural networks for routing, achieving better usability and survivability in mobile ad hoc network application scenarios. \rev{Geyer \emph{et al.} represent the network using graphs, and design a dedicated Graph-Query NN to address the distributed routing problem~\cite{geyer2018learning}. This novel architecture takes graphs as input and uses message passing between nodes in the graph, allowing it to operate with various network topologies. Pham \emph{et al.} shed light on routing protocols in knowledge-defined networking, using a Deep Deterministic Policy Gradient algorithm based on reinforcement learning \cite{pham2018deep}. Their agent takes traffic conditions as input and incorporates QoS into the reward function. Simulations show that their framework can effectively learn the correlations between traffic flows, which leads to better routing configurations.}\\

\noindent\textbf{Scheduling:} There are several studies that investigate scheduling with deep learning. Zhang \emph{et al.} introduce a deep Q learning-powered hybrid dynamic voltage and frequency scaling scheduling mechanism, to reduce the energy consumption in real-time systems (e.g. Wi-Fi, IoT, video applications)~\cite{zhang2017energy}. In their proposal, an AE is employed to \edit{approximate the Q function} and the framework performs experience replay \cite{schaul2015prioritized} to stabilize the training process and accelerate convergence. Simulations demonstrate that this method reduces by 4.2\% the energy consumption of a traditional Q learning based method. Similarly, the work in \cite{atallah2017deep} uses deep Q learning for scheduling in roadside communications networks. In particular, interactions between vehicular environments, including the sequence of actions, observations, and reward signals are formulated as an MDP. By approximating the Q value function, the agent learns a scheduling policy that achieves lower latency and busy time, and longer battery life, compared to traditional scheduling methods. 

More recently, Chinchali \emph{et al.} present a policy gradient based scheduler to optimize the cellular network traffic flow~\cite{chinchali2018}. Specifically, they cast the scheduling problem as a MDP and employ RF to predict network throughput, which is subsequently used as a component of a reward function. Evaluations with a realistic network simulator demonstrate that this proposal can dynamically adapt to traffic variations, which enables mobile networks to carry 14.7\% more data traffic, while outperforming heuristic schedulers by more than 2$\times$.  \edit{Wei \emph{et al.} address user scheduling and content caching simultaneously \cite{wei2018joint}. In particular, they train a DRL agent, consisting of an actor for deciding which base station should serve certain content, and whether to save the content. A critic is further employed to estimate the value function and deliver feedback to the actor. Simulations over a cluster of base stations show that the agent can yield low transmission delay.} \rev{Li \emph{et al.} shed light on resource allocation in a multi-user mobile computing scenario \cite{li2018deep}. They employ a deep Q learning framework to jointly optimize the offloading decision and computational resource  allocation, so as to minimize the sum cost of delay and energy consumption of all user equipment. Simulations show that their proposal can reduce the total cost of the system, as compared to fully-local, fully-offloading, and naive Q-learning approaches. }\\

\noindent\textbf{Resource Allocation:} Sun \emph{et al.} use a deep neural network to approximate the mapping between the input and output of the Weighted Minimum Mean Square Error resource allocation algorithm \cite{shi2011iteratively}, in interference-limited wireless network environments \cite{sun2017learning}. By effective imitation learning, the neural network approximation achieves close performance to that of its teacher. Deep learning has also been applied to cloud radio access networks, Xu \emph{et al.} employing deep Q learning to determine the on/off modes of remote radio heads given, the current mode and user demand \cite{xu2017deep3}. Comparisons with single base station association and fully coordinated association methods suggest that the proposed DRL controller allows the system to satisfy user demand while requiring significantly less energy. 

Ferreira \emph{et al.} employ deep State-Action-Reward-State-Action (SARSA) to address resource allocation management in cognitive communications \cite{ferreira2017multi}. By forecasting the effects of radio parameters, this framework avoids wasted trials of poor parameters, which reduces the computational resources required. \rev{In \cite{mennes2018neural}, Mennes \emph{et al.} employ MLPs to precisely forecast free slots prediction in a Multiple Frequencies Time Division Multiple Access (MF-TDMA) network, thereby achieving efficient scheduling. The authors conduct simulations with a network deployed in a 100$\times$100 room, showing that their solution can effectively reduces collisions by half. Zhou \emph{et al.} adopt LSTMs to predict traffic load at base stations in ultra dense networks \cite{8553651}. Based on the predictions, their method changes the resource allocation policy to avoid congestion, which leads to lower packet loss rates, and higher throughput and mean opinion scores.}\\
 
\noindent\textbf{Radio Control:} In \cite{naparstek2017deep}, the authors address the dynamic spectrum access problem in multichannel wireless network environments using deep reinforcement learning. In this setting, they incorporate an LSTM into a deep Q network, to maintain and memorize historical observations, allowing the architecture to perform precise state estimation, given partial observations. The training process is distributed to each user, which enables effective training parallelization and the learning of good policies for individual users. Experiments demonstrate that this framework achieves double the channel throughput, when compared to a benchmark method. \rev{Yu \emph{et al.} apply deep reinforcement learning to address challenges in wireless multiple access control \cite{yu2018deep2}, recognizing that in such tasks DRL agents are fast in terms of convergence and robust against non-optimal parameter settings. Li \emph{et al.} investigate power control for spectrum sharing in cognitive radios using DRL. In their design, a DQN agent is built to adjust the transmit power of a cognitive radio system, such that the overall signal-to-interference-plus-noise ratio is maximized.}

The work in \cite{o2016deep} sheds light on the radio control and signal detection problems. In particular, the authors introduce a radio signal search environment based on the Gym Reinforcement Learning platform. Their agent exhibits a steady learning process and is able to learn a radio signal search policy. \edit{Rutagemwa \emph{et al.} employ an RNN to perform traffic prediction, which can subsequently aid the dynamic spectrum assignment in mobile networks \cite{rutagemwa2018dynamic}. With accurate traffic forecasting, their proposal  improves the performance of spectrum sharing in dynamic wireless environments, as it attains near-optimal spectrum assignments.} \rev{In \cite{liu2018anti}, Liu \emph{et al.} approach the anti-jamming communications problem in dynamic and unknown environments with a DRL agent. Their system is based on a DQN with CNN, where the agent takes raw spectrum information as input and requires limited prior knowledge about the environment, in order to improve the overall throughput of the network in such adversarial circumstances.} 

\rev{Luong \emph{et al.} incorporate the blockchain technique into cognitive radio networking~\cite{luong2018joint}, employing a double DQN agent to maximize the number of successful transaction transmissions for secondary users, while minimizing the channel cost and transaction fees. Simulations show that the DQN method significantly outperforms na\:ive Q learning in terms of successful transactions, channel cost, and learning speed. DRL can further attack problems in the satellite communications domain. In \cite{ferreira2018multi}, Ferreira \emph{et al.} fuse multi-objective reinforcement learning \cite{ferreira2016multi} with deep neural networks to select among multiple radio transmitter settings while attempting to achieve multiple conflicting goals, in a dynamically changing satellite communications channel. Specifically, two set of NNs are employed to execute exploration and exploitation separately. This builds an ensembling system, with makes the framework more robust to the changing environment. Simulations demonstrate that their system can nearly optimize six different objectives (i.e. bit error rate, throughput, bandwidth, spectral efficiency, additional power consumption, and power efficiency), only with small performance errors compared to ideal solutions.}\\

\noindent \textbf{Other applications:} Deep learning is playing an important role in other network control problems as well. Mao \emph{et al.} develop the Pensieve system that generates adaptive video bit rate algorithms using deep reinforcement learning \cite{mao2017neural}. Specifically, Pensieve employs a state-of-the-art deep reinforcement learning algorithm A3C, which takes the bandwidth, bit rate and buffer size as input, and selects the bit rate that leads to the best expected return. The model is trained offline and deployed on an adaptive bit rate server, demonstrating that the system outperforms the best existing scheme by 12\%-25\% in terms of QoE. \edit{Liu \emph{et al.} apply deep Q learning to reduce the energy consumption in cellular networks \cite{liu2018deepnap}. They train an agent to dynamically switch on/off base stations based on traffic consumption in areas of interest. An action-wise experience replay mechanism is further designed for balancing different traffic behaviours. Experiments show that their proposal can significantly reduce the energy consumed by base stations, outperforming naive table-based Q learning approaches.} \rev{A control mechanism for unmanned aerial vehicles using DQN is proposed in \cite{liu2018energy}, where multiple objectives are targeted: maximizing energy efficiency, communications coverage, fairness and connectivity. The authors conduct extensive simulations in an virtual playground, showing that their agent is able to learn the dynamics of the environment, achieving superior performance over random and greedy control baselines.}

Kim and Kim link deep learning with the load balancing problem in IoT \cite{kim2017load}. The authors suggest that DBNs can effectively analyze network load and process structural configuration, thereby achieving efficient load balancing in IoT. \edit{Challita \emph{et al.} employ a deep reinforcement learning algorithm based on echo state networks to perform path planning for a cluster of unmanned aerial vehicles  \cite{challita2018deep}. Their proposal yields lower delay than a heuristic baseline. Xu \emph{et al.} employ a DRL agent to learn from network dynamics how to control traffic flow \cite{xu2018experience}. They advocate that DRL is suitable for this problem, as it performs remarkably well in handling dynamic environments and sophisticated state spaces. Simulations conducted over three network topologies confirm this viewpoint, as the DRL agent significantly reduces the delay, while providing throughput comparable to that of traditional approaches. \rev{Zhu \emph{et al.} employ the A3C algorithm to address the caching problem in mobile edge computing. Their method obtains superior cache hit ratios and traffic offloading  performance over three baselines caching methods. Several open challenges are also pointed out, which are worthy of future pursuit. The edge caching problem is also addressed in \cite{he2017software}, where He \emph{et al.} architect a DQN agent to perform dynamic orchestration of networking, caching, and computing. Their method facilitates high revenue to mobile virtual network operators. }} \\

\edit{\textbf{Lessons learned:} There exist three approaches to network control using deep learning i.e., reinforcement learning, imitation learning, and analysis-based control. Reinforcement learning requires to interact with the environment, trying different actions and obtaining feedback in order to improve. The agent will make mistakes during training, and usually needs a large number of steps of steps to become smart. Therefore, most works do not train the agent on the real infrastructure, as making mistakes usually can have serious consequences for the network. Instead, a simulator that mimics the real network environments is built and the agent is trained offline using that. This imposes high fidelity requirements on the simulator, as the agent can not work appropriately in an environment that is different from the one used for training. \rev{On the other hand, although DRL performs remarkable well in many applications, considerable amount of time and computing resources are required to train an usable agent. This should be considered in real-life implementation.}}

\edit{In contrast, the imitation learning mechanism ``learns by demonstration''. It requires a teacher that provides labels telling the agent what it should do under certain circumstances. In the networking context, this mechanism is usually employed to reduce the computational time \cite{mao2017routing}. Specifically, in some network application (e.g., routing), computing the optimal solution is time-consuming, which cannot satisfy the delay constraints of mobile network. To mitigate this, one can generate a large dataset offline, and use an NN agent to learn the optimal actions.}

\edit{Analysis-based control on the other hand, is suitable for problems were decisions cannot be based solely on the state of the network environment. One can use a NN to extract additional information (e.g. traffic forecasts), which subsequently aids decisions. For example, the dynamic spectrum assignment can benefit from the analysis-based control.}

\subsection{Deep Learning Driven Network Security}\label{sec:security}
With the increasing popularity of wireless connectivity, protecting users, network equipment and data from malicious attacks, unauthorized access and information leakage becomes crucial. Cyber security systems guard mobile devices and users through firewalls, anti-virus software, and Intrusion Detection Systems (IDS) \cite{buczak2016survey}. The firewall is an access security gateway that allows or blocks the uplink and downlink network traffic, based on pre-defined rules. Anti-virus software detects and removes computer viruses, worms and Trojans and malware. IDSs identify unauthorized and malicious activities, or rule violations in information systems. Each performs its own functions to protect network communication, central servers and edge devices.

\begin{table*}[h!]
\centering
\caption{A summary of work on deep learning driven network security.}
\label{tab:security}
\begin{tabular}{|c|C{2.1cm}|C{3.3cm}|C{6cm}|C{1.7cm}|C{1.7cm}|}
\hline
\textbf{Level}                        & \textbf{Reference}                                       & \textbf{Application}                                            & \textbf{Problem considered}                                                                                                           & \textbf{Learning paradigm} & \textbf{Model}                                           \\ \hline
\multirow{12}{*}{\edit{Infrastructure}} & Azar \emph{et al.} \cite{yousefi2017autoencoder}         & Cyber security applications                                     & Malware classification \& Denial of service, probing, remote to user \& user to root                                                  & Unsupervised \& supervised & Stacked AE                                               \\ \cline{2-6} 
                                      & Thing \cite{thing2017ieee}                               & IEEE 802.11 network anomaly detection and attack classification & Flooding, injection and impersonation attacks                                                                                          & Unsupervised \& supervised & Stacked AE                                               \\ \cline{2-6} 
                                      & Aminanto and Kim \cite{aminanto2016detecting}            & Wi-Fi impersonation attacks detection                            & Flooding, injection and impersonation attacks                                                                                          & Unsupervised \& supervised & MLP, AE                                                  \\ \cline{2-6} 
                                      & Feng \emph{et al.} \cite{feng2016anomaly}                & Spectrum anomaly detection                                      & Sudden signal-to-noise ratio changes in the communication channels                                                                                                        & Unsupervised \& supervised & AE                                                       \\ \cline{2-6} 
                                      & Khan \emph{et al.} \cite{khan2016distributed}            & Flooding attacks detection in wireless mesh networks             & Moderate and severe distributed flood attack                                                                                      & Supervised                 & MLP                                                      \\ \cline{2-6} 
                                      & Diro and Chilamkurti \cite{diro2017distributed}          & IoT distributed attacks detection                                & Denial of service, probing, remote to user \& user to root                                                                            & Supervised                 & MLP                                                      \\ \cline{2-6} 
                                      & Saied \emph{et al.} \cite{saied2016detection}            & Distributed denial of service attack detection                  & Known and unknown distributed denial of service attack                                                                                & Supervised                 & MLP                                                      \\ \cline{2-6} 
                                      & Martin \emph{et al.} \cite{lopez2017conditional}         & IoT intrusion detection                                         & Denial of service, probing, remote to user \& user to root                                                                            & Unsupervised \& supervised & Conditional VAE                                          \\ \cline{2-6} 
                                      & \edit{Hamedani \emph{et al.} \cite{hamedani2018reservoir}} & \edit{Attacks detection in delayed feedback networks}             & \edit{Attack detection in smart grids using reservoir computing}                                                                        & \edit{Supervised}            & \edit{MLP}                                                 \\ \cline{2-6} 
                                      & \edit{Luo and Nagarajany \cite{luo2018distributed}}        & \edit{Anomalies in WSNs}                                          & \edit{Spikes and burst recorded by temperature and relative humidity sensors}                                                           & \edit{Unsupervised}          & \edit{AE}                                                  \\ \cline{2-6} 
                                      & \edit{Das \emph{et al.} \cite{das2018deep}}                & \edit{IoT authentication}                                         & \edit{Long duration of signal imperfections}                                                                                            & \edit{Supervised}            & \edit{LSTM}                                                \\ \cline{2-6} 
                                      & \edit{Jiang \emph{et al.} \cite{jiang2018virtual}}         & \edit{MAC spoofing detection}                                & \edit{Packets from different hardware use same MAC address} & \edit{Supervised}            & \edit{CNN}                                                 \\ \cline{2-6} 
                                      & \rev{Jiang \emph{et al.} \cite{nguyen2018cyberattack}}         & \rev{Cyberattack detection in mobile cloud computing}                                & \rev{Various cyberattacks from 3 different datasets} & \rev{Unsupervised + supervised}            & \rev{RBM}                                                 \\ \hline
\multirow{12}{*}{\edit{Software}}       & Yuan \emph{et al.} \cite{yuan2014droid}                  & Android malware detection                                       & Apps in Contagio Mobile and Google Play Store                                                                                         & Unsupervised \& supervised & RBM                                                      \\ \cline{2-6} 
                                      & Yuan \emph{et al.} \cite{yuan2016droiddetector}          & Android malware detection                                       & Apps in Contagio Mobile, Google Play Store and Genome Project                                                                      & Unsupervised \& supervised & DBN                                                      \\ \cline{2-6} 
                                      & Su \emph{et al.} \cite{su2016deep}                       & Android malware detection                                       & Apps in Drebin, Android Malware Genome Project, the Contagio Community, and Google Play Store                                       & Unsupervised \& supervised & DBN + SVM                                                \\ \cline{2-6} 
                                      & Hou \emph{et al.} \cite{hou2016deep4maldroid}            & Android malware detection                                       & App samples from Comodo Cloud Security Center                                                                                         & Unsupervised \& supervised & Stacked AE                                               \\ \cline{2-6} 
                                      & Martinelli \cite{martinelli2017evaluating}               & Android malware detection                                       & Apps in Drebin, Android Malware Genome Project and Google Play Store                                                                  & Supersived                 & CNN                                                      \\ \cline{2-6} 
                                      & McLaughlin \emph{et al.} \cite{mclaughlin2017deep}       & Android malware detection                                       & Apps in Android Malware Genome project and Google Play Store                                                                          & Supersived                 & CNN                                                      \\ \cline{2-6} 
                                      & Chen \emph{et al.} \cite{chen2017deep123}                & Malicious application detection at the network edge             & Publicly-available malicious applications                                                                                             & Unsupervised \& supervised & RBM                                                      \\ \cline{2-6} 
                                      & Wang \emph{et al.} \cite{wang2017malware}                & Malware traffic classification                                  & Traffic extracted from 9 types of malware                                                                                             & Superivised                & CNN                                                      \\ \cline{2-6} 
                                      & Oulehla \emph{et al.} \cite{oulehla2016detection}        & Mobile botnet detection                                         & Client-server and hybrid botnets                                                                                                      & Unknown                    & Unknown                                                  \\ \cline{2-6} 
                                      & Torres \emph{et al.} \cite{torres2016analysis}           & Botnet detection                                                & Spam, HTTP and unknown traffic                                                                                                        & Superivised                & LSTM                                                     \\ \cline{2-6} 
                                      & Eslahi \emph{et al.} \cite{eslahi2016mobile}             & Mobile botnet detection                                         & HTTP botnet traffic                                                                                                                   & Superivised                & MLP                                                      \\ \cline{2-6} 
                                      & Alauthaman \emph{et al.} \cite{alauthaman2016p2p}        & Peer-to-peer botnet detection                                   & Waledac and Strom Bots                                                                                                                & Superivised                & MLP                                                      \\ \hline
\multirow{11}{*}{\edit{User privacy}}   & Shokri and Shmatikov \cite{shokri2015privacy}            & Privacy preserving deep learning                                & Avoiding sharing data in collaborative model training                                                                                 & Superivised                & MLP, CNN                                                 \\ \cline{2-6} 
                                      & Phong \emph{et al.} \cite{aono2017privacy}               & Privacy preserving deep learning                                & Addressing information leakage introduced in~\cite{shokri2015privacy}                                                                 & Supervised                 & MLP                                                      \\ \cline{2-6} 
                                      & Ossia \emph{et al.} \cite{ossia2017hybrid}               & Privacy-preserving mobile analytics                             & Offloading feature extraction from cloud                                                                                              & Supervised                 & CNN                                                      \\ \cline{2-6} 
                                      & Abadi \emph{et al.} \cite{abadi2016deep}                 & Deep learning with differential privacy                         & Preventing exposure of private information in training data                                                                           & Supervised                 & MLP                                                      \\ \cline{2-6} 
                                      & Osia \emph{et al.} \cite{osia2017privacy}                & Privacy-preserving personal model training                      & Offloading personal data from clouds                                                                                                  & Unsupervised \& supervised & MLP \\ \cline{2-6} 
                                      & Servia \emph{et al.} \cite{servia2017personal}           & Privacy-preserving model inference                              & Breaking down large models for privacy-preserving analytics                                                                           & Supervised                 & CNN                                                      \\ \cline{2-6} 
                                      & Hitaj \emph{et al.} \cite{hitaj2017deep}                 & Stealing information  from collaborative deep learning          & Breaking the ordinary and differentially private collaborative deep learning                                                          & Unsupervised               & GAN                                                      \\ \cline{2-6} 
                                      & Hitaj \emph{et al.} \cite{abadi2016deep}                 & Password guessing                                               & Generating passwords from leaked password set                                                                                         & Unsupervised               & GAN                                                      \\ \cline{2-6} 
                                      & Greydanus \cite{greydanus2017learning}                   & Enigma learning                                                 & Reconstructing functions of polyalphabetic cipher                                                                                     & Supervised                 & LSTM                                                     \\ \cline{2-6} 
                                      & Maghrebi \cite{maghrebi2016breaking}                     & Breaking cryptographic                                          & Side channel attacks                                                                                                                  & Supervised                 & MLP, AE, CNN, LSTM                                       \\ \cline{2-6} 
                                      & \edit{Liu \emph{et al.} \cite{liu2018genpass}}             & \edit{Password guessing}                                          & \edit{Employing adversarial generation to guess passwords }                                                                                 & \edit{Unsupervised}          & \edit{LSTM}                                                \\ \cline{2-6} 
                                      & \rev{Ning \emph{et al.} \cite{ning2018deepmag}}             & \rev{Mobile apps sniffing}                                          & \rev{Defending against mobile apps sniffing through noise injection}                                                                                 & \rev{Supervised}          & \rev{CNN}                                                \\ \cline{2-6} 
                                      & \rev{Wang \emph{et al.} \cite{wang2018not}}             & \rev{Private inference in mobile cloud}                                          & \rev{Computation offloading and privacy preserving for mobile inference}                                                                                 & \rev{Supervised}          & \rev{MLP, CNN}                                                \\ \hline
\end{tabular}
\end{table*}

Modern cyber security systems benefit increasingly from deep learning \cite{kwon2017survey}, since it can enable the system to \emph{(i)}~automatically learn signatures and patterns from experience and generalize to future intrusions (supervised learning); or \emph{(ii)} identify patterns that are clearly differed from regular behavior (unsupervised learning). This dramatically reduces the effort of pre-defined rules for discriminating intrusions. Beyond protecting networks from attacks,  deep learning can also be used for attack purposes, bringing huge potential to steal or crack user passwords or information. \edit{In this subsection, we review deep learning driven network security from three perspectives, namely infrastructure, software, and user privacy. Specifically, infrastructure level security work focuses on detecting anomalies that occur in the physical network and software level work is centred on identifying malware and botnets in mobile networks. From the user privacy perspective, we discuss methods to protect from how to protect against private information leakage, using deep learning. To our knowledge, no other reviews summarize these efforts.} We summarize these works in Table \ref{tab:security}. \\

\noindent\edit{\textbf{Infrastructure level security:} We mostly focus on anomaly detection at the infrastructure level, i.e. identifying network events (e.g., attacks, unexpected access and use of data) that do not conform to expected behaviors.} Many researchers exploit the outstanding unsupervised learning ability of AEs \cite{yousefi2017autoencoder}. For example, Thing investigates features of attacks and threats that exist in IEEE 802.11 networks \cite{thing2017ieee}. The author employs a stacked AE to categorize network traffic into 5 types (i.e. legitimate, flooding, injection and impersonation traffic), achieving 98.67\% overall accuracy. The AE is also exploited in \cite{aminanto2016detecting}, where Aminanto and Kim use an MLP and stacked AE for feature selection and extraction, demonstrating remarkable performance. Similarly, Feng \emph{et al.} use AEs to detect abnormal spectrum usage in wireless communications~\cite{feng2016anomaly}. Their experiments suggest that the detection accuracy can significantly benefit from the depth of AEs.

Distributed attack detection is also an important issue in mobile network security. Khan \emph{et al.} focus on detecting flooding attacks in wireless mesh networks \cite{khan2016distributed}. They simulate a wireless environment with 100 nodes, and artificially inject moderate and severe distributed flooding attacks, to generate a synthetic dataset. Their deep learning based methods achieve excellent false positive and false negative rates. Distributed attacks are also studied in \cite{diro2017distributed}, where the authors focus on an IoT scenario. Another work in \cite{saied2016detection} employs MLPs to detect distributed denial of service attacks. By characterizing typical patterns of attack incidents, the proposed model works well in detecting both known and unknown distributed denial of service attacks. \rev{More recently, Nguyen \emph{et al.} employ RBMs to classify cyberattacks in the mobile cloud in an online manner \cite{nguyen2018cyberattack}. Through unsupervised layer-wise pre-training and fine-tuning, their methods obtain over 90\% classification accuracy on three different datasets, significantly outperforming other machine learning approaches.}

Martin \emph{et al.} propose a conditional VAE to identify intrusion incidents in IoT \cite{lopez2017conditional}. In order to improve detection performance, their VAE infers missing features associated with incomplete measurements, which are common in IoT environments. The true data labels are embedded into the decoder layers to assist final classification. Evaluations on the well-known NSL-KDD dataset \cite{tavallaee2009detailed} demonstrate that their model achieves remarkable accuracy in identifying \edit{denial of service}, probing, remote to user and user to root attacks, outperforming traditional ML methods by 0.18 in terms of F1 score. \edit{Hamedani \emph{et al.} employ MLPs to detect malicious attacks in delayed feedback networks \cite{hamedani2018reservoir}. The proposal achieves more than 99\% accuracy over 10,000 simulations.}\\

\noindent\textbf{Software level security:} Nowadays, mobile devices are carrying considerable amount of private information. This information can be stolen and exploited by malicious apps installed on smartphones for ill-conceived purposes \cite{tam2017evolution}. Deep learning is being exploited for analyzing and detecting such threats.

Yuan \emph{et al.} use both labeled and unlabeled mobile apps to train an RBM \cite{yuan2014droid}. By learning from 300 samples, their model can classify Android malware with remarkable accuracy, outperforming traditional ML tools by up to 19\%. Their follow-up research in \cite{yuan2016droiddetector} named Droiddetector further improves the detection accuracy by 2\%. Similarly, Su \emph{et al.} analyze essential features of Android apps, namely requested permission, used permission, sensitive application programming interface calls, action and app components \cite{su2016deep}. They employ DBNs to extract features of malware and an SVM for classification, achieving high accuracy and only requiring 6 seconds per inference instance.

Hou \emph{et al.} attack the malware detection problem from a different perspective. Their research points out that signature-based detection is insufficient to deal with sophisticated Android malware \cite{hou2016deep4maldroid}. To address this problem, they propose the Component Traversal, which can automatically execute code routines to construct weighted directed graphs. By employing a Stacked AE for graph analysis, their framework Deep4MalDroid can accurately detect Android malware that intentionally repackages and obfuscates to bypass signatures and hinder analysis attempts to their inner operations. This work is followed by that of Martinelli \emph{et al.}, who exploit CNNs to discover the relationship between app types and extracted syscall traces from real mobile devices \cite{martinelli2017evaluating}. The CNN has also been used in \cite{mclaughlin2017deep}, where the authors draw inspiration from NLP and take the disassembled byte-code of an app as a text for analysis. Their experiments demonstrate that CNNs can effectively learn to detect sequences of opcodes that are indicative of malware. Chen \emph{et al.} incorporate location information into the detection framework and exploit an RBM for feature extraction and classification \cite{chen2017deep123}. Their proposal improves the performance of other ML methods.

 \edit{Botnets are another important threat to mobile networks.} A botnet is effectively a network that consists of machines compromised by bots. These machine are usually under the control of a botmaster who takes advantages of the bots to harm public services and systems \cite{rodriguez2013survey}. Detecting botnets is challenging and now becoming a pressing task in cyber security. 
Deep learning is playing an important role in this area. For example, Oulehla \emph{et al.} propose to employ neural networks to extract features from mobile botnet behaviors \cite{oulehla2016detection}. They design a parallel detection framework for identifying both client-server and hybrid botnets, and demonstrate encouraging performance. Torres \emph{et al.} investigate the common behavior patterns that botnets exhibit across their life cycle, using LSTMs \cite{torres2016analysis}. They employ both under-sampling  and  over-sampling to address the class imbalance between botnet and normal traffic in the dataset, which is common in anomaly detection problems. Similar issues are also studies in \cite{eslahi2016mobile} and \cite{alauthaman2016p2p}, where the authors use standard MLPs to perform mobile and peer-to-peer botnet detection respectively, achieving high overall accuracy.\\

\noindent\textbf{User privacy level:} Preserving user privacy during training and evaluating a deep neural network is another important research issue \cite{liu2016collaborative}. Initial research is conducted in \cite{shokri2015privacy}, where the authors enable user participation in the training and evaluation of a neural network, without sharing their input data. This allows to preserve individual's privacy while benefiting all users, as they collaboratively improve the model performance. Their framework is revisited and improved in \cite{aono2017privacy}, where another group of researchers employ additively homomorphic encryption, to address the information leakage problem ignored in \cite{shokri2015privacy}, without compromising model accuracy. This significantly boosts the security of the system.  \rev{More recently, Wang \emph{et al.} \cite{wang2018not} propose a framework called ARDEN to preserve users' privacy while reducing communication overhead in mobile-cloud deep learning applications. ARDEN partitions a NN across cloud and mobile devices, with heavy computation being conducted on the cloud and mobile devices performing only simple data transformation and perturbation, using a differentially private mechanism. This simultaneously guarantees user privacy, improves inference accuracy, and reduces resource consumption.}

Osia \emph{et al.} focus on privacy-preserving mobile analytics using deep learning. They design a client-server framework based on the Siamese architecture \cite{chopra2005learning}, which accommodates a feature extractor in mobile devices and correspondingly a classifier in the cloud \cite{ossia2017hybrid}. By offloading feature extraction from the cloud, their system offers strong privacy guarantees. An innovative work in \cite{abadi2016deep} implies that deep neural networks can be trained with differential privacy. The authors introduce a differentially private SGD to avoid disclosure of private information of training data. Experiments on two publicly-available image recognition datasets demonstrate that their algorithm is able to maintain users privacy, with a manageable cost in terms of complexity, efficiency, and performance. This approach is also useful for edge-based privacy filtering techniques such as Distributed One-class Learning~\cite{shamsabadi2018}. 

Servia \emph{et al.} consider training deep neural networks on distributed devices without violating privacy constraints \cite{servia2017personal}. Specifically, the authors retrain an initial model locally, tailored to individual users. This avoids transferring personal data to untrusted entities, hence user privacy is guaranteed. Osia \emph{et al.} focus on protecting user's personal data from the inferences' perspective. In particular, they break the entire deep neural network into a feature extractor (on the client side) and an analyzer (on the cloud side) to minimize the exposure of sensitive information. Through local processing of raw input data, sensitive personal information is transferred into abstract features, which avoids direct disclosure to the cloud.
Experiments on gender classification and emotion detection suggest that this framework can effectively preserve user privacy, while maintaining remarkable inference accuracy.

\edit{Deep learning has also been exploited for cyber attacks, including attempts to compromise private user information and guess passwords.}
In \cite{hitaj2017deep}, Hitaj \emph{et al.} suggest that learning a deep model collaboratively is not reliable. By training a GAN, their attacker is able to affect such learning process and lure the victims to disclose private information, by injecting fake training samples. Their GAN even successfully breaks the differentially private collaborative learning in \cite{abadi2016deep}. The authors further investigate the use of GANs for password guessing. In \cite{hitaj2017passgan}, they design PassGAN, which learns the distribution of a set of leaked passwords. Once trained on a dataset, PassGAN is able to match over 46\% of passwords in a different testing set, without user intervention or cryptography knowledge. This novel technique has potential to revolutionize current password guessing algorithms. 

Greydanus breaks a decryption rule using an LSTM network \cite{greydanus2017learning}. They treat decryption as a sequence-to-sequence translation task, and train a framework with large enigma pairs. The proposed LSTM demonstrates remarkable performance in learning polyalphabetic ciphers. Maghrebi \emph{et al.} exploit various deep learning models (i.e. MLP, AE, CNN, LSTM) to construct a precise profiling system and perform side channel key recovery attacks \cite{maghrebi2016breaking}. Surprisingly, deep learning based methods demonstrate overwhelming performance over other template machine learning attacks in terms of efficiency in  breaking both unprotected and protected Advanced Encryption Standard implementations. \rev{In \cite{ning2018deepmag}, Ning \emph{et al.} demonstrate that an attacker can use a CNN to infer with over 84\% accuracy what apps run on a smartphone and their usage, based on magnetometer or orientation data. The accuracy can increase to 98\% if motion sensors information is also taken into account, which jeopardizes user privacy. To mitigate this issue, the authors propose to inject Gaussian noise into the magnetometer and orientation data, which leads to a reduction in inference accuracy down to 15\%, thereby effectively mitigating the risk of privacy leakage.}\\

\edit{\textbf{Lessons learned:} Most deep learning based solutions focus on existing network attacks, yet new attacks emerge every day. As these new attacks may have different features and appear to behave `normally', old NN models may not easily detect them. Therefore, an effective deep learning technique should be able to \emph{(i)} rapidly transfer the knowledge of old attacks to detect newer ones; and \emph{(ii)} constantly absorb the features of newcomers and update the underlying model. Transfer learning and lifelong learning are strong candidates to address this problems, as we will discuss in Sec.\ref{sec:changing}. Research in this directions remains shallow, hence we expect more efforts in the future.}

\rev{Another issue to which attention should be paid is the fact that NNs are vulnerable to adversarial attacks. This has been briefly discussed in Sec.~\ref{sec:limit}. Although formal reports on this matter are lacking, hackers may exploit weaknesses in NN models and training procedures to perform attacks that subvert deep learning based cyber-defense systems. This is an important potential pitfall that should be considered in real implementations.}

\subsection{\edit{Deep Learning Driven Signal Processing}}
\edit{
Deep learning is also gaining increasing attention in signal processing, in applications including Multi-Input Multi-Output (MIMO) and modulation. MIMO has become a fundamental technique in current wireless communications, both in cellular and WiFi networks. By incorporating deep learning, MIMO performance is intelligently optimized based on environment conditions. Modulation recognition is also evolving to be more accurate, by taking advantage of deep learning. We give an overview of relevant work in this area in Table \ref{tab:mimo}.\\}

\begin{table*}[htb]
\centering 
\caption{\edit{A summary of deep learning driven signal processing.}}
\label{tab:mimo}
\color{black}
\begin{tabular}{|C{2CM}|C{3CM}|C{8CM}|C{3CM}|}
\hline
\textbf{Domain}                & \textbf{Reference}                                                & \textbf{Application}                                                & \textbf{Model}                   \\ \hline
\multirow{9}{*}{MIMO  systems} & Samuel \emph{et al.} \cite{samuel2017deep}                        & MIMO detection                                                      & MLP                              \\ \cline{2-4} 
                               & Yan \emph{et al.} \cite{yan2017signal}                            & Signal detection in a MIMO-OFDM system                              & AE+ELM                           \\ \cline{2-4} 
                               & Vieira \emph{et al.} \cite{vieira2017deep}                        & Massive MIMO fingerprint-based positioning                          & CNN                              \\ \cline{2-4} 
                               & Neumann \emph{et al.} \cite{neumann2017deep}                      & MIMO channel estimation                                             & CNN                              \\ \cline{2-4} 
                               & Wijaya \emph{et al.} \cite{wijaya2015intercell, wijaya2016neural} & Inter-cell interference cancellation and transmit power optimization & RBM                              \\ \cline{2-4} 
                               & O'Shea \emph{et al.} \cite{o2017deep}                             & Optimization of representations and encoding/decoding processes     & AE                               \\ \cline{2-4} 
                               & Borgerding \emph{et al.} \cite{borgerding2017amp}                 & Sparse linear inverse problem in MIMO                        & CNN                              \\ \cline{2-4} 
                               & Fujihashi \emph{et al.} \cite{fujihashi2018nonlinear}             & MIMO nonlinear equalization                                         & MLP                              \\\cline{2-4} 
                               & Huang \emph{et al.} \cite{huang2018deep}                          & Super-resolution channel and direction-of-arrival estimation        & MLP                              \\ \hline
\multirow{5}{*}{Modulation}    & Rajendran \emph{et al.} \cite{rajendran2017distributed}           & Automatic modulation classification                                 & LSTM                             \\ \cline{2-4} 
                               & West and O'Shea \cite{west2017deep}                               & Modulation recognition                                              & CNN, ResNet, Inception CNN, LSTM \\ \cline{2-4} 
                               & O'Shea \emph{et al.} \cite{o2016radio}                            & Modulation recognition                                              & Radio transformer network        \\ \cline{2-4} 
                               & O'Shea and Hoydis \cite{timothy2017introduction}                  & Modulation classification                                           & CNN                              \\ \cline{2-4} 
                               & Jagannath \emph{et al.} \cite{jagannath2018artificial}            & Modulation classification in a software defined radio testbed             & MLP                              \\ \hline
\multirow{7}{*}{Others}        & O'Shea \emph{et al.} \cite{o2016end}                              & Radio traffic sequence recognition                                  & LSTM                             \\ \cline{2-4} 
                               & O'Shea \emph{et el.} \cite{o2016learning}                         & Learning to communicate over an impaired channel                    & AE + radio transformer network   \\ \cline{2-4} 
                               & Ye \emph{et al.} \cite{ye2018power}                               & Channel estimation and signal detection in OFDM systsms.            & MLP                              \\ \cline{2-4} 
                               & Liang \emph{et al.} \cite{liang2018exploiting}                    & Channel decoding                                                    & CNN                              \\ \cline{2-4} 
                               & Lyu \emph{et al.} \cite{lyu2018performance}                       & NNs for channel decoding                          & MLP, CNN and RNN                 \\ \cline{2-4} 
                               & D{\"o}rner \emph{et al.} \cite{dorner2017deep}                    & Over-the-air communications system                          & AE                               \\ \cline{2-4}
                               & \rev{Liao \emph{et al.} \cite{liao2018rayleigh}}                   & \rev{Rayleigh fading channel prediction}                             & \rev{MLP}                         \\ \cline{2-4}
                               & \rev{Huang \emph{et al.} \cite{huang2018fully}}                   & \rev{Light-emitting diode (LED) visible light downlink error correction}                             & \rev{AE}                         \\\cline{2-4}
                               & \rev{Alkhateeb \emph{et al.} \cite{alkhateeb2018deep}}                   & \rev{Coordinated beamforming for highly-mobile millimeter wave systems}                             & \rev{MLP}                         \\ \cline{2-4}
                               & \rev{Gante \emph{et al.} \cite{gante2018beamformed}}                   & \rev{Millimeter wave positioning}                             & \rev{CNN}                         \\ \cline{2-4}
                               & \rev{Ye \emph{et al.} \cite{ye2018channel}}                   & \rev{Channel agnostic end-to-end learning based communication system}                             & \rev{Conditional GAN}                         \\\hline
\end{tabular}
\end{table*}

\noindent\edit{\textbf{MIMO Systems:} Samuel \emph{et al.} suggest that deep neural networks can be a good estimator of transmitted vectors in a MIMO channel. By unfolding a projected gradient descent method, they design an MLP-based detection network to perform binary MIMO detection \cite{samuel2017deep}. The Detection Network can be implemented on multiple channels after a single training. Simulations demonstrate that the proposed architecture achieves near-optimal accuracy, while requiring light computation without prior knowledge of Signal-to-Noise Ratio (SNR). Yan \emph{et al.} employ deep learning to solve a similar problem from a different perspective \cite{yan2017signal}. By considering the characteristic invariance of signals, they exploit an AE as a feature extractor, and subsequently use an Extreme Learning Machine (ELM) to classify signal sources in a MIMO orthogonal frequency division multiplexing (OFDM) system. Their proposal achieves higher detection accuracy than several traditional methods, while maintaining similar complexity.}

\edit{
Vieira \emph{et al.} show that massive MIMO channel measurements in cellular networks can be utilized for fingerprint-based inference of user positions \cite{vieira2017deep}. Specifically, they design CNNs with weight regularization to exploit the sparse and information-invariance of channel fingerprints, thereby achieving precise positions inference.  CNNs have also been employed for MIMO channel estimation. Neumann \emph{et al.} exploit the structure of the MIMO channel model to design a lightweight, approximated maximum likelihood estimator for a specific channel model \cite{neumann2017deep}. Their methods outperform traditional estimators in terms of computation cost and reduce the number of hyper-parameters to be tuned. A similar idea is implemented in \cite{ye2018power}, where Ye \emph{et al.} employ an MLP to perform channel estimation and signal detection in \mbox{OFDM systems.}}

\edit{
Wijaya \emph{et al.} consider applying deep learning to a different scenario \cite{wijaya2015intercell, wijaya2016neural}. The authors propose to use non-iterative neural networks to perform transmit power control at base stations, thereby preventing degradation of network performance due to inter-cell interference. The neural network is trained to estimate the optimal transmit power at every packet transmission, selecting that with the highest activation probability. Simulations demonstrate that the proposed framework significantly outperform the belief propagation algorithm that is routinely used for transmit power control in MIMO systems, while attaining a lower computational cost.}

\begin{figure*}[t]
\begin{center}
\includegraphics[width=0.9\textwidth]{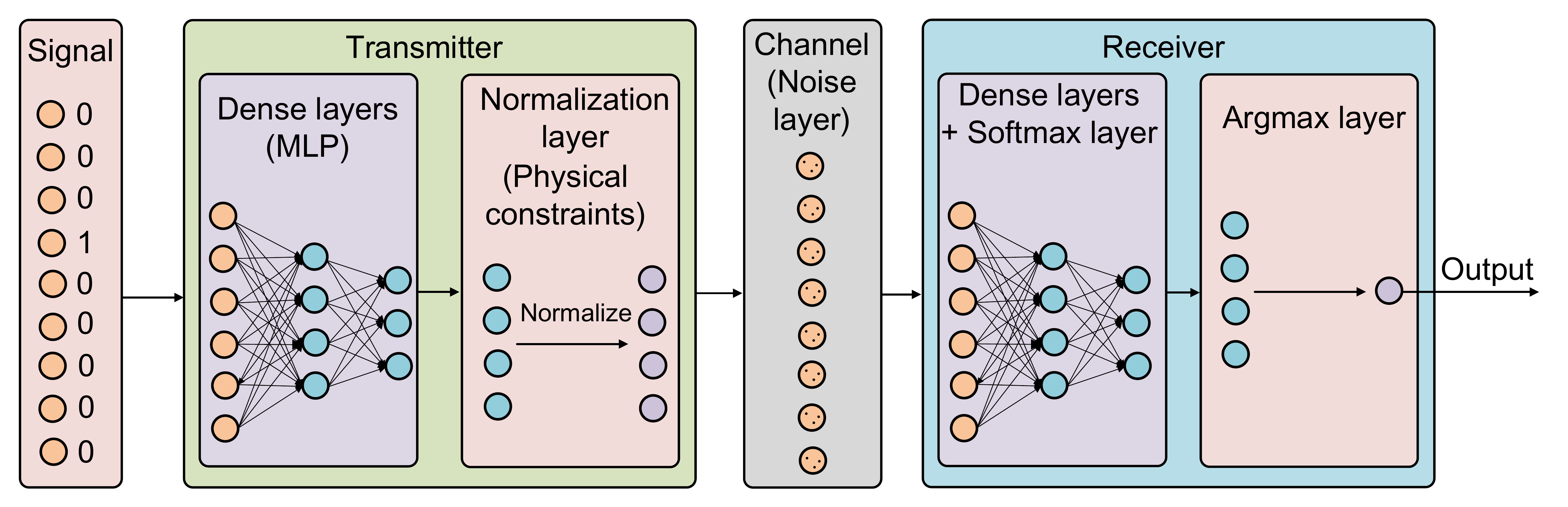}
\end{center}
\caption{\label{fig:physical} \rev{A communications system over an additive white Gaussian noise channel represented as an autoencoder.}}
\end{figure*}

\edit{
More recently, O'Shea \emph{et al.} bring deep learning to physical layer design \cite{o2017deep}. They incorporate an unsupervised deep AE into a single-user end-to-end MIMO system, to optimize representations and the encoding/decoding processes, for transmissions over a Rayleigh fading channel. \rev{We illustrate the adopted AE-based framework in Fig~\ref{fig:physical}. This design incorporates a transmitter consisting of an MLP followed by a normalization layer, which ensures that physical constraints on the signal are guaranteed. After transfer through an additive white Gaussian noise channel, a receiver employs another MLP to decode messages and select the one with the highest probability of occurrence. The system can be trained with an SGD algorithm in an end-to-end manner.} Experimental results show that the AE system outperforms the Space Time Block Code approach in terms of SNR by approximately 15 dB. In \cite{borgerding2017amp}, Borgerding \emph{et al.} propose to use deep learning to recover a sparse signal from noisy linear measurements in MIMO environments. The proposed scheme is evaluated on compressive random access and massive-MIMO channel estimation, where it achieves better accuracy over traditional algorithms and CNNs. \\}

\noindent\edit{
\textbf{Modulation:} 
West and O'Shea compare the modulation recognition accuracy of different deep learning architectures, including traditional CNN, ResNet, Inception CNN, and LSTM~\cite{west2017deep}. Their experiments suggest that the LSTM is the best candidate for modulation recognition, since it achieves the highest accuracy. Due to its superior performance, an LSTM is also employed for a similar task in \cite{rajendran2017distributed}. O'Shea \emph{et al.} then focus on tailoring deep learning architectures to radio properties. Their prior work is improved in \cite{o2016radio}, where they architect a novel deep radio transformer network for precise modulation recognition. Specifically, they introduce radio-domain specific parametric transformations into a spatial transformer network, which assists in the normalization of the received signal, thereby achieving superior performance. This framework also demonstrates automatic synchronization abilities, which reduces the dependency on traditional expert systems and expensive signal analytic processes. In \cite{timothy2017introduction}, O'Shea and Hoydis introduce several novel deep  learning applications for the network physical layer. They demonstrate a proof-of-concept where they employ a CNN for modulation classification and obtain satisfying accuracy.}\\

\noindent\edit{\textbf{Other signal processing applciations:} Deep learning is also adopted for radio signal analysis. In \cite{o2016end}, O'Shea \emph{et al.} employ an LSTM to replace sequence translation routines between radio transmitter and receiver. Although their framework works well in ideal environments, its performance drops significantly when introducing realistic channel effects. Later, the authors consider a different scenario in \cite{o2016learning}, where they exploit a regularized AE to enable reliable communications over an impaired channel. They further incorporate a radio transformer network for signal reconstruction at the decoder side, thereby achieving receiver synchronization. Simulations demonstrate that this approach is reliable and can be efficiently implemented.} 

\edit{In \cite{liang2018exploiting}, Liang \emph{et al.} exploit noise correlations to decode channels using a deep learning approach.  Specifically, they use a CNN to reduce channel noise estimation errors by learning the noise correlation. Experiments suggest that their framework can significantly improve the decoding performance. The decoding performance of MLPs , CNNs and RNNs is compared in \cite{lyu2018performance}. By conducting experiments in different setting, the obtained results suggest the RNN achieves the best decoding performance, nonetheless yielding the highest computational overhead.} \rev{Liao \emph{et al.} employ MLPs to perform accurate Rayleigh fading channel prediction~\cite{liao2018rayleigh}. The authors further equip their proposal with a sparse channel sample construction method to save system resources without compromising precision. Deep learning can further aid visible light communication. In \cite{huang2018fully}, Huang \emph{et al.} employ a deep learning based system for error correction in optical communications. Specifically, an AE is used in their work to perform dimension reduction on light-emitting diode (LED) visible light downlink, thereby maximizing the channel bandwidth . The proposal follows the theory in \cite{timothy2017introduction}, where O'Shea \emph{et al.} demonstrate that deep learning driven signal processing systems can perform as good as traditional encoding and/or modulation systems. }

\rev{Deep learning has been further adopted in solving millimeter wave beamforming. In \cite{alkhateeb2018deep}, Alkhateeb \emph{et al.} propose a millimeter wave communication system that utilizes MLPs to predict beamforming vectors from signals received from distributed base stations. By substituting a genie-aided solution with deep learning, their framework reduces the coordination overhead, enabling wide-coverage and low-latency beamforming. Similarly, Gante \emph{et al.} employ CNNs to infer the position of a device, given the received millimeter wave radiation. Their preliminary simulations show that the CNN-based system can achieve small estimation errors in a realistic outdoors scenario, significantly outperforming existing prediction approaches.}\\

\textbf{Lessons learned:} Deep learning is beginning to play an important role in signal processing applications and the performance demonstrated by early prototypes is remarkable. \rev{This is because deep learning can prove advantageous with regards to performance, complexity, and generalization capabilities. At this stage, research in this area is however incipient.} We can only expect that deep learning will become increasingly popular in this area.

\subsection{Emerging Deep Learning Applications in Mobile \mbox{Networks}}
In this part, we review work that builds upon deep learning in other mobile networking areas, which are beyond the scopes of the subjects discussed thus far. These emerging applications open several new research directions, as we discuss next. A summary of these works is given in Table~\ref{tab:emerging}.\\
 
\begin{table*}[htb]
\centering
\caption{A summary of emerging deep learning driven mobile network applications.}
\label{tab:emerging}
\begin{tabular}{|C{2.5cm}|C{3cm}|c|C{9cm}|}
\hline
\textbf{Reference}                               & \textbf{Application}                              & \textbf{Model}  & \edit{\textbf{Key contribution}}                                                                                                                                    \\ \hline
Gonzalez \emph{et al.} \cite{gonzalez2017network}  & Network data monetifzation                        & -               & \edit{A platform named Net2Vec to facilitate deep learning deployment in communication networks.}                                                            \\ \hline
Kaminski \emph{et al.} \cite{kaminski2017neural} & In-network computation for IoT                    & MLP             & \edit{Enables to perform collaborative data processing and reduces latency.}                                                                                          \\ \hline
Xiao \emph{et al.} \cite{xiao2017secure}         & Mobile crowdsensing                               & Deep Q learning & \edit{Mitigates vulnerabilities of mobile crowdsensing systems.}                                                                                \\ \hline
\edit{Luong \emph{et al.} \cite{luong2018optimal}} & \edit{Resource allocation for mobile blockchains}   & \edit{MLP}        & \edit{Employs deep learning to perform monotone transformations of miners'bids and outputs the allocation and conditional payment rules in optimal auctions. } \\ \hline
\edit{Gulati \emph{et al.} \cite{gulati2018deep}}  & \edit{Data dissemination in Internet of Vehicles (IoV)} & \edit{CNN}        & \edit{Investigates the relationship between data dissemination performance and social score, energy level, number of vehicles and their speed.}                   \\ \hline
\end{tabular}
\end{table*}

\noindent\textbf{Network Data Monetization:} Gonzalez \emph{et al.} employ unsupervised deep learning to generate real-time accurate user profiles \cite{gonzalez2017network} using an on-network machine learning platform called Net2Vec \cite{gonzalez2017net2vec}. Specifically, they analyze user browsing data in real time and generate user profiles using product categories. The profiles can be subsequently associated with the products that are of interest to the users and employed for online advertising. \\

\noindent\textbf{IoT In-Network Computation:} Instead of regarding IoT nodes as producers of data or the end consumers of processed information, Kaminski \emph{et al.} embed neural networks into an IoT deployment and allow the nodes to collaboratively process the data generated \cite{kaminski2017neural}. This enables low-latency communication, while offloading data storage and processing from the cloud. In particular, the authors map each hidden unit of a pre-trained neural network to a node in the IoT network, and investigate the optimal projection that leads to the minimum communication overhead. Their framework achieves functionality similar to in-network computation in WSNs and opens a new research directions in fog computing. \\

\noindent\textbf{Mobile Crowdsensing:}  \rev{Xiao \emph{et al.} investigate vulnerabilities facing crowdsensing in the mobile network context.} They argue that there exist malicious mobile users who intentionally provide false sensing data to servers, to save costs and preserve their privacy, which in turn can make mobile crowdsensings systems vulnerable \cite{xiao2017secure}. The authors model the server-users system as a Stackelberg game, where the server plays the role of a leader that is responsible for evaluating the sensing effort of individuals, by analyzing the accuracy of each sensing report. Users are paid based on the evaluation of their efforts, hence cheating users will be punished with zero reward. To design an optimal payment policy, the server employs a deep Q network, which derives knowledge from experience sensing reports, without requiring specific sensing models. Simulations demonstrate superior performance in terms of sensing quality, resilience to attacks, and server utility, as compared to traditional Q learning based and random payment strategies.\\


\noindent\edit{\textbf{Mobile Blockchain:} \rev{Substantial computing resource requirements and energy consumption limit the applicability of blockchain in mobile network environments.}
To mitigate this problem, Luong \emph{et al.} shed light on resource management in mobile blockchain networks based on optimal auction in \cite{luong2018optimal}.  They design an MLP to first conduct monotone transformations of the miners' bids and subsequently output the allocation scheme and conditional payment rules for each miner.  By running experiments with different settings, the results suggest the propsoed deep learning based framework can deliver much higher profit to edge computing service provider than the second-price auction baseline. }\\

\noindent\edit{\textbf{Internet of Vehicles (IoV):} Gulati \emph{et al.} extend the success of deep learning to IoV  \cite{gulati2018deep}. The authors design a deep learning-based content centric data dissemination approach that comprises three steps, namely \emph{(i)}~performing energy estimation on selected vehicles that are capable of data  dissemination; \emph{(ii)}~employing a Weiner process model to identify stable and reliable connections between vehicles;  and \emph{(iii)}~using a CNN to predict the social relationship among vehicles. Experiments unveil that the volume of data disseminated is positively related to social score, energy levels, and number of vehicles, while the speed of vehicles has negative impact on the connection probability.}\\

\textbf{Lessons learned:} \rev{The adoption of deep learning in the mobile and wireless networking domain is exciting and undoubtedly many advances are yet to appear. However, as discussed in Sec.~\ref{sec:limit},
deep learning solutions are not universal and may not be suitable for every problem. One should  rather regard deep learning as a powerful tool that can assist with fast and accurate inference, and facilitate the automation of some processes previously requiring human intervention. Nevertheless, deep learning algorithms will make mistakes, and their decisions might not be easy to interpret. In tasks that require high interpretability and low fault-tolerance, deep learning still has a long way to go, which also holds for the majority of ML algorithms.}


\section{Tailoring Deep Learning to Mobile Networks}\label{sec:tailor}
Although deep learning performs remarkably in many mobile networking areas, the No Free Lunch (NFL) theorem indicates that there is no single model that can work universally well in all problems \cite{wolpert1997no}. This implies that for any specific mobile and wireless networking problem, \edit{we may need to adapt different deep learning architectures to achieve the best performance.} In this section, we look at how to tailor deep learning to mobile networking applications from three perspectives, namely, mobile devices and systems, distributed data centers, and changing mobile network environments.

\subsection{Tailoring Deep Learning to Mobile Devices and Systems}
The ultra-low latency requirements of future 5G networks demand runtime efficiency from all operations performed by mobile systems. This also applies to deep learning driven applications. However, current mobile devices have limited hardware capabilities, which means that implementing complex deep learning architectures on such equipment may be computationally unfeasible, unless appropriate model tuning is performed. To address this issue, ongoing research improves existing deep learning architectures \cite{cheng2017survey}, such that the inference process does not violate latency or energy constraints \cite{lane2017squeezing, tang2017enabling}, nor raise any privacy concern \cite{wang2018deep}. We outline these works in Table \ref{tab:tailormobile} and discuss their key contributions next.
\begin{table*}[htb]
\centering
\caption{Summary of works on improving deep learning for mobile devices and systems.}
\label{tab:tailormobile}
\begin{tabular}{|c|C{11.5cm}|C{2cm}|}
\hline
\textbf{Reference}                                          & \textbf{Methods}                                                                 & \textbf{Target model} \\ \hline
Iandola \emph{et al.} \cite{iandola2017squeezenet}          & Filter size shrinking, reducing input channels and late downsampling             & CNN                   \\ \hline
Howard \emph{et al.} \cite{howard2017mobilenets}            & Depth-wise separable convolution                                                 & CNN                   \\ \hline
Zhang \emph{et al.} \cite{zhang2017shufflenet}              & Point-wise group convolution and channel shuffle                                 & CNN                   \\ \hline
Zhang \emph{et al.} \cite{zhang2017tucker}                  & Tucker decomposition                                                             & AE                    \\ \hline
Cao \emph{et al.} \cite{cao2017mobirnn}                     & Data parallelization by RenderScript                                             & RNN                   \\ \hline
Chen \emph{et al.} \cite{chen2016deep}                      & Space exploration for data reusability and kernel redundancy removal             & CNN                   \\ \hline
Rallapalli \emph{et al.} \cite{rallapalli2016very}          & Memory optimizations                                                             & CNN                   \\ \hline
Lane \emph{et al.} \cite{lane2016deepx}                     & Runtime layer compression and deep architecture decomposition                    & MLP, CNN              \\ \hline
Huynh \emph{et al.} \cite{huynh2017deepmon}                 & Caching, Tucker decomposition and computation offloading                         & CNN                   \\ \hline
Wu \emph{et al.} \cite{wu2016quantized}                     & Parameters quantization                                                          & CNN                   \\ \hline
Bhattacharya and Lane \cite{bhattacharya2016sparsification} & Sparsification of fully-connected layers and separation of convolutional kernels & MLP, CNN              \\ \hline
Georgiev \emph{et al.} \cite{georgiev2017low}               & Representation sharing                                                           & MLP                   \\ \hline
Cho and Brand \cite{cho2017mec}                             & Convolution operation optimization                                               & CNN                   \\ \hline
Guo and Potkonjak \cite{guo2017pruning}                     & Filters and classes pruning                                                      & CNN                   \\ \hline
Li \emph{et al.} \cite{li2017fitcnn}                        & Cloud assistance and incremental learning                                        & CNN                   \\ \hline
Zen \emph{et al.} \cite{zen2016fast}                        & Weight quantization                                                              & LSTM                  \\ \hline
Falcao \emph{et al.} \cite{falcao2017evaluation}            & Parallelization and memory sharing                                               & Stacked AE            \\ \hline
\rev{Fang \emph{et al.} \cite{fang2018nestdnn}}            & \rev{Model pruning and recovery
scheme   }                                            & \rev{CNN}           \\ \hline
\rev{Xu \emph{et al.} \cite{xu2018deepcache}}            & \rev{Reusable region lookup and reusable region propagation scheme}
                                               & \rev{CNN}           \\ \hline
\rev{Liu \emph{et al.} \cite{liu2018ondemand}}            & \rev{Using deep Q learning based optimizer to achieve appropriate balance between accuracy, latency, storage and energy consumption for deep NNs on mobile platforms}
                                               & \rev{CNN}           \\ \hline 
\rev{Chen \emph{et al.} \cite{chen2018tvm}}            & \rev{Machine learning based optimization system to  automatically explore and search for optimized tensor operators}
                                               & \rev{All NN architectures}           \\ \hline
\rev{Yao \emph{et al.} \cite{yao2018fastdeepiot}}            & \rev{Learning model execution time and performing the model compression}
                                               & \rev{MLP, CNN, GRU, and LSTM}           \\ \hline                                               
\end{tabular}
\end{table*}

\edit{Iandola \emph{et al.} design a compact architecture for embedded systems named SqueezeNet, which has similar accuracy to that of AlexNet \cite{krizhevsky2012imagenet}, a classical CNN, yet 50 times fewer parameters \cite{iandola2017squeezenet}. SqueezeNet is also based on CNNs, but its significantly smaller model size \emph{(i)} allows more efficiently training on distributed systems; \emph{(ii)} reduces the transmission overhead when updating the model at the client side; and \emph{(iii)} facilitates deployment on resource-limited embedded devices.} Howard \emph{et al.} extend this work and introduce an efficient family of streamlined CNNs called MobileNet, which uses depth-wise separable convolution operations to drastically reduce the number of computations required and the model size~\cite{howard2017mobilenets}. This new design can run with low latency and can satisfy the requirements of mobile and embedded vision applications. The authors further introduce two hyper-parameters to control the  width and resolution of multipliers, \edit{which can help strike an appropriate trade-off between accuracy and efficiency.} The ShuffleNet proposed by Zhang \emph{et al.} improves the accuracy of MobileNet by employing point-wise group convolution and channel shuffle, while retaining similar model complexity~\cite{zhang2017shufflenet}. In particular, the authors discover that more groups of convolution operations can reduce the computation requirements. 

Zhang \emph{et al.} focus on reducing the number of parameters of structures with fully-connected layers for mobile multimedia features learning \cite{zhang2017tucker}. This is achieved by applying Trucker decomposition to weight sub-tensors in the model, while maintaining decent reconstruction capability. The Trucker decomposition has also been employed in \cite{huynh2017deepmon}, where the authors seek to approximate a model with fewer parameters, in order to save memory. Mobile optimizations are further studied for RNN models. In \cite{cao2017mobirnn}, Cao \emph{et al.} use a mobile toolbox called RenderScript\footnote{Android Renderscript \url{https://developer.android.com/guide/topics/renderscript/compute.html.}} to parallelize specific data structures and enable mobile GPUs to perform computational accelerations. Their proposal reduces the latency when running RNN models on Android smartphones. Chen \emph{et al.} shed light on implementing CNNs on iOS mobile devices \cite{chen2016deep}. In particular, they reduce the model executions latency, through space exploration for data re-usability and kernel redundancy removal. The former alleviates the high bandwidth requirements of convolutional layers, while the latter reduces the memory and computational requirements, with negligible performance degradation. 

Rallapalli \emph{et al.} investigate offloading \emph{very deep} CNNs from clouds to edge devices, by employing memory optimization on both mobile CPUs and GPUs \cite{rallapalli2016very}. Their framework enables running at high speed deep CNNs with large memory requirements in mobile object detection applications. Lane \emph{et al.} develop a software accelerator, DeepX, to assist deep learning implementations on mobile devices. The proposed approach exploits two inference-time resource control algorithms, i.e., runtime layer compression and deep architecture decomposition \cite{lane2016deepx}. The runtime layer compression technique controls the memory and computation runtime during the inference phase, by extending model compression principles. This is important in mobile devices, since offloading the inference process to edge devices is more practical with current hardware platforms. Further, the deep architecture designs ``decomposition plans'' that seek to optimally allocate data and model operations to local and remote processors. By combining these two, DeepX enables maximizing energy and runtime efficiency, under given computation and memory constraints. \rev{Yao \emph{et al.} \cite{yao2018fastdeepiot} design a framework called FastDeepIoT, which fisrt learns the execution time of NN models on target devices, and subsequently conducts model compression to reduce the runtime without compromising the inference accuracy. Through this process, up to
78\% of execution time and 69\% of energy consumption is reduced, compared to state-of-the-art compression algorithms.}

\rev{More recently, Fang \emph{et al.} design a framework called NestDNN, to provide flexible resource-accuracy trade-offs on mobile devices \cite{fang2018nestdnn}. To this end, the NestDNN first adopts a model pruning and recovery scheme, which  translates deep NNs to single compact multi-capacity models. With this approach up to 4.22\% inference accuracy can be achieved with six mobile vision applications, at a 2.0$\times$ faster video frame processing rate and reducing energy consumption by 1.7$\times$. In \cite{xu2018deepcache}, Xu \emph{et al.} accelerate deep learning inference for mobile vision from the caching perspective. In particular, the proposed framework called DeepCache stores recent input frames as cache keys and recent feature maps for individual CNN layers as cache values. The authors further employ reusable region lookup and reusable region propagation, to enable a region matcher to only run once per input video frame and load cached feature maps at all layers inside the CNN. This reduces the inference time by 18\% and energy consumption by 20\% on average. Liu \emph{et al.} develop a usage-driven framework named AdaDeep, to select a combination of compression techniques for a specific deep NN on mobile platforms \cite{liu2018ondemand}. By using a deep Q learning optimizer, their proposal can achieve appropriate trade-offs between accuracy, latency, storage and energy consumption.} 

Beyond these works, researchers also successfully adapt deep learning architectures through other designs and sophisticated optimizations, such as parameters quantization \cite{wu2016quantized, zen2016fast}, sparsification and separation \cite{bhattacharya2016sparsification}, representation and memory sharing \cite{georgiev2017low, falcao2017evaluation}, convolution operation optimization \cite{cho2017mec}, pruning \cite{guo2017pruning}, cloud assistance \cite{li2017fitcnn} and compiler optimization \cite{chen2018tvm}. These techniques will be of great significance when embedding deep neural networks into mobile systems.


\subsection{Tailoring Deep Learning to Distributed Data Containers}
\begin{table*}[htb]
\centering
\caption{\edit{Summary of work on model and training parallelism for mobile systems and devices.}}
\label{tab:parallelism}
\color{black}
\begin{tabular}{|p{1.5cm}|C{2cm}|C{3cm}|C{6cm}|C{3cm}|}
\hline
\textbf{Parallelism Paradigm}         & \textbf{Reference}                                                  & \textbf{Target}                                                            & \textbf{Core Idea}                                                                                                                                                                                      & \textbf{Improvement}                                                \\ \hline
\multirow{4}{2cm}{Model parallelism}    & Dean \emph{et al.} \cite{dean2012large}                             & Very large deep neural networks in distributed systems.                    & Employs downpour SGD to support a large number of model replicas and a Sandblaster framework to support a variety of batch optimizations.                                                            & Up to 12$\times$ model training speed up, using 81 machines.      \\ \cline{2-5} 
                                      & Teerapittayanon \emph{et al.} \cite{teerapittayanon2017distributed} & Neural network on cloud and end devices.                                   & Maps a deep neural network to a distributed setting and jointly trains each individual section.                                                                                                            & Up to 20$\times$ reduction of communication cost.                      \\ \cline{2-5} 
                                      & De Coninck \emph{et al.} \cite{de2016distributed}                   & Neural networks on IoT devices.                                            & Distills from a pre-trained NN to obtain a smaller NN that performs classification on a subset of the entire space.                                                    & 10ms inference latency on a mobile device.            \\ \cline{2-5} 
                                      & Omidshafiei \emph{et al.} \cite{omidshafiei2017deep}                & Multi-task multi-agent reinforcement learning under partial observability. & Deep recurrent Q-networks \& cautiously-optimistic learners to approximate action-value function; decentralized concurrent experience replays trajectories to stabilize  training. & Near-optimal execution time.                    \\ \hline
\multirow{6}{2cm}{Training parallelism} & Recht \emph{et al.} \cite{recht2011hogwild}                         & Parallelized SGD.                                                          & Eliminates overhead associated with locking in distributed SGD.                                                                                                                                          & Up to 10$\times$ speed up in distributed training.               \\ \cline{2-5} 
                                      & Goyal \emph{et al.} \cite{goyal2017accurate}                        & Distributed synchronous SGD.                                               & Employs a hyper-parameter-free learning rule to adjust the learning rate and a warmup mechanism to address the early optimization problem.                                                               & Trains billions of images per day.                                   \\ \cline{2-5} 
                                      & Zhang \emph{et al.} \cite{zhang2016asynchronous}                    & Asynchronous distributed SGD.                                              & Combines the stochastic variance reduced gradient algorithm and a delayed proximal gradient algorithm.                                                                                                   & Up to 6$\times$ speed up                                        \\ \cline{2-5} 
                                      & Hardy \emph{et al.} \cite{hardy2017distributed}                     & Distributed deep learning on edge-devices.                                 & Compression technique (\emph{AdaComp}) to reduce ingress traffic at parameter severs.                                                                                                               & Up to 191$\times$ reduction in ingress traffic. \\ \cline{2-5} 
                                      & McMahan \emph{et al.} \cite{pmlr-v54-mcmahan17a}                    & Distributed training on mobile devices.                                    & Users collectively enjoy benefits of shared models trained with big data without centralized storage.                                                                    & Up to 64.3$\times$ training speedup.                      \\ \cline{2-5} 
                                      & Keith \emph{et al.} \cite{cryptoeprint:2017:281}                    & Data computation over mobile devices.                                      & Secure multi-party computation to obtain model parameters on distributed mobile devices.                                                                                               & Up to 1.98$\times$ communication expansion.                   \\ \hline
\end{tabular}
\end{table*}

\edit{Mobile systems generate and consume massive volumes of mobile data every day. This may involve similar content, but which is distributed around the world. Moving such data to centralized servers to perform model training and evaluation inevitably introduces communication and storage overheads, which does not scale. However, neglecting characteristics embedded in mobile data, which are associated with local culture, human mobility, geographical topology, etc., during model training can compromise the robustness of the model and implicitly the performance of the mobile network applications that build on such models. The solution is to offload model execution to distributed data centers or edge devices, to guarantee good performance, whilst alleviating the burden on the cloud.}

\edit{As such, one of the challenges facing parallelism, in the context of mobile networking, is that of training neural networks on a large number of mobile devices that are battery powered, have limited computational capabilities and in particular lack GPUs. The key goal of this paradigm is that of training with a large number of mobile CPUs at least as effective as with GPUs. The speed of training remains important, but becomes a secondary goal.}

\begin{figure*}[h!]
\begin{center}
\includegraphics[width=1\textwidth]{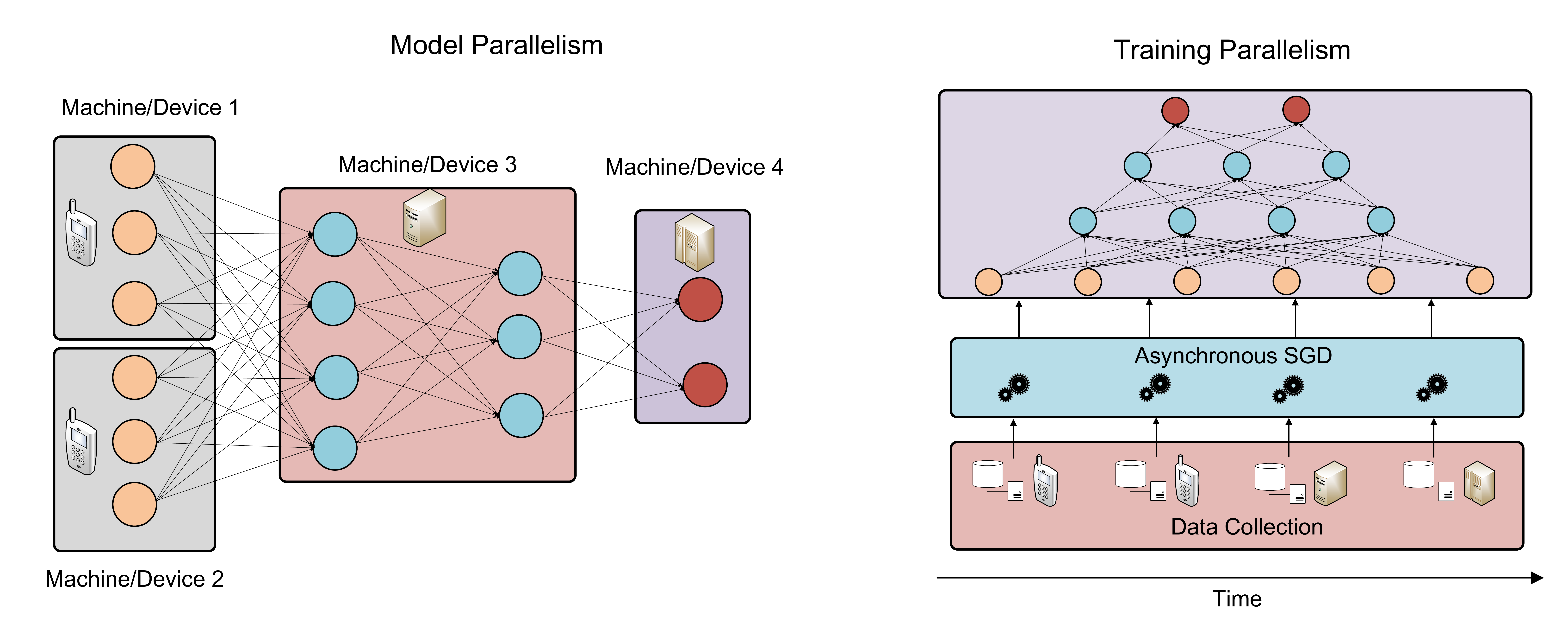}
\end{center}
\vspace*{-0,5em}
\caption{\label{fig:distributed} The underlying principles of model parallelism (left) and training parallelism (right).}
\end{figure*}

Generally, there are two routes to addressing this problem, namely, \emph{(i)} decomposing the model itself, to train (or make inference with) its components individually; or \emph{(ii)} scaling the training process to perform model update at different locations associated with data containers. Both schemes allow one to train a single model without requiring to centralize all data. \edit{We illustrate the principles of these two approaches in Fig. \ref{fig:distributed} and summarize the existing work  in Table~\ref{tab:parallelism}.}\\

\noindent \textbf{Model Parallelism.} Large-scale distributed deep learning is first studied in \cite{dean2012large}, where the authors develop a framework named \emph{DistBelief}, which enables training complex neural networks on thousands of machines. In their framework, the full model is partitioned into smaller components and distributed over various machines. Only nodes with edges (e.g. connections between layers) that cross boundaries between machines are required to communicate for parameters update and inference. This system further involves a parameter server, which enables each model replica to obtain the latest parameters during training. Experiments demonstrate that the proposed framework can be training significantly faster on a CPU cluster, compared to training on a single GPU, while achieving state-of-the-art classification performance on ImageNet~\cite{ILSVRC15}.

Teerapittayanon \emph{et al.} propose deep neural networks tailored to distributed systems, which include cloud servers, fog layers and geographically distributed devices \cite{teerapittayanon2017distributed}. The authors scale the overall neural network architecture and distribute its components hierarchically from cloud to end devices. The model exploits local aggregators and binary weights, to reduce computational storage, and communication overheads, while maintaining decent accuracy. Experiments on a multi-view multi-camera  dataset demonstrate that this proposal can perform efficient cloud-based training and local inference. Importantly, without violating latency constraints, the deep neural network obtains essential benefits associated with distributed systems, such as fault tolerance and privacy.

Coninck \emph{et al.} consider distributing deep learning over IoT for classification applications \cite{de2016distributed}. Specifically, they deploy a small neural network to local devices, to perform coarse classification, which enables fast response filtered data to be sent to central servers. If the local model fails to classify, the larger neural network in the cloud is activated to perform fine-grained classification. The overall architecture maintains good accuracy, while significantly reducing the latency typically introduced by large model inference.

Decentralized methods can also be applied to deep reinforcement learning. In \cite{omidshafiei2017deep}, Omidshafiei \emph{et al.} consider a multi-agent system with partial observability and limited communication, which is common in mobile systems. They combine a set of sophisticated methods and algorithms, including hysteresis learners, a deep recurrent Q network, concurrent experience replay trajectories and distillation, to enable multi-agent coordination using a single joint policy under a set of decentralized partially observable MDPs. Their framework can potentially play an important role in addressing control problems in distributed mobile systems. 
\\

\noindent \textbf{Training Parallelism} is also essential for mobile system, as mobile data usually come asynchronously from different sources. Training models effectively while maintaining consistency, fast convergence, and accuracy remains however challenging~\cite{gupta2016model}. 

A practical method to address this problem is to perform asynchronous SGD. The basic idea is to enable the server that maintains a model to accept delayed information (e.g. data, gradient updates) from workers. At each update iteration, the server only requires to wait for a smaller number of workers. This is essential for training a deep neural network over distributed machines in mobile systems. The asynchronous SGD is first studied in \cite{recht2011hogwild}, where the authors propose a lock-free parallel SGD named HOGWILD, which demonstrates significant faster convergence over locking counterparts. The Downpour SGD in \cite{dean2012large} improves the robustness of the training process when work nodes breakdown, as each model replica requests the latest version of the parameters. Hence a small number of machine failures will not have a significant impact on the training process. A similar idea has been employed in \cite{goyal2017accurate}, where Goyal \emph{et al.} investigate the usage of a set of techniques (i.e. learning rate adjustment, warm-up, batch normalization), which offer important insights into training large-scale deep neural networks on distributed systems. Eventually, their framework can train an network on ImageNet within 1 hour, which is impressive in comparison with traditional algorithms.

Zhang \emph{et al.} argue that most of asynchronous SGD algorithms suffer from slow convergence, due to the inherent variance of stochastic gradients \cite{zhang2016asynchronous}. They propose an improved SGD with variance reduction to speed up the convergence. Their algorithm outperforms other asynchronous SGD approaches in terms of convergence, when training deep neural networks on the Google Cloud Computing Platform. The asynchronous method has also been applied to deep reinforcement learning. In \cite{mnih2016asynchronous}, the authors create multiple environments, which allows agents to perform asynchronous updates to the main structure. The new A3C algorithm breaks the sequential dependency and speeds up the training of the traditional Actor-Critic algorithm significantly. In \cite{hardy2017distributed}, Hardy \emph{et al.} further study distributed deep learning over cloud and edge devices. In particular, they propose a training algorithm, \emph{AdaComp}, which allows to compress worker updates of the target model. This significantly reduce the communication overhead between cloud and edge, while retaining good fault tolerance.

Federated learning is an emerging parallelism approach that enables mobile devices to collaboratively learn a shared model, while retaining all training data on individual devices \cite{pmlr-v54-mcmahan17a, mcmahan2017federated}. Beyond offloading the training data from central servers, this approach performs model updates with a Secure Aggregation protocol \cite{cryptoeprint:2017:281}, which decrypts the average updates only if enough users have participated, without inspecting individual updates. 

\subsection{Tailoring Deep Learning to Changing Mobile Network Environments}\label{sec:changing}
\begin{figure*}[htb]
\begin{center}
\includegraphics[width=1\textwidth]{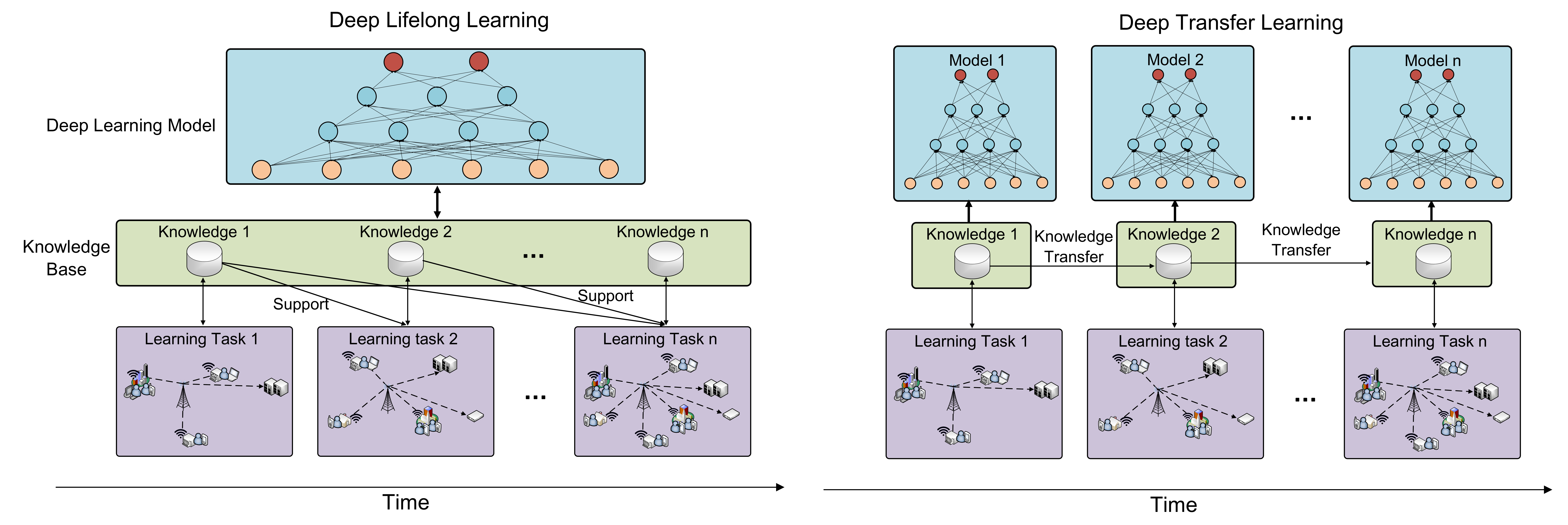}
\end{center}
\caption{\label{fig:changeable} The underlying principles of deep lifelong learning (left) and deep transfer learning (right). Lifelong learning retains the knowledge learned while transfer learning exploits labeled data of one domain to learn in a new target domain.}
\end{figure*}

Mobile network environments often exhibit changing patterns over time. For instance, the spatial distributions of mobile data traffic over a region may vary significantly between different times of the day \cite{furno2017joint}. Applying a deep learning model in changing mobile environments requires lifelong learning ability to continuously absorb new features, without forgetting old but essential patterns. Moreover, new smartphone-targeted viruses are spreading fast via mobile networks and may severely jeopardize users' privacy and business profits. These pose unprecedented challenges to current anomaly detection systems and anti-virus software, as such tools must react to new threats in a timely manner, using limited information. To this end, the model should have transfer learning ability, which can enable the fast transfer of knowledge from pre-trained models to different jobs or datasets. This will allow models to work well with limited threat samples (one-shot learning) or limited metadata descriptions of new threats (zero-shot learning). Therefore, both lifelong learning and transfer learning are essential for applications in ever changing mobile network environments. We illustrated these two learning paradigms in Fig. \ref{fig:changeable} and review essential research in this subsection.\\

\noindent \textbf{Deep Lifelong Learning} mimics human behaviors and seeks to build a machine that can continuously adapt to new environments, retain as much knowledge as possible from previous learning experience~\cite{chen2016lifelong}.
There exist several research efforts that adapt traditional deep learning to lifelong learning. For example, Lee \emph{et al.} propose a dual-memory deep learning architecture for lifelong learning of everyday human behaviors over non-stationary data streams \cite{lee2016dual}. To enable the pre-trained model to retain old knowledge while training with new data, their architecture includes two memory buffers, namely a deep memory and a fast memory. The deep memory is composed of several deep networks, which are built when the amount of data from an unseen distribution is accumulated and reaches a threshold. The fast memory component is a small neural network, which is updated immediately when coming across a new data sample. These two memory modules allow to perform continuous learning without forgetting old knowledge. Experiments on a non-stationary image data stream prove the effectiveness of this model, as it significantly outperforms other online deep learning algorithms. The memory mechanism has also been applied in \cite{graves2016hybrid}. In particular, the authors introduce a differentiable neural computer, which allows neural networks to dynamically read from and write to an external memory module. This enables lifelong lookup and forgetting of knowledge from external sources, as humans do.

Parisi \emph{et al.} consider a different lifelong learning scenario in \cite{parisi2017lifelong}. They abandon the memory modules in \cite{lee2016dual} and design a self-organizing architecture with recurrent neurons for processing time-varying patterns. A variant of the Growing When Required network is employed in each layer, to to predict neural activation sequences from the previous network layer. This allows learning time-vary correlations between inputs and labels, without requiring a predefined number of classes. Importantly, the framework is robust, as it has tolerance to missing and corrupted sample labels, which is common in mobile data.

Another interesting deep lifelong learning architecture is presented in \cite{tessler2017deep}, where Tessler \emph{et al.} build a DQN agent that can retain learned skills in playing the famous computer game Minecraft. The overall framework includes a pre-trained model, Deep Skill Network, which is trained a-priori on various sub-tasks of the game. When learning a new task, the old knowledge is maintained by incorporating reusable skills through a Deep Skill module, which consists of a Deep Skill Network array and a multi-skill distillation network. These allow the agent to selectively transfer knowledge to solve a new task. Experiments demonstrate that their proposal significantly outperforms traditional double DQNs in terms of accuracy and convergence. This technique has potential to be employed in solving mobile networking problems, as it can continuously acquire new knowledge.\\

\noindent \textbf{Deep Transfer Learning:} Unlike lifelong learning, transfer learning only seeks to use knowledge from a specific domain to aid learning in a target domain. Applying transfer learning can accelerate the new learning process, as the new task does not require to learn from scratch. This is essential to mobile network environments, as they require to agilely respond to new network patterns and threats. A number of important applications emerge in the computer network domain \cite{valente2017survey}, such as Web mining \cite{lopez2018deep}, caching \cite{bacstuug2015transfer} and base station sleep strategies \cite{li2014tact}. 

There exist two extreme transfer learning paradigms, namely one-shot learning and zero-shot learning. One-shot learning refers to a learning method that gains as much information as possible about a category from only one or a handful of samples, given a pre-trained model \cite{fei2006one}. On the other hand, zero-shot learning does not require any sample from a category \cite{palatucci2009zero}. It aims at learning a new distribution given meta description of the new category and correlations with existing training data. Though research towards deep one-shot learning \cite{rezende2016one, vinyals2016matching} and deep zero-shot learning \cite{changpinyo2016synthesized, oh2017zero} is in its infancy, both paradigms are very promising in detecting new threats or traffic patterns in mobile networks.

\section{Future Research Perspectives}\label{sec:future}
As deep learning is achieving increasingly promising results in the mobile networking domain, several important research issues remain to be addressed in the future. We conclude our survey by discussing these challenges and pinpointing key mobile networking research problems that could be tackled with novel deep learning tools. 

\subsection{Serving Deep Learning with Massive High-Quality Data}
Deep neural networks rely on massive and high-quality data to achieve good performance. When training a large and complex architecture, data volume and quality are very important, as deeper models usually have a huge set of parameters to be learned and configured. This issue remains true in mobile network applications. Unfortunately, unlike in other research areas such as computer vision and NLP, high-quality and large-scale labeled datasets still lack for mobile network applications, because service provides and operators keep the data collected confidential and are reluctant to release datasets. While this makes sense from a user privacy standpoint, to some extent it restricts the development of deep learning mechanisms for problems in the mobile networking domain. Moreover, mobile data collected by sensors and network equipment are frequently subject to loss, redundancy, mislabeling and class imbalance, and thus cannot be directly employed for training purpose.

To build intelligent 5G mobile network architecture, efficient and mature streamlining platforms for mobile data processing are in demand. This requires considerable amount of research efforts for data collection, transmission, cleaning, clustering, transformation, and annonymization. Deep learning applications in the mobile network area can only advance if researchers and industry stakeholder release more datasets, with a view to benefiting a wide range of communities.

\subsection{Deep Learning for Spatio-Temporal Mobile Data Mining}\label{sec:st-traffic}
\begin{figure}[t]
\begin{center}
\includegraphics[width=0.5\textwidth]{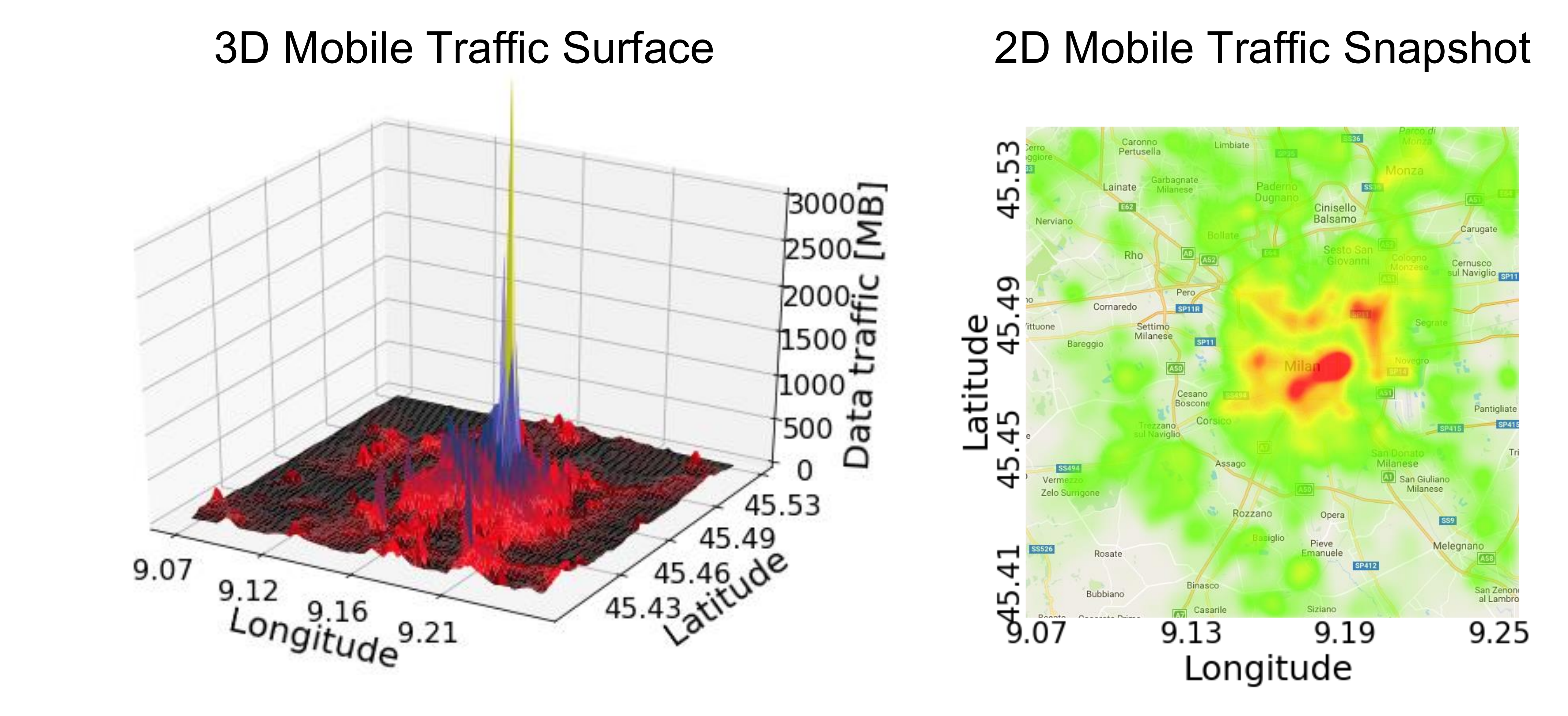}
\end{center}
\caption{\label{fig:mtraffic} Example of a 3D mobile traffic surface (left) and 2D projection (right) in  Milan, Italy. Figures adapted from \cite{chaoyun2017zipnet} using data from \cite{barlacchi2015multi}.}
\end{figure}

Accurate analysis of mobile traffic data over a geographical region is becoming increasingly essential for event localization, network resource allocation, context-based advertising and urban planning \cite{furno2017joint}. However, due to the mobility of smartphone users, the spatio-temporal distribution of mobile traffic \cite{wang2015understanding} and application popularity \cite{marquez2017apps} are difficult to understand (see the example city-scale traffic snapshot in Fig.~\ref{fig:mtraffic}).
Recent research suggests that data collected by mobile sensors (e.g. mobile traffic) over a city can be regarded as pictures taken by panoramic cameras, which provide a city-scale sensing system for urban surveillance \cite{liu2015urban}. These traffic sensing images enclose information associated with the movements of individuals \cite{naboulsi2016large}. 

\begin{figure}[htb]
\begin{center}
\includegraphics[width=0.5\textwidth]{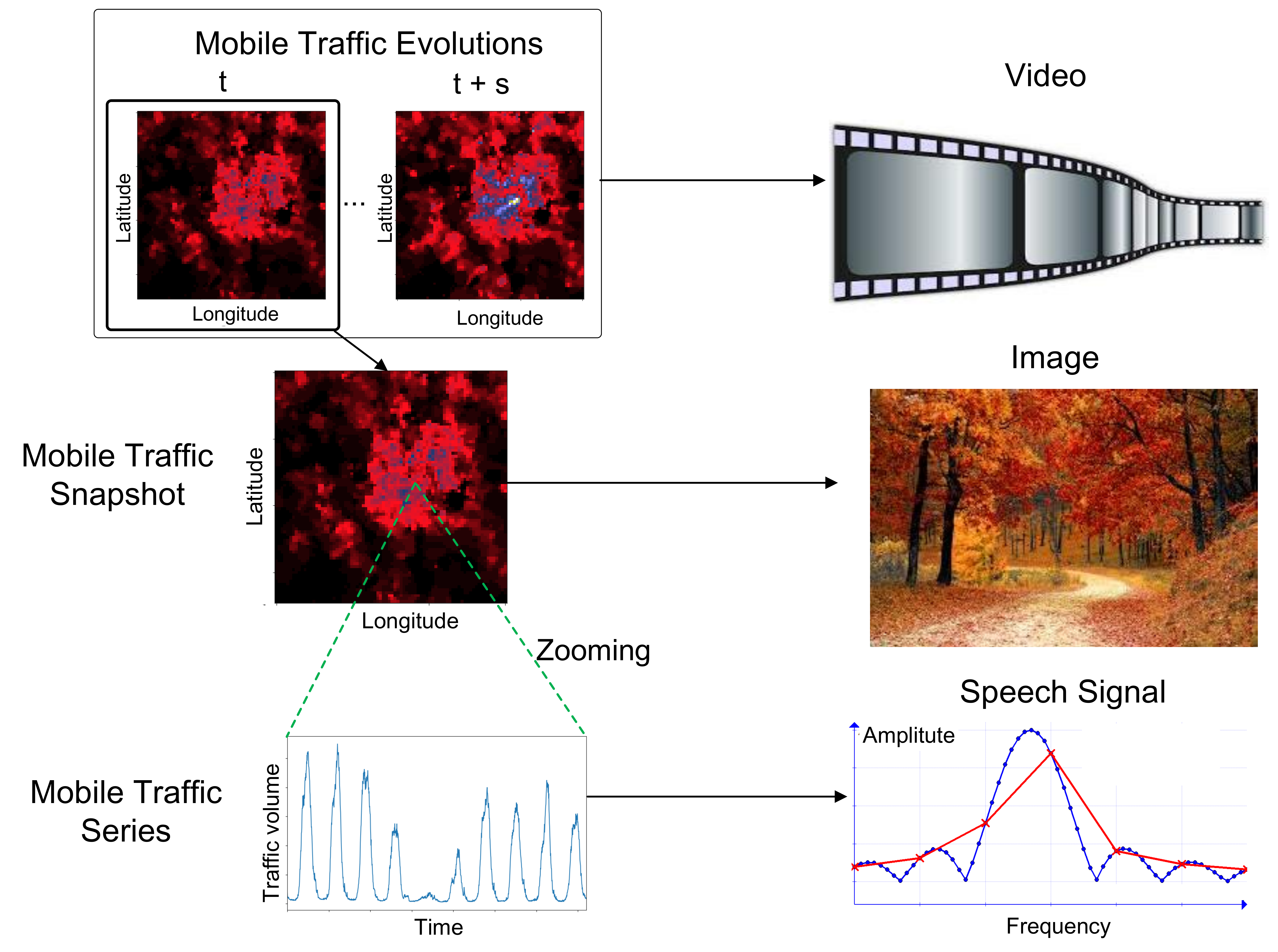}
\end{center}
\caption{\label{fig:compare} Analogies between mobile traffic data consumption in a city (left) and other types of data (right). }
\end{figure}

From both spatial and temporal dimensions perspective, we recognize that mobile traffic data have important similarity with videos or speech, which is an analogy made recently also in \cite{chaoyun2017zipnet} and exemplified in Fig.~\ref{fig:compare}.
Specifically, both videos and the large-scale evolution of mobile traffic are composed of sequences of ``frames''. Moreover, if we zoom into a small coverage area to measure long-term traffic consumption, we can observe that a single traffic consumption series looks similar to a natural language sequence. These observations suggest that, to some extent, well-established tools for computer vision (e.g. CNN) or NLP (e.g. RNN, LSTM) are promising candidate for mobile traffic analysis.

Beyond these similarity, we observe several properties of mobile traffic that makes it unique in comparison with images or language sequences. Namely,

\begin{enumerate}
\item The values of neighboring `pixels' in fine-grained traffic snapshots are not significantly different in general, while this happens quite often at the edges of natural images.
\item Single mobile traffic series usually exhibit some periodicity (both daily and weekly), yet this is not a feature seen among video pixels.
\item Due to user mobility, traffic consumption is more likely to stay or shift to neighboring cells in the near future, which is less likely to be seen in videos.
\end{enumerate}
Such spatio-temporal correlations in mobile traffic can be exploited as prior knowledge for model design. We recognize several unique advantages of employing deep learning for mobile traffic data mining:
\edit{
\begin{enumerate}
\item CNN structures work well in imaging applications, thus can also serve mobile traffic analysis tasks, given the analogies mentioned before.
\item LSTMs capture well temporal correlations in time series data such as natural language; hence this structure can also be adapted to traffic forecasting problems. 
\item GPU computing enables fast training of NNs and together with parallelization techniques can support low-latency mobile traffic analysis via deep learning tools.
\end{enumerate}
In essence, we expect deep learning tools tailored to mobile networking, will overcome the limitation of traditional regression and interpolation tools such as Exponential Smoothing~\cite{tikunov2007traffic}, Autoregressive Integrated Moving Average model~\cite{Kim2011}, or unifrom interpolation, which are commonly used in operational networks.}

\subsection{\rev{Deep learning for Geometric Mobile Data Mining}}
\begin{figure*}[t]
\begin{center}
\includegraphics[width=1\textwidth]{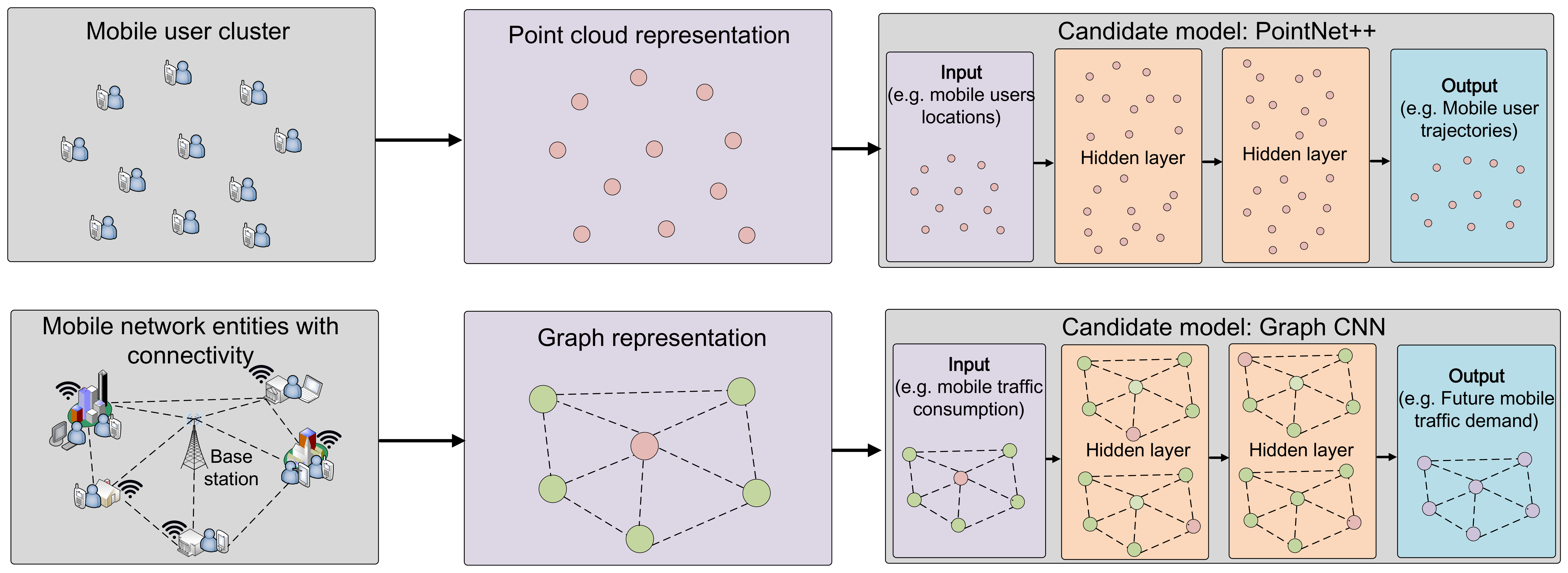}
\end{center}
\caption{\label{fig:geometric} \rev{Examples of mobile data with geometric properties (left), their geometric representations (middle) and their candidate models for analysis (right). PointNet++ could be used to infer user trajectories when fed with point cloud representations of user locations (above); A GraphCNN may be employed to forecast future mobile traffic demand at base station level (below).}}
\end{figure*}

\rev{As discussed in Sec.~\ref{sec:adv}, certain mobile data has important geometric properties. For instance, the location of mobile users or base stations along with the data carried can be viewed as point clouds in a 2D plane. If the temporal dimension is also added, this leads to a 3D point cloud representation, with either fixed or changing locations. In addition, the connectivity of mobile devices, routers, base stations, gateways, and so on can naturally construct a directed graph, where entities are represented as vertices, the links between them can be seen as edges, and data flows  may give direction to these edges. We show examples of geometric mobile data and their potential representations in Fig.~\ref{fig:geometric}. At the top of the figure a group of mobile users is represented as a point cloud. Likewise, mobile network entities (e.g. base station, gateway, users) are regarded as graphs below, following the rationale explained below. Due to the inherent complexity of such representations, traditional ML tools usually struggle to interpret geometric data and make reliable inferencess.}

\rev{In contrast, a variety of deep learning toolboxes for modelling geometric data exist, albeit not having been widely employed in mobile networking yet. For instance, PointNet \cite{charles2017pointnet} and the follow on PointNet++ \cite{qi2017pointnet} are the first solutions that employ deep learning for 3D point cloud applications, including classification and segmentation \cite{ioannidou2017deep}. We recognize that similar ideas can be applied to geometric mobile data analysis, such as clustering of mobile users or base stations, or user trajectory predictions. Further, deep learning for graphical data analysis is also evolving rapidly \cite{scarselli2009graph}. This is triggered by  research on Graph CNNs \cite{kipf2016semi}, which brings convolution concepts to graph-structured data. The applicability of Graph CNNs can be further extend to the temporal domain \cite{yuan2017temporal}. One possible application is the prediction of future traffic demand at individual base station level. We expect that such novel architectures will play an increasingly important role in network graph analysis and applications such as anomaly detection over a mobile network graph.}

\subsection{Deep Unsupervised Learning in Mobile Networks}
We observe that current deep learning practices in mobile networks largely employ supervised learning and reinforcement learning. However, as mobile networks generate considerable amounts of unlabeled data every day, data labeling is costly and requires domain-specific knowledge. To facilitate the analysis of raw mobile network data, unsupervised learning becomes essential in extracting insights from unlabeled data~\cite{usama2017unsupervised}, so as to optimize the mobile network functionality to improve QoE. 

The potential of a range of unsupervised deep learning tools including AE, RBM and GAN remains to be further explored. In general, these models require light feature engineering and are thus promising for learning from heterogeneous and unstructured mobile data. For instance, deep AEs work well for unsupervised anomaly detection~\cite{zhou2017anomaly}. Though less popular, RBMs can perform layer-wise unsupervised pre-training, which can accelerate the overall model training process. GANs are good at imitating data distributions, thus could be employed to mimic real mobile network environments. Recent research reveals that GANs can even protect communications by crafting custom cryptography to avoid eavesdropping~\cite{abadi2016learning}. All these tools require further research to fulfill their full potentials in the mobile networking domain.

\subsection{Deep Reinforcement Learning for Mobile Network Control}
Many mobile network control problems have been solved by constrained optimization, dynamic programming and game theory approaches. Unfortunately, these methods either make strong assumptions about the objective functions (e.g. function convexity) or data distribution (e.g. Gaussian or Poisson distributed), or suffer from high time and space complexity. As mobile networks become increasingly complex, such assumptions sometimes turn unrealistic. The objective functions are further affected by their increasingly large sets of variables, that pose severe computational and memory challenges to existing mathematical approaches.   

In contrast, deep reinforcement learning does not make strong assumptions about the target system. It employs function approximation, which explicitly addresses the problem of large state-action spaces, enabling reinforcement learning to scale to network control problems that were previously considered hard. Inspired by remarkable achievements in Atari~\cite{mnih2015human} and Go \cite{Silver1140} games, a number of researchers begin to explore DRL to solve complex network control problems, as we discussed in Sec. \ref{sec:control}. However, these works only scratch the surface and the potential of DRL to tackle mobile network control problems remains largely unexplored. For instance, as DeepMind trains a DRL agent to reduce Google's data centers cooling bill,\footnote{DeepMind AI Reduces Google Data Center Cooling Bill by 40\% \url{https://deepmind.com/blog/deepmind-ai-reduces-google-data-centre-cooling-bill-40/}} DRL could be exploited to extract rich features from cellular networks and enable intelligent on/off base stations switching, to reduce the infrastructure's energy footprint. Such exciting applications make us believe that advances in DRL that are yet to appear can revolutionize the autonomous control of future mobile networks.

\subsection{Summary}\label{sec:conclusion}
Deep learning is playing an increasingly important role in the mobile and wireless networking domain. In this paper, we provided a comprehensive survey of recent work that lies at the intersection between deep learning and mobile networking. We summarized both basic concepts and advanced principles of various deep learning models, then reviewed work specific to mobile networks across different application scenarios. We discussed how to tailor deep learning models to general mobile networking applications, an aspect overlooked by previous surveys. We concluded by pinpointing several open research issues and promising directions, which may lead to valuable future research results. Our hope is that this article will become a definite guide to researchers and practitioners interested in applying machine intelligence to complex problems in mobile network environments.

\section*{Acknowledgement}
We would like to thank Zongzuo Wang for sharing valuable insights on deep learning, which helped improving the quality of this paper. \edit{We also thank the anonymous reviewers, whose detailed and thoughtful feedback helped us give this survey more depth and a broader scope.} 


\ifCLASSOPTIONcaptionsoff
  \newpage
\fi

\bibliographystyle{unsrt}
\bibliography{references}

\begin{thebibliography}{100}

\bibitem{cisco2017}
{Cisco}.
\newblock {Cisco Visual Networking Index: Forecast and Methodology, 2016-2021},
  June 2017.

\bibitem{wang2015backhauling}
Ning Wang, Ekram Hossain, and Vijay~K Bhargava.
\newblock Backhauling {5G} small cells: A radio resource management
  perspective.
\newblock {\em IEEE Wireless Communications}, 22(5):41--49, 2015.

\bibitem{giust2015distributed}
Fabio Giust, Luca Cominardi, and Carlos~J Bernardos.
\newblock Distributed mobility management for future {5G} networks: overview
  and analysis of existing approaches.
\newblock {\em IEEE Communications Magazine}, 53(1):142--149, 2015.

\bibitem{agiwal2016next}
Mamta Agiwal, Abhishek Roy, and Navrati Saxena.
\newblock Next generation {5G} wireless networks: A comprehensive survey.
\newblock {\em IEEE Communications Surveys \& Tutorials}, 18(3):1617--1655,
  2016.

\bibitem{gupta2015survey}
Akhil Gupta and Rakesh~Kumar Jha.
\newblock A survey of {5G} network: Architecture and emerging technologies.
\newblock {\em IEEE access}, 3:1206--1232, 2015.

\bibitem{zheng2016big}
Kan Zheng, Zhe Yang, Kuan Zhang, Periklis Chatzimisios, Kan Yang, and Wei
  Xiang.
\newblock Big data-driven optimization for mobile networks toward {5G}.
\newblock {\em IEEE network}, 30(1):44--51, 2016.

\bibitem{jiang2017machine}
Chunxiao Jiang, Haijun Zhang, Yong Ren, Zhu Han, Kwang-Cheng Chen, and Lajos
  Hanzo.
\newblock Machine learning paradigms for next-generation wireless networks.
\newblock {\em IEEE Wireless Communications}, 24(2):98--105, 2017.

\bibitem{nguyen2017reinforcement}
Duong~D Nguyen, Hung~X Nguyen, and Langford~B White.
\newblock Reinforcement learning with network-assisted feedback for
  heterogeneous rat selection.
\newblock {\em IEEE Transactions on Wireless Communications}, 2017.

\bibitem{narudin2016evaluation}
Fairuz~Amalina Narudin, Ali Feizollah, Nor~Badrul Anuar, and Abdullah Gani.
\newblock Evaluation of machine learning classifiers for mobile malware
  detection.
\newblock {\em Soft Computing}, 20(1):343--357, 2016.

\bibitem{hsieh2017gaia}
Kevin Hsieh, Aaron Harlap, Nandita Vijaykumar, Dimitris Konomis, Gregory~R
  Ganger, Phillip~B Gibbons, and Onur Mutlu.
\newblock Gaia: Geo-distributed machine learning approaching {LAN} speeds.
\newblock In {\em USENIX Symposium on Networked Systems Design and
  Implementation (NSDI)}, pages 629--647, 2017.

\bibitem{xiao2017tux2}
Wencong Xiao, Jilong Xue, Youshan Miao, Zhen Li, Cheng Chen, Ming Wu, Wei Li,
  and Lidong Zhou.
\newblock Tux2: Distributed graph computation for machine learning.
\newblock In {\em USENIX Symposium on Networked Systems Design and
  Implementation (NSDI)}, pages 669--682, 2017.

\bibitem{anareport}
{Paolini, Monica and Fili, Senza }.
\newblock {Mastering Analytics: How to benefit from big data and network
  complexity: An Analyst Report}.
\newblock {\em RCR Wireless News}, 2017.

\bibitem{zhang2015convolutional}
Chaoyun Zhang, Pan Zhou, Chenghua Li, and Lijun Liu.
\newblock A convolutional neural network for leaves recognition using data
  augmentation.
\newblock In {\em Proc. IEEE International Conference on Pervasive Intelligence
  and Computing (PICOM)}, pages 2143--2150, 2015.

\bibitem{socher2012deep}
Richard Socher, Yoshua Bengio, and Christopher~D Manning.
\newblock Deep learning for {NLP} (without magic).
\newblock In {\em Tutorial Abstracts of ACL 2012}, pages 5--5. Association for
  Computational Linguistics.

\bibitem{specialissue}
{IEEE Network special issue: Exploring Deep Learning for Efficient and Reliable
  Mobile Sensing}.
\newblock
  \url{http://www.comsoc.org/netmag/cfp/exploring-deep-learning-efficient-and-reliable-mobile-sensing},
  2017.
\newblock [Online; accessed 14-July-2017].

\bibitem{wang2017machine}
Mowei Wang, Yong Cui, Xin Wang, Shihan Xiao, and Junchen Jiang.
\newblock Machine learning for networking: Workflow, advances and
  opportunities.
\newblock {\em IEEE Network}, 32(2):92--99, 2018.

\bibitem{alsheikh2016mobile}
Mohammad~Abu Alsheikh, Dusit Niyato, Shaowei Lin, Hwee-Pink Tan, and Zhu Han.
\newblock Mobile big data analytics using deep learning and {Apache Spark}.
\newblock {\em IEEE network}, 30(3):22--29, 2016.

\bibitem{goodfellow2016deep}
Ian Goodfellow, Yoshua Bengio, and Aaron Courville.
\newblock {\em Deep learning}.
\newblock MIT press, 2016.

\bibitem{mnih2015human}
Volodymyr Mnih, Koray Kavukcuoglu, David Silver, Andrei~A. Rusu, Joel Veness,
  Marc~G. Bellemare, Alex Graves, Martin Riedmiller, Andreas~K. Fidjeland,
  Georg Ostrovski, Stig Petersen, Charles Beattie, Amir Sadik, Ioannis
  Antonoglou, Helen King, Dharshan Kumaran, Daan Wierstra, Shane Legg, and
  Demis Hassabis.
\newblock Human-level control through deep reinforcement learning.
\newblock {\em Nature}, 518(7540):529--533, 2015.

\bibitem{lecun2015deep}
Yann LeCun, Yoshua Bengio, and Geoffrey Hinton.
\newblock Deep learning.
\newblock {\em Nature}, 521(7553):436--444, 2015.

\bibitem{schmidhuber2015deep}
J{\"u}rgen Schmidhuber.
\newblock Deep learning in neural networks: {An} overview.
\newblock {\em Neural networks}, 61:85--117, 2015.

\bibitem{liu2017survey}
Weibo Liu, Zidong Wang, Xiaohui Liu, Nianyin Zeng, Yurong Liu, and Fuad~E
  Alsaadi.
\newblock A survey of deep neural network architectures and their applications.
\newblock {\em Neurocomputing}, 234:11--26, 2017.

\bibitem{deng2014deep}
Li~Deng, Dong Yu, et~al.
\newblock Deep learning: methods and applications.
\newblock {\em Foundations and Trends{\textregistered} in Signal Processing},
  7(3--4):197--387, 2014.

\bibitem{deng2014tutorial}
Li~Deng.
\newblock A tutorial survey of architectures, algorithms, and applications for
  deep learning.
\newblock {\em APSIPA Transactions on Signal and Information Processing}, 3,
  2014.

\bibitem{Pouyanfar:2018:SDL:3271482.3234150}
\textcolor{black}{Samira} \color{black}Pouyanfar, Saad Sadiq, Yilin Yan, Haiman
  Tian, Yudong Tao, Maria~Presa Reyes, Mei-Ling Shyu, Shu-Ching Chen, and S.~S.
  Iyengar.
\newblock {\color{black}A Survey on Deep Learning: Algorithms, Techniques, and
  Applications}.
\newblock {\em {\color{black}ACM Computing Surveys (CSUR)}},
  {\color{black}51}({\color{black}5}):{\color{black}92:1--92:36},
  {\color{black}2018}\color{black}.

\bibitem{kai2017brief}
Kai Arulkumaran, Marc~Peter Deisenroth, Miles Brundage, and Anil~Anthony
  Bharath.
\newblock Deep reinforcement learning: A brief survey.
\newblock {\em IEEE Signal Processing Magazine}, 34(6):26--38, 2017.

\bibitem{hussein2017imitation}
Ahmed Hussein, Mohamed~Medhat Gaber, Eyad Elyan, and Chrisina Jayne.
\newblock Imitation learning: A survey of learning methods.
\newblock {\em ACM Computing Surveys (CSUR)}, 50(2):21:1--21:35,
  2017\color{black}.

\bibitem{chen2014big}
Xue-Wen Chen and Xiaotong Lin.
\newblock Big data deep learning: challenges and perspectives.
\newblock {\em IEEE access}, 2:514--525, 2014.

\bibitem{najafabadi2015deep}
Maryam~M Najafabadi, Flavio Villanustre, Taghi~M Khoshgoftaar, Naeem Seliya,
  Randall Wald, and Edin Muharemagic.
\newblock Deep learning applications and challenges in big data analytics.
\newblock {\em Journal of Big Data}, 2(1):1, 2015.

\bibitem{hordri2017systematic}
NF~Hordri, A~Samar, SS~Yuhaniz, and SM~Shamsuddin.
\newblock A systematic literature review on features of deep learning in big
  data analytics.
\newblock {\em International Journal of Advances in Soft Computing \& Its
  Applications}, 9(1), 2017.

\bibitem{gheisari2017survey}
Mehdi Gheisari, Guojun Wang, and Md~Zakirul~Alam Bhuiyan.
\newblock A survey on deep learning in big data.
\newblock In {\em Proc. IEEE International Conference on Computational Science
  and Engineering (CSE) and Embedded and Ubiquitous Computing (EUC)}, volume~2,
  pages 173--180, 2017.

\bibitem{zhang2017deep}
Shuai Zhang, Lina Yao, and Aixin Sun.
\newblock Deep learning based recommender system: A survey and new
  perspectives.
\newblock {\em arXiv preprint arXiv:1707.07435}, 2017\color{black}.

\bibitem{yu2017networking}
Shui Yu, Meng Liu, Wanchun Dou, Xiting Liu, and Sanming Zhou.
\newblock Networking for big data: A survey.
\newblock {\em IEEE Communications Surveys \& Tutorials}, 19(1):531--549, 2017.

\bibitem{alsheikh2014machine}
Mohammad~Abu Alsheikh, Shaowei Lin, Dusit Niyato, and Hwee-Pink Tan.
\newblock Machine learning in wireless sensor networks: Algorithms, strategies,
  and applications.
\newblock {\em IEEE Communications Surveys \& Tutorials}, 16(4):1996--2018,
  2014.

\bibitem{tsai2014data}
Chun-Wei Tsai, Chin-Feng Lai, Ming-Chao Chiang, Laurence~T Yang, et~al.
\newblock Data mining for {Internet} of things: A survey.
\newblock {\em IEEE Communications Surveys and Tutorials}, 16(1):77--97, 2014.

\bibitem{cheng2017exploiting}
Xiang Cheng, Luoyang Fang, Xuemin Hong, and Liuqing Yang.
\newblock Exploiting mobile big data: Sources, features, and applications.
\newblock {\em IEEE Network}, 31(1):72--79, 2017.

\bibitem{bkassiny2013survey}
Mario Bkassiny, Yang Li, and Sudharman~K Jayaweera.
\newblock A survey on machine-learning techniques in cognitive radios.
\newblock {\em IEEE Communications Surveys \& Tutorials}, 15(3):1136--1159,
  2013.

\bibitem{andrews2014will}
Jeffrey~G Andrews, Stefano Buzzi, Wan Choi, Stephen~V Hanly, Angel Lozano,
  Anthony~CK Soong, and Jianzhong~Charlie Zhang.
\newblock What will {5G} be?
\newblock {\em IEEE Journal on selected areas in communications},
  32(6):1065--1082, 2014.

\bibitem{panwar2016survey}
Nisha Panwar, Shantanu Sharma, and Awadhesh~Kumar Singh.
\newblock A survey on {5G}: The next generation of mobile communication.
\newblock {\em Physical Communication}, 18:64--84, 2016.

\bibitem{elijah2016comprehensive}
Olakunle Elijah, Chee~Yen Leow, Tharek~Abdul Rahman, Solomon Nunoo, and
  Solomon~Zakwoi Iliya.
\newblock A comprehensive survey of pilot contamination in massive {MIMO--5G}
  system.
\newblock {\em IEEE Communications Surveys \& Tutorials}, 18(2):905--923, 2016.

\bibitem{buzzi2016survey}
Stefano Buzzi, I~Chih-Lin, Thierry~E Klein, H~Vincent Poor, Chenyang Yang, and
  Alessio Zappone.
\newblock A survey of energy-efficient techniques for {5G} networks and
  challenges ahead.
\newblock {\em IEEE Journal on Selected Areas in Communications},
  34(4):697--709, 2016.

\bibitem{peng2015system}
Mugen Peng, Yong Li, Zhongyuan Zhao, and Chonggang Wang.
\newblock System architecture and key technologies for {5G} heterogeneous cloud
  radio access networks.
\newblock {\em IEEE network}, 29(2):6--14, 2015.

\bibitem{niu2015survey}
Yong Niu, Yong Li, Depeng Jin, Li~Su, and Athanasios~V Vasilakos.
\newblock A survey of millimeter wave communications (mmwave) for {5G}:
  opportunities and challenges.
\newblock {\em Wireless Networks}, 21(8):2657--2676, 2015.

\bibitem{foukas2017network}
Xenofon Foukas, Georgios Patounas, Ahmed Elmokashfi, and Mahesh~K Marina.
\newblock Network slicing in {5G}: Survey and challenges.
\newblock {\em IEEE Communications Magazine}, 55(5):94--100, 2017.

\bibitem{taleb2017multi}
Tarik Taleb, Konstantinos Samdanis, Badr Mada, Hannu Flinck, Sunny Dutta, and
  Dario Sabella.
\newblock On multi-access edge computing: A survey of the emerging {5G} network
  edge architecture \& orchestration.
\newblock {\em IEEE Communications Surveys \& Tutorials}, 2017.

\bibitem{mach2017mobile}
Pavel Mach and Zdenek Becvar.
\newblock Mobile edge computing: A survey on architecture and computation
  offloading.
\newblock {\em IEEE Communications Surveys \& Tutorials}, 2017.

\bibitem{mao2017survey}
Yuyi Mao, Changsheng You, Jun Zhang, Kaibin Huang, and Khaled~B Letaief.
\newblock A survey on mobile edge computing: The communication perspective.
\newblock {\em IEEE Communications Surveys \& Tutorials}, 2017.

\bibitem{wang2017data}
Ying Wang, Peilong Li, Lei Jiao, Zhou Su, Nan Cheng, Xuemin~Sherman Shen, and
  Ping Zhang.
\newblock A data-driven architecture for personalized {QoE} management in {5G}
  wireless networks.
\newblock {\em IEEE Wireless Communications}, 24(1):102--110, 2017.

\bibitem{han2015mobile}
Qilong Han, Shuang Liang, and Hongli Zhang.
\newblock Mobile cloud sensing, big data, and {5G} networks make an intelligent
  and smart world.
\newblock {\em IEEE Network}, 29(2):40--45, 2015.

\bibitem{singh2017survey}
Sukhdeep Singh, Navrati Saxena, Abhishek Roy, and HanSeok Kim.
\newblock A survey on {5G} network technologies from social perspective.
\newblock {\em IETE Technical Review}, 34(1):30--39, 2017.

\bibitem{chen20175g}
Min Chen, Jun Yang, Yixue Hao, Shiwen Mao, and Kai Hwang.
\newblock A {5G} cognitive system for healthcare.
\newblock {\em Big Data and Cognitive Computing}, 1(1):2, 2017.

\bibitem{chen2015energy}
Xianfu Chen, Jinsong Wu, Yueming Cai, Honggang Zhang, and Tao Chen.
\newblock {Energy-efficiency oriented traffic offloading in wireless networks:
  A brief survey and a learning approach for heterogeneous cellular networks}.
\newblock {\em IEEE Journal on Selected Areas in Communications},
  33(4):627--640, {2015}\color{black}.

\bibitem{wu2016big}
Jinsong Wu, Song Guo, Jie Li, and Deze Zeng.
\newblock Big data meet green challenges: big data toward green applications.
\newblock {\em IEEE Systems Journal}, 10(3):888--900, 2016\color{black}.

\bibitem{buda2016can}
Teodora~Sandra Buda, Haytham Assem, Lei Xu, Danny Raz, Udi Margolin, Elisha
  Rosensweig, Diego~R Lopez, Marius-Iulian Corici, Mikhail Smirnov, Robert
  Mullins, et~al.
\newblock Can machine learning aid in delivering new use cases and scenarios in
  {5G}?
\newblock In {\em IEEE/IFIP Network Operations and Management Symposium
  (NOMS)}, pages 1279--1284, 2016.

\bibitem{imran2014challenges}
Ali Imran, Ahmed Zoha, and Adnan Abu-Dayya.
\newblock {Challenges in 5G: how to empower SON with big data for enabling 5G}.
\newblock {\em IEEE Network}, 28(6):27--33, 2014.

\bibitem{keshavamurthy2016conceptual}
Bharath Keshavamurthy and Mohammad Ashraf.
\newblock Conceptual design of proactive {SONs} based on the big data framework
  for {5G} cellular networks: A novel machine learning perspective facilitating
  a shift in the son paradigm.
\newblock In {\em Proc. IEEE International Conference on System Modeling \&
  Advancement in Research Trends (SMART)}, pages 298--304, 2016.

\bibitem{valente2017survey}
Paulo Valente~Klaine, Muhammad~Ali Imran, Oluwakayode Onireti, and Richard~Demo
  Souza.
\newblock A survey of machine learning techniques applied to self organizing
  cellular networks.
\newblock {\em IEEE Communications Surveys and Tutorials}, 2017.

\bibitem{li2017intelligent}
Rongpeng Li, Zhifeng Zhao, Xuan Zhou, Guoru Ding, Yan Chen, Zhongyao Wang, and
  Honggang Zhang.
\newblock Intelligent {5G}: When cellular networks meet artificial
  intelligence.
\newblock {\em IEEE Wireless communications}, 24(5):175--183, 2017.

\bibitem{bui2017survey}
Nicola Bui, Matteo Cesana, S~Amir Hosseini, Qi~Liao, Ilaria Malanchini, and
  Joerg Widmer.
\newblock A survey of anticipatory mobile networking: Context-based
  classification, prediction methodologies, and optimization techniques.
\newblock {\em IEEE Communications Surveys \& Tutorials}, 19(3):1790--1821,
  2017.

\bibitem{kasnesis2017changing}
Panagiotis Kasnesis, Charalampos Patrikakis, and Iakovos Venieris.
\newblock Changing the game of mobile data analysis with deep learning.
\newblock {\em IT Professional}, 2017.

\bibitem{cheng2017mobile}
Xiang Cheng, Luoyang Fang, Liuqing Yang, and Shuguang Cui.
\newblock Mobile big data: the fuel for data-driven wireless.
\newblock {\em IEEE Internet of Things Journal}, 4(5):1489--1516, 2017.

\bibitem{wang2017big}
Lidong Wang and Randy Jones.
\newblock Big data analytics for network intrusion detection: A survey.
\newblock {\em International Journal of Networks and Communications},
  7(1):24--31, 2017.

\bibitem{kato2017deep}
Nei Kato, Zubair~Md Fadlullah, Bomin Mao, Fengxiao Tang, Osamu Akashi, Takeru
  Inoue, and Kimihiro Mizutani.
\newblock The deep learning vision for heterogeneous network traffic control:
  proposal, challenges, and future perspective.
\newblock {\em IEEE Wireless Communications}, 24(3):146--153, 2017.

\bibitem{zorzi2015cognition}
Michele Zorzi, Andrea Zanella, Alberto Testolin, Michele De~Filippo De~Grazia,
  and Marco Zorzi.
\newblock Cognition-based networks: A new perspective on network optimization
  using learning and distributed intelligence.
\newblock {\em IEEE Access}, 3:1512--1530, 2015.

\bibitem{fadlullah2017state}
Zubair Fadlullah, Fengxiao Tang, Bomin Mao, Nei Kato, Osamu Akashi, Takeru
  Inoue, and Kimihiro Mizutani.
\newblock State-of-the-art deep learning: Evolving machine intelligence toward
  tomorrow's intelligent network traffic control systems.
\newblock {\em IEEE Communications Surveys \& Tutorials}, 19(4):2432--2455,
  2017.

\bibitem{mohammadi2018deep}
Mehdi Mohammadi, Ala Al-Fuqaha, Sameh Sorour, and Mohsen Guizani.
\newblock {Deep Learning for IoT Big Data and Streaming Analytics: A Survey}.
\newblock {\em IEEE Communications Surveys \& Tutorials}, 2018\color{black}.

\bibitem{ahad2016neural}
Nauman Ahad, Junaid Qadir, and Nasir Ahsan.
\newblock Neural networks in wireless networks: Techniques, applications and
  guidelines.
\newblock {\em Journal of Network and Computer Applications}, 68:1--27, 2016.

\bibitem{mao2018deep}
Qian Mao, Fei Hu, and Qi~Hao.
\newblock Deep learning for intelligent wireless networks: A comprehensive
  survey.
\newblock {\em IEEE Communications Surveys \& Tutorials}, 2018\color{black}.

\bibitem{luong2018applications}
\textcolor{black}{Nguyen Cong} \color{black}Luong, Dinh~Thai Hoang, Shimin
  Gong, Dusit Niyato, Ping Wang, Ying-Chang Liang, and Dong~In Kim.
\newblock {\color{black}Applications of Deep Reinforcement Learning in
  Communications and Networking: A Survey}.
\newblock {\em {\color{black}arXiv preprint arXiv:1810.07862}},
  {\color{black}2018}\color{black}.

\bibitem{zhou2017intelligent2}
Xiangwei Zhou, Mingxuan Sun, Ye~Geoffrey Li, and Biing-Hwang Juang.
\newblock {Intelligent Wireless Communications Enabled by Cognitive Radio and
  Machine Learning}.
\newblock {\em arXiv preprint arXiv:1710.11240}, 2017\color{black}.

\bibitem{chen2017machine2}
Mingzhe Chen, Ursula Challita, Walid Saad, Changchuan Yin, and M{\'e}rouane
  Debbah.
\newblock Machine learning for wireless networks with artificial intelligence:
  A tutorial on neural networks.
\newblock {\em arXiv preprint arXiv:1710.02913}, 2017\color{black}.

\bibitem{gharaibeh2017smart}
Ammar Gharaibeh, Mohammad~A Salahuddin, Sayed~Jahed Hussini, Abdallah
  Khreishah, Issa Khalil, Mohsen Guizani, and Ala Al-Fuqaha.
\newblock Smart cities: A survey on data management, security, and enabling
  technologies.
\newblock {\em IEEE Communications Surveys \& Tutorials}, 19(4):2456--2501,
  2017.

\bibitem{lane2015can}
Nicholas~D Lane and Petko Georgiev.
\newblock Can deep learning revolutionize mobile sensing?
\newblock In {\em Proc. 16th ACM International Workshop on Mobile Computing
  Systems and Applications}, pages 117--122, 2015.

\bibitem{ota2017deep}
Kaoru Ota, Minh~Son Dao, Vasileios Mezaris, and Francesco~GB De~Natale.
\newblock Deep learning for mobile multimedia: A survey.
\newblock {\em ACM Transactions on Multimedia Computing, Communications, and
  Applications (TOMM)}, 13(3s):34, 2017.

\bibitem{mishra2018detailed}
Preeti Mishra, Vijay Varadharajan, Uday Tupakula, and Emmanuel~S Pilli.
\newblock A detailed investigation and analysis of using machine learning
  techniques for intrusion detection.
\newblock {\em IEEE Communications Surveys \& Tutorials}, 2018\color{black}.

\bibitem{li2017deeprl}
Yuxi Li.
\newblock Deep reinforcement learning: An overview.
\newblock {\em arXiv preprint arXiv:1701.07274}, 2017.

\bibitem{chen2018deep0}
Longbiao Chen, Dingqi Yang, Daqing Zhang, Cheng Wang, Jonathan Li, et~al.
\newblock Deep mobile traffic forecast and complementary base station
  clustering for {C-RAN} optimization.
\newblock {\em Journal of Network and Computer Applications}, 121:59--69, 2018.

\bibitem{mnih2016asynchronous}
Volodymyr Mnih, Adria~Puigdomenech Badia, Mehdi Mirza, Alex Graves, Timothy
  Lillicrap, Tim Harley, David Silver, and Koray Kavukcuoglu.
\newblock Asynchronous methods for deep reinforcement learning.
\newblock In {\em Proc. International Conference on Machine Learning (ICML)},
  pages 1928--1937, 2016.

\bibitem{arjovsky2017wasserstein}
Martin Arjovsky, Soumith Chintala, and L{\'e}on Bottou.
\newblock Wasserstein generative adversarial networks.
\newblock In {\em Proc. International Conference on Machine Learning}, pages
  214--223, 2017.

\bibitem{damianou2013deep2}
Andreas Damianou and Neil Lawrence.
\newblock Deep {Gaussian} processes.
\newblock In {\em Artificial Intelligence and Statistics}, pages 207--215,
  2013\color{black}.

\bibitem{garnelo2018neural}
Marta Garnelo, Jonathan Schwarz, Dan Rosenbaum, Fabio Viola, Danilo~J Rezende,
  SM~Eslami, and Yee~Whye Teh.
\newblock Neural processes.
\newblock {\em arXiv preprint arXiv:1807.01622}, 2018\color{black}.

\bibitem{zhou2017deep2}
Zhi-Hua Zhou and Ji~Feng.
\newblock Deep forest: towards an alternative to deep neural networks.
\newblock In {\em Proc. 26th International Joint Conference on Artificial
  Intelligence}, pages 3553--3559. AAAI Press, 2017\color{black}.

\bibitem{mcculloch:1943}
W.~McCulloch and W.~Pitts.
\newblock A logical calculus of the ideas immanent in nervous activity.
\newblock {\em Bulletin of Mathematical Biophysics}, (5).

\bibitem{williams1986learning}
David~E Rumelhart, Geoffrey~E Hinton, and Ronald~J Williams.
\newblock Learning representations by back-propagating errors.
\newblock {\em Nature}, 323(6088):533, 1986.

\bibitem{lecun1995convolutional}
Yann LeCun, Yoshua Bengio, et~al.
\newblock Convolutional networks for images, speech, and time series.
\newblock {\em The handbook of brain theory and neural networks},
  3361(10):1995, 1995.

\bibitem{krizhevsky2012imagenet}
Alex Krizhevsky, Ilya Sutskever, and Geoffrey~E Hinton.
\newblock Imagenet classification with deep convolutional neural networks.
\newblock In {\em Advances in neural information processing systems}, pages
  1097--1105, 2012.

\bibitem{Domingos:2012}
Pedro Domingos.
\newblock A few useful things to know about machine learning.
\newblock {\em Communications of the ACM}, 55(10):78--87, 2012.

\bibitem{tsang2005core}
Ivor~W Tsang, James~T Kwok, and Pak-Ming Cheung.
\newblock {Core vector machines: Fast {SVM} training on very large data sets}.
\newblock {\em Journal of Machine Learning Research}, 6:363--392, 2005.

\bibitem{rasmussen2006gaussian}
Carl~Edward Rasmussen and Christopher~KI Williams.
\newblock {\em Gaussian processes for machine learning}, volume~1.
\newblock MIT press Cambridge, 2006.

\bibitem{le2008representational}
Nicolas Le~Roux and Yoshua Bengio.
\newblock Representational power of restricted boltzmann machines and deep
  belief networks.
\newblock {\em Neural computation}, 20(6):1631--1649, 2008.

\bibitem{goodfellow2014generative}
Ian Goodfellow, Jean Pouget-Abadie, Mehdi Mirza, Bing Xu, David Warde-Farley,
  Sherjil Ozair, Aaron Courville, and Yoshua Bengio.
\newblock Generative adversarial nets.
\newblock In {\em Advances in neural information processing systems}, pages
  2672--2680, 2014.

\bibitem{schroff2015facenet}
Florian Schroff, Dmitry Kalenichenko, and James Philbin.
\newblock Facenet: A unified embedding for face recognition and clustering.
\newblock In {\em Proc. IEEE Conference on Computer Vision and Pattern
  Recognition}, pages 815--823, 2015.

\bibitem{kingma2014semi}
Diederik~P Kingma, Shakir Mohamed, Danilo~Jimenez Rezende, and Max Welling.
\newblock Semi-supervised learning with deep generative models.
\newblock In {\em Advances in Neural Information Processing Systems}, pages
  3581--3589, 2014.

\bibitem{stewart2017label}
Russell Stewart and Stefano Ermon.
\newblock Label-free supervision of neural networks with physics and domain
  knowledge.
\newblock In {\em Proc. National Conference on Artificial Intelligence (AAAI)},
  pages 2576--2582, 2017.

\bibitem{rezende2016one}
Danilo Rezende, Ivo Danihelka, Karol Gregor, Daan Wierstra, et~al.
\newblock One-shot generalization in deep generative models.
\newblock In {\em Proc. International Conference on Machine Learning (ICML)},
  pages 1521--1529, 2016.

\bibitem{socher2013zero}
Richard Socher, Milind Ganjoo, Christopher~D Manning, and Andrew Ng.
\newblock Zero-shot learning through cross-modal transfer.
\newblock In {\em Advances in neural information processing systems}, pages
  935--943, 2013.

\bibitem{georgiev2017low}
Petko Georgiev, Sourav Bhattacharya, Nicholas~D Lane, and Cecilia Mascolo.
\newblock Low-resource multi-task audio sensing for mobile and embedded devices
  via shared deep neural network representations.
\newblock {\em Proc. ACM on Interactive, Mobile, Wearable and Ubiquitous
  Technologies (IMWUT)}, 1(3):50, 2017.

\bibitem{monti2017geometric}
\textcolor{black}{Federico} \color{black}Monti, Davide Boscaini, Jonathan
  Masci, Emanuele Rodola, Jan Svoboda, and Michael~M Bronstein.
\newblock {\color{black}Geometric deep learning on graphs and manifolds using
  mixture model {CNNs}}.
\newblock In {\em {\color{black}Proc. IEEE Conference on Computer Vision and
  Pattern Recognition (CVPR)}}, volume~{\color{black}1}, page~{\color{black}3},
  {\color{black}2017}\color{black}.

\bibitem{le2004geometric}
\textcolor{black}{Brigitte} \color{black}Le Roux and Henry Rouanet.
\newblock {\em {\color{black}Geometric data analysis: from correspondence
  analysis to structured data analysis}}.
\newblock {\color{black}Springer Science \& Business Media},
  {\color{black}2004}\color{black}.

\bibitem{qi2017pointnet}
\textcolor{black}{Charles R.} \color{black}Qi, Li~Yi, Hao Su, and Leonidas~J
  Guibas.
\newblock {\color{black}PointNet++: Deep hierarchical feature learning on point
  sets in a metric space}.
\newblock In {\em {\color{black}Advances in Neural Information Processing
  Systems}}, pages~{\color{black}5099--5108}, {\color{black}2017}\color{black}.

\bibitem{kipf2016semi}
\textcolor{black}{Thomas N} \color{black}Kipf and Max Welling.
\newblock {\color{black}Semi-supervised classification with graph convolutional
  networks}.
\newblock In {\em {\color{black}Proc. International Conference on Learning
  Representations (ICLR)}}, {\color{black}2017}\color{black}.

\bibitem{wang2018spatio}
\textcolor{black}{Xu} \color{black}Wang, Zimu Zhou, Fu~Xiao, Kai Xing, Zheng
  Yang, Yunhao Liu, and Chunyi Peng.
\newblock {\color{black}Spatio-temporal analysis and prediction of cellular
  traffic in metropolis}.
\newblock {\em {\color{black}IEEE Transactions on Mobile Computing}},
  {\color{black}2018}\color{black}.

\bibitem{nguyen2015deep}
\textcolor{black}{Anh} \color{black}Nguyen, Jason Yosinski, and Jeff Clune.
\newblock \color{black}{Deep neural networks are easily fooled: High confidence
  predictions for unrecognizable images}.
\newblock In {\em {\color{black}Proc. IEEE Conference on Computer Vision and
  Pattern Recognition}}, pages~{\color{black}427--436},
  {\color{black}2015}\color{black}.

\bibitem{behzadan2017vulnerability}
\textcolor{black}{Vahid} \color{black}Behzadan and Arslan Munir.
\newblock {\color{black}Vulnerability of deep reinforcement learning to policy
  induction attacks}.
\newblock In {\em {\color{black}Proc. International Conference on Machine
  Learning and Data Mining in Pattern Recognition}},
  pages~{\color{black}262--275}. {\color{black}Springer},
  {\color{black}2017}\color{black}.

\bibitem{madani2018robustness}
\textcolor{black}{Pooria} \color{black}Madani and Natalija Vlajic.
\newblock {\color{black}Robustness of deep autoencoder in intrusion detection
  under adversarial contamination}.
\newblock In {\em {\color{black}Proc. 5th ACM Annual Symposium and Bootcamp on
  Hot Topics in the Science of Security}}, page~{\color{black}1},
  {\color{black}2018}\color{black}.

\bibitem{bau2017network}
\textcolor{black}{David} \color{black}Bau, Bolei Zhou, Aditya Khosla, Aude
  Oliva, and Antonio Torralba.
\newblock {\color{black}Network Dissection: Quantifying Interpretability of
  Deep Visual Representations}.
\newblock In {\em {\color{black}Proc. IEEE Conference on Computer Vision and
  Pattern Recognition (CVPR)}}, pages~{\color{black}3319--3327},
  {\color{black}2017}\color{black}.

\bibitem{wu2017beyond}
\textcolor{black}{Mike} \color{black}Wu, Michael~C Hughes, Sonali Parbhoo,
  Maurizio Zazzi, Volker Roth, and Finale Doshi-Velez.
\newblock {\color{black}Beyond sparsity: Tree regularization of deep models for
  interpretability}.
\newblock {\color{black}2018}\color{black}.

\bibitem{chakraborty2017interpretability}
\textcolor{black}{Supriyo} \color{black}Chakraborty, Richard Tomsett, Ramya
  Raghavendra, Daniel Harborne, Moustafa Alzantot, Federico Cerutti, Mani
  Srivastava, Alun Preece, Simon Julier, Raghuveer~M Rao, et~al.
\newblock {\color{black}Interpretability of deep learning models: a survey of
  results}.
\newblock In {\em {\color{black}Proc. IEEE Smart World Congress Workshop:
  DAIS}}, {\color{black}2017}\color{black}.

\bibitem{perez2017effectiveness}
\textcolor{black}{Luis} \color{black}Perez and Jason Wang.
\newblock {\color{black}The effectiveness of data augmentation in image
  classification using deep learning}.
\newblock {\em {\color{black}arXiv preprint arXiv:1712.04621}},
  {\color{black}2017}\color{black}.

\bibitem{liu2017progressive}
Chenxi Liu, Barret Zoph, Jonathon Shlens, Wei Hua, Li-Jia Li, Li~Fei-Fei, Alan
  Yuille, Jonathan Huang, and Kevin Murphy.
\newblock Progressive neural architecture search.
\newblock {\em arXiv preprint arXiv:1712.00559}, 2017\color{black}.

\bibitem{zhang2016deep3}
Wei Zhang, Kan Liu, Weidong Zhang, Youmei Zhang, and Jason Gu.
\newblock Deep neural networks for wireless localization in indoor and outdoor
  environments.
\newblock {\em Neurocomputing}, 194:279--287, 2016.

\bibitem{ordonez2016deep}
Francisco~Javier Ord{\'o}{\~n}ez and Daniel Roggen.
\newblock Deep convolutional and {LSTM} recurrent neural networks for
  multimodal wearable activity recognition.
\newblock {\em Sensors}, 16(1):115, 2016.

\bibitem{de2016distributed}
Elias De~Coninck, Tim Verbelen, Bert Vankeirsbilck, Steven Bohez, Pieter
  Simoens, Piet Demeester, and Bart Dhoedt.
\newblock Distributed neural networks for {Internet of Things}: the big-little
  approach.
\newblock In {\em Internet of Things. IoT Infrastructures: Second International
  Summit, IoT 360$^{\circ}$ 2015, Rome, Italy, October 27-29, Revised Selected
  Papers, Part II}, pages 484--492. Springer, 2016.

\bibitem{jouppi2017datacenter}
Norman~P Jouppi, Cliff Young, Nishant Patil, David Patterson, Gaurav Agrawal,
  Raminder Bajwa, Sarah Bates, Suresh Bhatia, Nan Boden, Al~Borchers, et~al.
\newblock In-datacenter performance analysis of a tensor processing unit.
\newblock In {\em ACM/IEEE 44th Annual International Symposium on Computer
  Architecture (ISCA)}, pages 1--12, 2017.

\bibitem{nickolls2008scalable}
John Nickolls, Ian Buck, Michael Garland, and Kevin Skadron.
\newblock Scalable parallel programming with {CUDA}.
\newblock {\em Queue}, 6(2):40--53, 2008.

\bibitem{chetlur2014cudnn}
Sharan Chetlur, Cliff Woolley, Philippe Vandermersch, Jonathan Cohen, John
  Tran, Bryan Catanzaro, and Evan Shelhamer.
\newblock {cuDNN}: Efficient primitives for deep learning.
\newblock {\em arXiv preprint arXiv:1410.0759}, 2014.

\bibitem{tensorflow2015-whitepaper}
Mart{\'\i}n Abadi, Paul Barham, Jianmin Chen, Zhifeng Chen, Andy Davis, Jeffrey
  Dean, Matthieu Devin, Sanjay Ghemawat, Geoffrey Irving, Michael Isard, et~al.
\newblock {TensorFlow}: A system for large-scale machine learning.
\newblock In {\em USENIX Symposium on Operating Systems Design and
  Implementation (OSDI)}, volume~16, pages 265--283, 2016.

\bibitem{2016arXiv160502688short}
{Theano Development Team}.
\newblock {Theano: A {Python} framework for fast computation of mathematical
  expressions}.
\newblock {\em arXiv e-prints}, abs/1605.02688, May 2016.

\bibitem{jia2014caffe}
Yangqing Jia, Evan Shelhamer, Jeff Donahue, Sergey Karayev, Jonathan Long, Ross
  Girshick, Sergio Guadarrama, and Trevor Darrell.
\newblock Caffe: Convolutional architecture for fast feature embedding.
\newblock {\em arXiv preprint arXiv:1408.5093}, 2014.

\bibitem{torch}
R.~Collobert, K.~Kavukcuoglu, and C.~Farabet.
\newblock Torch7: A {Matlab}-like environment for machine learning.
\newblock In {\em Proc. BigLearn, NIPS Workshop}, 2011.

\bibitem{gokhale2014240}
Vinayak Gokhale, Jonghoon Jin, Aysegul Dundar, Berin Martini, and Eugenio
  Culurciello.
\newblock A 240 {G}-ops/s mobile coprocessor for deep neural networks.
\newblock In {\em Proc. IEEE Conference on Computer Vision and Pattern
  Recognition Workshops}, pages 682--687, 2014.

\bibitem{ncnn}
{ncnn -- a high-performance neural network inference framework optimized for
  the mobile platform }.
\newblock \url{https://github.com/Tencent/ncnn}, 2017.
\newblock [Online; accessed 25-July-2017].

\bibitem{huawei2017kirin}
{Huawei announces the {Kirin} 970- new flagship {SoC} with {AI} capabilities}.
\newblock
  \url{http://www.androidauthority.com/huawei-announces-kirin-970-797788/},
  2017.
\newblock [Online; accessed 01-Sep-2017].

\bibitem{coreml}
{Core ML: Integrate machine learning models into your app.}
\newblock \url{https://developer.apple.com/documentation/coreml}, 2017.
\newblock [Online; accessed 25-July-2017].

\bibitem{sutskever2013importance}
Ilya Sutskever, James Martens, George~E Dahl, and Geoffrey~E Hinton.
\newblock On the importance of initialization and momentum in deep learning.
\newblock {\em Proc. international conference on machine learning (ICML)},
  28:1139--1147, 2013.

\bibitem{dean2012large}
Jeffrey Dean, Greg Corrado, Rajat Monga, Kai Chen, Matthieu Devin, Mark Mao,
  Andrew Senior, Paul Tucker, Ke~Yang, Quoc~V Le, et~al.
\newblock Large scale distributed deep networks.
\newblock In {\em Advances in neural information processing systems}, pages
  1223--1231, 2012.

\bibitem{kingma2015adam}
Diederik Kingma and Jimmy Ba.
\newblock {Adam: A method for stochastic optimization}.
\newblock In {\em Proc. International Conference on Learning Representations
  (ICLR)}, 2015.

\bibitem{kraska2013mlbase}
Tim Kraska, Ameet Talwalkar, John~C Duchi, Rean Griffith, Michael~J Franklin,
  and Michael~I Jordan.
\newblock {MLbase}: A distributed machine-learning system.
\newblock In {\em CIDR}, volume~1, pages 2--1, 2013.

\bibitem{chilimbi2014project}
Trishul~M Chilimbi, Yutaka Suzue, Johnson Apacible, and Karthik Kalyanaraman.
\newblock Project adam: Building an efficient and scalable deep learning
  training system.
\newblock In {\em USENIX Symposium on Operating Systems Design and
  Implementation (OSDI)}, volume~14, pages 571--582, 2014.

\bibitem{cui2016geeps}
Henggang Cui, Hao Zhang, Gregory~R Ganger, Phillip~B Gibbons, and Eric~P Xing.
\newblock Geeps: Scalable deep learning on distributed {GPUs} with a
  {GPU}-specialized parameter server.
\newblock In {\em Proc. Eleventh ACM European Conference on Computer Systems},
  page~4, 2016.

\bibitem{Lineaat8084}
Xing Lin, Yair Rivenson, Nezih~T. Yardimci, Muhammed Veli, Yi~Luo, Mona
  Jarrahi, and Aydogan Ozcan.
\newblock All-optical machine learning using diffractive deep neural networks.
\newblock {\em Science}, 2018\color{black}.

\bibitem{spring2017scalable}
Ryan Spring and Anshumali Shrivastava.
\newblock Scalable and sustainable deep learning via randomized hashing.
\newblock {\em Proc. ACM SIGKDD Conference on Knowledge Discovery and Data
  Mining}, 2017.

\bibitem{mirhoseini2017device}
Azalia Mirhoseini, Hieu Pham, Quoc~V Le, Benoit Steiner, Rasmus Larsen, Yuefeng
  Zhou, Naveen Kumar, Mohammad Norouzi, Samy Bengio, and Jeff Dean.
\newblock Device placement optimization with reinforcement learning.
\newblock {\em Proc. International Conference on Machine Learning}, 2017.

\bibitem{xing2015petuum}
Eric~P Xing, Qirong Ho, Wei Dai, Jin~Kyu Kim, Jinliang Wei, Seunghak Lee, Xun
  Zheng, Pengtao Xie, Abhimanu Kumar, and Yaoliang Yu.
\newblock Petuum: A new platform for distributed machine learning on big data.
\newblock {\em IEEE Transactions on Big Data}, 1(2):49--67, 2015.

\bibitem{moritz2018ray}
\textcolor{black}{Philipp} \color{black}Moritz, Robert Nishihara, Stephanie
  Wang, Alexey Tumanov, Richard Liaw, Eric Liang, Melih Elibol, Zongheng Yang,
  William Paul, Michael~I Jordan, et~al.
\newblock {\color{black}Ray: A Distributed Framework for Emerging {AI}
  Applications}.
\newblock In {\em {\color{black}Proc. 13th USENIX Symposium on Operating
  Systems Design and Implementation (OSDI)}}, pages~{\color{black}561--577},
  {\color{black}2018}\color{black}.

\bibitem{alzantot2017rstensorflow}
Moustafa Alzantot, Yingnan Wang, Zhengshuang Ren, and Mani~B Srivastava.
\newblock {RSTensorFlow}: {GPU} enabled tensorflow for deep learning on
  commodity {Android} devices.
\newblock In {\em Proc. 1st ACM International Workshop on Deep Learning for
  Mobile Systems and Applications}, pages 7--12, 2017.

\bibitem{tensorlayer}
Hao Dong, Akara Supratak, Luo Mai, Fangde Liu, Axel Oehmichen, Simiao Yu, and
  Yike Guo.
\newblock {TensorLayer}: A versatile library for efficient deep learning
  development.
\newblock In {\em Proc. ACM on Multimedia Conference}, MM '17, pages
  1201--1204, 2017.

\bibitem{paszke2017automatic}
Adam Paszke, Sam Gross, Soumith Chintala, Gregory Chanan, Edward Yang, Zachary
  DeVito, Zeming Lin, Alban Desmaison, Luca Antiga, and Adam Lerer.
\newblock Automatic differentiation in {PyTorch}.
\newblock 2017.

\bibitem{chen2015mxnet}
Tianqi Chen, Mu~Li, Yutian Li, Min Lin, Naiyan Wang, Minjie Wang, Tianjun Xiao,
  Bing Xu, Chiyuan Zhang, and Zheng Zhang.
\newblock Mxnet: A flexible and efficient machine learning library for
  heterogeneous distributed systems.
\newblock {\em arXiv preprint arXiv:1512.01274}, 2015.

\bibitem{ruder2016overview}
Sebastian Ruder.
\newblock An overview of gradient descent optimization algorithms.
\newblock {\em arXiv preprint arXiv:1609.04747}, 2016.

\bibitem{zeiler2012adadelta}
Matthew~D Zeiler.
\newblock {ADADELTA}: an adaptive learning rate method.
\newblock {\em arXiv preprint arXiv:1212.5701}, 2012\color{black}.

\bibitem{dozat2016incorporating}
Timothy Dozat.
\newblock Incorporating {Nesterov} momentum into {Adam}.
\newblock 2016\color{black}.

\bibitem{andrychowicz2016learning}
Marcin Andrychowicz, Misha Denil, Sergio Gomez, Matthew~W Hoffman, David Pfau,
  Tom Schaul, and Nando de~Freitas.
\newblock Learning to learn by gradient descent by gradient descent.
\newblock In {\em Advances in Neural Information Processing Systems}, pages
  3981--3989, 2016.

\bibitem{szegedy2015going}
Christian Szegedy, Wei Liu, Yangqing Jia, Pierre Sermanet, Scott Reed, Dragomir
  Anguelov, Dumitru Erhan, Vincent Vanhoucke, and Andrew Rabinovich.
\newblock Going deeper with convolutions.
\newblock In {\em Proc. IEEE conference on computer vision and pattern
  recognition}, pages 1--9, 2015.

\bibitem{stable2018zhou}
Yingxue Zhou, Sheng Chen, and Arindam Banerjee.
\newblock Stable gradient descent.
\newblock In {\em Proc. Conference on Uncertainty in Artificial Intelligence},
  2018\color{black}.

\bibitem{wen2017terngrad}
Wei Wen, Cong Xu, Feng Yan, Chunpeng Wu, Yandan Wang, Yiran Chen, and Hai Li.
\newblock {TernGrad}: Ternary gradients to reduce communication in distributed
  deep learning.
\newblock In {\em Advances in neural information processing systems}, 2017.

\bibitem{bonomi2014fog}
Flavio Bonomi, Rodolfo Milito, Preethi Natarajan, and Jiang Zhu.
\newblock Fog computing: A platform for internet of things and analytics.
\newblock In {\em Big Data and Internet of Things: A Roadmap for Smart
  Environments}, pages 169--186. Springer, 2014.

\bibitem{mao2017modnn}
Jiachen Mao, Xiang Chen, Kent~W Nixon, Christopher Krieger, and Yiran Chen.
\newblock {MoDNN}: Local distributed mobile computing system for deep neural
  network.
\newblock In {\em Proc. IEEE Design, Automation \& Test in Europe Conference \&
  Exhibition (DATE)}, pages 1396--1401, 2017.

\bibitem{mukherjee2018survey}
Mithun Mukherjee, Lei Shu, and Di~Wang.
\newblock Survey of fog computing: Fundamental, network applications, and
  research challenges.
\newblock {\em IEEE Communications Surveys \& Tutorials}, 2018\color{black}.

\bibitem{vaquero2014finding}
Luis~M Vaquero and Luis Rodero-Merino.
\newblock Finding your way in the fog: Towards a comprehensive definition of
  fog computing.
\newblock {\em ACM SIGCOMM Computer Communication Review}, 44(5):27--32,
  2014\color{black}.

\bibitem{aazam2018offloading}
\textcolor{black}{Mohammad} \color{black}Aazam, Sherali Zeadally, and Khaled~A
  Harras.
\newblock {\color{black}Offloading in fog computing for {IoT}: Review, enabling
  technologies, and research opportunities}.
\newblock {\em {\color{black}Future Generation Computer Systems}},
  {\color{black}87}:{\color{black}278--289}, {\color{black}2018}\color{black}.

\bibitem{Buyya2018Manifesto}
\textcolor{black}{Rajkumar} \color{black}Buyya, Satish~Narayana Srirama,
  Giuliano Casale, Rodrigo Calheiros, Yogesh Simmhan, Blesson Varghese, Erol
  Gelenbe, Bahman Javadi, Luis~Miguel Vaquero, Marco A.~S. Netto, Adel~Nadjaran
  Toosi, Maria~Alejandra Rodriguez, Ignacio~M. Llorente, Sabrina De Capitani~Di
  Vimercati, Pierangela Samarati, Dejan Milojicic, Carlos Varela, Rami Bahsoon,
  Marcos Dias~De Assuncao, Omer Rana, Wanlei Zhou, Hai Jin, Wolfgang Gentzsch,
  Albert~Y. Zomaya, and Haiying Shen.
\newblock {\color{black}A Manifesto for Future Generation Cloud Computing:
  Research Directions for the Next Decade}.
\newblock {\em {\color{black}ACM Computing Survey}},
  {\color{black}51}({\color{black}5}):{\color{black}105:1--105:38},
  {\color{black}2018}\color{black}.

\bibitem{sze2017efficient}
\textcolor{black}{Vivienne} \color{black}Sze, Yu-Hsin Chen, Tien-Ju Yang, and
  Joel~S Emer.
\newblock {\color{black}Efficient processing of deep neural networks: A
  tutorial and survey}.
\newblock {\em {\color{black}Proc. of the IEEE}},
  {\color{black}105}({\color{black}12}):{\color{black}2295--2329},
  {\color{black}2017}\color{black}.

\bibitem{bang201714}
Suyoung Bang, Jingcheng Wang, Ziyun Li, Cao Gao, Yejoong Kim, Qing Dong, Yen-Po
  Chen, Laura Fick, Xun Sun, Ron Dreslinski, et~al.
\newblock 14.7 a 288$\mu$w programmable deep-learning processor with 270kb
  on-chip weight storage using non-uniform memory hierarchy for mobile
  intelligence.
\newblock In {\em Proc. IEEE International Conference on Solid-State Circuits
  (ISSCC)}, pages 250--251, 2017.

\bibitem{akopyan2016design}
Filipp Akopyan.
\newblock Design and tool flow of {IBM}'s truenorth: an ultra-low power
  programmable neurosynaptic chip with 1 million neurons.
\newblock In {\em Proc. International Symposium on Physical Design}, pages
  59--60. ACM, 2016.

\bibitem{latifi2016cnndroid}
Seyyed~Salar Latifi~Oskouei, Hossein Golestani, Matin Hashemi, and Soheil
  Ghiasi.
\newblock Cnndroid: {GPU}-accelerated execution of trained deep convolutional
  neural networks on {A}ndroid.
\newblock In {\em Proc. Proc. ACM on Multimedia Conference}, pages 1201--1205,
  2016.

\bibitem{cortes2017adanet}
Corinna Cortes, Xavi Gonzalvo, Vitaly Kuznetsov, Mehryar Mohri, and Scott Yang.
\newblock Adanet: Adaptive structural learning of artificial neural networks.
\newblock {\em Proc. International Conference on Machine Learning (ICML)},
  2017.

\bibitem{hinton2006fast}
Geoffrey~E Hinton, Simon Osindero, and Yee-Whye Teh.
\newblock A fast learning algorithm for deep belief nets.
\newblock {\em Neural computation}, 18(7):1527--1554, 2006.

\bibitem{lee2009convolutional}
Honglak Lee, Roger Grosse, Rajesh Ranganath, and Andrew~Y Ng.
\newblock Convolutional deep belief networks for scalable unsupervised learning
  of hierarchical representations.
\newblock In {\em Proc. 26th ACM annual international conference on machine
  learning}, pages 609--616, 2009.

\bibitem{vincent2010stacked}
Pascal Vincent, Hugo Larochelle, Isabelle Lajoie, Yoshua Bengio, and
  Pierre-Antoine Manzagol.
\newblock Stacked denoising autoencoders: Learning useful representations in a
  deep network with a local denoising criterion.
\newblock {\em Journal of Machine Learning Research}, 11(Dec):3371--3408, 2010.

\bibitem{kingma2014auto}
Diederik~P Kingma and Max Welling.
\newblock {Auto-encoding variational bayes}.
\newblock In {\em Proc. International Conference on Learning Representations
  (ICLR)}, 2014.

\bibitem{he2016deep}
Kaiming He, Xiangyu Zhang, Shaoqing Ren, and Jian Sun.
\newblock Deep residual learning for image recognition.
\newblock In {\em Proc. IEEE conference on computer vision and pattern
  recognition}, pages 770--778, 2016.

\bibitem{ji20133d}
Shuiwang Ji, Wei Xu, Ming Yang, and Kai Yu.
\newblock {3D} convolutional neural networks for human action recognition.
\newblock {\em IEEE transactions on pattern analysis and machine intelligence},
  35(1):221--231, 2013.

\bibitem{huang2017densely}
Gao Huang, Zhuang Liu, Kilian~Q Weinberger, and Laurens van~der Maaten.
\newblock Densely connected convolutional networks.
\newblock {\em IEEE Conference on Computer Vision and Pattern Recognition},
  2017.

\bibitem{gers1999learning}
Felix~A Gers, J{\"u}rgen Schmidhuber, and Fred Cummins.
\newblock Learning to forget: Continual prediction with {LSTM}.
\newblock 1999.

\bibitem{sutskever2014sequence}
Ilya Sutskever, Oriol Vinyals, and Quoc~V Le.
\newblock Sequence to sequence learning with neural networks.
\newblock In {\em Advances in neural information processing systems}, pages
  3104--3112, 2014.

\bibitem{xingjian2015convolutional}
Shi Xingjian, Zhourong Chen, Hao Wang, Dit-Yan Yeung, Wai-Kin Wong, and
  Wang-chun Woo.
\newblock Convolutional {LSTM} network: A machine learning approach for
  precipitation nowcasting.
\newblock In {\em Advances in neural information processing systems}, pages
  802--810, 2015.

\bibitem{qi2017loss}
Guo-Jun Qi.
\newblock Loss-sensitive generative adversarial networks on {Lipschitz}
  densities.
\newblock {\em arXiv preprint arXiv:1701.06264}, 2017.

\bibitem{brock2018large}
\textcolor{black}{Andrew} \color{black}Brock, Jeff Donahue, and Karen Simonyan.
\newblock {\color{black}Large scale {GAN} training for high fidelity natural
  image synthesis}.
\newblock {\em {\color{black}arXiv preprint arXiv:1809.11096}},
  {\color{black}2018}\color{black}.

\bibitem{silver2016mastering}
David Silver, Aja Huang, Chris~J Maddison, Arthur Guez, Laurent Sifre, George
  Van Den~Driessche, Julian Schrittwieser, Ioannis Antonoglou, Veda
  Panneershelvam, Marc Lanctot, et~al.
\newblock Mastering the game of {Go} with deep neural networks and tree search.
\newblock {\em Nature}, 529(7587):484--489, 2016.

\bibitem{hessel2017rainbow}
Matteo Hessel, Joseph Modayil, Hado Van~Hasselt, Tom Schaul, Georg Ostrovski,
  Will Dabney, Dan Horgan, Bilal Piot, Mohammad Azar, and David Silver.
\newblock Rainbow: Combining improvements in deep reinforcement learning.
\newblock 2017\color{black}.

\bibitem{schulman2017proximal}
John Schulman, Filip Wolski, Prafulla Dhariwal, Alec Radford, and Oleg Klimov.
\newblock Proximal policy optimization algorithms.
\newblock {\em arXiv preprint arXiv:1707.06347}, 2017\color{black}.

\bibitem{collobert2004links}
Ronan Collobert and Samy Bengio.
\newblock Links between perceptrons, {MLP}s and {SVM}s.
\newblock In {\em Proc. twenty-first ACM international conference on Machine
  learning}, page~23, 2004.

\bibitem{glorot2011deep}
Xavier Glorot, Antoine Bordes, and Yoshua Bengio.
\newblock Deep sparse rectifier neural networks.
\newblock In {\em Proc. fourteenth international conference on artificial
  intelligence and statistics}, pages 315--323, 2011\color{black}.

\bibitem{klambauer2017self}
G{\"u}nter Klambauer, Thomas Unterthiner, Andreas Mayr, and Sepp Hochreiter.
\newblock Self-normalizing neural networks.
\newblock In {\em Advances in Neural Information Processing Systems}, pages
  971--980, 2017\color{black}.

\bibitem{ioffe2015batch}
\textcolor{black}{Sergey} \color{black}Ioffe and Christian Szegedy.
\newblock {\color{black}Batch Normalization: Accelerating Deep Network Training
  by Reducing Internal Covariate Shift}.
\newblock In {\em {\color{black}Proc. International Conference on Machine
  Learning}}, pages~{\color{black}448--456}, {\color{black}2015}\color{black}.

\bibitem{hinton2002training}
Geoffrey~E Hinton.
\newblock Training products of experts by minimizing contrastive divergence.
\newblock {\em Neural computation}, 14(8):1771--1800, 2002.

\bibitem{casella1992explaining}
George Casella and Edward~I George.
\newblock Explaining the {Gibbs} sampler.
\newblock {\em The American Statistician}, 46(3):167--174, 1992.

\bibitem{kuremoto2014forecast}
Takashi Kuremoto, Masanao Obayashi, Kunikazu Kobayashi, Takaomi Hirata, and
  Shingo Mabu.
\newblock Forecast chaotic time series data by {DBNs}.
\newblock In {\em Proc. 7th IEEE International Congress on Image and Signal
  Processing (CISP)}, pages 1130--1135, 2014.

\bibitem{dauphin2013stochastic}
Yann Dauphin and Yoshua Bengio.
\newblock Stochastic ratio matching of {RBMs} for sparse high-dimensional
  inputs.
\newblock In {\em Advances in Neural Information Processing Systems}, pages
  1340--1348, 2013.

\bibitem{sainath2011making}
Tara~N Sainath, Brian Kingsbury, Bhuvana Ramabhadran, Petr Fousek, Petr Novak,
  and Abdel-rahman Mohamed.
\newblock Making deep belief networks effective for large vocabulary continuous
  speech recognition.
\newblock In {\em Proc. IEEE Workshop on Automatic Speech Recognition and
  Understanding (ASRU)}, pages 30--35, 2011.

\bibitem{bengio2009learning}
Yoshua Bengio et~al.
\newblock Learning deep architectures for {AI}.
\newblock {\em Foundations and trends{\textregistered} in Machine Learning},
  2(1):1--127, 2009.

\bibitem{sakurada2014anomaly}
Mayu Sakurada and Takehisa Yairi.
\newblock Anomaly detection using autoencoders with nonlinear dimensionality
  reduction.
\newblock In {\em Proc. Workshop on Machine Learning for Sensory Data Analysis
  (MLSDA)}, page~4. ACM, 2014.

\bibitem{nicolau2016hybrid}
Miguel Nicolau, James McDermott, et~al.
\newblock A hybrid autoencoder and density estimation model for anomaly
  detection.
\newblock In {\em Proc. International Conference on Parallel Problem Solving
  from Nature}, pages 717--726. Springer, 2016.

\bibitem{thing2017ieee}
Vrizlynn~LL Thing.
\newblock {IEEE} 802.11 network anomaly detection and attack classification: A
  deep learning approach.
\newblock In {\em Proc. IEEE Wireless Communications and Networking Conference
  (WCNC)}, pages 1--6, 2017.

\bibitem{mao2017routing}
Bomin Mao, Zubair~Md Fadlullah, Fengxiao Tang, Nei Kato, Osamu Akashi, Takeru
  Inoue, and Kimihiro Mizutani.
\newblock Routing or computing? the paradigm shift towards intelligent computer
  network packet transmission based on deep learning.
\newblock {\em IEEE Transactions on Computers}, 66(11):1946--1960, 2017.

\bibitem{radu2016towards}
Valentin Radu, Nicholas~D Lane, Sourav Bhattacharya, Cecilia Mascolo, Mahesh~K
  Marina, and Fahim Kawsar.
\newblock Towards multimodal deep learning for activity recognition on mobile
  devices.
\newblock In {\em Proc. ACM International Joint Conference on Pervasive and
  Ubiquitous Computing: Adjunct}, pages 185--188, 2016.

\bibitem{radu2018multimodal}
Valentin Radu, Catherine Tong, Sourav Bhattacharya, Nicholas~D Lane, Cecilia
  Mascolo, Mahesh~K Marina, and Fahim Kawsar.
\newblock Multimodal deep learning for activity and context recognition.
\newblock {\em Proc. ACM on Interactive, Mobile, Wearable and Ubiquitous
  Technologies (IMWUT)}, 1(4):157, 2018.

\bibitem{raghavendra2016learning}
Ramachandra Raghavendra and Christoph Busch.
\newblock Learning deeply coupled autoencoders for smartphone based robust
  periocular verification.
\newblock In {\em Proc. IEEE International Conference on Image Processing
  (ICIP)}, pages 325--329, 2016.

\bibitem{li2016wavelet}
Jing Li, Jingyuan Wang, and Zhang Xiong.
\newblock Wavelet-based stacked denoising autoencoders for cell phone base
  station user number prediction.
\newblock In {\em Proc. IEEE International Conference on Internet of Things
  (iThings) and IEEE Green Computing and Communications (GreenCom) and IEEE
  Cyber, Physical and Social Computing (CPSCom) and IEEE Smart Data
  (SmartData)}, pages 833--838, 2016.

\bibitem{ILSVRC15}
Olga Russakovsky, Jia Deng, Hao Su, Jonathan Krause, Sanjeev Satheesh, Sean Ma,
  Zhiheng Huang, Andrej Karpathy, Aditya Khosla, Michael Bernstein,
  Alexander~C. Berg, and Li~Fei-Fei.
\newblock {ImageNet Large Scale Visual Recognition Challenge}.
\newblock {\em International Journal of Computer Vision (IJCV)},
  115(3):211--252, 2015.

\bibitem{active2017yunho}
Junmo~Kim Yunho~Jeon.
\newblock Active convolution: Learning the shape of convolution for image
  classification.
\newblock In {\em Proc. IEEE Conference on Computer Vision and Pattern
  Recognition}, 2017.

\bibitem{dai2017deformable}
Jifeng Dai, Haozhi Qi, Yuwen Xiong, Yi~Li, Guodong Zhang, Han Hu, and Yichen
  Wei.
\newblock Deformable convolutional networks.
\newblock In {\em Proc. IEEE International Conference on Computer Vision},
  pages 764--773, 2017.

\bibitem{Zhu2018morel}
\textcolor{black}{Xizhou} \color{black}Zhu, Han Hu, Stephen Lin, and Jifeng
  Dai.
\newblock {\color{black}Deformable ConvNets v2: More Deformable, Better
  Results}.
\newblock {\em {\color{black}arXiv preprint arXiv:1811.11168}},
  {\color{black}2018}\color{black}.

\bibitem{bengio1994learning}
Yoshua Bengio, Patrice Simard, and Paolo Frasconi.
\newblock Learning long-term dependencies with gradient descent is difficult.
\newblock {\em IEEE transactions on neural networks}, 5(2):157--166, 1994.

\bibitem{graves2013hybrid}
Alex Graves, Navdeep Jaitly, and Abdel-rahman Mohamed.
\newblock Hybrid speech recognition with deep bidirectional {LSTM}.
\newblock In {\em Proc. IEEE Workshop on Automatic Speech Recognition and
  Understanding (ASRU)}, pages 273--278, 2013.

\bibitem{johnson2016supervised}
Rie Johnson and Tong Zhang.
\newblock Supervised and semi-supervised text categorization using {LSTM} for
  region embeddings.
\newblock In {\em Proc. International Conference on Machine Learning (ICML)},
  pages 526--534, 2016.

\bibitem{goodfellow2016nips}
\textcolor{black}{Ian} \color{black}Goodfellow.
\newblock {\color{black}{NIPS} 2016 tutorial: Generative adversarial networks}.
\newblock {\em {\color{black}arXiv preprint arXiv:1701.00160}},
  {\color{black}2016}\color{black}.

\bibitem{ledig2017photo}
Christian Ledig, Lucas Theis, Ferenc Husz{\'a}r, Jose Caballero, Andrew
  Cunningham, Alejandro Acosta, Andrew Aitken, Alykhan Tejani, Johannes Totz,
  Zehan Wang, et~al.
\newblock Photo-realistic single image super-resolution using a generative
  adversarial network.
\newblock In {\em Proc. IEEE Conference on Computer Vision and Pattern
  Recognition}, 2017.

\bibitem{li2017perceptual}
Jianan Li, Xiaodan Liang, Yunchao Wei, Tingfa Xu, Jiashi Feng, and Shuicheng
  Yan.
\newblock Perceptual generative adversarial networks for small object
  detection.
\newblock In {\em Proc. IEEE Conference on Computer Vision and Pattern
  Recognition}, 2017.

\bibitem{li2017generative}
Yijun Li, Sifei Liu, Jimei Yang, and Ming-Hsuan Yang.
\newblock Generative face completion.
\newblock In {\em IEEE Conference on Computer Vision and Pattern Recognition},
  2017.

\bibitem{gawlowicz2018ns3}
\textcolor{black}{Piotr} \color{black}Gaw{\l}owicz and Anatolij Zubow.
\newblock {\color{black}{NS3-Gym}: Extending {OpenAI Gym} for Networking
  Research}.
\newblock {\em {\color{black}arXiv preprint arXiv:1810.03943}},
  {\color{black}2018}\color{black}.

\bibitem{gu2016continuous}
Shixiang Gu, Timothy Lillicrap, Ilya Sutskever, and Sergey Levine.
\newblock Continuous deep {Q}-learning with model-based acceleration.
\newblock In {\em Proc. International Conference on Machine Learning}, pages
  2829--2838, 2016.

\bibitem{moravvcik2017deepstack}
Matej Morav{\v{c}}{\'\i}k, Martin Schmid, Neil Burch, Viliam Lis{\`y}, Dustin
  Morrill, Nolan Bard, Trevor Davis, Kevin Waugh, Michael Johanson, and Michael
  Bowling.
\newblock Deepstack: Expert-level artificial intelligence in heads-up no-limit
  poker.
\newblock {\em Science}, 356(6337):508--513, 2017.

\bibitem{levine2016learning}
Sergey Levine, Peter Pastor, Alex Krizhevsky, Julian Ibarz, and Deirdre
  Quillen.
\newblock Learning hand-eye coordination for robotic grasping with deep
  learning and large-scale data collection.
\newblock {\em The International Journal of Robotics Research},
  37(4-5):421--436, 2018.

\bibitem{sallab2017deep}
Ahmad~EL Sallab, Mohammed Abdou, Etienne Perot, and Senthil Yogamani.
\newblock Deep reinforcement learning framework for autonomous driving.
\newblock {\em Electronic Imaging}, (19):70--76, 2017.

\bibitem{li2014tact}
Rongpeng Li, Zhifeng Zhao, Xianfu Chen, Jacques Palicot, and Honggang Zhang.
\newblock Tact: A transfer actor-critic learning framework for energy saving in
  cellular radio access networks.
\newblock {\em IEEE Transactions on Wireless Communications}, 13(4):2000--2011,
  2014.

\bibitem{al2015application}
Hasan~AA Al-Rawi, Ming~Ann Ng, and Kok-Lim~Alvin Yau.
\newblock Application of reinforcement learning to routing in distributed
  wireless networks: a review.
\newblock {\em Artificial Intelligence Review}, 43(3):381--416, 2015.

\bibitem{liu2015reinforcement}
Yan-Jun Liu, Li~Tang, Shaocheng Tong, CL~Philip Chen, and Dong-Juan Li.
\newblock Reinforcement learning design-based adaptive tracking control with
  less learning parameters for nonlinear discrete-time {MIMO} systems.
\newblock {\em IEEE Transactions on Neural Networks and Learning Systems},
  26(1):165--176, 2015.

\bibitem{pierucci2016neural}
Laura Pierucci and Davide Micheli.
\newblock A neural network for quality of experience estimation in mobile
  communications.
\newblock {\em IEEE MultiMedia}, 23(4):42--49, 2016.

\bibitem{gwon2014inferring}
Youngjune~L Gwon and HT~Kung.
\newblock Inferring origin flow patterns in wi-fi with deep learning.
\newblock In {\em Proc. 11th IEEE International Conference on Autonomic
  Computing (ICAC))}, pages 73--83, 2014.

\bibitem{nie2017network}
Laisen Nie, Dingde Jiang, Shui Yu, and Houbing Song.
\newblock Network traffic prediction based on deep belief network in wireless
  mesh backbone networks.
\newblock In {\em Proc. IEEEWireless Communications and Networking Conference
  (WCNC)}, pages 1--5, 2017.

\bibitem{moyo2015generalization}
Vusumuzi Moyo et~al.
\newblock The generalization ability of artificial neural networks in
  forecasting {TCP/IP} traffic trends: How much does the size of learning rate
  matter?
\newblock {\em International Journal of Computer Science and Application},
  2015.

\bibitem{wangspatiotemporal}
Jing Wang, Jian Tang, Zhiyuan Xu, Yanzhi Wang, Guoliang Xue, Xing Zhang, and
  Dejun Yang.
\newblock Spatiotemporal modeling and prediction in cellular networks: A big
  data enabled deep learning approach.
\newblock In {\em Proc. 36th Annual IEEE International Conference on Computer
  Communications (INFOCOM)}, 2017.

\bibitem{zhang2017long}
Chaoyun Zhang and Paul Patras.
\newblock Long-term mobile traffic forecasting using deep spatio-temporal
  neural networks.
\newblock In {\em Proc. Eighteenth ACM International Symposium on Mobile Ad Hoc
  Networking and Computing}, pages 231--240, 2018.

\bibitem{chaoyun2017zipnet}
Chaoyun Zhang, Xi~Ouyang, and Paul Patras.
\newblock {ZipNet-GAN}: Inferring fine-grained mobile traffic patterns via a
  generative adversarial neural network.
\newblock In {\em Proc. 13th ACM Conference on Emerging Networking Experiments
  and Technologies}, 2017.

\bibitem{huang2017study}
Chih-Wei Huang, Chiu-Ti Chiang, and Qiuhui Li.
\newblock A study of deep learning networks on mobile traffic forecasting.
\newblock In {\em Proc. 28th IEEE Annual International Symposium on Personal,
  Indoor, and Mobile Radio Communications (PIMRC)}, pages 1--6,
  2017\color{black}.

\bibitem{zhang2018citywide}
Chuanting Zhang, Haixia Zhang, Dongfeng Yuan, and Minggao Zhang.
\newblock Citywide cellular traffic prediction based on densely connected
  convolutional neural networks.
\newblock {\em IEEE Communications Letters}, 2018\color{black}.

\bibitem{navabi2018predicting}
Shiva Navabi, Chenwei Wang, Ozgun~Y Bursalioglu, and Haralabos Papadopoulos.
\newblock Predicting wireless channel features using neural networks.
\newblock {\em arXiv preprint arXiv:1802.00107}, 2018\color{black}.

\bibitem{wang2015applications}
Zhanyi Wang.
\newblock The applications of deep learning on traffic identification.
\newblock {\em BlackHat USA}, 2015.

\bibitem{wang2017end}
Wei Wang, Ming Zhu, Jinlin Wang, Xuewen Zeng, and Zhongzhen Yang.
\newblock End-to-end encrypted traffic classification with one-dimensional
  convolution neural networks.
\newblock In {\em Proc. IEEE International Conference on Intelligence and
  Security Informatics}, 2017.

\bibitem{lotfollahi2017deep}
Mohammad Lotfollahi, Ramin Shirali, Mahdi~Jafari Siavoshani, and Mohammdsadegh
  Saberian.
\newblock Deep packet: A novel approach for encrypted traffic classification
  using deep learning.
\newblock {\em arXiv preprint arXiv:1709.02656}, 2017.

\bibitem{wang2017malware}
Wei Wang, Ming Zhu, Xuewen Zeng, Xiaozhou Ye, and Yiqiang Sheng.
\newblock Malware traffic classification using convolutional neural network for
  representation learning.
\newblock In {\em Proc. IEEE International Conference on Information Networking
  (ICOIN)}, pages 712--717, 2017.

\bibitem{liang2016mercury}
Victor~C Liang, Richard~TB Ma, Wee~Siong Ng, Li~Wang, Marianne Winslett, Huayu
  Wu, Shanshan Ying, and Zhenjie Zhang.
\newblock Mercury: Metro density prediction with recurrent neural network on
  streaming {CDR} data.
\newblock In {\em Proc. IEEE 32nd International Conference on Data Engineering
  (ICDE)}, pages 1374--1377, 2016.

\bibitem{felbo2016using}
Bjarke Felbo, P{\aa}l Sunds{\o}y, Alex'Sandy' Pentland, Sune Lehmann, and
  Yves-Alexandre de~Montjoye.
\newblock Using deep learning to predict demographics from mobile phone
  metadata.
\newblock In {\em Proc. workshop track of the International Conference on
  Learning Representations (ICLR)}, 2016.

\bibitem{chen2017comprehensive}
Nai~Chun Chen, Wanqin Xie, Roy~E Welsch, Kent Larson, and Jenny Xie.
\newblock Comprehensive predictions of tourists' next visit location based on
  call detail records using machine learning and deep learning methods.
\newblock In {\em Proc. IEEE International Congress on Big Data (BigData
  Congress)}, pages 1--6, 2017.

\bibitem{lin2017deep}
Ziheng Lin, Mogeng Yin, Sidney Feygin, Madeleine Sheehan, Jean-Francois
  Paiement, and Alexei Pozdnoukhov.
\newblock Deep generative models of urban mobility.
\newblock {\em IEEE Transactions on Intelligent Transportation Systems}, 2017.

\bibitem{xu2017large}
Chang Xu, Kuiyu Chang, Khee-Chin Chua, Meishan Hu, and Zhenxiang Gao.
\newblock Large-scale {Wi-Fi} hotspot classification via deep learning.
\newblock In {\em Proc. 26th International Conference on World Wide Web
  Companion}, pages 857--858. International World Wide Web Conferences Steering
  Committee, 2017.

\bibitem{meng2018qoe}
Qianyu Meng, Kun Wang, Bo~Liu, Toshiaki Miyazaki, and Xiaoming He.
\newblock {QoE}-based big data analysis with deep learning in pervasive edge
  environment.
\newblock In {\em Proc. IEEE International Conference on Communications (ICC)},
  pages 1--6, 2018.

\bibitem{fang2018mobile}
\textcolor{black}{Luoyang} \color{black}Fang, Xiang Cheng, Haonan Wang, and
  Liuqing Yang.
\newblock {\color{black}Mobile Demand Forecasting via Deep Graph-Sequence
  Spatiotemporal Modeling in Cellular Networks}.
\newblock {\em {\color{black}IEEE Internet of Things Journal}},
  {\color{black}2018}\color{black}.

\bibitem{luo2018channel}
\textcolor{black}{Changqing} \color{black}Luo, Jinlong Ji, Qianlong Wang, Xuhui
  Chen, and Pan Li.
\newblock {\color{black}Channel state information prediction for {5G} wireless
  communications: A deep learning approach}.
\newblock {\em {\color{black}IEEE Transactions on Network Science and
  Engineering}}, {\color{black}2018}\color{black}.

\bibitem{8553650}
\textcolor{black}{Peng} \color{black}Li, Zhikui Chen, Laurence~T. Yang, Jing
  Gao, Qingchen Zhang, and M.~Jamal Deen.
\newblock {\color{black}An Improved Stacked Auto-Encoder for Network Traffic
  Flow Classification}.
\newblock {\em {\color{black}IEEE Network}},
  {\color{black}32}({\color{black}6}):{\color{black}22--27},
  {\color{black}2018}\color{black}.

\bibitem{feng2018deeptp}
\textcolor{black}{Jie} \color{black}Feng, Xinlei Chen, Rundong Gao, Ming Zeng,
  and Yong Li.
\newblock {\color{black}{DeepTP}: An End-to-End Neural Network for Mobile
  Cellular Traffic Prediction}.
\newblock {\em {\color{black}IEEE Network}},
  {\color{black}32}({\color{black}6}):{\color{black}108--115},
  {\color{black}2018}\color{black}.

\bibitem{zhu2018deep}
\textcolor{black}{Hao} \color{black}Zhu, Yang Cao, Wei Wang, Tao Jiang, and Shi
  Jin.
\newblock {\color{black}Deep Reinforcement Learning for Mobile Edge Caching:
  Review, New Features, and Open Issues}.
\newblock {\em {\color{black}IEEE Network}},
  {\color{black}32}({\color{black}6}):{\color{black}50--57},
  {\color{black}2018}\color{black}.

\bibitem{liu2016poster}
Sicong Liu and Junzhao Du.
\newblock Poster: Mobiear-building an environment-independent acoustic sensing
  platform for the deaf using deep learning.
\newblock In {\em Proc. 14th ACM Annual International Conference on Mobile
  Systems, Applications, and Services Companion}, pages 50--50, 2016.

\bibitem{sicong2017ubiear}
Liu Sicong, Zhou Zimu, Du~Junzhao, Shangguan Longfei, Jun Han, and Xin Wang.
\newblock Ubiear: Bringing location-independent sound awareness to the
  hard-of-hearing people with smartphones.
\newblock {\em Proc. ACM Interactive, Mobile, Wearable and Ubiquitous
  Technologies (IMWUT)}, 1(2):17, 2017.

\bibitem{jindal2016integrating}
Vasu Jindal.
\newblock Integrating mobile and cloud for {PPG} signal selection to monitor
  heart rate during intensive physical exercise.
\newblock In {\em Proc. ACM International Workshop on Mobile Software
  Engineering and Systems}, pages 36--37, 2016.

\bibitem{kim2016deep}
Edward Kim, Miguel Corte-Real, and Zubair Baloch.
\newblock A deep semantic mobile application for thyroid cytopathology.
\newblock In {\em Medical Imaging 2016: PACS and Imaging Informatics: Next
  Generation and Innovations}, volume 9789, page 97890A. International Society
  for Optics and Photonics, 2016.

\bibitem{sathyanarayana2016sleep}
Aarti Sathyanarayana, Shafiq Joty, Luis Fernandez-Luque, Ferda Ofli, Jaideep
  Srivastava, Ahmed Elmagarmid, Teresa Arora, and Shahrad Taheri.
\newblock Sleep quality prediction from wearable data using deep learning.
\newblock {\em JMIR mHealth and uHealth}, 4(4), 2016.

\bibitem{li2017personal}
Honggui Li and Maria Trocan.
\newblock Personal health indicators by deep learning of smart phone sensor
  data.
\newblock In {\em Proc. 3rd IEEE International Conference on Cybernetics
  (CYBCONF)}, pages 1--5, 2017.

\bibitem{hosseini2017deep}
Mohammad-Parsa Hosseini, Tuyen~X Tran, Dario Pompili, Kost Elisevich, and Hamid
  Soltanian-Zadeh.
\newblock Deep learning with edge computing for localization of
  epileptogenicity using multimodal rs-{fMRI} and {EEG} big data.
\newblock In {\em Proc. IEEE International Conference on Autonomic Computing
  (ICAC)}, pages 83--92, 2017.

\bibitem{stamate2017deep}
Cosmin Stamate, George~D Magoulas, Stefan K{\"u}ppers, Effrosyni Nomikou,
  Ioannis Daskalopoulos, Marco~U Luchini, Theano Moussouri, and George Roussos.
\newblock Deep learning parkinson's from smartphone data.
\newblock In {\em Proc. IEEE International Conference on Pervasive Computing
  and Communications (PerCom)}, pages 31--40, 2017.

\bibitem{quisel2017collecting}
Tom Quisel, Luca Foschini, Alessio Signorini, and David~C Kale.
\newblock Collecting and analyzing millions of mhealth data streams.
\newblock In {\em Proc. 23rd ACM SIGKDD International Conference on Knowledge
  Discovery and Data Mining}, pages 1971--1980, 2017.

\bibitem{khan2017deep}
Usman~Mahmood Khan, Zain Kabir, Syed~Ali Hassan, and Syed~Hassan Ahmed.
\newblock A deep learning framework using passive {WiFi} sensing for
  respiration monitoring.
\newblock In {\em Proc. IEEE Global Communications Conference (GLOBECOM)},
  pages 1--6, 2017.

\bibitem{li2016deepcham}
Dawei Li, Theodoros Salonidis, Nirmit~V Desai, and Mooi~Choo Chuah.
\newblock Deepcham: Collaborative edge-mediated adaptive deep learning for
  mobile object recognition.
\newblock In {\em Proc. IEEE/ACM Symposium on Edge Computing (SEC)}, pages
  64--76, 2016.

\bibitem{tobias2016convolutional}
Luis Tob{\'\i}as, Aur{\'e}lien Ducournau, Fran{\c{c}}ois Rousseau, Gr{\'e}goire
  Mercier, and Ronan Fablet.
\newblock Convolutional neural networks for object recognition on mobile
  devices: A case study.
\newblock In {\em Proc. 23rd IEEE International Conference on Pattern
  Recognition (ICPR)}, pages 3530--3535, 2016.

\bibitem{pouladzadeh2017mobile}
Parisa Pouladzadeh and Shervin Shirmohammadi.
\newblock Mobile multi-food recognition using deep learning.
\newblock {\em ACM Transactions on Multimedia Computing, Communications, and
  Applications (TOMM)}, 13(3s):36, 2017.

\bibitem{tanno2016deepfoodcam}
Ryosuke Tanno, Koichi Okamoto, and Keiji Yanai.
\newblock {DeepFoodCam}: A {DCNN}-based real-time mobile food recognition
  system.
\newblock In {\em Proc. 2nd ACM International Workshop on Multimedia Assisted
  Dietary Management}, pages 89--89, 2016.

\bibitem{kuhad2015using}
Pallavi Kuhad, Abdulsalam Yassine, and Shervin Shimohammadi.
\newblock Using distance estimation and deep learning to simplify calibration
  in food calorie measurement.
\newblock In {\em Proc. IEEE International Conference on Computational
  Intelligence and Virtual Environments for Measurement Systems and
  Applications (CIVEMSA)}, pages 1--6, 2015.

\bibitem{teng2016facial}
Teng Teng and Xubo Yang.
\newblock Facial expressions recognition based on convolutional neural networks
  for mobile virtual reality.
\newblock In {\em Proc. 15th ACM SIGGRAPH Conference on Virtual-Reality
  Continuum and Its Applications in Industry-Volume 1}, pages 475--478, 2016.

\bibitem{rao2017mobile}
Jinmeng Rao, Yanjun Qiao, Fu~Ren, Junxing Wang, and Qingyun Du.
\newblock A mobile outdoor augmented reality method combining deep learning
  object detection and spatial relationships for geovisualization.
\newblock {\em Sensors}, 17(9):1951, 2017.

\bibitem{zeng2014convolutional}
Ming Zeng, Le~T Nguyen, Bo~Yu, Ole~J Mengshoel, Jiang Zhu, Pang Wu, and Joy
  Zhang.
\newblock Convolutional neural networks for human activity recognition using
  mobile sensors.
\newblock In {\em Proc. 6th IEEE International Conference on Mobile Computing,
  Applications and Services (MobiCASE)}, pages 197--205, 2014.

\bibitem{almaslukh2017effective}
Bandar Almaslukh, Jalal AlMuhtadi, and Abdelmonim Artoli.
\newblock An effective deep autoencoder approach for online smartphone-based
  human activity recognition.
\newblock {\em International Journal of Computer Science and Network Security
  (IJCSNS)}, 17(4):160, 2017.

\bibitem{li2016deep22}
Xinyu Li, Yanyi Zhang, Ivan Marsic, Aleksandra Sarcevic, and Randall~S Burd.
\newblock Deep learning for {RFID}-based activity recognition.
\newblock In {\em Proc. 14th ACM Conference on Embedded Network Sensor Systems
  CD-ROM}, pages 164--175, 2016.

\bibitem{bhattacharya2016smart}
Sourav Bhattacharya and Nicholas~D Lane.
\newblock From smart to deep: Robust activity recognition on smartwatches using
  deep learning.
\newblock In {\em Proc. IEEE International Conference on Pervasive Computing
  and Communication Workshops (PerCom Workshops)}, pages 1--6, 2016.

\bibitem{antoniou2016general}
Antreas Antoniou and Plamen Angelov.
\newblock A general purpose intelligent surveillance system for mobile devices
  using deep learning.
\newblock In {\em Proc. IEEE International Joint Conference on Neural Networks
  (IJCNN)}, pages 2879--2886, 2016.

\bibitem{wang2016interacting}
Saiwen Wang, Jie Song, Jaime Lien, Ivan Poupyrev, and Otmar Hilliges.
\newblock Interacting with {Soli}: Exploring fine-grained dynamic gesture
  recognition in the radio-frequency spectrum.
\newblock In {\em Proc. 29th ACM Annual Symposium on User Interface Software
  and Technology}, pages 851--860, 2016.

\bibitem{gao2016ihear}
Yang Gao, Ning Zhang, Honghao Wang, Xiang Ding, Xu~Ye, Guanling Chen, and
  Yu~Cao.
\newblock ihear food: Eating detection using commodity bluetooth headsets.
\newblock In {\em Proc. IEEE First International Conference on Connected
  Health: Applications, Systems and Engineering Technologies (CHASE)}, pages
  163--172, 2016.

\bibitem{zhu2015using}
Jindan Zhu, Amit Pande, Prasant Mohapatra, and Jay~J Han.
\newblock Using deep learning for energy expenditure estimation with wearable
  sensors.
\newblock In {\em Proc. 17th IEEE International Conference on E-health
  Networking, Application \& Services (HealthCom)}, pages 501--506, 2015.

\bibitem{sundsoy2016deep}
P{\aa}l Sunds{\o}y, Johannes Bjelland, B~Reme, A~Iqbal, and Eaman Jahani.
\newblock Deep learning applied to mobile phone data for individual income
  classification.
\newblock {\em ICAITA doi}, 10, 2016.

\bibitem{chen2015deep}
Yuqing Chen and Yang Xue.
\newblock A deep learning approach to human activity recognition based on
  single accelerometer.
\newblock In {\em Proc. IEEE International Conference on Systems, Man, and
  Cybernetics (SMC)}, pages 1488--1492, 2015.

\bibitem{ha2016convolutional}
Sojeong Ha and Seungjin Choi.
\newblock Convolutional neural networks for human activity recognition using
  multiple accelerometer and gyroscope sensors.
\newblock In {\em Proc. IEEE International Joint Conference on Neural Networks
  (IJCNN)}, pages 381--388, 2016.

\bibitem{edel2016binarized}
Marcus Edel and Enrico K{\"o}ppe.
\newblock Binarized-{BLSTM-RNN} based human activity recognition.
\newblock In {\em Proc. IEEE International Conference on Indoor Positioning and
  Indoor Navigation (IPIN)}, pages 1--7, 2016.

\bibitem{xue2018appdna}
\textcolor{black}{Shuangshuang} \color{black}Xue, Lan Zhang, Anran Li,
  Xiang-Yang Li, Chaoyi Ruan, and Wenchao Huang.
\newblock {\color{black}{AppDNA}: App behavior profiling via graph-based deep
  learning}.
\newblock In {\em {\color{black}Proc. IEEE Conference on Computer
  Communications}}, pages~{\color{black}1475--1483},
  {\color{black}2018}\color{black}.

\bibitem{liu2018finding}
\textcolor{black}{Huiqi} \color{black}Liu, Xiang-Yang Li, Lan Zhang, Yaochen
  Xie, Zhenan Wu, Qian Dai, Ge~Chen, and Chunxiao Wan.
\newblock {\color{black}Finding the stars in the fireworks: Deep understanding
  of motion sensor fingerprint}.
\newblock In {\em {\color{black}Proc. IEEE Conference on Computer
  Communications}}, pages~{\color{black}126--134},
  {\color{black}2018}\color{black}.

\bibitem{okita2017recognition}
Tsuyoshi Okita and Sozo Inoue.
\newblock Recognition of multiple overlapping activities using compositional
  cnn-lstm model.
\newblock In {\em Proc. ACM International Joint Conference on Pervasive and
  Ubiquitous Computing and Proc. ACM International Symposium on Wearable
  Computers}, pages 165--168. ACM, 2017.

\bibitem{mittal2016spotgarbage}
Gaurav Mittal, Kaushal~B Yagnik, Mohit Garg, and Narayanan~C Krishnan.
\newblock Spotgarbage: smartphone app to detect garbage using deep learning.
\newblock In {\em Proc. ACM International Joint Conference on Pervasive and
  Ubiquitous Computing}, pages 940--945, 2016.

\bibitem{seidenari2017deep}
Lorenzo Seidenari, Claudio Baecchi, Tiberio Uricchio, Andrea Ferracani, Marco
  Bertini, and Alberto~Del Bimbo.
\newblock Deep artwork detection and retrieval for automatic context-aware
  audio guides.
\newblock {\em ACM Transactions on Multimedia Computing, Communications, and
  Applications (TOMM)}, 13(3s):35, 2017.

\bibitem{zeng2017mobiledeeppill}
Xiao Zeng, Kai Cao, and Mi~Zhang.
\newblock Mobiledeeppill: A small-footprint mobile deep learning system for
  recognizing unconstrained pill images.
\newblock In {\em Proc. 15th ACM Annual International Conference on Mobile
  Systems, Applications, and Services}, pages 56--67, 2017.

\bibitem{zoudeepsense}
Han Zou, Yuxun Zhou, Jianfei Yang, Hao Jiang, Lihua Xie, and Costas~J Spanos.
\newblock Deepsense: Device-free human activity recognition via autoencoder
  long-term recurrent convolutional network.
\newblock 2018\color{black}.

\bibitem{zeng2017mobile}
Xiao Zeng.
\newblock Mobile sensing through deep learning.
\newblock In {\em Proc. Workshop on MobiSys Ph. D. Forum}, pages 5--6. ACM,
  2017.

\bibitem{wang2015phasefi}
Xuyu Wang, Lingjun Gao, and Shiwen Mao.
\newblock {PhaseFi}: Phase fingerprinting for indoor localization with a deep
  learning approach.
\newblock In {\em Proc. IEEE Global Communications Conference (GLOBECOM)},
  pages 1--6, 2015.

\bibitem{wang2016csi}
Xuyu Wang, Lingjun Gao, and Shiwen Mao.
\newblock {CSI} phase fingerprinting for indoor localization with a deep
  learning approach.
\newblock {\em IEEE Internet of Things Journal}, 3(6):1113--1123, 2016.

\bibitem{feng2018evaluation}
Chunhai Feng, Sheheryar Arshad, Ruiyun Yu, and Yonghe Liu.
\newblock Evaluation and improvement of activity detection systems with
  recurrent neural network.
\newblock In {\em Proc. IEEE International Conference on Communications (ICC)},
  pages 1--6, 2018\color{black}.

\bibitem{cao2017deepmood}
Bokai Cao, Lei Zheng, Chenwei Zhang, Philip~S Yu, Andrea Piscitello, John
  Zulueta, Olu Ajilore, Kelly Ryan, and Alex~D Leow.
\newblock Deepmood: Modeling mobile phone typing dynamics for mood detection.
\newblock In {\em Proc. 23rd ACM SIGKDD International Conference on Knowledge
  Discovery and Data Mining}, pages 747--755, 2017.

\bibitem{ran2018deepdecision}
Xukan Ran, Haoliang Chen, Xiaodan Zhu, Zhenming Liu, and Jiasi Chen.
\newblock Deepdecision: A mobile deep learning framework for edge video
  analytics.
\newblock In {\em Proc. IEEE International Conference on Computer
  Communications}, 2018\color{black}.

\bibitem{siri}
Siri Team.
\newblock {Deep Learning for Siri's Voice: On-device Deep Mixture Density
  Networks for Hybrid Unit Selection Synthesis}.
\newblock \url{https://machinelearning.apple.com/2017/08/06/siri-voices.html},
  2017.
\newblock [Online; accessed 16-Sep-2017].

\bibitem{mcgraw2016personalized}
Ian McGraw, Rohit Prabhavalkar, Raziel Alvarez, Montse~Gonzalez Arenas,
  Kanishka Rao, David Rybach, Ouais Alsharif, Ha{\c{s}}im Sak, Alexander
  Gruenstein, Fran{\c{c}}oise Beaufays, et~al.
\newblock Personalized speech recognition on mobile devices.
\newblock In {\em Proc. IEEE International Conference on Acoustics, Speech and
  Signal Processing (ICASSP)}, pages 5955--5959, 2016.

\bibitem{prabhavalkar2016compression}
Rohit Prabhavalkar, Ouais Alsharif, Antoine Bruguier, and Lan McGraw.
\newblock On the compression of recurrent neural networks with an application
  to {LVCSR} acoustic modeling for embedded speech recognition.
\newblock In {\em Proc. IEEE International Conference on Acoustics, Speech and
  Signal Processing (ICASSP)}, pages 5970--5974, 2016.

\bibitem{yoshioka2015ntt}
Takuya Yoshioka, Nobutaka Ito, Marc Delcroix, Atsunori Ogawa, Keisuke
  Kinoshita, Masakiyo Fujimoto, Chengzhu Yu, Wojciech~J Fabian, Miquel Espi,
  Takuya Higuchi, et~al.
\newblock The {NTT CHiME}-3 system: Advances in speech enhancement and
  recognition for mobile multi-microphone devices.
\newblock In {\em Proc. IEEE Workshop on Automatic Speech Recognition and
  Understanding (ASRU)}, pages 436--443, 2015.

\bibitem{ruan2016speech}
Sherry Ruan, Jacob~O Wobbrock, Kenny Liou, Andrew Ng, and James Landay.
\newblock Speech is 3x faster than typing for english and mandarin text entry
  on mobile devices.
\newblock {\em arXiv preprint arXiv:1608.07323}, 2016.

\bibitem{ignatov2017dslr}
Andrey Ignatov, Nikolay Kobyshev, Radu Timofte, Kenneth Vanhoey, and Luc
  Van~Gool.
\newblock {DSLR}-quality photos on mobile devices with deep convolutional
  networks.
\newblock In {\em IEEE International Conference on Computer Vision (ICCV)},
  2017.

\bibitem{lu2017demo}
Zongqing Lu, Noor Felemban, Kevin Chan, and Thomas La~Porta.
\newblock Demo abstract: On-demand information retrieval from videos using deep
  learning in wireless networks.
\newblock In {\em Proc. IEEE/ACM Second International Conference on
  Internet-of-Things Design and Implementation (IoTDI)}, pages 279--280, 2017.

\bibitem{lee2016reducing}
Jemin Lee, Jinse Kwon, and Hyungshin Kim.
\newblock Reducing distraction of smartwatch users with deep learning.
\newblock In {\em Proc. 18th ACM International Conference on Human-Computer
  Interaction with Mobile Devices and Services Adjunct}, pages 948--953, 2016.

\bibitem{vu2016transportation}
Toan~H Vu, Le~Dung, and Jia-Ching Wang.
\newblock Transportation mode detection on mobile devices using recurrent nets.
\newblock In {\em Proc. ACM on Multimedia Conference}, pages 392--396, 2016.

\bibitem{fang2017learning}
Shih-Hau Fang, Yu-Xaing Fei, Zhezhuang Xu, and Yu~Tsao.
\newblock Learning transportation modes from smartphone sensors based on deep
  neural network.
\newblock {\em IEEE Sensors Journal}, 17(18):6111--6118, 2017.

\bibitem{zhao2018rf}
Mingmin Zhao, Yonglong Tian, Hang Zhao, Mohammad~Abu Alsheikh, Tianhong Li,
  Rumen Hristov, Zachary Kabelac, Dina Katabi, and Antonio Torralba.
\newblock {RF}-based {3D} skeletons.
\newblock In {\em Proc. ACM Conference of the ACM Special Interest Group on
  Data Communication (SIGCOMM)}, pages 267--281, 2018\color{black}.

\bibitem{katevas2017practical}
Kleomenis Katevas, Ilias Leontiadis, Martin Pielot, and Joan Serr{\`a}.
\newblock Practical processing of mobile sensor data for continual deep
  learning predictions.
\newblock In {\em Proc. 1st ACM International Workshop on Deep Learning for
  Mobile Systems and Applications}, 2017.

\bibitem{yao2017deepsense}
Shuochao Yao, Shaohan Hu, Yiran Zhao, Aston Zhang, and Tarek Abdelzaher.
\newblock Deepsense: A unified deep learning framework for time-series mobile
  sensing data processing.
\newblock In {\em Proc. 26th International Conference on World Wide Web}, pages
  351--360. International World Wide Web Conferences Steering Committee, 2017.

\bibitem{ohara2017detecting}
Kazuya Ohara, Takuya Maekawa, and Yasuyuki Matsushita.
\newblock Detecting state changes of indoor everyday objects using {Wi-Fi}
  channel state information.
\newblock {\em Proc. ACM on Interactive, Mobile, Wearable and Ubiquitous
  Technologies (IMWUT)}, 1(3):88, 2017.

\bibitem{liu2017deep}
Wu~Liu, Huadong Ma, Heng Qi, Dong Zhao, and Zhineng Chen.
\newblock Deep learning hashing for mobile visual search.
\newblock {\em EURASIP Journal on Image and Video Processing}, (1):17, 2017.

\bibitem{ouyang2016deepspace}
Xi~Ouyang, Chaoyun Zhang, Pan Zhou, and Hao Jiang.
\newblock {DeepSpace}: An online deep learning framework for mobile big data to
  understand human mobility patterns.
\newblock {\em arXiv preprint arXiv:1610.07009}, 2016.

\bibitem{yang2017neural}
Hua Yang, Zhimei Li, and Zhiyong Liu.
\newblock Neural networks for {MANET AODV}: an optimization approach.
\newblock {\em Cluster Computing}, pages 1--9, 2017.

\bibitem{song2016deeptransport}
Xuan Song, Hiroshi Kanasugi, and Ryosuke Shibasaki.
\newblock {DeepTransport}: Prediction and simulation of human mobility and
  transportation mode at a citywide level.
\newblock In {\em Proc. International Joint Conference on Artificial
  Intelligence}, pages 2618--2624, 2016.

\bibitem{zhang2017deep123}
Junbo Zhang, Yu~Zheng, and Dekang Qi.
\newblock Deep spatio-temporal residual networks for citywide crowd flows
  prediction.
\newblock In {\em Proc. National Conference on Artificial Intelligence (AAAI)},
  2017.

\bibitem{subramanian2014implementation}
J~Venkata Subramanian and M~Abdul~Karim Sadiq.
\newblock Implementation of artificial neural network for mobile movement
  prediction.
\newblock {\em Indian Journal of science and Technology}, 7(6):858--863, 2014.

\bibitem{ezema2017artificial}
Longinus~S Ezema and Cosmas~I Ani.
\newblock Artificial neural network approach to mobile location estimation in
  {GSM} network.
\newblock {\em International Journal of Electronics and Telecommunications},
  63(1):39--44, 2017.

\bibitem{shao2018depedo}
Wenhua Shao, Haiyong Luo, Fang Zhao, Cong Wang, Antonino Crivello, and
  Muhammad~Zahid Tunio.
\newblock {DePedo}: Anti periodic negative-step movement pedometer with deep
  convolutional neural networks.
\newblock In {\em Proc. IEEE International Conference on Communications (ICC)},
  pages 1--6, 2018\color{black}.

\bibitem{yayeh2018mobility}
\textcolor{black}{Yirga} \color{black}Yayeh, Hsin-piao Lin, Getaneh Berie,
  Abebe~Belay Adege, Lei Yen, and Shiann-Shiun Jeng.
\newblock {\color{black}Mobility prediction in mobile {Ad-hoc} network using
  deep learning}.
\newblock In {\em {\color{black}Proc. IEEE International Conference on Applied
  System Invention (ICASI)}}, pages~{\color{black}1203--1206},
  {\color{black}2018}\color{black}.

\bibitem{chen2016learning2}
\textcolor{black}{Quanjun} \color{black}Chen, Xuan Song, Harutoshi Yamada, and
  Ryosuke Shibasaki.
\newblock {\color{black}Learning Deep Representation from Big and Heterogeneous
  Data for Traffic Accident Inference}.
\newblock In {\em {\color{black}Proc. National Conference on Artificial
  Intelligence (AAAI)}}, pages~{\color{black}338--344},
  {\color{black}2016}\color{black}.

\bibitem{song2017deepmob}
\textcolor{black}{Xuan} \color{black}Song, Ryosuke Shibasaki, Nicholos~Jing
  Yuan, Xing Xie, Tao Li, and Ryutaro Adachi.
\newblock {\color{black}{DeepMob}: learning deep knowledge of human emergency
  behavior and mobility from big and heterogeneous data}.
\newblock {\em {\color{black}ACM Transactions on Information Systems (TOIS)}},
  {\color{black}35}({\color{black}4}):{\color{black}41},
  {\color{black}2017}\color{black}.

\bibitem{yao2017trajectory}
\textcolor{black}{Di} \color{black}Yao, Chao Zhang, Zhihua Zhu, Jianhui Huang,
  and Jingping Bi.
\newblock {\color{black}Trajectory clustering via deep representation
  learning}.
\newblock In {\em {\color{black}Proc. IEEE International Joint Conference on
  Neural Networks (IJCNN)}}, pages~{\color{black}3880--3887},
  {\color{black}2017}\color{black}.

\bibitem{liu2018urban}
\textcolor{black}{Zhidan} \color{black}Liu, Zhenjiang Li, Kaishun Wu, and
  Mo~Li.
\newblock {\color{black}Urban Traffic Prediction from Mobility Data Using Deep
  Learning}.
\newblock {\em {\color{black}IEEE Network}},
  {\color{black}32}({\color{black}4}):{\color{black}40--46},
  {\color{black}2018}\color{black}.

\bibitem{wickramasuriya2017base}
Dilranjan~S Wickramasuriya, Calvin~A Perumalla, Kemal Davaslioglu, and
  Richard~D Gitlin.
\newblock Base station prediction and proactive mobility management in virtual
  cells using recurrent neural networks.
\newblock In {\em Proc. IEEE Wireless and Microwave Technology Conference
  (WAMICON)}, pages 1--6.

\bibitem{tkavcik2016neural}
\textcolor{black}{Jan} \color{black}Tka{\v{c}}{\'\i}k and Pavel Kord{\'\i}k.
\newblock {\color{black}Neural turing machine for sequential learning of human
  mobility patterns}.
\newblock In {\em {\color{black}Proc. IEEE International Joint Conference on
  Neural Networks (IJCNN)}}, pages~{\color{black}2790--2797},
  {\color{black}2016}\color{black}.

\bibitem{kim2018method}
\textcolor{black}{Dong Yup} \color{black}Kim and Ha~Yoon Song.
\newblock {\color{black}Method of predicting human mobility patterns using deep
  learning}.
\newblock {\em {\color{black}Neurocomputing}},
  {\color{black}280}:{\color{black}56--64}, {\color{black}2018}\color{black}.

\bibitem{jiang2018deepurbanmomentum}
\textcolor{black}{Renhe} \color{black}Jiang, Xuan Song, Zipei Fan, Tianqi Xia,
  Quanjun Chen, Satoshi Miyazawa, and Ryosuke Shibasaki.
\newblock {\color{black}{DeepUrbanMomentum}: An Online Deep-Learning System for
  Short-Term Urban Mobility Prediction}.
\newblock In {\em {\color{black}Proc. National Conference on Artificial
  Intelligence (AAAI)}}, {\color{black}2018}\color{black}.

\bibitem{wang2018deep1231}
\textcolor{black}{Chujie} \color{black}Wang, Zhifeng Zhao, Qi~Sun, and Honggang
  Zhang.
\newblock {\color{black}Deep Learning-based Intelligent Dual Connectivity for
  Mobility Management in Dense Network}.
\newblock {\em {\color{black}arXiv preprint arXiv:1806.04584}},
  {\color{black}2018}\color{black}.

\bibitem{jiang2018deep}
\textcolor{black}{Renhe} \color{black}Jiang, Xuan Song, Zipei Fan, Tianqi Xia,
  Quanjun Chen, Qi~Chen, and Ryosuke Shibasaki.
\newblock {\color{black}Deep {ROI}-Based Modeling for Urban Human Mobility
  Prediction}.
\newblock {\em {\color{black}Proc. ACM on Interactive, Mobile, Wearable and
  Ubiquitous Technologies (IMWUT)}},
  {\color{black}2}({\color{black}1}):{\color{black}14},
  {\color{black}2018}\color{black}.

\bibitem{feng2018deepmove}
\textcolor{black}{Jie} \color{black}Feng, Yong Li, Chao Zhang, Funing Sun,
  Fanchao Meng, Ang Guo, and Depeng Jin.
\newblock {\color{black}{DeepMove}: Predicting Human Mobility with Attentional
  Recurrent Networks}.
\newblock In {\em {\color{black}Proc. World Wide Web Conference on World Wide
  Web}}, pages~{\color{black}1459--1468}, {\color{black}2018}\color{black}.

\bibitem{wang2015deepfi}
Xuyu Wang, Lingjun Gao, Shiwen Mao, and Santosh Pandey.
\newblock {DeepFi}: Deep learning for indoor fingerprinting using channel state
  information.
\newblock In {\em Proc. IEEE Wireless Communications and Networking Conference
  (WCNC)}, pages 1666--1671, 2015.

\bibitem{wang2017cifi}
Xuyu Wang, Xiangyu Wang, and Shiwen Mao.
\newblock {CiFi}: Deep convolutional neural networks for indoor localization
  with 5 {GHz} {Wi-Fi}.
\newblock In {\em Proc. IEEE International Conference on Communications (ICC)},
  pages 1--6, 2017.

\bibitem{wang2017biloc}
Xuyu Wang, Lingjun Gao, and Shiwen Mao.
\newblock {BiLoc}: Bi-modal deep learning for indoor localization with
  commodity {5GHz WiFi}.
\newblock {\em IEEE Access}, 5:4209--4220, 2017.

\bibitem{nowicki2017low}
Micha{\l} Nowicki and Jan Wietrzykowski.
\newblock Low-effort place recognition with {WiFi} fingerprints using deep
  learning.
\newblock In {\em Proc. International Conference Automation}, pages 575--584.
  Springer, 2017.

\bibitem{zhang2016device}
Xiao Zhang, Jie Wang, Qinghua Gao, Xiaorui Ma, and Hongyu Wang.
\newblock Device-free wireless localization and activity recognition with deep
  learning.
\newblock In {\em Proc. IEEE International Conference on Pervasive Computing
  and Communication Workshops (PerCom Workshops)}, pages 1--5, 2016.

\bibitem{wang2017device}
Jie Wang, Xiao Zhang, Qinhua Gao, Hao Yue, and Hongyu Wang.
\newblock Device-free wireless localization and activity recognition: A deep
  learning approach.
\newblock {\em IEEE Transactions on Vehicular Technology}, 66(7):6258--6267,
  2017.

\bibitem{mohammadi2017semi}
Mehdi Mohammadi, Ala Al-Fuqaha, Mohsen Guizani, and Jun-Seok Oh.
\newblock Semi-supervised deep reinforcement learning in support of {IoT} and
  smart city services.
\newblock {\em IEEE Internet of Things Journal}, 2017.

\bibitem{anzum2018zone}
Nafisa Anzum, Syeda~Farzia Afroze, and Ashikur Rahman.
\newblock Zone-based indoor localization using neural networks: A view from a
  real testbed.
\newblock In {\em Proc. IEEE International Conference on Communications (ICC)},
  pages 1--7, 2018\color{black}.

\bibitem{wang2018deepml}
Xuyu Wang, Zhitao Yu, and Shiwen Mao.
\newblock {DeepML}: Deep {LSTM} for indoor localization with smartphone
  magnetic and light sensors.
\newblock In {\em Proc. IEEE International Conference on Communications (ICC)},
  pages 1--6, 2018\color{black}.

\bibitem{kumar2016indoor}
Anil Kumar Tirumala~Ravi Kumar, Bernd Sch{\"a}ufele, Daniel Becker, Oliver
  Sawade, and Ilja Radusch.
\newblock Indoor localization of vehicles using deep learning.
\newblock In {\em Proc. 17th IEEE International Symposium on World of Wireless,
  Mobile and Multimedia Networks (WoWMoM)}, pages 1--6.

\bibitem{zhengj2016mobile}
Zejia Zhengj and Juyang Weng.
\newblock Mobile device based outdoor navigation with on-line learning neural
  network: A comparison with convolutional neural network.
\newblock In {\em Proc. IEEE Conference on Computer Vision and Pattern
  Recognition Workshops}, pages 11--18, 2016.

\bibitem{vieira2017deep}
Joao Vieira, Erik Leitinger, Muris Sarajlic, Xuhong Li, and Fredrik Tufvesson.
\newblock Deep convolutional neural networks for massive {MIMO}
  fingerprint-based positioning.
\newblock In {\em Proc. 28th IEEE Annual International Symposium on Personal,
  Indoor and Mobile Radio Communications}, 2017.

\bibitem{wang2017csi}
\textcolor{black}{Xuyu} \color{black}Wang, Lingjun Gao, Shiwen Mao, and Santosh
  Pandey.
\newblock {\color{black}{CSI}-based fingerprinting for indoor localization: A
  deep learning approach}.
\newblock {\em {\color{black}IEEE Transactions on Vehicular Technology}},
  {\color{black}66}({\color{black}1}):{\color{black}763--776},
  {\color{black}2017}\color{black}.

\bibitem{chen2017confi}
\textcolor{black}{Hao} \color{black}Chen, Yifan Zhang, Wei Li, Xiaofeng Tao,
  and Ping Zhang.
\newblock {\color{black}{ConFi}: Convolutional neural networks based indoor
  {wi-fi} localization using channel state information}.
\newblock {\em {\color{black}IEEE Access}},
  {\color{black}5}:{\color{black}18066--18074},
  {\color{black}2017}\color{black}.

\bibitem{shokry2018deeploc}
\textcolor{black}{Ahmed} \color{black}Shokry, Marwan Torki, and Moustafa
  Youssef.
\newblock {\color{black}{DeepLoc}: A ubiquitous accurate and low-overhead
  outdoor cellular localization system}.
\newblock In {\em {\color{black}Proc. 26th ACM SIGSPATIAL International
  Conference on Advances in Geographic Information Systems}},
  pages~{\color{black}339--348}, {\color{black}2018}\color{black}.

\bibitem{zhou2018device}
\textcolor{black}{Rui} \color{black}Zhou, Meng Hao, Xiang Lu, Mingjie Tang, and
  Yang Fu.
\newblock {\color{black}Device-Free Localization Based on {CSI} Fingerprints
  and Deep Neural Networks}.
\newblock In {\em {\color{black}Proc. 15th Annual IEEE International Conference
  on Sensing, Communication, and Networking (SECON)}},
  pages~{\color{black}1--9}, {\color{black}2018}\color{black}.

\bibitem{zhang2017deeppositioning}
\textcolor{black}{Wei} \color{black}Zhang, Rahul Sengupta, John Fodero, and
  Xiaolin Li.
\newblock {\color{black}{DeepPositioning}: Intelligent Fusion of Pervasive
  Magnetic Field and {WiFi} Fingerprinting for Smartphone Indoor Localization
  via Deep Learning}.
\newblock In {\em {\color{black}Proc. 16th IEEE International Conference on
  Machine Learning and Applications (ICMLA)}}, pages~{\color{black}7--13},
  {\color{black}2017}\color{black}.

\bibitem{adege2018applying}
\textcolor{black}{Abebe} \color{black}Adege, Hsin-Piao Lin, Getaneh Tarekegn,
  and Shiann-Shiun Jeng.
\newblock {\color{black}Applying Deep Neural Network {(DNN)} for Robust Indoor
  Localization in Multi-Building Environment}.
\newblock {\em {\color{black}Applied Sciences}},
  {\color{black}8}({\color{black}7}):{\color{black}1062},
  {\color{black}2018}\color{black}.

\bibitem{ibrahim2018cnn}
\textcolor{black}{Mai} \color{black}Ibrahim, Marwan Torki, and Mustafa
  ElNainay.
\newblock {\color{black}{CNN} based Indoor Localization using {RSS}
  Time-Series}.
\newblock In {\em {\color{black}Proc. IEEE Symposium on Computers and
  Communications (ISCC)}}, pages~{\color{black}1044--1049},
  {\color{black}2018}\color{black}.

\bibitem{niitsoo2018convolutional}
\textcolor{black}{Arne} \color{black}Niitsoo, Thorsten Edelh{\"a}u$\beta$er,
  and Christopher Mutschler.
\newblock {\color{black}Convolutional neural networks for position estimation
  in {TDoA}-based locating systems}.
\newblock In {\em {\color{black}Proc. IEEE International Conference on Indoor
  Positioning and Indoor Navigation (IPIN)}}, pages~{\color{black}1--8},
  {\color{black}2018}\color{black}.

\bibitem{wang2018deep6}
\textcolor{black}{Xuyu} \color{black}Wang, Xiangyu Wang, and Shiwen Mao.
\newblock {\color{black}Deep convolutional neural networks for indoor
  localization with {CSI} images}.
\newblock {\em {\color{black}IEEE Transactions on Network Science and
  Engineering}}, {\color{black}2018}\color{black}.

\bibitem{xiao20173}
\textcolor{black}{Chao} \color{black}Xiao, Daiqin Yang, Zhenzhong Chen, and
  Guang Tan.
\newblock {\color{black}{3-D BLE} indoor localization based on denoising
  autoencoder}.
\newblock {\em {\color{black}IEEE Access}},
  {\color{black}5}:{\color{black}12751--12760},
  {\color{black}2017}\color{black}.

\bibitem{Hsu:2017:ZIS:3139486.3130924}
\textcolor{black}{Chen-Yu} \color{black}Hsu, Aayush Ahuja, Shichao Yue, Rumen
  Hristov, Zachary Kabelac, and Dina Katabi.
\newblock {\color{black}Zero-Effort In-Home Sleep and Insomnia Monitoring Using
  Radio Signals}.
\newblock {\em {\color{black}Proc. ACM Interactive, Mobile, Wearable and
  Ubiquitous Technologies (IMWUT)}}, {\color{black}1}({\color{black}3}),
  {\color{black}sep} {\color{black}2017}\color{black}.

\bibitem{guan2017high}
\textcolor{black}{Weipeng} \color{black}Guan, Yuxiang Wu, Canyu Xie, Hao Chen,
  Ye~Cai, and Yingcong Chen.
\newblock {\color{black}High-precision approach to localization scheme of
  visible light communication based on artificial neural networks and modified
  genetic algorithms}.
\newblock {\em {\color{black}Optical Engineering}},
  {\color{black}56}({\color{black}10}):{\color{black}106103},
  {\color{black}2017}\color{black}.

\bibitem{chuang2014effective}
Po-Jen Chuang and Yi-Jun Jiang.
\newblock Effective neural network-based node localisation scheme for wireless
  sensor networks.
\newblock {\em IET Wireless Sensor Systems}, 4(2):97--103, 2014.

\bibitem{bernas2015fully}
Marcin Bernas and Bart{\l}omiej P{\l}aczek.
\newblock Fully connected neural networks ensemble with signal strength
  clustering for indoor localization in wireless sensor networks.
\newblock {\em International Journal of Distributed Sensor Networks},
  11(12):403242, 2015.

\bibitem{payal2015analysis}
Ashish Payal, Chandra~Shekhar Rai, and BV~Ramana Reddy.
\newblock Analysis of some feedforward artificial neural network training
  algorithms for developing localization framework in wireless sensor networks.
\newblock {\em Wireless Personal Communications}, 82(4):2519--2536, 2015.

\bibitem{dong2017range}
Yuhan Dong, Zheng Li, Rui Wang, and Kai Zhang.
\newblock Range-based localization in underwater wireless sensor networks using
  deep neural network.
\newblock In {\em IPSN}, pages 321--322, 2017.

\bibitem{yan2016real}
Xiaofei Yan, Hong Cheng, Yandong Zhao, Wenhua Yu, Huan Huang, and Xiaoliang
  Zheng.
\newblock Real-time identification of smoldering and flaming combustion phases
  in forest using a wireless sensor network-based multi-sensor system and
  artificial neural network.
\newblock {\em Sensors}, 16(8):1228, 2016.

\bibitem{wang2017temperature}
Baowei Wang, Xiaodu Gu, Li~Ma, and Shuangshuang Yan.
\newblock Temperature error correction based on {BP} neural network in
  meteorological wireless sensor network.
\newblock {\em International Journal of Sensor Networks}, 23(4):265--278, 2017.

\bibitem{lee2017deep222}
Ki-Seong Lee, Sun-Ro Lee, Youngmin Kim, and Chan-Gun Lee.
\newblock Deep learning--based real-time query processing for wireless sensor
  network.
\newblock {\em International Journal of Distributed Sensor Networks}, 13(5),
  2017.

\bibitem{li2016adaptive}
Jiakai Li and Gursel Serpen.
\newblock Adaptive and intelligent wireless sensor networks through neural
  networks: an illustration for infrastructure adaptation through hopfield
  network.
\newblock {\em Applied Intelligence}, 45(2):343--362, 2016.

\bibitem{khorasani2017energy}
Fereshteh Khorasani and Hamid~Reza Naji.
\newblock Energy efficient data aggregation in wireless sensor networks using
  neural networks.
\newblock {\em International Journal of Sensor Networks}, 24(1):26--42, 2017.

\bibitem{li2015distributed}
Chunlin Li, Xiaofu Xie, Yuejiang Huang, Hong Wang, and Changxi Niu.
\newblock Distributed data mining based on deep neural network for wireless
  sensor network.
\newblock {\em International Journal of Distributed Sensor Networks},
  11(7):157453, 2015.

\bibitem{luo2018distributed}
Tie Luo and Sai~G Nagarajany.
\newblock Distributed anomaly detection using autoencoder neural networks in
  {WSN for IoT}.
\newblock In {\em Proc. IEEE International Conference on Communications (ICC)},
  pages 1--6, 2018\color{black}.

\bibitem{kumar2019machine}
\textcolor{black}{D Praveen} \color{black}Kumar, Tarachand Amgoth, and Chandra
  Sekhara~Rao Annavarapu.
\newblock {\color{black}Machine learning algorithms for wireless sensor
  networks: A survey}.
\newblock {\em {\color{black}Information Fusion}},
  {\color{black}49}:{\color{black}1--25}, {\color{black}2019}\color{black}.

\bibitem{heydari2017reduce}
\textcolor{black}{Nasim} \color{black}Heydari and Behrouz Minaei-Bidgoli.
\newblock {\color{black}Reduce energy consumption and send secure data wireless
  multimedia sensor networks using a combination of techniques for multi-layer
  watermark and deep learning}.
\newblock {\em {\color{black}International Journal of Computer Science and
  Network Security (IJCSNS)}},
  {\color{black}17}({\color{black}2}):{\color{black}98--105},
  {\color{black}2017}\color{black}.

\bibitem{phoemphon2018hybrid}
\textcolor{black}{Songyut} \color{black}Phoemphon, Chakchai So-In, and
  Dusit~Tao Niyato.
\newblock {\color{black}A hybrid model using fuzzy logic and an extreme
  learning machine with vector particle swarm optimization for wireless sensor
  network localization}.
\newblock {\em {\color{black}Applied Soft Computing}},
  {\color{black}65}:{\color{black}101--120}, {\color{black}2018}\color{black}.

\bibitem{banihashemian2018new}
\textcolor{black}{Seyed Saber} \color{black}Banihashemian, Fazlollah Adibnia,
  and Mehdi~A Sarram.
\newblock {\color{black}A new range-free and storage-efficient localization
  algorithm using neural networks in wireless sensor networks}.
\newblock {\em {\color{black}Wireless Personal Communications}},
  {\color{black}98}({\color{black}1}):{\color{black}1547--1568},
  {\color{black}2018}\color{black}.

\bibitem{sun2017wnn}
\textcolor{black}{Wei} \color{black}Sun, Wei Lu, Qiyue Li, Liangfeng Chen,
  Daoming Mu, and Xiaojing Yuan.
\newblock {\color{black}{WNN-LQE}: Wavelet-Neural-Network-Based Link Quality
  Estimation for Smart Grid {WSNs}}.
\newblock {\em {\color{black}IEEE Access}},
  {\color{black}5}:{\color{black}12788--12797},
  {\color{black}2017}\color{black}.

\bibitem{kang2018novel}
\textcolor{black}{Jiheon} \color{black}Kang, Youn-Jong Park, Jaeho Lee,
  Soo-Hyun Wang, and Doo-Seop Eom.
\newblock {\color{black}Novel Leakage Detection by Ensemble {CNN-SVM} and
  Graph-based Localization in Water Distribution Systems}.
\newblock {\em {\color{black}IEEE Transactions on Industrial Electronics}},
  {\color{black}65}({\color{black}5}):{\color{black}4279--4289},
  {\color{black}2018}\color{black}.

\bibitem{mehmood2017eldc}
\textcolor{black}{Amjad} \color{black}Mehmood, Zhihan Lv, Jaime Lloret, and
  Muhammad~Muneer Umar.
\newblock {\color{black}{ELDC}: An artificial neural network based
  energy-efficient and robust routing scheme for pollution monitoring in WSNs}.
\newblock {\em {\color{black}IEEE Transactions on Emerging Topics in
  Computing}}, {\color{black}2017}\color{black}.

\bibitem{alsheikh2016rate}
\textcolor{black}{Mohammad Abu} \color{black}Alsheikh, Shaowei Lin, Dusit
  Niyato, and Hwee-Pink Tan.
\newblock {\color{black}Rate-distortion balanced data compression for wireless
  sensor networks}.
\newblock {\em {\color{black}IEEE Sensors Journal}},
  {\color{black}16}({\color{black}12}):{\color{black}5072--5083},
  {\color{black}2016}\color{black}.

\bibitem{el2016robust}
\textcolor{black}{Ahmad} \color{black}El Assaf, Slim Zaidi, Sofi{\`e}ne Affes,
  and Nahi Kandil.
\newblock {\color{black}Robust {ANNs-Based WSN} Localization in the Presence of
  Anisotropic Signal Attenuation}.
\newblock {\em {\color{black}IEEE Wireless Communications Letters}},
  {\color{black}5}({\color{black}5}):{\color{black}504--507},
  {\color{black}2016}\color{black}.

\bibitem{wang2017deep3}
\textcolor{black}{Yuzhi} \color{black}Wang, Anqi Yang, Xiaoming Chen, Pengjun
  Wang, Yu~Wang, and Huazhong Yang.
\newblock {\color{black}A deep learning approach for blind drift calibration of
  sensor networks}.
\newblock {\em {\color{black}IEEE Sensors Journal}},
  {\color{black}17}({\color{black}13}):{\color{black}4158--4171},
  {\color{black}2017}\color{black}.

\bibitem{jia2018continuous}
\textcolor{black}{Zhenhua} \color{black}Jia, Xinmeng Lyu, Wuyang Zhang,
  Richard~P Martin, Richard~E Howard, and Yanyong Zhang.
\newblock {\color{black}Continuous Low-Power Ammonia Monitoring Using Long
  Short-Term Memory Neural Networks}.
\newblock In {\em {\color{black}Proc. 16th ACM Conference on Embedded Networked
  Sensor Systems}}, pages~{\color{black}224--236},
  {\color{black}2018}\color{black}.

\bibitem{liu2017deep222}
Lu~Liu, Yu~Cheng, Lin Cai, Sheng Zhou, and Zhisheng Niu.
\newblock Deep learning based optimization in wireless network.
\newblock In {\em Proc. IEEE International Conference on Communications (ICC)},
  pages 1--6, 2017.

\bibitem{subramanian2016poster}
Shivashankar Subramanian and Arindam Banerjee.
\newblock Poster: Deep learning enabled {M2M} gateway for network optimization.
\newblock In {\em Proc. 14th ACM Annual International Conference on Mobile
  Systems, Applications, and Services Companion}, pages 144--144, 2016.

\bibitem{he2017optimization}
Ying He, Chengchao Liang, F~Richard Yu, Nan Zhao, and Hongxi Yin.
\newblock Optimization of cache-enabled opportunistic interference alignment
  wireless networks: A big data deep reinforcement learning approach.
\newblock In {\em Proc. 2017 IEEE International Conference on Communications
  (ICC)}, pages 1--6.

\bibitem{he2017deep3}
Ying He, Zheng Zhang, F~Richard Yu, Nan Zhao, Hongxi Yin, Victor~CM Leung, and
  Yanhua Zhang.
\newblock Deep reinforcement learning-based optimization for cache-enabled
  opportunistic interference alignment wireless networks.
\newblock {\em IEEE Transactions on Vehicular Technology}, 2017.

\bibitem{mismar2017deep}
Faris~B Mismar and Brian~L Evans.
\newblock Deep reinforcement learning for improving downlink mmwave
  communication performance.
\newblock {\em arXiv preprint arXiv:1707.02329}, 2017.

\bibitem{wang2018handover}
Zhi Wang, Lihua Li, Yue Xu, Hui Tian, and Shuguang Cui.
\newblock Handover optimization via asynchronous multi-user deep reinforcement
  learning.
\newblock In {\em Proc. IEEE International Conference on Communications (ICC)},
  pages 1--6, 2018\color{black}.

\bibitem{chen2018heterogeneous}
Ziqi Chen and David~B Smith.
\newblock Heterogeneous machine-type communications in cellular networks:
  Random access optimization by deep reinforcement learning.
\newblock In {\em Proc. IEEE International Conference on Communications (ICC)},
  pages 1--6, 2018\color{black}.

\bibitem{chen2018auto}
Li~Chen, Justinas Lingys, Kai Chen, and Feng Liu.
\newblock {AuTO}: scaling deep reinforcement learning for datacenter-scale
  automatic traffic optimization.
\newblock In {\em Proc. ACM Conference of the ACM Special Interest Group on
  Data Communication (SIGCOMM)}, pages 191--205, 2018\color{black}.

\bibitem{lee2017classification}
YangMin Lee.
\newblock Classification of node degree based on deep learning and routing
  method applied for virtual route assignment.
\newblock {\em Ad Hoc Networks}, 58:70--85, 2017.

\bibitem{tang2017removing}
Fengxiao Tang, Bomin Mao, Zubair~Md Fadlullah, Nei Kato, Osamu Akashi, Takeru
  Inoue, and Kimihiro Mizutani.
\newblock On removing routing protocol from future wireless networks: A
  real-time deep learning approach for intelligent traffic control.
\newblock {\em IEEE Wireless Communications}, 2017.

\bibitem{zhang2017energy}
Qingchen Zhang, Man Lin, Laurence~T Yang, Zhikui Chen, and Peng Li.
\newblock Energy-efficient scheduling for real-time systems based on deep
  {Q}-learning model.
\newblock {\em IEEE Transactions on Sustainable Computing}, 2017.

\bibitem{atallah2017deep}
Ribal Atallah, Chadi Assi, and Maurice Khabbaz.
\newblock Deep reinforcement learning-based scheduling for roadside
  communication networks.
\newblock In {\em Proc. 15th IEEE International Symposium on Modeling and
  Optimization in Mobile, Ad Hoc, and Wireless Networks (WiOpt)}, pages 1--8,
  2017.

\bibitem{chinchali2018}
Sandeep Chinchali, Pan Hu, Tianshu Chu, Manu Sharma, Manu Bansal, Rakesh Misra,
  Marco Pavone, and Katti Sachin.
\newblock Cellular network traffic scheduling with deep reinforcement learning.
\newblock In {\em Proc. National Conference on Artificial Intelligence (AAAI)},
  2018.

\bibitem{wei2018joint}
Yifei Wei, Zhiqiang Zhang, F~Richard Yu, and Zhu Han.
\newblock Joint user scheduling and content caching strategy for mobile edge
  networks using deep reinforcement learning.
\newblock In {\em Proc. IEEE International Conference on Communications
  Workshops (ICC Workshops)}, 2018\color{black}.

\bibitem{sun2017learning}
Haoran Sun, Xiangyi Chen, Qingjiang Shi, Mingyi Hong, Xiao Fu, and Nikos~D
  Sidiropoulos.
\newblock Learning to optimize: Training deep neural networks for wireless
  resource management.
\newblock In {\em Proc. 18th IEEE International Workshop on Signal Processing
  Advances in Wireless Communications (SPAWC)}, pages 1--6, 2017.

\bibitem{xu2017deep3}
Zhiyuan Xu, Yanzhi Wang, Jian Tang, Jing Wang, and Mustafa~Cenk Gursoy.
\newblock A deep reinforcement learning based framework for power-efficient
  resource allocation in cloud {RANs}.
\newblock In {\em Proc. 2017 IEEE International Conference on Communications
  (ICC)}, pages 1--6.

\bibitem{ferreira2017multi}
Paulo Victor~R Ferreira, Randy Paffenroth, Alexander~M Wyglinski, Timothy~M
  Hackett, Sven~G Bil{\'e}n, Richard~C Reinhart, and Dale~J Mortensen.
\newblock Multi-objective reinforcement learning-based deep neural networks for
  cognitive space communications.
\newblock In {\em Proc. Cognitive Communications for Aerospace Applications
  Workshop (CCAA)}, pages 1--8. IEEE, 2017.

\bibitem{ye2018deep12}
Hao Ye and Geoffrey~Ye Li.
\newblock Deep reinforcement learning for resource allocation in {V2V}
  communications.
\newblock In {\em Proc. IEEE International Conference on Communications (ICC)},
  pages 1--6, 2018\color{black}.

\bibitem{challita2018proactive}
Ursula Challita, Li~Dong, and Walid Saad.
\newblock Proactive resource management for {LTE} in unlicensed spectrum: A
  deep learning perspective.
\newblock {\em IEEE Transactions on Wireless Communications},
  2018\color{black}.

\bibitem{naparstek2017deep}
Oshri Naparstek and Kobi Cohen.
\newblock Deep multi-user reinforcement learning for dynamic spectrum access in
  multichannel wireless networks.
\newblock In {\em Proc. IEEE Global Communications Conference}, pages 1--7,
  2017.

\bibitem{o2016deep}
Timothy~J O'Shea and T~Charles Clancy.
\newblock Deep reinforcement learning radio control and signal detection with
  {KeRLym}, a gym {RL} agent.
\newblock {\em arXiv preprint arXiv:1605.09221}, 2016.

\bibitem{wijaya2015intercell}
Michael~Andri Wijaya, Kazuhiko Fukawa, and Hiroshi Suzuki.
\newblock Intercell-interference cancellation and neural network transmit power
  optimization for {MIMO} channels.
\newblock In {\em Proc. IEEE 82nd Vehicular Technology Conference (VTC Fall)},
  pages 1--5, 2015.

\bibitem{rutagemwa2018dynamic}
Humphrey Rutagemwa, Amir Ghasemi, and Shuo Liu.
\newblock Dynamic spectrum assignment for land mobile radio with deep recurrent
  neural networks.
\newblock In {\em Proc. IEEE International Conference on Communications
  Workshops (ICC Workshops)}, 2018\color{black}.

\bibitem{wijaya2016neural}
Michael~Andri Wijaya, Kazuhiko Fukawa, and Hiroshi Suzuki.
\newblock Neural network based transmit power control and interference
  cancellation for {MIMO} small cell networks.
\newblock {\em IEICE Transactions on Communications}, 99(5):1157--1169, 2016.

\bibitem{mao2017neural}
Hongzi Mao, Ravi Netravali, and Mohammad Alizadeh.
\newblock Neural adaptive video streaming with pensieve.
\newblock In {\em Proc. Conference of the ACM Special Interest Group on Data
  Communication (SIGCOMM)}, pages 197--210. ACM, 2017.

\bibitem{oda2017design}
Tetsuya Oda, Ryoichiro Obukata, Makoto Ikeda, Leonard Barolli, and Makoto
  Takizawa.
\newblock Design and implementation of a simulation system based on deep
  {Q}-network for mobile actor node control in wireless sensor and actor
  networks.
\newblock In {\em Proc. 31st IEEE International Conference on Advanced
  Information Networking and Applications Workshops (WAINA)}, pages 195--200,
  2017.

\bibitem{oda2017performance}
Tetsuya Oda, Donald Elmazi, Miralda Cuka, Elis Kulla, Makoto Ikeda, and Leonard
  Barolli.
\newblock Performance evaluation of a deep {Q}-network based simulation system
  for actor node mobility control in wireless sensor and actor networks
  considering three-dimensional environment.
\newblock In {\em Proc. International Conference on Intelligent Networking and
  Collaborative Systems}, pages 41--52. Springer, 2017.

\bibitem{kim2017load}
Hye-Young Kim and Jong-Min Kim.
\newblock A load balancing scheme based on deep-learning in {IoT}.
\newblock {\em Cluster Computing}, 20(1):873--878, 2017.

\bibitem{challita2018deep}
Ursula Challita, Walid Saad, and Christian Bettstetter.
\newblock Deep reinforcement learning for interference-aware path planning of
  cellular connected {UAVs}.
\newblock In {\em Proc. IEEE International Conference on Communications (ICC)},
  2018\color{black}.

\bibitem{luo2018online}
Changqing Luo, Jinlong Ji, Qianlong Wang, Lixing Yu, and Pan Li.
\newblock Online power control for {5G} wireless communications: A deep
  {Q}-network approach.
\newblock In {\em Proc. IEEE International Conference on Communications (ICC)},
  pages 1--6, 2018\color{black}.

\bibitem{yu2018deep2}
Yiding Yu, Taotao Wang, and Soung~Chang Liew.
\newblock Deep-reinforcement learning multiple access for heterogeneous
  wireless networks.
\newblock In {\em Proc. IEEE International Conference on Communications (ICC)},
  pages 1--7, 2018\color{black}.

\bibitem{xu2018experience}
Zhiyuan Xu, Jian Tang, Jingsong Meng, Weiyi Zhang, Yanzhi Wang, Chi~Harold Liu,
  and Dejun Yang.
\newblock Experience-driven networking: A deep reinforcement learning based
  approach.
\newblock In {\em Proc. IEEE International Conference on Computer
  Communications}, 2018\color{black}.

\bibitem{liu2018deepnap}
Jingchu Liu, Bhaskar Krishnamachari, Sheng Zhou, and Zhisheng Niu.
\newblock {DeepNap}: Data-driven base station sleeping operations through deep
  reinforcement learning.
\newblock {\em IEEE Internet of Things Journal}, 2018\color{black}.

\bibitem{zhao2018deep}
\textcolor{black}{Zhifeng} \color{black}Zhao, Rongpeng Li, Qi~Sun, Yangchen
  Yang, Xianfu Chen, Minjian Zhao, Honggang Zhang, et~al.
\newblock {\color{black}Deep Reinforcement Learning for Network Slicing}.
\newblock {\em {\color{black}arXiv preprint arXiv:1805.06591}},
  {\color{black}2018}\color{black}.

\bibitem{li2018deep}
\textcolor{black}{Ji} \color{black}Li, Hui Gao, Tiejun Lv, and Yueming Lu.
\newblock \color{black}{Deep reinforcement learning based computation
  offloading and resource allocation for {MEC}}.
\newblock In {\em {\color{black}Proc. IEEE Wireless Communications and
  Networking Conference (WCNC)}}, pages~{\color{black}1--6},
  {\color{black}2018}\color{black}.

\bibitem{mennes2018neural}
\textcolor{black}{Ruben} \color{black}Mennes, Miguel Camelo, Maxim Claeys, and
  Steven Latr{\'e}.
\newblock {\color{black}A neural-network-based {MF-TDMA MAC} scheduler for
  collaborative wireless networks}.
\newblock In {\em {\color{black}Proc. IEEE Wireless Communications and
  Networking Conference (WCNC)}}, pages~{\color{black}1--6}, 2018\color{black}.

\bibitem{8553651}
\textcolor{black}{Yibo} \color{black}Zhou, Zubair~Md. Fadlullah, Bomin Mao, and
  Nei Kato.
\newblock {\color{black}A Deep-Learning-Based Radio Resource Assignment
  Technique for {5G} Ultra Dense Networks}.
\newblock {\em {\color{black}IEEE Network}},
  {\color{black}32}({\color{black}6}):{\color{black}28--34},
  {\color{black}2018}\color{black}.

\bibitem{mao2017tensor}
\textcolor{black}{Bomin} \color{black}Mao, Zubair~Md Fadlullah, Fengxiao Tang,
  Nei Kato, Osamu Akashi, Takeru Inoue, and Kimihiro Mizutani.
\newblock {\color{black}A Tensor Based Deep Learning Technique for Intelligent
  Packet Routing}.
\newblock In {\em {\color{black}Proc. IEEE Global Communications Conference}},
  pages~{\color{black}1--6}. {\color{black}IEEE},
  {\color{black}2017}\color{black}.

\bibitem{geyer2018learning}
\textcolor{black}{Fabien} \color{black}Geyer and Georg Carle.
\newblock {\color{black}Learning and Generating Distributed Routing Protocols
  Using Graph-Based Deep Learning}.
\newblock In {\em {\color{black}Proc. ACM Workshop on Big Data Analytics and
  Machine Learning for Data Communication Networks}},
  pages~{\color{black}40--45}, {\color{black}2018}\color{black}.

\bibitem{luong2018joint}
\textcolor{black}{Nguyen Cong} \color{black}Luong, Tran~The Anh, Huynh
  Thi~Thanh Binh, Dusit Niyato, Dong~In Kim, and Ying-Chang Liang.
\newblock {\color{black}Joint Transaction Transmission and Channel Selection in
  Cognitive Radio Based Blockchain Networks: A Deep Reinforcement Learning
  Approach}.
\newblock {\em {\color{black}arXiv preprint arXiv:1810.10139}},
  {\color{black}2018}\color{black}.

\bibitem{li2017intelligent2}
\textcolor{black}{Xingjian} \color{black}Li, Jun Fang, Wen Cheng, Huiping Duan,
  Zhi Chen, and Hongbin Li.
\newblock {\color{black}Intelligent Power Control for Spectrum Sharing in
  Cognitive Radios: A Deep Reinforcement Learning Approach}.
\newblock {\em {\color{black}IEEE access}}, {\color{black}2018}\color{black}.

\bibitem{lee2018deep}
\textcolor{black}{Woongsup} \color{black}Lee, Minhoe Kim, and Dong-Ho Cho.
\newblock {\color{black}Deep Learning Based Transmit Power Control in Underlaid
  Device-to-Device Communication}.
\newblock {\em {\color{black}IEEE Systems Journal}},
  {\color{black}2018}\color{black}.

\bibitem{liu2018energy}
\textcolor{black}{Chi Harold} \color{black}Liu, Zheyu Chen, Jian Tang, Jie Xu,
  and Chengzhe Piao.
\newblock {\color{black}Energy-efficient {UAV} control for effective and fair
  communication coverage: A deep reinforcement learning approach}.
\newblock {\em {\color{black}IEEE Journal on Selected Areas in
  Communications}},
  {\color{black}36}({\color{black}9}):{\color{black}2059--2070},
  {\color{black}2018}\color{black}.

\bibitem{he2017software}
\textcolor{black}{Ying} \color{black}He, F~Richard Yu, Nan Zhao, Victor~CM
  Leung, and Hongxi Yin.
\newblock {\color{black}Software-defined networks with mobile edge computing
  and caching for smart cities: A big data deep reinforcement learning
  approach}.
\newblock {\em {\color{black}IEEE Communications Magazine}},
  {\color{black}55}({\color{black}12}):{\color{black}31--37},
  {\color{black}2017}\color{black}.

\bibitem{liu2018anti}
\textcolor{black}{Xin} \color{black}Liu, Yuhua Xu, Luliang Jia, Qihui Wu, and
  Alagan Anpalagan.
\newblock {\color{black}Anti-jamming Communications Using Spectrum Waterfall: A
  Deep Reinforcement Learning Approach}.
\newblock {\em {\color{black}IEEE Communications Letters}},
  {\color{black}22}({\color{black}5}):{\color{black}998--1001},
  {\color{black}2018}\color{black}.

\bibitem{pham2018deep}
\textcolor{black}{Quang Tran Anh} \color{black}Pham, Yassine Hadjadj-Aoul, and
  Abdelkader Outtagarts.
\newblock {\color{black}Deep Reinforcement Learning based {QoS}-aware Routing
  in Knowledge-defined networking}.
\newblock In {\em {\color{black}Proc. Qshine EAI International Conference on
  Heterogeneous Networking for Quality, Reliability, Security and Robustness}},
  pages~{\color{black}1--13}, {\color{black}2018}\color{black}.

\bibitem{ferreira2016multi}
\textcolor{black}{Paulo} \color{black}Ferreira, Randy Paffenroth, Alexander
  Wyglinski, Timothy~M Hackett, Sven Bil{\'e}n, Richard Reinhart, and Dale
  Mortensen.
\newblock Multi-objective reinforcement learning for cognitive radio--based
  satellite communications.
\newblock In {\em Proc. 34th AIAA International Communications Satellite
  Systems Conference}, page 5726, 2016\color{black}.

\bibitem{yousefi2017autoencoder}
Mahmood Yousefi-Azar, Vijay Varadharajan, Len Hamey, and Uday Tupakula.
\newblock Autoencoder-based feature learning for cyber security applications.
\newblock In {\em Proc. IEEE International Joint Conference on Neural Networks
  (IJCNN)}, pages 3854--3861, 2017.

\bibitem{aminanto2016detecting}
Muhamad~Erza Aminanto and Kwangjo Kim.
\newblock Detecting impersonation attack in {WiFi} networks using deep learning
  approach.
\newblock In {\em Proc. International Workshop on Information Security
  Applications}, pages 136--147. Springer, 2016.

\bibitem{feng2016anomaly}
Qingsong Feng, Zheng Dou, Chunmei Li, and Guangzhen Si.
\newblock Anomaly detection of spectrum in wireless communication via deep
  autoencoder.
\newblock In {\em Proc. International Conference on Computer Science and its
  Applications}, pages 259--265. Springer, 2016.

\bibitem{khan2016distributed}
Muhammad~Altaf Khan, Shafiullah Khan, Bilal Shams, and Jaime Lloret.
\newblock Distributed flood attack detection mechanism using artificial neural
  network in wireless mesh networks.
\newblock {\em Security and Communication Networks}, 9(15):2715--2729, 2016.

\bibitem{diro2017distributed}
Abebe~Abeshu Diro and Naveen Chilamkurti.
\newblock Distributed attack detection scheme using deep learning approach for
  {Internet of Things}.
\newblock {\em Future Generation Computer Systems}, 2017.

\bibitem{saied2016detection}
Alan Saied, Richard~E Overill, and Tomasz Radzik.
\newblock Detection of known and unknown {DDoS} attacks using artificial neural
  networks.
\newblock {\em Neurocomputing}, 172:385--393, 2016.

\bibitem{lopez2017conditional}
Manuel Lopez-Martin, Belen Carro, Antonio Sanchez-Esguevillas, and Jaime
  Lloret.
\newblock Conditional variational autoencoder for prediction and feature
  recovery applied to intrusion detection in {IoT}.
\newblock {\em Sensors}, 17(9):1967, 2017.

\bibitem{hamedani2018reservoir}
Kian Hamedani, Lingjia Liu, Rachad Atat, Jinsong Wu, and Yang Yi.
\newblock Reservoir computing meets smart grids: attack detection using delayed
  feedback networks.
\newblock {\em IEEE Transactions on Industrial Informatics}, 14(2):734--743,
  2018\color{black}.

\bibitem{das2018deep}
Rajshekhar Das, Akshay Gadre, Shanghang Zhang, Swarun Kumar, and Jose~MF Moura.
\newblock A deep learning approach to {IoT} authentication.
\newblock In {\em Proc. IEEE International Conference on Communications (ICC)},
  pages 1--6, 2018\color{black}.

\bibitem{jiang2018virtual}
Peng Jiang, Hongyi Wu, Cong Wang, and Chunsheng Xin.
\newblock Virtual {MAC} spoofing detection through deep learning.
\newblock In {\em Proc. IEEE International Conference on Communications (ICC)},
  pages 1--6, 2018\color{black}.

\bibitem{yuan2014droid}
Zhenlong Yuan, Yongqiang Lu, Zhaoguo Wang, and Yibo Xue.
\newblock {Droid-Sec}: deep learning in {Android} malware detection.
\newblock In {\em ACM SIGCOMM Computer Communication Review}, volume~44, pages
  371--372, 2014.

\bibitem{yuan2016droiddetector}
Zhenlong Yuan, Yongqiang Lu, and Yibo Xue.
\newblock Droiddetector: {Android} malware characterization and detection using
  deep learning.
\newblock {\em Tsinghua Science and Technology}, 21(1):114--123, 2016.

\bibitem{su2016deep}
Xin Su, Dafang Zhang, Wenjia Li, and Kai Zhao.
\newblock A deep learning approach to {Android} malware feature learning and
  detection.
\newblock In {\em IEEE Trustcom/BigDataSE/ISPA}, pages 244--251, 2016.

\bibitem{hou2016deep4maldroid}
Shifu Hou, Aaron Saas, Lifei Chen, and Yanfang Ye.
\newblock {Deep4MalDroid}: A deep learning framework for {Android} malware
  detection based on linux kernel system call graphs.
\newblock In {\em Proc. IEEE/WIC/ACM International Conference on Web
  Intelligence Workshops (WIW)}, pages 104--111, 2016.

\bibitem{martinelli2017evaluating}
Fabio Martinelli, Fiammetta Marulli, and Francesco Mercaldo.
\newblock Evaluating convolutional neural network for effective mobile malware
  detection.
\newblock {\em Procedia Computer Science}, 112:2372--2381, 2017.

\bibitem{nguyen2018cyberattack}
\textcolor{black}{Khoi Khac} \color{black}Nguyen, Dinh~Thai Hoang, Dusit
  Niyato, Ping Wang, Diep Nguyen, and Eryk Dutkiewicz.
\newblock {\color{black}Cyberattack detection in mobile cloud computing: A deep
  learning approach}.
\newblock In {\em {\color{black}Proc. IEEE Wireless Communications and
  Networking Conference (WCNC)}}, pages~{\color{black}1--6},
  {\color{black}2018}\color{black}.

\bibitem{mclaughlin2017deep}
Niall McLaughlin, Jesus Martinez~del Rincon, BooJoong Kang, Suleiman Yerima,
  Paul Miller, Sakir Sezer, Yeganeh Safaei, Erik Trickel, Ziming Zhao, Adam
  Doup{\'e}, et~al.
\newblock Deep android malware detection.
\newblock In {\em Proc. Seventh ACM on Conference on Data and Application
  Security and Privacy}, pages 301--308, 2017.

\bibitem{chen2017deep123}
Yuanfang Chen, Yan Zhang, and Sabita Maharjan.
\newblock Deep learning for secure mobile edge computing.
\newblock {\em arXiv preprint arXiv:1709.08025}, 2017.

\bibitem{oulehla2016detection}
Milan Oulehla, Zuzana~Kom{\'\i}nkov{\'a} Oplatkov{\'a}, and David Malanik.
\newblock Detection of mobile botnets using neural networks.
\newblock In {\em Proc. IEEE Future Technologies Conference (FTC)}, pages
  1324--1326, 2016.

\bibitem{torres2016analysis}
Pablo Torres, Carlos Catania, Sebastian Garcia, and Carlos~Garcia Garino.
\newblock An analysis of recurrent neural networks for botnet detection
  behavior.
\newblock In {\em Proc. IEEE Biennial Congress of Argentina (ARGENCON)}, pages
  1--6, 2016.

\bibitem{eslahi2016mobile}
Meisam Eslahi, Moslem Yousefi, Maryam~Var Naseri, YM~Yussof, NM~Tahir, and
  H~Hashim.
\newblock Mobile botnet detection model based on retrospective pattern
  recognition.
\newblock {\em International Journal of Security and Its Applications},
  10(9):39--+, 2016.

\bibitem{alauthaman2016p2p}
Mohammad Alauthaman, Nauman Aslam, Li~Zhang, Rafe Alasem, and MA~Hossain.
\newblock A p2p botnet detection scheme based on decision tree and adaptive
  multilayer neural networks.
\newblock {\em Neural Computing and Applications}, pages 1--14, 2016.

\bibitem{shokri2015privacy}
Reza Shokri and Vitaly Shmatikov.
\newblock Privacy-preserving deep learning.
\newblock In {\em Proc. 22nd ACM SIGSAC conference on computer and
  communications security}, pages 1310--1321, 2015.

\bibitem{aono2017privacy}
Yoshinori Aono, Takuya Hayashi, Lihua Wang, Shiho Moriai, et~al.
\newblock Privacy-preserving deep learning: Revisited and enhanced.
\newblock In {\em Proc. International Conference on Applications and Techniques
  in Information Security}, pages 100--110. Springer, 2017.

\bibitem{ossia2017hybrid}
Seyed~Ali Ossia, Ali~Shahin Shamsabadi, Ali Taheri, Hamid~R Rabiee, Nic Lane,
  and Hamed Haddadi.
\newblock A hybrid deep learning architecture for privacy-preserving mobile
  analytics.
\newblock {\em arXiv preprint arXiv:1703.02952}, 2017.

\bibitem{abadi2016deep}
Mart{\'\i}n Abadi, Andy Chu, Ian Goodfellow, H~Brendan McMahan, Ilya Mironov,
  Kunal Talwar, and Li~Zhang.
\newblock Deep learning with differential privacy.
\newblock In {\em Proc. ACM SIGSAC Conference on Computer and Communications
  Security}, pages 308--318. ACM, 2016.

\bibitem{osia2017privacy}
Seyed~Ali Osia, Ali~Shahin Shamsabadi, Ali Taheri, Kleomenis Katevas, Hamed
  Haddadi, and Hamid~R Rabiee.
\newblock Private and scalable personal data analytics using a hybrid
  edge-cloud deep learning.
\newblock {\em IEEE Computer Magazine Special Issue on Mobile and Embedded Deep
  Learning}, 2018.

\bibitem{servia2017personal}
Sandra Servia-Rodriguez, Liang Wang, Jianxin~R Zhao, Richard Mortier, and Hamed
  Haddadi.
\newblock Personal model training under privacy constraints.
\newblock In {\em Proc. 3rd ACM/IEEE International Conference on
  Internet-of-Things Design and Implementation}, Apr 2018.

\bibitem{hitaj2017deep}
Briland Hitaj, Giuseppe Ateniese, and Fernando Perez-Cruz.
\newblock Deep models under the {GAN}: information leakage from collaborative
  deep learning.
\newblock In {\em Proc. ACM SIGSAC Conference on Computer and Communications
  Security}, pages 603--618, 2017.

\bibitem{greydanus2017learning}
Sam Greydanus.
\newblock Learning the enigma with recurrent neural networks.
\newblock {\em arXiv preprint arXiv:1708.07576}, 2017.

\bibitem{maghrebi2016breaking}
Houssem Maghrebi, Thibault Portigliatti, and Emmanuel Prouff.
\newblock Breaking cryptographic implementations using deep learning
  techniques.
\newblock In {\em Proc. International Conference on Security, Privacy, and
  Applied Cryptography Engineering}, pages 3--26. Springer, 2016.

\bibitem{liu2018genpass}
Yunyu Liu, Zhiyang Xia, Ping Yi, Yao Yao, Tiantian Xie, Wei Wang, and Ting Zhu.
\newblock G{ENPass}: A general deep learning model for password guessing with
  {PCFG} rules and adversarial generation.
\newblock In {\em Proc. IEEE International Conference on Communications (ICC)},
  pages 1--6, 2018\color{black}.

\bibitem{ning2018deepmag}
\textcolor{black}{Rui} \color{black}Ning, Cong Wang, ChunSheng Xin, Jiang Li,
  and Hongyi Wu.
\newblock Deepmag: sniffing mobile apps in magnetic field through deep
  convolutional neural networks.
\newblock In {\em Proc. IEEE International Conference on Pervasive Computing
  and Communications (PerCom)}, pages 1--10, 2018\color{black}.

\bibitem{o2017deep}
Timothy~J O'Shea, Tugba Erpek, and T~Charles Clancy.
\newblock Deep learning based {MIMO} communications.
\newblock {\em arXiv preprint arXiv:1707.07980}, 2017.

\bibitem{borgerding2017amp}
Mark Borgerding, Philip Schniter, and Sundeep Rangan.
\newblock {AMP}-inspired deep networks for sparse linear inverse problems.
\newblock {\em IEEE Transactions on Signal Processing}, 2017.

\bibitem{fujihashi2018nonlinear}
Takuya Fujihashi, Toshiaki Koike-Akino, Takashi Watanabe, and Philip~V Orlik.
\newblock Nonlinear equalization with deep learning for multi-purpose visual
  {MIMO} communications.
\newblock In {\em Proc. IEEE International Conference on Communications (ICC)},
  pages 1--6, 2018\color{black}.

\bibitem{rajendran2017distributed}
Sreeraj Rajendran, Wannes Meert, Domenico Giustiniano, Vincent Lenders, and
  Sofie Pollin.
\newblock Deep learning models for wireless signal classification with
  distributed low-cost spectrum sensors.
\newblock {\em IEEE Transactions on Cognitive Communications and Networking},
  2018.

\bibitem{west2017deep}
Nathan~E West and Tim O'Shea.
\newblock Deep architectures for modulation recognition.
\newblock In {\em Proc. IEEE International Symposium on Dynamic Spectrum Access
  Networks (DySPAN)}, pages 1--6, 2017.

\bibitem{o2016radio}
Timothy~J O'Shea, Latha Pemula, Dhruv Batra, and T~Charles Clancy.
\newblock Radio transformer networks: Attention models for learning to
  synchronize in wireless systems.
\newblock In {\em Proc. 50th Asilomar Conference on Signals, Systems and
  Computers}, pages 662--666, 2016.

\bibitem{gante2018beamformed}
\textcolor{black}{Jo{\~a}o} \color{black}Gante, Gabriel Falc{\~a}o, and Leonel
  Sousa.
\newblock {\color{black}Beamformed Fingerprint Learning for Accurate Millimeter
  Wave Positioning}.
\newblock {\em {\color{black}arXiv preprint arXiv:1804.04112}},
  {\color{black}2018}\color{black}.

\bibitem{alkhateeb2018deep}
\textcolor{black}{Ahmed} \color{black}Alkhateeb, Sam Alex, Paul Varkey, Ying
  Li, Qi~Qu, and Djordje Tujkovic.
\newblock {\color{black}Deep Learning Coordinated Beamforming for Highly-Mobile
  Millimeter Wave Systems}.
\newblock {\em {\color{black}IEEE Access}},
  {\color{black}6}:{\color{black}37328--37348},
  {\color{black}2018}\color{black}.

\bibitem{neumann2017deep}
David Neumann, Wolfgang Utschick, and Thomas Wiese.
\newblock Deep channel estimation.
\newblock In {\em Proc. 21th International ITG Workshop on Smart Antennas},
  pages 1--6. VDE, 2017.

\bibitem{samuel2017deep}
Neev Samuel, Tzvi Diskin, and Ami Wiesel.
\newblock Deep {MIMO} detection.
\newblock {\em arXiv preprint arXiv:1706.01151}, 2017.

\bibitem{yan2017signal}
Xin Yan, Fei Long, Jingshuai Wang, Na~Fu, Weihua Ou, and Bin Liu.
\newblock Signal detection of {MIMO-OFDM} system based on auto encoder and
  extreme learning machine.
\newblock In {\em Proc. IEEE International Joint Conference on Neural Networks
  (IJCNN)}, pages 1602--1606, 2017.

\bibitem{timothy2017introduction}
Timothy O'Shea and Jakob Hoydis.
\newblock An introduction to deep learning for the physical layer.
\newblock {\em IEEE Transactions on Cognitive Communications and Networking},
  3(4):563--575, 2017.

\bibitem{jagannath2018artificial}
Jithin Jagannath, Nicholas Polosky, Daniel O'Connor, Lakshmi~N Theagarajan,
  Brendan Sheaffer, Svetlana Foulke, and Pramod~K Varshney.
\newblock Artificial neural network based automatic modulation classification
  over a software defined radio testbed.
\newblock In {\em Proc. IEEE International Conference on Communications (ICC)},
  pages 1--6, 2018\color{black}.

\bibitem{o2016end}
Timothy~J O'Shea, Seth Hitefield, and Johnathan Corgan.
\newblock End-to-end radio traffic sequence recognition with recurrent neural
  networks.
\newblock In {\em Proc. IEEE Global Conference on Signal and Information
  Processing (GlobalSIP)}, pages 277--281, 2016.

\bibitem{o2016learning}
Timothy~J O'Shea, Kiran Karra, and T~Charles Clancy.
\newblock Learning to communicate: Channel auto-encoders, domain specific
  regularizers, and attention.
\newblock In {\em Proc. IEEE International Symposium on Signal Processing and
  Information Technology (ISSPIT)}, pages 223--228, 2016.

\bibitem{ye2018power}
Hao Ye, Geoffrey~Ye Li, and Biing-Hwang Juang.
\newblock Power of deep learning for channel estimation and signal detection in
  {OFDM} systems.
\newblock {\em IEEE Wireless Communications Letters}, 7(1):114--117,
  2018\color{black}.

\bibitem{liang2018exploiting}
Fei Liang, Cong Shen, and Feng Wu.
\newblock Exploiting noise correlation for channel decoding with convolutional
  neural networks.

\bibitem{lyu2018performance}
Wei Lyu, Zhaoyang Zhang, Chunxu Jiao, Kangjian Qin, and Huazi Zhang.
\newblock Performance evaluation of channel decoding with deep neural networks.
\newblock In {\em Proc. IEEE International Conference on Communications (ICC)},
  pages 1--6, 2018\color{black}.

\bibitem{dorner2017deep}
Sebastian D{\"o}rner, Sebastian Cammerer, Jakob Hoydis, and Stephan ten Brink.
\newblock Deep learning based communication over the air.
\newblock {\em IEEE Journal of Selected Topics in Signal Processing},
  12(1):132--143, 2018.

\bibitem{liao2018rayleigh}
\textcolor{black}{Run-Fa} \color{black}Liao, Hong Wen, Jinsong Wu, Huanhuan
  Song, Fei Pan, and Lian Dong.
\newblock {\color{black}The Rayleigh Fading Channel Prediction via Deep
  Learning}.
\newblock {\em {\color{black}Wireless Communications and Mobile Computing}},
  {\color{black}2018}\color{black}.

\bibitem{huang2018deep}
\textcolor{black}{Huang} \color{black}Hongji, Yang Jie, Song Yiwei, Huang Hao,
  and Gui Guan.
\newblock {\color{black}Deep Learning for Super-Resolution Channel Estimation
  and {DOA} Estimation based Massive {MIMO} System}.
\newblock {\em {\color{black}IEEE Transactions on Vehicular Technology}},
  {\color{black}2018}\color{black}.

\bibitem{huang2018fully}
\textcolor{black}{Sihao} \color{black}Huang and Haowen Lin.
\newblock {\color{black}{Fully optical spacecraft communications: implementing
  an omnidirectional PV-cell receiver and 8 Mb/s LED visible light downlink
  with deep learning error correction}}.
\newblock {\em {\color{black}IEEE Aerospace and Electronic Systems Magazine}},
  {\color{black}33}({\color{black}4}):{\color{black}16--22},
  {\color{black}2018}\color{black}.

\bibitem{gonzalez2017network}
Roberto Gonzalez, Alberto Garcia-Duran, Filipe Manco, Mathias Niepert, and
  Pelayo Vallina.
\newblock Network data monetization using {Net2Vec}.
\newblock In {\em Proc. ACM SIGCOMM Posters and Demos}, pages 37--39, 2017.

\bibitem{kaminski2017neural}
Nichoas Kaminski, Irene Macaluso, Emanuele Di~Pascale, Avishek Nag, John Brady,
  Mark Kelly, Keith Nolan, Wael Guibene, and Linda Doyle.
\newblock A neural-network-based realization of in-network computation for the
  {Internet of Things}.
\newblock In {\em Proc. IEEE International Conference on Communications (ICC)},
  pages 1--6, 2017.

\bibitem{xiao2017secure}
Liang Xiao, Yanda Li, Guoan Han, Huaiyu Dai, and H~Vincent Poor.
\newblock A secure mobile crowdsensing game with deep reinforcement learning.
\newblock {\em IEEE Transactions on Information Forensics and Security}, 2017.

\bibitem{luong2018optimal}
Nguyen~Cong Luong, Zehui Xiong, Ping Wang, and Dusit Niyato.
\newblock Optimal auction for edge computing resource management in mobile
  blockchain networks: A deep learning approach.
\newblock In {\em Proc. IEEE International Conference on Communications (ICC)},
  pages 1--6, 2018\color{black}.

\bibitem{gulati2018deep}
Amuleen Gulati, Gagangeet~Singh Aujla, Rajat Chaudhary, Neeraj Kumar, and
  Mohammad~S Obaidat.
\newblock Deep learning-based content centric data dissemination scheme for
  {Internet of Vehicles}.
\newblock In {\em Proc. IEEE International Conference on Communications (ICC)},
  pages 1--6, 2018\color{black}.

\bibitem{ahmed2018recent}
\textcolor{black}{Ejaz} \color{black}Ahmed, Ibrar Yaqoob, Ibrahim Abaker~Targio
  Hashem, Junaid Shuja, Muhammad Imran, Nadra Guizani, and Sheikh~Tahir Bakhsh.
\newblock {\color{black}Recent Advances and Challenges in Mobile Big Data}.
\newblock {\em {\color{black}IEEE Communications Magazine}},
  {\color{black}56}({\color{black}2}):{\color{black}102--108},
  {\color{black}2018}\color{black}.

\bibitem{yazti2014mobile}
Demetrios~Zeinalipour Yazti and Shonali Krishnaswamy.
\newblock Mobile big data analytics: research, practice, and opportunities.
\newblock In {\em Proc. 15th IEEE International Conference on Mobile Data
  Management (MDM)}, volume~1, pages 1--2, 2014.

\bibitem{naboulsi2016large}
Diala Naboulsi, Marco Fiore, Stephane Ribot, and Razvan Stanica.
\newblock Large-scale mobile traffic analysis: a survey.
\newblock {\em IEEE Communications Surveys \& Tutorials}, 18(1):124--161, 2016.

\bibitem{ngiam2011multimodal}
Jiquan Ngiam, Aditya Khosla, Mingyu Kim, Juhan Nam, Honglak Lee, and Andrew~Y
  Ng.
\newblock Multimodal deep learning.
\newblock In {\em Proc. 28th international conference on machine learning
  (ICML)}, pages 689--696, 2011.

\bibitem{alawe2018improving}
\textcolor{black}{Imad} \color{black}Alawe, Adlen Ksentini, Yassine
  Hadjadj-Aoul, and Philippe Bertin.
\newblock {\color{black}Improving traffic forecasting for {5G} core network
  scalability: A Machine Learning approach}.
\newblock {\em {\color{black}IEEE Network}},
  {\color{black}32}({\color{black}6}):{\color{black}42--49},
  {\color{black}2018}\color{black}.

\bibitem{aceto2018mobile}
Giuseppe Aceto, Domenico Ciuonzo, Antonio Montieri, and Antonio Pescap{\'e}.
\newblock Mobile encrypted traffic classification using deep learning.
\newblock In {\em Proc. 2nd IEEE Network Traffic Measurement and Analysis
  Conference}, 2018\color{black}.

\bibitem{al2015internet}
Ala Al-Fuqaha, Mohsen Guizani, Mehdi Mohammadi, Mohammed Aledhari, and Moussa
  Ayyash.
\newblock Internet of things: A survey on enabling technologies, protocols, and
  applications.
\newblock {\em IEEE Communications Surveys \& Tutorials}, 17(4):2347--2376,
  2015.

\bibitem{seneviratne2017survey}
Suranga Seneviratne, Yining Hu, Tham Nguyen, Guohao Lan, Sara Khalifa, Kanchana
  Thilakarathna, Mahbub Hassan, and Aruna Seneviratne.
\newblock A survey of wearable devices and challenges.
\newblock {\em IEEE Communications Surveys \& Tutorials}, 19(4):2573--2620,
  2017.

\bibitem{li2018learning}
He~Li, Kaoru Ota, and Mianxiong Dong.
\newblock Learning {IoT} in edge: Deep learning for the {Internet of Things}
  with edge computing.
\newblock {\em IEEE Network}, 32(1):96--101, 2018.

\bibitem{lane2015early}
Nicholas~D Lane, Sourav Bhattacharya, Petko Georgiev, Claudio Forlivesi, and
  Fahim Kawsar.
\newblock An early resource characterization of deep learning on wearables,
  smartphones and internet-of-things devices.
\newblock In {\em Proc. ACM International Workshop on Internet of Things
  towards Applications}, pages 7--12, 2015.

\bibitem{ravi2017deep}
Daniele Rav{\`\i}, Charence Wong, Fani Deligianni, Melissa Berthelot, Javier
  Andreu-Perez, Benny Lo, and Guang-Zhong Yang.
\newblock Deep learning for health informatics.
\newblock {\em IEEE journal of biomedical and health informatics}, 21(1):4--21,
  2017.

\bibitem{miotto2017deep}
Riccardo Miotto, Fei Wang, Shuang Wang, Xiaoqian Jiang, and Joel~T Dudley.
\newblock Deep learning for healthcare: review, opportunities and challenges.
\newblock {\em Briefings in Bioinformatics}, 2017.

\bibitem{ronao2016human}
Charissa~Ann Ronao and Sung-Bae Cho.
\newblock Human activity recognition with smartphone sensors using deep
  learning neural networks.
\newblock {\em Expert Systems with Applications}, 59:235--244, 2016.

\bibitem{wang2017deep}
Jindong Wang, Yiqiang Chen, Shuji Hao, Xiaohui Peng, and Lisha Hu.
\newblock Deep learning for sensor-based activity recognition: A survey.
\newblock {\em Pattern Recognition Letters}, 2018.

\bibitem{ran2017delivering}
Xukan Ran, Haoliang Chen, Zhenming Liu, and Jiasi Chen.
\newblock Delivering deep learning to mobile devices via offloading.
\newblock In {\em Proc. ACM Workshop on Virtual Reality and Augmented Reality
  Network}, pages 42--47, 2017.

\bibitem{vyas2017survey}
Vishakha~V Vyas, KH~Walse, and RV~Dharaskar.
\newblock A survey on human activity recognition using smartphone.
\newblock {\em International Journal}, 5(3), 2017.

\bibitem{zen2014deep}
Heiga Zen and Andrew Senior.
\newblock Deep mixture density networks for acoustic modeling in statistical
  parametric speech synthesis.
\newblock In {\em Proc. IEEE International Conference on Acoustics, Speech and
  Signal Processing (ICASSP)}, pages 3844--3848, 2014.

\bibitem{zhao2016urban}
Kai Zhao, Sasu Tarkoma, Siyuan Liu, and Huy Vo.
\newblock Urban human mobility data mining: An overview.
\newblock In {\em Proc. IEEE International Conference on Big Data (Big Data)},
  pages 1911--1920, 2016.

\bibitem{yang2017neural2}
Cheng Yang, Maosong Sun, Wayne~Xin Zhao, Zhiyuan Liu, and Edward~Y Chang.
\newblock A neural network approach to jointly modeling social networks and
  mobile trajectories.
\newblock {\em ACM Transactions on Information Systems (TOIS)}, 35(4):36, 2017.

\bibitem{graves2014neural}
\textcolor{black}{Alex} \color{black}Graves, Greg Wayne, and Ivo Danihelka.
\newblock {\color{black}Neural turing machines}.
\newblock {\em {\color{black}arXiv preprint arXiv:1410.5401}},
  {\color{black}2014}\color{black}.

\bibitem{xia2017indoor}
Shixiong Xia, Yi~Liu, Guan Yuan, Mingjun Zhu, and Zhaohui Wang.
\newblock Indoor fingerprint positioning based on {Wi-Fi}: An overview.
\newblock {\em ISPRS International Journal of Geo-Information}, 6(5):135, 2017.

\bibitem{davidson2016survey}
Pavel Davidson and Robert Pich{\'e}.
\newblock A survey of selected indoor positioning methods for smartphones.
\newblock {\em IEEE Communications Surveys \& Tutorials}, 19(2):1347--1370,
  2017.

\bibitem{xiao2016survey}
\textcolor{black}{Jiang} \color{black}Xiao, Zimu Zhou, Youwen Yi, and Lionel~M
  Ni.
\newblock A survey on wireless indoor localization from the device perspective.
\newblock {\em ACM Computing Surveys (CSUR)}, 49(2):25, 2016\color{black}.

\bibitem{xiao2012fifs}
Jiang Xiao, Kaishun Wu, Youwen Yi, and Lionel~M Ni.
\newblock {FIFS}: Fine-grained indoor fingerprinting system.
\newblock In {\em Proc. 21st International Conference on Computer
  Communications and Networks (ICCCN)}, pages 1--7, 2012.

\bibitem{youssef2005horus}
Moustafa Youssef and Ashok Agrawala.
\newblock The {H}orus {WLAN} location determination system.
\newblock In {\em Proc. 3rd ACM international conference on Mobile systems,
  applications, and services}, pages 205--218, 2005.

\bibitem{brunato2005statistical}
Mauro Brunato and Roberto Battiti.
\newblock Statistical learning theory for location fingerprinting in wireless
  {LANs}.
\newblock {\em Computer Networks}, 47(6):825--845, 2005.

\bibitem{ho2016generative}
Jonathan Ho and Stefano Ermon.
\newblock Generative adversarial imitation learning.
\newblock In {\em Advances in Neural Information Processing Systems}, pages
  4565--4573, 2016.

\bibitem{zorzi2016cobanets}
Michele Zorzi, Andrea Zanella, Alberto Testolin, Michele De~Filippo De~Grazia,
  and Marco Zorzi.
\newblock {COBANETS}: A new paradigm for cognitive communications systems.
\newblock In {\em Proc. IEEE International Conference on Computing, Networking
  and Communications (ICNC)}, pages 1--7, 2016.

\bibitem{roopaei2017deep}
Mehdi Roopaei, Paul Rad, and Mo~Jamshidi.
\newblock Deep learning control for complex and large scale cloud systems.
\newblock {\em Intelligent Automation \& Soft Computing}, pages 1--3, 2017.

\bibitem{ferreira2018multi}
\textcolor{black}{Paulo Victor R} \color{black}Ferreira, Randy Paffenroth,
  Alexander~M Wyglinski, Timothy~M Hackett, Sven~G Bil{\'e}n, Richard~C
  Reinhart, and Dale~J Mortensen.
\newblock {\color{black}Multi-objective Reinforcement Learning for Cognitive
  Satellite Communications using Deep Neural Network Ensembles}.
\newblock {\em {\color{black}IEEE Journal on Selected Areas in
  Communications}}, {\color{black}2018}\color{black}.

\bibitem{schaul2015prioritized}
Tom Schaul, John Quan, Ioannis Antonoglou, and David Silver.
\newblock Prioritized experience replay.
\newblock In {\em Proc. International Conference on Learning Representations
  (ICLR)}, 2016.

\bibitem{shi2011iteratively}
Qingjiang Shi, Meisam Razaviyayn, Zhi-Quan Luo, and Chen He.
\newblock An iteratively weighted {MMSE} approach to distributed sum-utility
  maximization for a {MIMO} interfering broadcast channel.
\newblock {\em IEEE Transactions on Signal Processing}, 59(9):4331--4340, 2011.

\bibitem{buczak2016survey}
Anna~L Buczak and Erhan Guven.
\newblock A survey of data mining and machine learning methods for cyber
  security intrusion detection.
\newblock {\em IEEE Communications Surveys \& Tutorials}, 18(2):1153--1176,
  2016.

\bibitem{wang2018not}
\textcolor{black}{Ji} \color{black}Wang, Jianguo Zhang, Weidong Bao, Xiaomin
  Zhu, Bokai Cao, and Philip~S Yu.
\newblock {\color{black}Not just privacy: Improving performance of private deep
  learning in mobile cloud}.
\newblock In {\em {\color{black}Proc. 24th ACM SIGKDD International Conference
  on Knowledge Discovery \& Data Mining}}, pages~{\color{black}2407--2416},
  {\color{black}2018}\color{black}.

\bibitem{kwon2017survey}
Donghwoon Kwon, Hyunjoo Kim, Jinoh Kim, Sang~C Suh, Ikkyun Kim, and Kuinam~J
  Kim.
\newblock A survey of deep learning-based network anomaly detection.
\newblock {\em Cluster Computing}, pages 1--13, 2017.

\bibitem{tavallaee2009detailed}
Mahbod Tavallaee, Ebrahim Bagheri, Wei Lu, and Ali~A Ghorbani.
\newblock A detailed analysis of the kdd cup 99 data set.
\newblock In {\em Proc. IEEE Symposium on Computational Intelligence for
  Security and Defense Applications}, pages 1--6, 2009.

\bibitem{tam2017evolution}
Kimberly Tam, Ali Feizollah, Nor~Badrul Anuar, Rosli Salleh, and Lorenzo
  Cavallaro.
\newblock The evolution of android malware and {Android} analysis techniques.
\newblock {\em ACM Computing Surveys (CSUR)}, 49(4):76, 2017.

\bibitem{rodriguez2013survey}
Rafael~A Rodr{\'\i}guez-G{\'o}mez, Gabriel Maci{\'a}-Fern{\'a}ndez, and Pedro
  Garc{\'\i}a-Teodoro.
\newblock Survey and taxonomy of botnet research through life-cycle.
\newblock {\em ACM Computing Surveys (CSUR)}, 45(4):45, 2013.

\bibitem{liu2016collaborative}
Menghan Liu, Haotian Jiang, Jia Chen, Alaa Badokhon, Xuetao Wei, and Ming-Chun
  Huang.
\newblock A collaborative privacy-preserving deep learning system in
  distributed mobile environment.
\newblock In {\em Proc. IEEE International Conference on Computational Science
  and Computational Intelligence (CSCI)}, pages 192--197, 2016.

\bibitem{chopra2005learning}
Sumit Chopra, Raia Hadsell, and Yann LeCun.
\newblock Learning a similarity metric discriminatively, with application to
  face verification.
\newblock In {\em Proc. IEEE Conference on Computer Vision and Pattern
  Recognition (CVPR)}, volume~1, pages 539--546, 2005.

\bibitem{shamsabadi2018}
Ali~Shahin Shamsabadi, Hamed Haddadi, and Andrea Cavallaro.
\newblock Distributed one-class learning.
\newblock In {\em Proc. IEEE International Conference on Image Processing
  (ICIP)}, 2018.

\bibitem{hitaj2017passgan}
Briland Hitaj, Paolo Gasti, Giuseppe Ateniese, and Fernando Perez-Cruz.
\newblock Pass{GAN}: A deep learning approach for password guessing.
\newblock {\em arXiv preprint arXiv:1709.00440}, 2017.

\bibitem{ye2018channel}
\textcolor{black}{Hao} \color{black}Ye, Geoffrey~Ye Li, Biing-Hwang~Fred Juang,
  and Kathiravetpillai Sivanesan.
\newblock {\color{black}Channel agnostic end-to-end learning based
  communication systems with conditional {GAN}}.
\newblock {\em {\color{black}arXiv preprint arXiv:1807.00447}},
  {\color{black}2018}\color{black}.

\bibitem{gonzalez2017net2vec}
Roberto Gonzalez, Filipe Manco, Alberto Garcia-Duran, Jose Mendes, Felipe
  Huici, Saverio Niccolini, and Mathias Niepert.
\newblock {Net2Vec}: Deep learning for the network.
\newblock In {\em Proc. ACM Workshop on Big Data Analytics and Machine Learning
  for Data Communication Networks}, pages 13--18, 2017.

\bibitem{wolpert1997no}
David~H Wolpert and William~G Macready.
\newblock No free lunch theorems for optimization.
\newblock {\em IEEE transactions on evolutionary computation}, 1(1):67--82,
  1997.

\bibitem{cheng2017survey}
Yu~Cheng, Duo Wang, Pan Zhou, and Tao Zhang.
\newblock Model compression and acceleration for deep neural networks: The
  principles, progress, and challenges.
\newblock {\em IEEE Signal Processing Magazine}, 35(1):126--136, 2018.

\bibitem{lane2017squeezing}
Nicholas~D Lane, Sourav Bhattacharya, Akhil Mathur, Petko Georgiev, Claudio
  Forlivesi, and Fahim Kawsar.
\newblock Squeezing deep learning into mobile and embedded devices.
\newblock {\em IEEE Pervasive Computing}, 16(3):82--88, 2017.

\bibitem{tang2017enabling}
Jie Tang, Dawei Sun, Shaoshan Liu, and Jean-Luc Gaudiot.
\newblock Enabling deep learning on {IoT} devices.
\newblock {\em Computer}, 50(10):92--96, 2017.

\bibitem{wang2018deep}
\textcolor{black}{Ji} \color{black}Wang, Bokai Cao, Philip Yu, Lichao Sun,
  Weidong Bao, and Xiaomin Zhu.
\newblock {\color{black}Deep learning towards mobile applications}.
\newblock In {\em {\color{black}Proc. 38th IEEE International Conference on
  Distributed Computing Systems (ICDCS)}}, pages~{\color{black}1385--1393},
  {\color{black}2018}\color{black}.

\bibitem{iandola2017squeezenet}
Forrest~N Iandola, Song Han, Matthew~W Moskewicz, Khalid Ashraf, William~J
  Dally, and Kurt Keutzer.
\newblock {SqueezeNet}: {AlexNet}-level accuracy with 50x fewer parameters and<
  0.5 {MB} model size.
\newblock In {\em Proc. International Conference on Learning Representations
  (ICLR)}, 2017.

\bibitem{howard2017mobilenets}
Andrew~G Howard, Menglong Zhu, Bo~Chen, Dmitry Kalenichenko, Weijun Wang,
  Tobias Weyand, Marco Andreetto, and Hartwig Adam.
\newblock Mobilenets: Efficient convolutional neural networks for mobile vision
  applications.
\newblock {\em arXiv preprint arXiv:1704.04861}, 2017.

\bibitem{zhang2017shufflenet}
Xiangyu Zhang, Xinyu Zhou, Mengxiao Lin, and Jian Sun.
\newblock {ShuffleNet}: An extremely efficient convolutional neural network for
  mobile devices.
\newblock In {\em The IEEE Conference on Computer Vision and Pattern
  Recognition (CVPR)}, June 2018.

\bibitem{zhang2017tucker}
Qingchen Zhang, Laurence~T Yang, Xingang Liu, Zhikui Chen, and Peng Li.
\newblock A tucker deep computation model for mobile multimedia feature
  learning.
\newblock {\em ACM Transactions on Multimedia Computing, Communications, and
  Applications (TOMM)}, 13(3s):39, 2017.

\bibitem{cao2017mobirnn}
Qingqing Cao, Niranjan Balasubramanian, and Aruna Balasubramanian.
\newblock {MobiRNN}: Efficient recurrent neural network execution on mobile
  {GPU}.
\newblock In {\em Proc. 1st ACM International Workshop on Deep Learning for
  Mobile Systems and Applications}, pages 1--6, 2017.

\bibitem{chen2016deep}
Chun-Fu Chen, Gwo~Giun Lee, Vincent Sritapan, and Ching-Yung Lin.
\newblock Deep convolutional neural network on {iOS} mobile devices.
\newblock In {\em Proc. IEEE International Workshop on Signal Processing
  Systems (SiPS)}, pages 130--135, 2016.

\bibitem{rallapalli2016very}
S~Rallapalli, H~Qiu, A~Bency, S~Karthikeyan, R~Govindan, B~Manjunath, and
  R~Urgaonkar.
\newblock Are very deep neural networks feasible on mobile devices.
\newblock {\em IEEE Transactions on Circuits and Systems for Video Technology},
  2016.

\bibitem{lane2016deepx}
Nicholas~D Lane, Sourav Bhattacharya, Petko Georgiev, Claudio Forlivesi, Lei
  Jiao, Lorena Qendro, and Fahim Kawsar.
\newblock {DeepX}: A software accelerator for low-power deep learning inference
  on mobile devices.
\newblock In {\em Proc. 15th ACM/IEEE International Conference on Information
  Processing in Sensor Networks (IPSN)}, pages 1--12, 2016.

\bibitem{huynh2017deepmon}
Loc~N Huynh, Rajesh~Krishna Balan, and Youngki Lee.
\newblock {DeepMon}: Building mobile {GPU} deep learning models for continuous
  vision applications.
\newblock In {\em Proc. 15th ACM Annual International Conference on Mobile
  Systems, Applications, and Services}, pages 186--186, 2017.

\bibitem{wu2016quantized}
Jiaxiang Wu, Cong Leng, Yuhang Wang, Qinghao Hu, and Jian Cheng.
\newblock Quantized convolutional neural networks for mobile devices.
\newblock In {\em Proc. IEEE Conference on Computer Vision and Pattern
  Recognition}, pages 4820--4828, 2016.

\bibitem{bhattacharya2016sparsification}
Sourav Bhattacharya and Nicholas~D Lane.
\newblock Sparsification and separation of deep learning layers for constrained
  resource inference on wearables.
\newblock In {\em Proc. 14th ACM Conference on Embedded Network Sensor Systems
  CD-ROM}, pages 176--189, 2016.

\bibitem{cho2017mec}
Minsik Cho and Daniel Brand.
\newblock {MEC}: Memory-efficient convolution for deep neural network.
\newblock In {\em Proc. International Conference on Machine Learning (ICML)},
  pages 815--824, 2017.

\bibitem{guo2017pruning}
Jia Guo and Miodrag Potkonjak.
\newblock Pruning filters and classes: Towards on-device customization of
  convolutional neural networks.
\newblock In {\em Proc. 1st ACM International Workshop on Deep Learning for
  Mobile Systems and Applications}, pages 13--17, 2017.

\bibitem{li2017fitcnn}
Shiming Li, Duo Liu, Chaoneng Xiang, Jianfeng Liu, Yingjian Ling, Tianjun Liao,
  and Liang Liang.
\newblock Fitcnn: A cloud-assisted lightweight convolutional neural network
  framework for mobile devices.
\newblock In {\em Proc. 23rd IEEE International Conference on Embedded and
  Real-Time Computing Systems and Applications (RTCSA)}, pages 1--6, 2017.

\bibitem{zen2016fast}
Heiga Zen, Yannis Agiomyrgiannakis, Niels Egberts, Fergus Henderson, and
  Przemys{\l}aw Szczepaniak.
\newblock Fast, compact, and high quality {LSTM-RNN} based statistical
  parametric speech synthesizers for mobile devices.
\newblock {\em arXiv preprint arXiv:1606.06061}, 2016.

\bibitem{falcao2017evaluation}
Gabriel Falcao, Lu{\'\i}s~A Alexandre, J~Marques, Xavier Fraz{\~a}o, and Joao
  Maria.
\newblock On the evaluation of energy-efficient deep learning using stacked
  autoencoders on mobile gpus.
\newblock In {\em Proc. 25th IEEE Euromicro International Conference on
  Parallel, Distributed and Network-based Processing}, pages 270--273, 2017.

\bibitem{fang2018nestdnn}
\textcolor{black}{Biyi} \color{black}Fang, Xiao Zeng, and Mi~Zhang.
\newblock {\color{black}{NestDNN}: Resource-Aware Multi-Tenant On-Device Deep
  Learning for Continuous Mobile Vision}.
\newblock In {\em {\color{black}Proc. 24th ACM Annual International Conference
  on Mobile Computing and Networking}}, pages~{\color{black}115--127},
  {\color{black}2018}\color{black}.

\bibitem{xu2018deepcache}
\textcolor{black}{Mengwei} \color{black}Xu, Mengze Zhu, Yunxin Liu,
  Felix~Xiaozhu Lin, and Xuanzhe Liu.
\newblock {\color{black}{DeepCache}: Principled Cache for Mobile Deep Vision}.
\newblock In {\em {\color{black}Proc. 24th ACM Annual International Conference
  on Mobile Computing and Networking}}, pages~{\color{black}129--144},
  {\color{black}2018}\color{black}.

\bibitem{liu2018ondemand}
\textcolor{black}{Sicong} \color{black}Liu, Yingyan Lin, Zimu Zhou, Kaiming
  Nan, Hui Liu, and Junzhao Du.
\newblock {\color{black}On-Demand Deep Model Compression for Mobile Devices: A
  Usage-Driven Model Selection Framework}.
\newblock In {\em {\color{black}Proc. 16th ACM Annual International Conference
  on Mobile Systems, Applications, and Services}},
  pages~{\color{black}389--400}, {\color{black}2018}\color{black}.

\bibitem{chen2018tvm}
\textcolor{black}{Tianqi} \color{black}Chen, Thierry Moreau, Ziheng Jiang,
  Lianmin Zheng, Eddie Yan, Haichen Shen, Meghan Cowan, Leyuan Wang, Yuwei Hu,
  Luis Ceze, Carlos Guestrin, and Arvind Krishnamurthy.
\newblock {\color{black}{TVM}: An Automated End-to-End Optimizing Compiler for
  Deep Learning}.
\newblock In {\em {\color{black}13th USENIX Symposium on Operating Systems
  Design and Implementation (OSDI 18)}}, pages~{\color{black}578--594},
  {\color{black}2018}\color{black}.

\bibitem{yao2018fastdeepiot}
\textcolor{black}{Shuochao} \color{black}Yao, Yiran Zhao, Huajie Shao,
  ShengZhong Liu, Dongxin Liu, Lu~Su, and Tarek Abdelzaher.
\newblock {\color{black}{FastDeepIoT}: Towards Understanding and Optimizing
  Neural Network Execution Time on Mobile and Embedded Devices}.
\newblock In {\em {\color{black}Proc. 16th ACM Conference on Embedded Networked
  Sensor Systems}}, pages~{\color{black}278--291},
  {\color{black}2018}\color{black}.

\bibitem{teerapittayanon2017distributed}
Surat Teerapittayanon, Bradley McDanel, and HT~Kung.
\newblock Distributed deep neural networks over the cloud, the edge and end
  devices.
\newblock In {\em Proc. 37th IEEE International Conference on Distributed
  Computing Systems (ICDCS)}, pages 328--339, 2017.

\bibitem{omidshafiei2017deep}
Shayegan Omidshafiei, Jason Pazis, Christopher Amato, Jonathan~P How, and John
  Vian.
\newblock Deep decentralized multi-task multi-agent reinforcement learning
  under partial observability.
\newblock In {\em Proc. International Conference on Machine Learning (ICML)},
  pages 2681--2690, 2017.

\bibitem{recht2011hogwild}
Benjamin Recht, Christopher Re, Stephen Wright, and Feng Niu.
\newblock Hogwild: A lock-free approach to parallelizing stochastic gradient
  descent.
\newblock In {\em Advances in neural information processing systems}, pages
  693--701, 2011.

\bibitem{goyal2017accurate}
Priya Goyal, Piotr Doll{\'a}r, Ross Girshick, Pieter Noordhuis, Lukasz
  Wesolowski, Aapo Kyrola, Andrew Tulloch, Yangqing Jia, and Kaiming He.
\newblock Accurate, large {Minibatch SGD}: Training imagenet in 1 hour.
\newblock {\em arXiv preprint arXiv:1706.02677}, 2017.

\bibitem{zhang2016asynchronous}
ShuaiZheng Ruiliang~Zhang and JamesT Kwok.
\newblock Asynchronous distributed semi-stochastic gradient optimization.
\newblock In {\em Proc. National Conference on Artificial Intelligence (AAAI)},
  2016.

\bibitem{hardy2017distributed}
Corentin Hardy, Erwan Le~Merrer, and Bruno Sericola.
\newblock Distributed deep learning on edge-devices: feasibility via adaptive
  compression.
\newblock In {\em Proc. 16th IEEE International Symposium on Network Computing
  and Applications (NCA)}, pages 1--8, 2017.

\bibitem{pmlr-v54-mcmahan17a}
Brendan McMahan, Eider Moore, Daniel Ramage, Seth Hampson, and Blaise~Aguera
  y~Arcas.
\newblock Communication-efficient learning of deep networks from decentralized
  data.
\newblock In {\em Proc. 20th International Conference on Artificial
  Intelligence and Statistics}, volume~54, pages 1273--1282, Fort Lauderdale,
  FL, USA, 20--22 Apr 2017.

\bibitem{cryptoeprint:2017:281}
Keith Bonawitz, Vladimir Ivanov, Ben Kreuter, Antonio Marcedone, H.~Brendan
  McMahan, Sarvar Patel, Daniel Ramage, Aaron Segal, and Karn Seth.
\newblock Practical secure aggregation for privacy preserving machine learning.
\newblock Cryptology ePrint Archive, Report 2017/281, 2017.
\newblock \url{https://eprint.iacr.org/2017/281}.

\bibitem{gupta2016model}
Suyog Gupta, Wei Zhang, and Fei Wang.
\newblock Model accuracy and runtime tradeoff in distributed deep learning: A
  systematic study.
\newblock In {\em Proc. IEEE 16th International Conference on Data Mining
  (ICDM)}, pages 171--180, 2016.

\bibitem{mcmahan2017federated}
B~McMahan and Daniel Ramage.
\newblock Federated learning: Collaborative machine learning without
  centralized training data.
\newblock {\em Google Research Blog}, 2017.

\bibitem{furno2017joint}
Angelo Fumo, Marco Fiore, and Razvan Stanica.
\newblock Joint spatial and temporal classification of mobile traffic demands.
\newblock In {\em Proc. IEEE Conference on Computer Communications}, pages
  1--9, 2017.

\bibitem{chen2016lifelong}
Zhiyuan Chen and Bing Liu.
\newblock Lifelong machine learning.
\newblock {\em Synthesis Lectures on Artificial Intelligence and Machine
  Learning}, 10(3):1--145, 2016.

\bibitem{lee2016dual}
Sang-Woo Lee, Chung-Yeon Lee, Dong-Hyun Kwak, Jiwon Kim, Jeonghee Kim, and
  Byoung-Tak Zhang.
\newblock Dual-memory deep learning architectures for lifelong learning of
  everyday human behaviors.
\newblock In {\em Proc. International Joint Conferences on Artificial
  Intelligence}, pages 1669--1675, 2016.

\bibitem{graves2016hybrid}
Alex Graves, Greg Wayne, Malcolm Reynolds, Tim Harley, Ivo Danihelka, Agnieszka
  Grabska-Barwi{\'n}ska, Sergio~G{\'o}mez Colmenarejo, Edward Grefenstette,
  Tiago Ramalho, John Agapiou, et~al.
\newblock Hybrid computing using a neural network with dynamic external memory.
\newblock {\em Nature}, 538(7626):471--476, 2016.

\bibitem{parisi2017lifelong}
German~I Parisi, Jun Tani, Cornelius Weber, and Stefan Wermter.
\newblock Lifelong learning of human actions with deep neural network
  self-organization.
\newblock {\em Neural Networks}, 2017.

\bibitem{tessler2017deep}
Chen Tessler, Shahar Givony, Tom Zahavy, Daniel~J Mankowitz, and Shie Mannor.
\newblock A deep hierarchical approach to lifelong learning in {Minecraft}.
\newblock In {\em Proc. National Conference on Artificial Intelligence (AAAI)},
  pages 1553--1561, 2017.

\bibitem{lopez2018deep}
Daniel L{\'o}pez-S{\'a}nchez, Ang{\'e}lica~Gonz{\'a}lez Arrieta, and Juan~M
  Corchado.
\newblock Deep neural networks and transfer learning applied to multimedia web
  mining.
\newblock In {\em Proc. 14th International Conference Distributed Computing and
  Artificial Intelligence}, volume 620, page 124. Springer, 2018.

\bibitem{bacstuug2015transfer}
Ejder Ba{\c{s}}tu{\u{g}}, Mehdi Bennis, and M{\'e}rouane Debbah.
\newblock A transfer learning approach for cache-enabled wireless networks.
\newblock In {\em Proc. 13th IEEE International Symposium on Modeling and
  Optimization in Mobile, Ad Hoc, and Wireless Networks (WiOpt)}, pages
  161--166, 2015.

\bibitem{fei2006one}
Li~Fei-Fei, Rob Fergus, and Pietro Perona.
\newblock One-shot learning of object categories.
\newblock {\em IEEE transactions on pattern analysis and machine intelligence},
  28(4):594--611, 2006.

\bibitem{palatucci2009zero}
Mark Palatucci, Dean Pomerleau, Geoffrey~E Hinton, and Tom~M Mitchell.
\newblock Zero-shot learning with semantic output codes.
\newblock In {\em Advances in neural information processing systems}, pages
  1410--1418, 2009.

\bibitem{vinyals2016matching}
Oriol Vinyals, Charles Blundell, Tim Lillicrap, Daan Wierstra, et~al.
\newblock Matching networks for one shot learning.
\newblock In {\em Advances in Neural Information Processing Systems}, pages
  3630--3638, 2016.

\bibitem{changpinyo2016synthesized}
Soravit Changpinyo, Wei-Lun Chao, Boqing Gong, and Fei Sha.
\newblock Synthesized classifiers for zero-shot learning.
\newblock In {\em Proc. IEEE Conference on Computer Vision and Pattern
  Recognition}, pages 5327--5336, 2016.

\bibitem{oh2017zero}
Junhyuk Oh, Satinder Singh, Honglak Lee, and Pushmeet Kohli.
\newblock Zero-shot task generalization with multi-task deep reinforcement
  learning.
\newblock In {\em Proc. International Conference on Machine Learning (ICML)},
  pages 2661--2670, 2017.

\bibitem{wang2015understanding}
Huandong Wang, Fengli Xu, Yong Li, Pengyu Zhang, and Depeng Jin.
\newblock Understanding mobile traffic patterns of large scale cellular towers
  in urban environment.
\newblock In {\em Proc. ACM Internet Measurement Conference}, pages 225--238,
  2015.

\bibitem{marquez2017apps}
Cristina Marquez, Marco Gramaglia, Marco Fiore, Albert Banchs, Cezary
  Ziemlicki, and Zbigniew Smoreda.
\newblock Not all {Apps} are created equal: Analysis of spatiotemporal
  heterogeneity in nationwide mobile service usage.
\newblock In {\em Proc. 13th ACM Conference on Emerging Networking Experiments
  and Technologies}, 2017.

\bibitem{barlacchi2015multi}
Gianni Barlacchi, Marco De~Nadai, Roberto Larcher, Antonio Casella, Cristiana
  Chitic, Giovanni Torrisi, Fabrizio Antonelli, Alessandro Vespignani, Alex
  Pentland, and Bruno Lepri.
\newblock A multi-source dataset of urban life in the city of {Milan} and the
  province of {Trentino}.
\newblock {\em Scientific data}, 2, 2015.

\bibitem{liu2015urban}
Liang Liu, Wangyang Wei, Dong Zhao, and Huadong Ma.
\newblock Urban resolution: New metric for measuring the quality of urban
  sensing.
\newblock {\em IEEE Transactions on Mobile Computing}, 14(12):2560--2575, 2015.

\bibitem{tikunov2007traffic}
Denis Tikunov and Toshikazu Nishimura.
\newblock Traffic prediction for mobile network using holt-winter's exponential
  smoothing.
\newblock In {\em Proc. 15th IEEE International Conference on Software,
  Telecommunications and Computer Networks}, pages 1--5, 2007.

\bibitem{Kim2011}
Hyun-Woo Kim, Jun-Hui Lee, Yong-Hoon Choi, Young-Uk Chung, and Hyukjoon Lee.
\newblock {Dynamic bandwidth provisioning using ARIMA-based traffic forecasting
  for Mobile WiMAX}.
\newblock {\em Computer Communications}, 34(1):99--106, 2011.

\bibitem{charles2017pointnet}
\textcolor{black}{R Qi} \color{black}Charles, Hao Su, Mo~Kaichun, and
  Leonidas~J Guibas.
\newblock Pointnet: Deep learning on point sets for {3D} classification and
  segmentation.
\newblock In {\em {\color{black}Proc. IEEE Conference on Computer Vision and
  Pattern Recognition (CVPR)}}, pages~{\color{black}77--85},
  {\color{black}2017}\color{black}.

\bibitem{ioannidou2017deep}
Anastasia Ioannidou, Elisavet Chatzilari, Spiros Nikolopoulos, and Ioannis
  Kompatsiaris.
\newblock Deep learning advances in computer vision with {3D} data: A survey.
\newblock {\em ACM Computing Surveys (CSUR)}, 50(2):20, 2017.

\bibitem{scarselli2009graph}
\textcolor{black}{Franco} \color{black}Scarselli, Marco Gori, Ah~Chung Tsoi,
  Markus Hagenbuchner, and Gabriele Monfardini.
\newblock {\color{black}The graph neural network model}.
\newblock {\em {\color{black}IEEE Transactions on Neural Networks}},
  {\color{black}20}({\color{black}1}):{\color{black}61--80},
  {\color{black}2009}\color{black}.

\bibitem{yuan2017temporal}
\textcolor{black}{Yuan} \color{black}Yuan, Xiaodan Liang, Xiaolong Wang,
  Dit-Yan Yeung, and Abhinav Gupta.
\newblock Temporal dynamic graph {LSTM} for action-driven video object
  detection.
\newblock In {\em Proc. IEEE International Conference on Computer Vision
  (ICCV)}, pages 1819--1828, 2017\color{black}.

\bibitem{usama2017unsupervised}
Muhammad Usama, Junaid Qadir, Aunn Raza, Hunain Arif, Kok-Lim~Alvin Yau, Yehia
  Elkhatib, Amir Hussain, and Ala Al-Fuqaha.
\newblock Unsupervised machine learning for networking: Techniques,
  applications and research challenges.
\newblock {\em arXiv preprint arXiv:1709.06599}, 2017.

\bibitem{zhou2017anomaly}
Chong Zhou and Randy~C Paffenroth.
\newblock Anomaly detection with robust deep autoencoders.
\newblock In {\em Proc. 23rd ACM SIGKDD International Conference on Knowledge
  Discovery and Data Mining}, pages 665--674, 2017.

\bibitem{abadi2016learning}
Mart{\'\i}n Abadi and David~G Andersen.
\newblock Learning to protect communications with adversarial neural
  cryptography.
\newblock In {\em Proc. International Conference on Learning Representations
  (ICLR)}, 2017\color{black}.

\bibitem{Silver1140}
\textcolor{black}{David} \color{black}Silver, Thomas Hubert, Julian
  Schrittwieser, Ioannis Antonoglou, Matthew Lai, Arthur Guez, Marc Lanctot,
  Laurent Sifre, Dharshan Kumaran, Thore Graepel, Timothy Lillicrap, Karen
  Simonyan, and Demis Hassabis.
\newblock \color{black}{A} general reinforcement learning algorithm that
  masters chess, shogi, and {Go} through self-play.
\newblock {\em Science}, 362(6419):1140--1144, 2018\color{black}.

\end{thebibliography}

\end{document}